# TOWARDS ROBUST QUANTUM COMPUTATION

Debbie W. Leung

July 2000

I certify that I have read this dissertation and that in my opinion it is fully adequate, in scope and quality, as a dissertation for the degree of Doctor of Philosophy.

_______________________________
Yoshihisa Yamamoto
(Principal Advisor)

I certify that I have read this dissertation and that in my opinion it is fully adequate, in scope and quality, as a dissertation for the degree of Doctor of Philosophy.

_______________________________
Isaac L. Chuang
(Almaden Research Center, IBM)
(Co-Advisor)

I certify that I have read this dissertation and that in my opinion it is fully adequate, in scope and quality, as a dissertation for the degree of Doctor of Philosophy.

_______________________________
Thomas M. Cover
(Department of Electrical Engineering)

I certify that I have read this dissertation and that in my opinion it is fully adequate, in scope and quality, as a dissertation for the degree of Doctor of Philosophy.

_______________________________
Stephen Shenker

Approved for the University Committee on Graduate Studies:

_______________________________

# Abstract

It has been shown in the past two decades that quantum devices can achieve useful information processing tasks impossible in classical devices. Meanwhile, the development of quantum error correcting codes and fault-tolerant computation methods has also set down a solid theoretical promise to overcome decoherence. However, to-date, we have not been able to apply such quantum information processing tasks in real life, because the theoretically modest requirements in these schemes still present daunting experimental challenges.

This Dissertation aims at reducing the resources required for various schemes related to simple and robust quantum computation, focusing on quantum error correcting codes and solution NMR quantum computation. A variety of techniques have been developed, including high rate quantum codes for specific noise processes, relaxed criteria for quantum error correction, systematic construction of fault-tolerant gates, techniques in quantum process tomography, techniques in bulk mixed state computation and efficient schemes to selectively implement coupled logic gates using naturally occurring Hamiltonians. A detailed experimental study of quantum error correcting code in NMR is also presented. The goal is to get ready tools and techniques that may apply to some useful candidate systems for implementing quantum computation in the near future.

# Acknowledgements

I am greatly indebted to my advisors, Prof. Yoshihisa Yamamoto and Dr. Ike Chuang for their patient guidance and generous support during the completion of this Dissertation. Professor Yamamoto has provided a wonderfully challenging research environment and has greatly encouraged independence in my work. Ike has introduced me to the subject of quantum information, and has greatly motivated my interest in many different topics. He generously sacrifices his time to teach, to advise, and to help. He has also actively provided opportunities for me to learn from various resources. I feel extremely fortunate to be able to work with both Prof. Yamamoto and Ike.

I would like to thank Prof. Tom Cover, Prof. Steve Shenker, and Prof. Charlie Marcus who kindly serve as reading committee members for this Dissertation, and Prof. Vaughan Pratt who chair the defense of this Dissertation.

I have also been partially supported by Prof. James Harris, and the managers Nabil Amer and Bill Risk from IBM. This Dissertation would not have existed if not for the generous financial support from ARO, DARPA, ERATO, IBM, JST-ICORP, and NTT.

Yurika Peterman, Aihui Lin, Mayumi Hakkaku, Gail Chun-Creech, Dawn Hyde, Marcia Keating and Kathleen Guan have given me lots of adminstrative help. I am particularly grateful to Yurika and Aihui for their caring acts.

I benefit greatly from interacting with other members in the research groups of Prof. Yamamoto and Ike. Orly Alter, Hui Cao, Jungsang Kim, Seema Lathi, Robert Liu, and Fumiko Yamaguchi have given much needed advice in handling the difficulties being a graduate student.

I would like to thank the following friends and colleagues. I am grateful to Hoi-Kwong

Lo and Hoi-Fung Chau for being extremely kind. They have given helpful comments and advice, and I enjoy learning from and working with them. Hoi-Kwong has made a decisive influence on my choosing quantum information as the dissertation subject. Michael Nielsen's work on understanding quantum operations, and Daniel Gottesman's work on quantum error correcting codes have inspired a lot of the current dissertation work. Discussions with them are always pleasant and enlightening. A significant portion of this Dissertation is based on their work. Mark Kubinec, Prof. Alex Pines, Mark Sherwood, Shigeki Takeuchi, and Nino Yannoni have been great teachers of NMR. It will be difficult to list all those I have learnt from and those who have shown exceptional kindness during my dissertation work. I will give a partial list: Dorit Aharonov, Dave Bacon, Charles Bennett, David DiVincenzo, Arthur Ekert, Chris Fuchs, Lov Grover, Pawel Horedecki, Richard Josza, Julie Kempe, Daniel Lidar, Asher Peres, John Preskill, John Smolin, Barbara Terhal, and Ashish Thapliyal. Following the suggestion of Chris Fuchs, special thanks are given to Peter Shor *for funding*.

The friendship from Tom, Chris and Ben-GoG, Kian-meng, Isabel and Eric, Ah Gay, Ah Yip, Ah Shing, Kenny, Honsing, Calvin, Sam, Alice and Helios, Janice and Robert, Tiffany and Norris, Lauren, Joshua, and Marcus are deeply appreciated.

Last but certainly not least, I would like to thank my parents and siblings O.

# Contents

# List of Tables

# List of Figures

# Part I

# Introduction

# Chapter 1

# Motivation

## 1.1 Time frame

The two important but distant subjects of quantum mechanics and the theory of computation were first put together by Benioff, Feynman and Deutsch back in the mid 1980's [14, 51, 44]. In the following decade, many interesting results and intriguing possibilities were found, including various cryptographic and communication protocols [15, 18, 16] and computation problems which exhibit a gap between the computation power of classical and quantum models [46, 106]. These were followed by an extended three-year-long excitement brought by the discovery of the fast factoring and search algorithms [103, 61], refined notions of primitives of computation [45, 12, 47, 13], and the development of quantum error correcting codes [105, 108] which can be used to achieve reliable quantum computation [11, 48, 68, 74, 57, 96, 104].

The work described in this Dissertation started in April, 1996, in the peak of the "theoretical excitement". Quantum information is a very interesting subject, not only because one can attempt to engineer faster computation devices, but one is facing challenges which will change our fundamental understanding of many subjects, including quantum mechanics, the theory of computation, cryptography, and information theory.

Meanwhile, our theoretical development was not well matched by experimental progress. While theorists proved that quantum computation can be made arbitrarily reliable if the elementary error probability is below some optimistic threshold value of about $10^{-4}$, and if coding are performed, taking about hundreds of quantum bits to encode one, experimentalists heroically prepared a couple of quantum bits and demonstrated non-trivial logic operations. It is believed that most of the obstacles are technical, and will be resolved with time.

## 1.2   Overview of the Dissertation

Most of the work in this Dissertation attempts to narrow the gap between theory and experiment. The focus is mostly theoretical. Various approaches are taken, which revolve around more efficient methods to make quantum computation reliable. One approach is to view quantum error correction as a trade-off between resources and reliability, and the goal is to minimize the resources needed to achieve a certain level of reliability. Another approach is to consider the trade-off between resources and the computation power. This is motivated by the surprising realization of solution NMR quantum computation towards the end of 1996. In this implementation, difficult experimental requirements are waived by various techniques which compromise experimental feasibility and the computation capability. Focusing on the subjects of quantum error correction and solution NMR quantum computation, techniques were developed in this Dissertation which may apply to a wide range of candidate systems. Specifically, the following results were obtained.

1. Quantum error correcting codes for specific noise processes which achieve better rates than any codes correcting for general errors [31, 78].

2. Relaxed sufficient criteria for quantum error correction which admit more efficient quantum codes [78].

3. A systematic construction of fault-tolerant logic operations which unifies known constructions and enlarges the set of primitives [115].

4. An alternative recipe to perform quantum process tomography, the complete characterization of quantum processes useful for designing better codes and gates (unpublished).

5. An efficient scheme to perform selective logic operations using some non-selective evolution as a primitive [77].

6. An experimental realization of a quantum code in NMR which was a first demonstration of noise reduction by coding [79]. A thorough characterization of systematic errors on the NMR system was also made.

Contributions were also made in the development of various techniques used in NMR quantum computation, including a particular hybrid scheme for state labeling (unpublished), a modification of the quantum process tomography procedure to the NMR system [27] and a study of the computation power of bulk systems with imperfect initial states [114].

Results on quantum error correction for specific noise processes were obtained in collaboration with I. Chuang, M. Nielsen, and Y. Yamamoto. Results on fault-tolerant gate constructions and the computation power of bulk systems were obtained in collaboration with X. Zhou and I. Chuang. The efficient scheme to perform selective logic operations was co-developed with I. Chuang, F. Yamaguchi, and Y. Yamamoto. Similar result was independently reported in [66]. The experimental study of quantum error correction was a joint effort with L. Vandersypen, I. Chuang, X. Zhou, M. Sherwood, C. Yannoni, and M. Kubinec. Related result was independently reported in [41]. The ongoing study of quantum process tomography in NMR is done in collaboration with A. Childs and I. Chuang.

This Dissertation is structured into three parts.

- Part I contains the current overview chapter, and Chapter 2, which reviews useful background for this Dissertation. Quantum circuits, universal sets of quantum logic gates, the formalism of quantum operations, quantum process tomography, and some common noise processing are covered. Item 4 is included as part of the discussion in quantum process tomography.

- Part II focuses on quantum error correction and consists of three chapters. Chapter 3 reviews the major results in the development of quantum error correction. It covers classical error correction, the general theory of quantum error correction, the stabilizer formalism, fault-tolerant quantum computation and the threshold theorem.

Chapter 4 describes quantum codes for specific noise processes and the relaxed criteria for quantum error correction.  Chapter 5 describes the systematic construction of fault-tolerant gates.

- Part III focuses on NMR quantum computation, and also consists of three chapters. Chapter 6 describes the elements in NMR quantum computation.  The problem of initial state preparation, process tomography, and bulk computation will be described, and some partial solutions will be given.  Chapter 7 describes the efficient scheme to select specific coupling in NMR and related systems.  Chapter 8 describes the experiment on quantum error correction.

Review materials in Chapters 2 and 3 are primarily based on [88, 28, 85, 32, 83, 20, 95, 89, 56, 96, 104].  Section 4.2.5 in Chapter 4 is based on [56].  The primitives described in Section 5.2 are due to Bennett and Gottesman.  Section 6.3.1 is based on [86].  Many results presented in Chapter 6 are due to Chuang.

The problem in building quantum computation devices is not just decoherence, but rather the many simultaneous requirements of quantum information processing which contradict each other.  It is important to (1) prepare the system in a fiducial initial state, to (2) perform a universal set of logic gates, to (3) implement measurement which will correctly read out the computational results, and to (4) maintain long coherence times.  Throughout the past few years, more and more candidate implementations have been proposed, each presents some novel solution to part of the listed requirements and faces more difficulties in other requirements.  It is the aim of this Dissertation to make ready some of the theoretical techniques which can be adapted to a wide range of candidate systems.

# Chapter 2

# Fundamental concepts

## 2.1 The Basics

The fundamental unit of classical information is a *bit*, a random variable that takes value on $\{0, 1\}$. It can also be viewed as a 1-dimensional vector over $Z_2$ (the integers modulo 2). An $n$-bit classical string takes value on a set of $2^n$ possible states. It is also an $n$-dimensional vector over $Z_2$. In contrast, the fundamental unit of quantum information, a quantum bit or a *qubit* is a 2-dimensional unit vector over the complex field C. A state of $n$ qubits is a vector over a $2^n$-dimensional Hilbert space which is a *tensor product* of $n$ 2-dimensional Hilbert spaces. Following the Dirac notation, a vector or a "ket" is denoted by $|\cdot\rangle$. The dual of a vector or a "bra" is denoted by $\langle\cdot|$. $\{|0\rangle, |1\rangle\}$ usually stands for the basis of a qubit. This basis is often called the *computational basis*. The state of an arbitrary qubit is given by a vector $|\psi\rangle = a|0\rangle + b|1\rangle$ with *norm square* $\langle\psi|\psi\rangle = 1$ to represent some "total probability". An $n$-qubit state is likewise a vector over basis states with $n$-bit labels, following the tensor product structure of the $n$-qubit Hilbert space. The *conjugate basis* is defined as $|\pm\rangle = \frac{1}{\sqrt{2}}(|0\rangle \pm |1\rangle)$. We may use 2-dimensional subspaces in a larger $n$-qubit space to represent qubits. To distinguish these subspaces from the constituent qubits of the larger space, we call the embedded ones "logical qubits" and the constituent ones "physical qubits".

The vector $|\psi\rangle$ is a *pure* state. As the overall phase is irrelevant, a pure state is more precisely represented by a *projector* $|\psi\rangle\langle\psi|$. In contrast, a *mixed* state is a *distribution* or an *ensemble* of pure states. For example, the state is $|\psi_k\rangle$ with probability $p_k$. Mathematically, a mixed state is a *convex combination* of projectors. For example, the above distribution is represented by $\rho = \sum_k p_k|\psi_k\rangle\langle\psi_k|$ which is called a *density matrix*. This definition follows naturally if the expectation of an operator $O$ is generalized from $\langle\psi|O|\psi\rangle$ to $\text{Tr}(O\rho)$. It is immediate that any density matrix is *positive* with unit trace.

In a usual quantum information processing task, one prepares some initial state, say $|\psi\rangle$ or $\rho$, and applies a sequence of unitary operations and measurements. A unitary operation $U$ evolves $|\psi\rangle$ to $U|\psi\rangle$ and $\rho$ to $U\rho U^\dagger$. Unitary operations are quantum analogs of "gates". Quantum mechanics admits *projective measurements* of hermitian operators. Each measurement outcome is some eigenvalue of the measured operators. The measurement projects the original state onto the eigenspace corresponding to the measurement outcome. ($|\psi\rangle \to P|\psi\rangle$ or $\rho \to P\rho P$ where the projector $P$ depends on the outcome). Measuring $\sigma_z$ (to be defined in the next section) projects $|\psi\rangle = a|0\rangle + b|1\rangle$ onto $|0\rangle$ or $|1\rangle$. We postpone the discussion on non-unitary evolution and generalized measurements until Section 2.3.

## 2.2   Quantum circuits

The quantum circuit is both a model and a representation of quantum information processing. We will discuss the usual conventions and introduce common circuit elements in this section. Consider the following example of a quantum circuit.

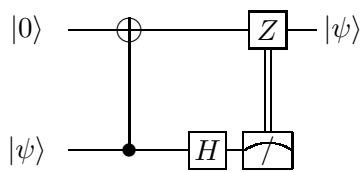

Figure 2.1: An example of a quantum circuit

The following conventions are used throughout the Dissertation:

- The horizontal and vertical directions schematically represent changing time and

space.

- Time runs from left to right.

- Horizontal single-lines represent quantum registers.

- Input states are at the far left, and output states are at the far right.

- Boxes enclosing capital letters represent gates. Some gates are given by more special symbols, such as the symbol connecting two registers with a $\oplus$ on one and a $\bullet$ on the other.

- The meter represents a projective measurement along the computational basis.

- The double line represents *classical information*. An operation connected to a measurement box by a double line is performed *conditioned* on the measurement outcome being $|1\rangle$.

We now define some important quantum gates:

The Pauli matrices are defined in the computational basis as:

$$\sigma_x = \begin{bmatrix} 0 & 1 \\ 1 & 0 \end{bmatrix} \quad \sigma_y = \begin{bmatrix} 0 & -i \\ i & 0 \end{bmatrix} \quad \sigma_z = \begin{bmatrix} 1 & 0 \\ 0 & -1 \end{bmatrix} \tag{2.1}$$

where the basis is ordered as $(|0\rangle, |1\rangle)$. We often use the shorthands:

$$X = \sigma_x, \quad Z = \sigma_z, \quad \text{and} \quad Y = XZ = -i\sigma_y = \begin{bmatrix} 0 & -1 \\ 1 & 0 \end{bmatrix}. \tag{2.2}$$

"$I$" stands for the identity matrix ($2 \times 2$ unless otherwise stated). We may use $\sigma_{0,1,2,3}$ to stand for $I, \sigma_{x,y,z}$. For a system of $n$ qubits, $\sigma_i^{(j)}$ denotes a $\sigma_i$ acting on the $j$-th qubit (and an $I$ acting on every other qubit). The Pauli group over $n$ qubits, $\mathcal{G}_n$, is generated by $\sigma_i^{(j)}$ for $i = 1, 2, 3$ and $j = 1, \cdots, n$. We may refer to the smaller real subgroup generated by $X$ and $Z$ also as the Pauli group, in which case the usage will be made explicit.

For a fixed unitary operator $U$, the mapping $O \to UOU^\dagger$ is called the *conjugation by* $U$. We say that $U$ takes $O$ to $UOU^\dagger$.

The *Clifford group* consists of operations which take $\mathcal{G}_n$ to $\mathcal{G}_n$. Since conjugation is reversible, the Clifford group is a subgroup of the permutation group of $\mathcal{G}_n$. The Clifford group can be generated by $P$, $H$, and CNOT (see Section 5.8 in [56]), defined as follows.

• The *phase gate* $P$ is a $\pi/2$ phase rotation which equals $\sqrt{Z}$:

$$P = \begin{bmatrix} 1 & 0 \\ 0 & i \end{bmatrix} \tag{2.3}$$

• The *Hadamard gate* $H$ is a $\pi$ rotation about the axis $\frac{1}{\sqrt{2}}(X + Z)$:

$$H = \frac{1}{\sqrt{2}} \begin{bmatrix} 1 & 1 \\ 1 & -1 \end{bmatrix} \tag{2.4}$$

• The controlled-NOT (CNOT) acts on two qubits, called the control and the target bits. When the control bit is $|1\rangle$, CNOT applies an $X$ to the target. If the first and second qubits represent the control and the target bits respectively, the CNOT has matrix representation:

$$\begin{bmatrix} 1 & 0 & 0 & 0 \\ 0 & 1 & 0 & 0 \\ 0 & 0 & 0 & 1 \\ 0 & 0 & 1 & 0 \end{bmatrix} \tag{2.5}$$

We may write CNOT$_{ab}$ where $a$ and $b$ are labels for the control and the target bits. The Clifford group over $\mathcal{G}_n$ is generated by $P^{(i)}$, $H^{(i)}$ and CNOT$_{ij}$ for $i, j \in \{1, \cdots, n\}$.

There are many other important gates. The *rotation operator* of angle $\theta$ about the axis $\hat{\eta}$ is defined as

$$R_\eta(\theta) = e^{-i\theta\hat{\eta}\cdot\vec{\sigma}/2} = \cos\frac{\theta}{2}I - i\sin\frac{\theta}{2}\hat{\eta}\cdot\vec{\sigma}. \tag{2.6}$$

Up to an overall phase factor, Eq. (2.6) represents the most general *one-qubit* unitary operation. Other gates of interest include the Toffoli gate (controlled-controlled-NOT), the Fredkin gate (controlled-SWAP), and the $\pi/8$ gate $Z^{\frac{1}{4}} = \text{Diag}(1, \sqrt{i})$.

The circuit symbols for some of these important gates are shown in Fig. 2.2.

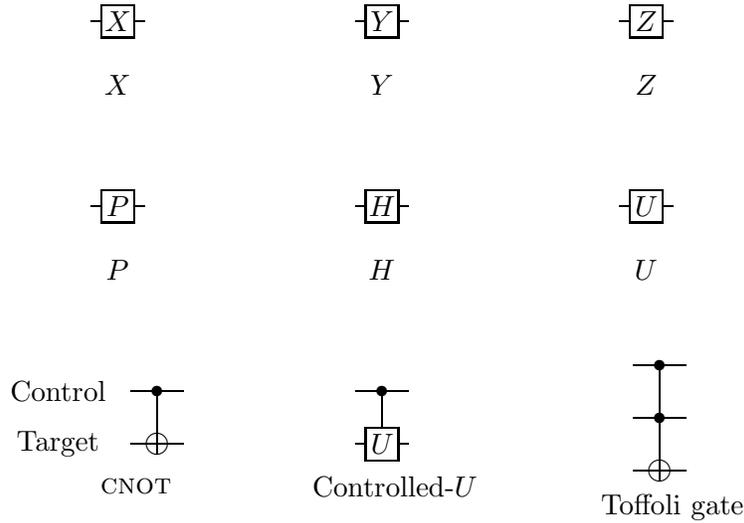

Figure 2.2: Symbols for quantum gates (courtesy of [56]).

We conclude this section with a discussion on the *universality* of quantum gates. It is known that classical circuits can be built with just Toffoli gates, or gates from the set {NOT, OR, AND}. These are examples of *universal* sets of gates for classical computation. Therefore, an arbitrary circuit can be built from certain elementary components. Similarly, a set of gates is universal for quantum computation if it generates a *dense* set in the set of all unitary operations. In other words, an arbitrary unitary operation can be *approximated* to arbitrary accuracy by compositing gates from the set. An important universal set of gates for quantum computation is the set of all single qubit operations and CNOT. Another important universal set of gates is {CNOT, $P$, $H$, Toffoli}. Note that the first set is continuous, while the second set is discrete.

The proofs of universality are out of scope of this Dissertation. Interested readers can refer to the original papers [97, 45, 12, 47, 81, 13] or an informative review in [88]. We outline the main ideas. An arbitrary $d \times d$ unitary matrix is a product of at most $d(d-1)/2$ matrices each acting on two basis states. [1] Each two-level gate on $n$ qubits ($d = 2^n$) is a product of $\mathcal{O}(n)$ operations each having $(n-1)$ control bits and one target bit. Such operations can in turns be expressed as $\mathcal{O}(n)$ CNOTs and single qubit operations, proving the universality of the first set. Note that the continuous set generates all unitary operations exactly. In contrast, the discrete set generates a *dense* subset of the unitary operations. In particular, the discrete set can approximate any single qubit gates (and therefore any gates). This is based on two additional facts. First of all, each single qubit gate can be expressed as at most three rotations about two non-parallel axes. Second, iterating a fixed irrational angle rotation can approximate an arbitrary angle rotation about the same axis. This is the crucial step in which a continuum of gates is generated by a discrete set. The irrational angle rotation can be implemented using $H$, $P$ and the Toffoli gate, completing the proof. This particular construction can approximate an arbitrary angle of rotation to accuracy $\epsilon$ using $\approx \mathcal{O}(1/\epsilon)$ elementary gates. Consider a circuit of $N$ gates taken from the first universal set. To approximate the circuit to an overall accuracy $\epsilon$, each gate has to be accurate to $\epsilon/N$ [19]. Therefore, $\mathcal{O}(N^2/\epsilon)$ gates from the discrete set are required. A more efficient construction by Solovay (unpublished) and Kitaev [69] requires only $\mathcal{O}(N \log^c(N/\epsilon))$ steps where $c \approx 2 - 4$.

---

[1] This is reminiscent of decomposition of a permutation into swaps or transpositions.

Finally, we remark that even though composition from the continuous universal set is exact and composition from the discrete set is approximate, gates are inevitably inaccurate in real life, making such distinction irrelevant. Even more counter-intuitively, the discrete set can be more useful for achieving accuracy when the elementary gates are imperfect, as we will see in Chapter 3.

## 2.3 Quantum operations

In this section, we extend our discussion on closed quantum systems and unitary evolution to open quantum systems and their evolution described by the formalism of quantum operations. We will see that an open quantum system can be modeled as part of a larger, closed quantum system undergoing unitary evolution. Surprisingly this simple model is extremely general – it can describe any physically reasonable dynamics to be defined. Such formalism will be used to describe noise processes and non-unitary computation procedures which are the main subjects of this Dissertation.

In Section 2.3.1, we will derive three equivalent approaches to describe open quantum systems: an axiomatic approach, a subsystem evolution or system-reservoir coupling approach and an operator sum representation approach which has a stochastic process interpretation. These equivalent approaches describe a class of possible dynamics called *quantum operations*. In Section 2.3.2, we will describe two methods to perform *quantum process tomography*, the procedure to completely characterize a quantum process. Some common noise processes are described in Section 2.4, using the formalism developed. The traditional description of open quantum systems in terms of master equations is given in Appendix A.1.

Section 2.3.1 is based on [85, 28]. The interpretation of a crucial result in [28] is original. Section 2.3.2 is based on [32] and original (unpublished) work.

### 2.3.1 Equivalent approaches to quantum operations

The common scenario depicted in Fig. 2.3 is a physically motivated example of quantum operations:

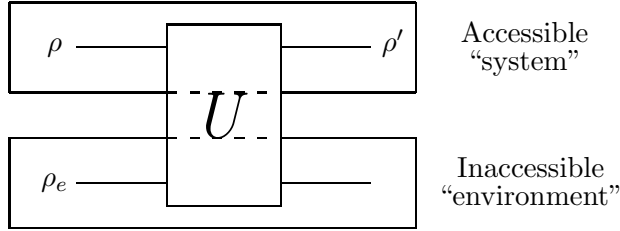

Figure 2.3: Circuit for system-environment coupling

We assume that the system and environment are initially in a product state $\rho \otimes \rho_e$. Due to some system-environment interaction, the density matrix of the combined system is evolved unitarily to $\rho_c = U(\rho \otimes \rho_e)U^\dagger$. Suppose $\rho'$ is the density matrix for the system when the environment is inaccessible ($\rho'$ is called the *reduced* density matrix) . How does $\rho_c$ determine $\rho'$? For consistency, $\rho'$ and $\rho_c$ should predict the same expectation value of any *system observable $O$*. Therefore, $\mathrm{Tr}_s(O\rho') = \mathrm{Tr}_{se}\left[ (O \otimes I)\rho_c \right]$. This is true only if

$$\rho' = \mathrm{Tr}_e\left[ U\rho_c U^\dagger \right] = \mathrm{Tr}_e\left[ U(\rho \otimes \rho_e)U^\dagger \right] \tag{2.7}$$

where $\text{Tr}_e$ denotes the *partial trace* [2] on the environment. The partial trace is equivalent to measuring the environment in an arbitrary basis and *forgetting* the result, so that the system is in a statistical mixture of states corresponding to the inaccessible measurement outcomes. $\rho'$ in Eq. (2.7) defines a mapping $\mathcal{E}$ on the system density matrices, which is written as $\rho' = \mathcal{E}(\rho)$.

The system-environment coupling scenario can be generalized to the first approach of quantum operations, considered as state changes in a subsystem (see Fig. 2.4).

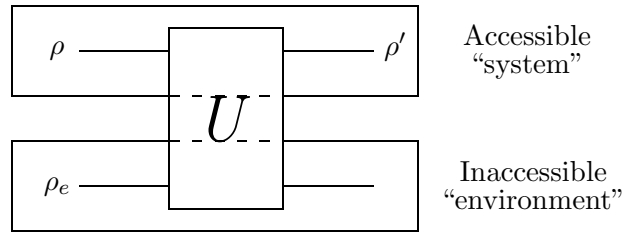

Figure 2.4: Circuit for system-environment coupling

The system interacts with some ancillary system and some subspace labeled by "$o$" is discarded. We allow state changes which are conditioned on some measurement results in the discarded subsystem. Mathematically, the process is described as

$$\mathcal{E}(\rho) = \text{Tr}_o \left[ U(\rho \otimes \rho_a)U^\dagger(I \otimes P_o) \right] \tag{2.8}$$

where $P_o$ is a projector acting on the discarded subsystem. The probability of having the final state is given by $\text{Tr}(\mathcal{E}(\rho)) \leq 1$ which can be less than unity. The output density matrix is given by $\rho' = \frac{\mathcal{E}(\rho)}{\text{Tr}(\mathcal{E}(\rho))}$. It is important to absorb the normalization into $\rho'$ to keep $\mathcal{E}$ linear. Having described the approach of quantum operations based on subsystem evolution, we now show that such formalism is equivalent to two others. A summary of these three approaches and their relations is shown in Fig. 2.5.

---

[2] Consider two systems $A$ and $B$. The partial trace over system $B$ is a linear operation on the composite system $AB$ defined by $\text{Tr}_B \left[ |a_1\rangle\langle a_2| \otimes |b_1\rangle\langle b_2| \right] \equiv |a_1\rangle\langle a_2| \; \text{Tr}(|b_1\rangle\langle b_2|)$ where $|a_i\rangle$ and $|b_i\rangle$ for $i = 1, 2$ are vectors in $A$ and $B$.

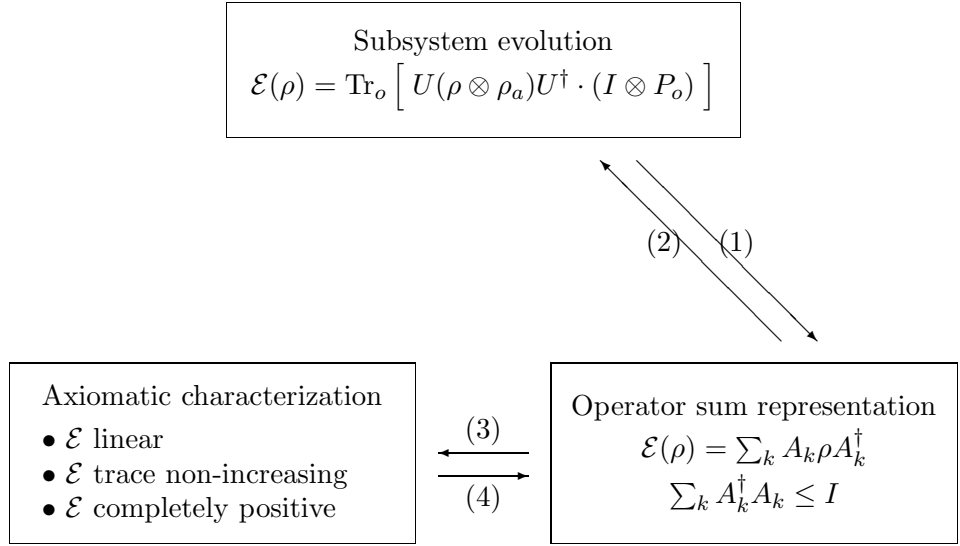

Figure 2.5: Three equivalent approaches to quantum operations

To elaborate on the highly non-trivial result depicted in Fig. 2.5, we need to define some notations. Let $\mathcal{H}_1$ and $\mathcal{H}_2$ be the input and output Hilbert spaces for the quantum operation, and $\mathcal{B}(\mathcal{H}_i)$ be the set of bounded operators acting on $\mathcal{H}_i$. Figure 2.5 asserts that the three classes of mappings from $\mathcal{B}(\mathcal{H}_1)$ to $\mathcal{B}(\mathcal{H}_2)$ with the following characterizations are the same:

1. Subsystem evolutions which are given by:

$$\mathcal{E}(\rho) = \mathrm{Tr}_o \left[ U(\rho \otimes \rho_a) U^\dagger (I_2 \otimes P_o) \right] \tag{2.9}$$

where $\rho \in \mathcal{B}(\mathcal{H}_1)$ is the input density matrix, $\rho_a \in \mathcal{B}(\mathcal{H}_a)$ is the initial ancilla density matrix, $I_2$ is the identity operator in $\mathcal{B}(\mathcal{H}_2)$, and $P_o$ is a projector acting on the discarded system (labeled by "$o$") which is the complement of $\mathcal{H}_2$ in $\mathcal{H}_1 \otimes \mathcal{H}_a$.

2. State changes described by some *operator sum representation* [101, 28, 75]

$$\mathcal{E}(\rho) = \sum_k A_k \rho A_k^\dagger \tag{2.10}$$

$$\text{where} \quad \sum_k A_k^\dagger A_k \leq I \tag{2.11}$$

An input state becomes a mixture of output states each resulting from the action of an $A_k$. These are linear operations from $\mathcal{H}_1$ to $\mathcal{H}_2$ which are analogous to a set of probable "events". The $A_k$ are sometimes called "Kraus operators". In this Dissertation, $A_k$ are referred to as the *operation elements* of $\mathcal{E}$. When the input state is $\rho$, the probability of the "event" $A_k$ is given by $\text{Tr}(A_k^\dagger A_k \rho)$. Equation (2.11) restricts the total probability to be no greater than unity.

3. The set of mappings from $\mathcal{B}(\mathcal{H}_1)$ to $\mathcal{B}(\mathcal{H}_2)$ which satisfy three axioms: linearity, *complete positivity* and trace non-increasing. Complete positivity is defined by the following:

> *Definition 1*: An operator $A$ is *positive* if all eigenvalues are non-negative.

> *Definition 2*: A linear map $\mathcal{M}$ acting on operators is *positive* if $A \geq 0 \Rightarrow$ $\mathcal{M}(A) \geq 0$.

> *Definition 3*: A linear map $\mathcal{M}$ on $\mathcal{H}_1$ is *completely positive* if, for any ancillary Hilbert space $\mathcal{H}_a$ and for all operators $\tilde{A}$ on $\mathcal{H}_1 \otimes \mathcal{H}_a$, $\tilde{A} \geq 0 \Rightarrow$ $(\mathcal{M} \otimes \mathcal{I}_a)(\tilde{A}) \geq 0$, where $\mathcal{I}_a$ is the identity operation on $\mathcal{B}(\mathcal{H}_a)$.

The physical significance of complete positivity is the following. A quantum operation takes a density matrix to another and therefore has to be a positive map. Moreover, if an input state is initially entangled with some ancillary system, and $\mathcal{E}$ acts on the system while the identity operation acts on the ancillary system, the density matrix representing the final combined state has to be positive.

There are important implications of the equivalence. Any physically reasonable process (such as defined by the three basic axioms) can always be described by some operator sum representation or be *modeled* by some coupling with an external system followed by discarding part of the combined system. The fact that continuous distortion of a quantum state has an operator sum representation is important in quantum error correction, as we will see in subsequent chapters.

After examining the physical contents of the three approaches, we now proceed to prove their equivalence. The proofs are ordered as given in Fig. 2.5.

**Proof of (1):** Suppose

$$\mathcal{E}(\rho) = \mathrm{Tr}_o \left[ U(\rho \otimes \rho_a)U^\dagger(I_2 \otimes P_o) \right] \tag{2.12}$$

We want to find an operator sum representation for $\mathcal{E}$. We write $\rho_a = \sum_l \lambda_l |i_l\rangle\langle i_l|$ where $|i_l\rangle$ and $\lambda_l$ are the eigenstates and the corresponding (non-negative) eigenvalues of $\rho_a$. Let $|j_k\rangle$ be the eigenvectors of $P_o$, ordered such that $P_o = \sum_{k=1}^K |j_k\rangle\langle j_k|$. Performing the partial trace on the basis $\{|j_k\rangle\}$, Eq. (2.12) can be rewritten as:

$$\mathcal{E}(\rho) = \sum_{k=1}^K \sum_l \lambda_l \langle j_k|U \left[ \rho \otimes |i_l\rangle\langle i_l| \right] U^\dagger|j_k\rangle \tag{2.13}$$

$$= \sum_{k=1}^K \sum_l A_{lk}\rho A_{lk}^\dagger \quad \text{where} \quad A_{lk} = \langle j_k|U|i_l\rangle \tag{2.14}$$

Note that $A_{lk} = \langle j_k|U|i_l\rangle$ is indeed an operator mapping $\mathcal{B}(\mathcal{H}_1)$ to $\mathcal{B}(\mathcal{H}_2)$. It remains to show $\sum_{k=1}^K \sum_l A_{lk}^\dagger A_{lk} \le I$. For any state $|\psi\rangle$,

$$\langle\psi| \left[ \sum_{k=1}^K \sum_l A_{lk}^\dagger A_{lk} \right] |\psi\rangle = \mathrm{Tr} \left[ \sum_{k=1}^K \sum_l A_{lk}|\psi\rangle\langle\psi|A_{lk}^\dagger \right] = \mathrm{Tr}(\mathcal{E}(|\psi\rangle\langle\psi|)) \le 1 \tag{2.15}$$

where the last inequality is immediate from Eq. (2.12). This completes the proof that any subsystem evolution can be described by an operator sum representation.

**Proof of (2):** Let $\mathcal{E}(\rho) = \sum_{k=1}^N A_k\rho A_k^\dagger$ be given, with $\sum_{k=1}^N A_k^\dagger A_k \le I$. Let $A_0 = \sqrt{I - \sum_{k=1}^N A_k^\dagger A_k}$. Let $\mathcal{H}_a$ be some $N+1$ dimensional space with basis $\{|k\rangle\}$. Consider an operator $U$ which acts on $\mathcal{H}_1 \otimes \mathcal{H}_a$ according to

$$U(|\psi\rangle|0\rangle) = \sum_{k=0}^N (A_k|\psi\rangle)|k\rangle \qquad \forall|\psi\rangle. \tag{2.16}$$

At this point, $U$ may not be unitary. However, from Eq. (2.16),

$$
\left[\ \langle\psi_1|\langle 0|U^\dagger\ \right]\left[\ U|\psi_2\rangle|0\rangle\ \right] \quad = \quad \sum_{kk'=0}^{N}\left[\ (\langle\psi_1|A_k^\dagger)\langle k|\ \right]\left[\ (A_{k'}|\psi_2\rangle)|k'\rangle\ \right] \tag{2.17}
$$

$$
= \quad \sum_{kk'=0}^{N}\langle\psi_1|A_k^\dagger A_{k'}|\psi_2\rangle\langle k|k'\rangle \tag{2.18}
$$

$$
= \quad \langle\psi_1|\sum_{k=0}^{N}A_k^\dagger A_k|\psi_2\rangle \tag{2.19}
$$

$$
= \quad \langle\psi_1|\psi_2\rangle\langle 0|0\rangle \tag{2.20}
$$

Therefore, $U$ preserves the inner product on the subspace spanned by $|\psi\rangle|0\rangle$ and can be extended to a unitary operation on the whole space. Let $P = \sum_{k=1}^{N}|k\rangle\langle k|$. Then,

$$
\mathrm{Tr}_a\left[\ U(|\psi\rangle\langle\psi|\otimes|0\rangle\langle 0|)U^\dagger(I\otimes P)\ \right] \tag{2.21}
$$

$$
= \quad \sum_{k,k',k''=1}^{N}\langle k|\left[\ (A_{k'}|\psi\rangle)|k'\rangle(\langle\psi|A_{k''}^\dagger)\langle k''|\ \right]|k\rangle \tag{2.22}
$$

$$
= \quad \sum_{k=1}^{N}A_k|\psi\rangle\langle\psi|A_k^\dagger \tag{2.23}
$$

Since Eqs. (2.21)-(2.23) hold for all pure states in $\mathcal{H}_1$, they also hold for mixed states. Therefore, Eq. (2.21) is indeed a correct system-environment coupling model having the same dynamics as the given operator sum representation.

**Proof of (3):** $\mathcal{E}(\rho) = \sum_k A_k\rho A_k^\dagger$ with $\sum_k A_k^\dagger A_k \leq I$ is obviously linear and trace non-increasing. Let $\tilde{\rho}$ be a positive operator in $\mathcal{B}(\mathcal{H}_1\otimes\mathcal{H}_a)$ where $\mathcal{H}_a$ is *any* finite dimensional Hilbert space. By hypothesis, $(\mathcal{E}\otimes\mathcal{I}_a)(\tilde{\rho}) = \sum_k(A_k\otimes I_a)\ \tilde{\rho}\ (A_k\otimes I_a)^\dagger$. Since $\tilde{\rho}$ is a convex sum of outer-products, $(\mathcal{E}\otimes\mathcal{I}_a)(\tilde{\rho})$ is also a positive sum of outer-products and is positive. Hence, $\mathcal{E}$ is completely positive.

**Proof of (4):** This is the most non-trivial part of the result. This ingenious result was proved by Choi in [28]. The present proof is essentially the same proof, but re-casted in a completely different language. A related but slightly different proof can be found in Chapter 3 of [85].

Suppose $\mathcal{E}$ is a completely positive linear map from $\mathcal{B}(\mathcal{H}_1)$ to $\mathcal{B}(\mathcal{H}_2)$. Our goal is to show that $\mathcal{E}$ can be expressed as in Eq. (2.10). For concreteness, let $n_i$ be the dimension of $\mathcal{H}_i$.

Consider the mapping $\mathcal{I} \otimes \mathcal{E}$ from $\mathcal{B}(\mathcal{H}_1 \otimes \mathcal{H}_1)$ to $\mathcal{B}(\mathcal{H}_1 \otimes \mathcal{H}_2)$. In particular, consider the action of $\mathcal{I} \otimes \mathcal{E}$ on the following operator in $\mathcal{B}(\mathcal{H}_1 \otimes \mathcal{H}_1)$:

$$Y = \sum_{i,j=1}^{n_1} |i\rangle\langle j| \otimes |i\rangle\langle j| \tag{2.24}$$

where $\{|i\rangle\}_{i=1,\cdots,n_1}$ is a basis for $\mathcal{H}_1$. Note that $Y = n_1|\Phi\rangle\langle\Phi|$ where $|\Phi\rangle$ is the maximally entangled state $|\Phi\rangle = \frac{1}{\sqrt{n_1}}\sum_i |i\rangle \otimes |i\rangle$. $Y$ is therefore positive. $Y$ can be explicitly written as:

$$Y = \begin{bmatrix}
1 & 0 & \cdot & 0 & 0 & 1 & \cdot & 0 & \cdot & \cdot & \cdot & \cdot & 0 & 0 & \cdot & 1 \\
0 & 0 & \cdot & 0 & 0 & 0 & \cdot & 0 & \cdot & \cdot & \cdot & \cdot & 0 & 0 & \cdot & 0 \\
\cdot & \cdot & \cdot & \cdot & & & & & & & & & \cdot & \cdot & \cdot & \cdot \\
0 & 0 & \cdot & 0 & 0 & 0 & \cdot & 0 & \cdot & \cdot & \cdot & \cdot & 0 & 0 & \cdot & 0 \\
0 & 0 & \cdot & 0 & 0 & 0 & \cdot & 0 & \cdot & \cdot & \cdot & \cdot & 0 & 0 & \cdot & 0 \\
1 & 0 & \cdot & 0 & 0 & 1 & \cdot & 0 & \cdot & \cdot & \cdot & \cdot & 0 & 0 & \cdot & 1 \\
\cdot & \cdot & \cdot & \cdot & & & & & & & & & \cdot & \cdot & \cdot & \cdot \\
0 & 0 & \cdot & 0 & 0 & 0 & \cdot & 0 & \cdot & \cdot & \cdot & \cdot & 0 & 0 & \cdot & 0 \\
\cdot & \cdot & \cdot & \cdot & \cdot & \cdot & \cdot & \cdot & \cdot & \cdot & \cdot & \cdot & \cdot & \cdot & \cdot & \cdot \\
\cdot & \cdot & \cdot & \cdot & \cdot & \cdot & \cdot & \cdot & \cdot & \cdot & \cdot & \cdot & \cdot & \cdot & \cdot & \cdot \\
\cdot & \cdot & \cdot & \cdot & \cdot & \cdot & \cdot & \cdot & \cdot & \cdot & \cdot & \cdot & \cdot & \cdot & \cdot & \cdot \\
\cdot & \cdot & \cdot & \cdot & \cdot & \cdot & \cdot & \cdot & \cdot & \cdot & \cdot & \cdot & \cdot & \cdot & \cdot & \cdot \\
0 & 0 & 0 & 0 & 0 & 0 & 0 & 0 & \cdot & \cdot & \cdot & \cdot & 0 & 0 & 0 & 0 \\
0 & \cdot & \cdot & \cdot & 0 & \cdot & \cdot & \cdot & \cdot & \cdot & \cdot & \cdot & 0 & \cdot & \cdot & \cdot \\
0 & 0 & 0 & 0 & 0 & 0 & 0 & 0 & \cdot & \cdot & \cdot & \cdot & 0 & 0 & 0 & 0 \\
1 & 0 & 0 & 0 & 0 & 1 & 0 & 0 & \cdot & \cdot & \cdot & \cdot & 0 & 0 & 0 & 1
\end{bmatrix} \tag{2.25}$$

which is an $n_1 \times n_1$ array of $n_1 \times n_1$ matrices. The $(i,j)$ block is exactly $|i\rangle\langle j|$. Hence,

$(\mathcal{I} \otimes \mathcal{E})(Y) =$

$$
\begin{bmatrix}
\mathcal{E}\begin{pmatrix} 1 & 0 & \cdot & 0 \\ 0 & 0 & \cdot & 0 \\ \cdot & \cdot & \cdot & \cdot \\ 0 & 0 & \cdot & 0 \end{pmatrix} & \mathcal{E}\begin{pmatrix} 0 & 1 & \cdot & 0 \\ 0 & 0 & \cdot & 0 \\ \cdot & \cdot & \cdot & \cdot \\ 0 & 0 & \cdot & 0 \end{pmatrix} & \begin{matrix} \cdot & \cdot & \cdot & \cdot \end{matrix} & \mathcal{E}\begin{pmatrix} 0 & 0 & \cdot & 1 \\ 0 & 0 & \cdot & 0 \\ \cdot & \cdot & \cdot & \cdot \\ 0 & 0 & \cdot & 0 \end{pmatrix} \\[2em]
\mathcal{E}\begin{pmatrix} 0 & 0 & \cdot & 0 \\ 1 & 0 & \cdot & 0 \\ \cdot & \cdot & \cdot & \cdot \\ 0 & 0 & \cdot & 0 \end{pmatrix} & \mathcal{E}\begin{pmatrix} 0 & 0 & \cdot & 0 \\ 0 & 1 & \cdot & 0 \\ \cdot & \cdot & \cdot & \cdot \\ 0 & 0 & \cdot & 0 \end{pmatrix} & \begin{matrix} \cdot & \cdot & \cdot & \cdot \end{matrix} & \mathcal{E}\begin{pmatrix} 0 & 0 & \cdot & 0 \\ 0 & 0 & \cdot & 1 \\ \cdot & \cdot & \cdot & \cdot \\ 0 & 0 & \cdot & 0 \end{pmatrix} \\[2em]
\begin{matrix} \cdot & \cdot & \cdot \\ \cdot & \cdot & \cdot \\ \cdot & \cdot & \cdot \\ \cdot & \cdot & \cdot \end{matrix} & \begin{matrix} \cdot & \cdot & \cdot \\ \cdot & \cdot & \cdot \\ \cdot & \cdot & \cdot \\ \cdot & \cdot & \cdot \end{matrix} & \begin{matrix} \cdot & \cdot & \cdot \\ \cdot & \cdot & \cdot \\ \cdot & \cdot & \cdot \\ \cdot & \cdot & \cdot \end{matrix} & \begin{matrix} \cdot & \cdot & \cdot \\ \cdot & \cdot & \cdot \\ \cdot & \cdot & \cdot \\ \cdot & \cdot & \cdot \end{matrix} \\[2em]
\mathcal{E}\begin{pmatrix} 0 & 0 & 0 & 0 \\ 0 & \cdot & \cdot & \cdot \\ 0 & 0 & 0 & 0 \\ 1 & 0 & 0 & 0 \end{pmatrix} & \mathcal{E}\begin{pmatrix} 0 & 0 & 0 & 0 \\ 0 & \cdot & \cdot & \cdot \\ 0 & 0 & 0 & 0 \\ 0 & 1 & 0 & 0 \end{pmatrix} & \begin{matrix} \cdot & \cdot & \cdot & \cdot \end{matrix} & \mathcal{E}\begin{pmatrix} 0 & 0 & 0 & 0 \\ 0 & \cdot & \cdot & \cdot \\ 0 & 0 & 0 & 0 \\ 0 & 0 & 0 & 1 \end{pmatrix}
\end{bmatrix}
$$

$(2.26)$

which is an $n_1 \times n_1$ array of $n_2 \times n_2$ matrices. The $(i, j)$ block is exactly $\mathcal{E}(|i\rangle\langle j|)$.

We now express $(\mathcal{I} \otimes \mathcal{E})(Y)$ in a manner completely independent of Eq. (2.26). Since $Y$ is positive and $\mathcal{E}$ is completely positive, $(\mathcal{I} \otimes \mathcal{E})(Y)$ is positive, and can be expressed as $(\mathcal{I} \otimes \mathcal{E})(Y) = \sum_{l=1}^{n_1 n_2} |a_k\rangle\langle a_k|$; the $|a_k\rangle$ are the eigenvectors of $(\mathcal{I} \otimes \mathcal{E})(Y)$, normalized to $\sqrt{\lambda_k}$, where $\lambda_k$ are the positive eigenvalues of $(\mathcal{I} \otimes \mathcal{E})(Y)$. We can write $(\mathcal{I} \otimes \mathcal{E})(Y)$ as:

where the column represents $|a_k\rangle$ and the row represents $\langle a_k|$. We divide the column into $n_1$ segments each of length $n_2$, and define a matrix $A_k$ with the $i$-th column being the $i$-th segment, so that the $i$-th segment is exactly $A_k|i\rangle$. Therefore,

$$(\mathcal{I} \otimes \mathcal{E})(Y) = \sum_k \begin{bmatrix} A_k|1\rangle\langle 1|A_k^\dagger & A_k|1\rangle\langle 2|A_k^\dagger & \cdots & A_k|1\rangle\langle n_1|A_k^\dagger \\ A_k|2\rangle\langle 1|A_k^\dagger & A_k|2\rangle\langle 2|A_k^\dagger & \cdots & A_k|2\rangle\langle n_1|A_k^\dagger \\ \cdots & \cdots & \cdots & \cdots \\ A_k|n_1\rangle\langle 1|A_k^\dagger & A_k|n_1\rangle\langle 2|A_k^\dagger & \cdots & A_k|n_1\rangle\langle n_1|A_k^\dagger \end{bmatrix} \tag{2.27}$$

Comparing Eqs. (2.26) and (2.27), $\mathcal{E}(\rho) = \sum_k A_k \rho A_k^\dagger \ \forall \rho$. Using this expression for $\mathcal{E}$ and the fact that $\mathcal{E}$ is trace non-increasing, $\sum_k A_k^\dagger A_k \leq I$. This completes the proof.

We will primarily be using the operator sum representation approach for quantum operations. We conclude this section with a discussion on the freedom of the operation elements. The operation elements $A_k$ are not unique for a quantum operation $\mathcal{E}$. The classic example is the *phase flip channel* on a qubit, given by $\mathcal{E}(\rho) = (\rho + Z\rho Z)/2 = |0\rangle\langle0|\rho|0\rangle\langle0| + |1\rangle\langle1|\rho|1\rangle\langle1|$, which turns all off-diagonal elements in $\rho$ to 0. It means that randomly applying $I$ or $Z$ is the same as measuring the qubit without knowing the result! These two processes or interpretations are indistinguishable if we have no access to the agent coupled to the system. This example can be generalized:

   *Theorem 1*: Consider two sets of operation elements $\{A_k\}$ and $\{B_l\}$, where the $A_k$ are independent. Then, the two sets generate the same quantum operation iff $B_l = \sum_k u_{lk} A_k$ where $u_{lk}$ are entries of an isometric matrix. [3] This is called the "freedom in the operation elements" for a quantum operation.

   **Proof:** Following the proof of (4), the theorem is the same as saying that $\sum_k |a_k\rangle\langle a_k| = \sum_l |b_l\rangle\langle b_l|$ iff $|b_l\rangle = \sum_k u_{lk}|a_k\rangle$, where $|a_k\rangle$ and $A_k$ are related as in the proof of (4) and similarly for $|b_l\rangle$ and $B_l$.

   **Sufficiency:** If $|b_l\rangle = \sum_k u_{lk}|a_k\rangle$, then, $\sum_l |b_l\rangle\langle b_l| = \sum_{lkk'} u_{lk}u_{lk'}^*|a_k\rangle\langle a_{k'}| = \sum_{kk'}(\sum_l u_{lk}u_{k'l}^\dagger)|a_k\rangle\langle a_{k'}| = \sum_k |a_k\rangle\langle a_k|$.

   **Necessity:** If $\sum_k |a_k\rangle\langle a_k| = \sum_l |b_l\rangle\langle b_l|$, We consider the spans of $\{|a_k\rangle\}$ and $\{|b_l\rangle\}$. For any $|a\rangle$ which is orthogonal to all $|a_k\rangle$, $\sum_l \langle a|b_l\rangle\langle b_l|a\rangle = 0$ by hypothesis, and $|a\rangle$ is orthogonal to all $|b_l\rangle$. Therefore, the span of $\{|b_l\rangle\}$ is in the span of $\{|a_k\rangle\}$, and $|b_l\rangle = \sum_k c_{lk}|a_k\rangle$ for some constants $c_{lk}$. Re-applying this to the hypothesis, $\sum_{lkk'} c_{lk}c_{lk'}^*|a_k\rangle\langle a_{k'}| = \sum_k |a_k\rangle\langle a_k|$. Using the linear independence of $|a_k\rangle$ (and thus that of $|a_k\rangle\langle a_{k'}|$), $\sum_l c_{lk}c_{lk'}^* = \delta_{kk'}$ and the $c_{lk}$ are entries of an isometric matrix. This completes the proof.

Note that we can similar argue that the span of $\{|a_k\rangle\}$ is in that of $\{|b_l\rangle\}$. Therefore a set of independent operation elements has minimal cardinality. Moreover, such independent sets are "canonical": every other set generating the same operation can be expressed in terms of the canonical ones by Theorem 1.

---

[3] A matrix $u$ is isometric if $u^\dagger u = I$ ($u^\dagger u = I \Leftrightarrow uu^\dagger = I$). Square isometric matrices are unitary.

The above proof also applies to the freedom in decomposing a positive matrix into a sum of outer products. The implication is that, different ensembles of states resulting in the same density matrix are indistinguishable, without access to the information outside the system.

This degree of freedom in the operation elements is also manifest in the subsystem evolution approach. This is exactly the freedom in unitarily evolving the discarded system, which cannot affect the quantum operation as observed in the system only. The proof is straightforward, and we only state the result using Fig. 2.6.

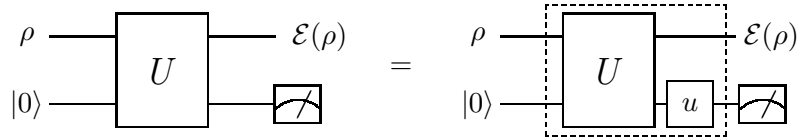

Figure 2.6: The degree of freedom in the operation elements, from the system-environment coupling approach. In the left, we construct a system-environment coupling model for $\{A_k\}$ as described in the proof of (2), and retain all the definitions made before. In the right is the construction based on $B_l = \sum_k u_{lk} A_k$. In other words, the dashed box is the evolution required if we apply the construction in the proof of (2) to the set $\{B_l\}$.

### 2.3.2 Quantum process tomography

Quantum process tomography is a procedure by which an unknown quantum operation of a system can be fully experimentally characterized. Complete determination of a quantum operation is important for characterizing our devices, such as quantum logic gates and quantum channels, as well as for understanding our noise sources which may lead to better correction techniques.

Traditionally, such characterization revolves around semi-classical concepts such as coupling strengths, relaxation rates, and phase coherence times [82, 52]. However, it is a simple exercise following the discussion in the previous section that there are $16^n - 4^n$ degrees of freedom in the quantum operations on an $n$-qubit system. Therefore, traditional techniques are insufficient for the purpose of quantum process tomography.

Methods for performing quantum process tomography were first reported in [32, 94].

In this section, we summarize the method described in [32] which is used in Part III. A completely different method will also be given, based on the proof of (4) in the previous section. This alternative method is original.

The basic assumptions in quantum process tomography are as follows. The unknown quantum operation, $\mathcal{E}$, is given as an "oracle" or a "blackbox" we call upon without knowing its internal mechanism. We prepare certain input states and *measure* the corresponding output density matrices to learn about $\mathcal{E}$ systematically. The task to measure the density matrix of a quantum system is called *quantum state tomography*, and will be described in detail in Part III. We simply assume the ability to do so for now. We want to obtain an operator sum representation for $\mathcal{E}$ to be used in our various applications. Therefore, the question is, how to convert experimental measurements on density matrices to knowledge of the operation elements?

## Method I [32]

Let the unknown quantum operation be $\mathcal{E}(\rho) = \sum_k A_k \rho A_k^\dagger$. The crucial observation is to transform the information carried by the operation elements to the coefficients in the "$\chi$-representation" to be defined. For simplicity, the input and output Hilbert spaces are both equal to $\mathcal{H}$ and have $n$ dimensions. The procedure is as follows:

1. Choose a *fixed* basis $\{B_m\}_{m=1,\cdots,n^2}$ for the operators acting on $\mathcal{H}$. Express each $A_k$ as $A_k = \sum_m c_{km} B_m$, and rewrite the operator sum representation as

$$\mathcal{E}(\rho) = \sum_{mnk} c_{km} c_{kn}^* B_m \rho B_n^\dagger = \sum_{mn} \chi_{mn} B_m \rho B_n^\dagger \qquad (2.28)$$

   where $\chi_{mn} = \sum_k c_{km} c_{kn}^*$. Equation (2.28) is called a $\chi$-representation of $\mathcal{E}$. Note that the information on $\mathcal{E}$ lies with the coefficients $\chi_{mn}$ rather than the fixed operators $B_m$.

2. Choose a fixed basis $\{\rho_i\}_{i=1,\cdots,n^2}$ for the density matrices. The $\rho_i$ correspond to physical states.

3. For each $i$, obtain $\mathcal{E}(\rho_i)$ in terms of $\rho_j$ in two different ways.

- Experimentally, prepare $\rho_i$, apply $\mathcal{E}$ and measure $\mathcal{E}(\rho_i)$. Express the outcome as $\mathcal{E}(\rho_i) = \sum_j \rho_j \lambda_{ij}$.

- Theoretically, express $B_m \rho_i B_n^\dagger = \sum_j \kappa_{ij}^{mn} \rho_j$, so that $\mathcal{E}(\rho) = \sum_{mn} \chi_{mn} B_m \rho B_n^\dagger = \sum_{mn} \kappa_{ij}^{mn} \chi_{mn} \rho_j$.

4. Comparing the experimental and theoretical results, and using the fact that $\{\rho_j\}$ is a basis, we obtain $\lambda_{ij} = \sum_{mn} \kappa_{ij}^{mn} \chi_{mn}$. Considering $\kappa$ as a matrix with double indices $ij$ and $mn$, we have $\vec{\lambda} = \kappa \vec{\chi}$. Having found $\vec{\lambda}$ experimentally, we invert $\kappa$ to obtain $\vec{\chi}$.

5. Note that $\chi$ is positive by definition. Therefore, $\chi = WDW^\dagger$ where $W$ is unitary and $D$ is positive and diagonal. Putting $\chi_{mn} = \sum_k W_{mk} D_{kk} W_{kn}^\dagger$ into Eq. (2.28), we obtain *an* operator sum representation with operation elements $\tilde{A}_k = \sum_m W_{mk} \sqrt{D_{kk}} B_m$.

Note that the resulting set of operation elements is canonical.

## Method II

The second method follows almost immediately from the proof of (4) in Section 2.3. We retain all the notations defined in the proof of (4). The crucial observation is that $Y$ and $(\mathcal{I} \otimes \mathcal{E})(Y)$ in the proof of (4) correspond to physical states $|\Phi\rangle\langle\Phi|$ and $(\mathcal{I} \otimes \mathcal{E})(|\Phi\rangle\langle\Phi|)$ which can be prepared and measured. The procedure is to:

1. Take two copies of $\mathcal{H}$ and prepare the maximally entangled state $|\Phi\rangle$.

2. Subject one system to the action of $\mathcal{E}$, while not doing anything to the other.

3. Measure the combined output density matrix $(\mathcal{I} \otimes \mathcal{E})(|\Phi\rangle\langle\Phi|) = \frac{1}{n}(\mathcal{I} \otimes \mathcal{E})(Y)$, multiply by $n$ and decompose as $\sum_k |a_k\rangle\langle a_k|$. Divide $|a_k\rangle$ (of length $n^2$) into $n$ equal segments. $A_k$ is the matrix having the $i$-th segment as its $i$-th column.

This recipe is interesting because the mathematical proof that a quantum operation has an operator sum representation can be directly be translated to an experimental procedure for finding it.

**Comparison**

Some interesting comparisons can be made on the two methods. Method I involves a $\chi$-representation as an intermediate step, which is unnecessary in Method II (the density matrix measured in Method II is directly related to the operation elements). Method I also involves a basis of physical states $\rho_i$, which cannot be chosen to be orthonormal, and will complicate the analysis. In contrast, only one physical input state is required for Method II, which automatically contains the information $\mathcal{E}(|i\rangle\langle j|)$ for the unphysical orthonormal basis $|i\rangle\langle j|$. However, Method II requires the preparation of a maximally entangled state and the ability to isolate one of the two systems prepared.

The two methods consume equivalent amount of resources, which is determined by the number of degrees of freedom in the quantum operation. In general, to measure an $n \times n$ density matrix, $n^2$ *ensemble* measurements are needed, requiring $\approx \mathcal{O}(n^2)$ steps. For method I, to determine $n^2$ $n \times n$ density matrices requires $\approx \mathcal{O}(n^4)$ steps. For method II, to determine one $n^2 \times n^2$ density matrix also requires $\approx \mathcal{O}(n^4)$ steps. In both cases, the number of steps is of the same order as the number of degrees of freedom in the quantum operation, therefore, both procedures are optimal in some sense.

## 2.4   Common noise processes

As an application of the quantum operation formalism, we describe some important noise processes using the operator sum representation. The traditional description of some of these processes using master equations is given in Appendix A.1. Noise processes are often called noisy channels, following the model communication problem:

$$\rho \;\rule[0.5ex]{1.5em}{0.4pt}\; \boxed{\mathcal{E}} \;\rule[0.5ex]{1.5em}{0.4pt}\; \mathcal{E}(\rho)$$

Figure 2.7: A quantum channel

We will consider primarily noise processes on a single qubit. In this case, the density matrix can always be written as

$$\rho = \frac{1}{2}(I + \vec{r}\cdot\vec{\sigma}) \tag{2.29}$$

where $\vec{r}$ is a real vector in the unit ball and is often called the Bloch vector. In fact, the convention for single qubit rotation in Eq. (2.6) is chosen to correspond to the rotation of the Bloch vector. The effects of a quantum process $\mathcal{E}$ can easily be visualized by comparing the Bloch vectors for $\rho$ and $\mathcal{E}(\rho)$.

**Phase Damping**

Phase damping can be described by a channel in which a $Z$ occurs with probability $p \in [0, 1]$. The operator sum representation is given by

$$\mathcal{E}_{PD}(\rho) = (1 - p)\rho + pZ\rho Z \,. \tag{2.30}$$

Consequently,

$$(r_x, r_y, r_z) \rightarrow ((1 - 2p)r_x, (1 - 2p)r_y, r_z) \tag{2.31}$$

resulting in the loss of coherence between different basis states. The maximum amount of decoherence occurs at $p = 1/2$.

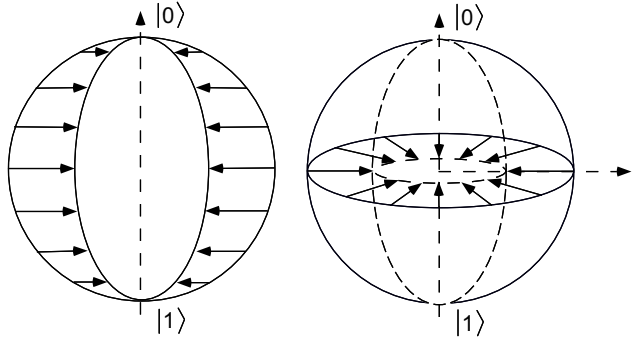

Figure 2.8: Trajectories of different points on the Bloch sphere under the effect of phase damping. Points move along perpendiculars to the $\hat{z}$-axis at rates proportional to the distances to the $\hat{z}$-axis. As a result, the Bloch sphere turns into an ellipsoid.

The same process can arise from many different physical situations. For example, if the qubit is measured with probability $2p$,

$$\mathcal{E}_M(\rho) = (1 - 2p)\rho + 2p \left[ \ket{0}\bra{0} \rho \ket{0}\bra{0} + \ket{1}\bra{1} \rho \ket{1}\bra{1} \right] \,, \tag{2.32}$$

which is just $\mathcal{E}_{PD}(\rho)$ for $0 \leq p \leq 1/2$. Phase damping can arise from even more apparently different process. For example, if some random phase shift $P_\theta = e^{-i\theta Z/2}$ occurs, the density matrix is evolved as

$$\rho = \begin{bmatrix} a & b \\ b^* & c \end{bmatrix} \rightarrow P_\theta \rho P_\theta^\dagger = \begin{bmatrix} a & be^{-i\theta} \\ b^* e^{i\theta} & c \end{bmatrix}. \tag{2.33}$$

For a noise process in which $\theta$ is a random walk with independent and identically distributed steps, after a time $t$, the density matrix resulting from averaging over $\theta$ is

$$\begin{bmatrix} a & be^{-\lambda t} \\ b^* e^{-\lambda t} & c \end{bmatrix}, \tag{2.34}$$

for some constant $\lambda$.

The important implication is that, we may model a complicated real life process by much simpler ones without losing the essential physical content. We will see later how this fact is used in quantum error correction.

**Random Pauli channel**

In a random Pauli channel, an "error" $X$, $Y$ or $Z$ can happen each with probability $p/3$. The operator sum representation is given by

$$\mathcal{E}_{RP}(\rho) = (1-p)\rho + p/3 \left( X\rho X + Y\rho Y + Z\rho Z \right). \tag{2.35}$$

The random Pauli channel is also called the depolarizing channel, because $\mathcal{E}_{RP}$ is same as swapping $\rho$ with $I/2$ with probability $4p/3$! The result is a *symmetric* shrinking of the *Bloch sphere* towards the origin.

In the *generalized random Pauli channel*, different errors can occur with different probabilities, in which case, the *Bloch sphere* still shrinks towards the origin, but asymmetrically.

**Unital processes**

All the channels described so far fix the identity. These are examples of unital processes, defined by $\mathcal{E}(I) = I$. Trace-preserving unital processes are the quantum analogs of the classical doubly stochastic processes, and have many interesting properties [87]. For instance, the von-Neuman entropy of a state (defined to be $\text{Tr}(\rho \log \rho)$) can only increase under unital processes. *Random unitary processes* of the form

$$\mathcal{E}(\rho) = \sum_k p_k U_k \rho U_k^\dagger \tag{2.36}$$

where $U_k$ are unitary, are unital. The converse does not hold, except for qubit unital processes. An original proof is given in Appendix B.1.

Non-unital processes play an important role in quantum information processing, because they provide the only means to reduce entropy and prepare initial states.

**Amplitude damping**

Amplitude damping is a process by which energy is lost to a zero temperature environment. It is defined by

$$\mathcal{E}_{AD}(\rho) = \sum_{k=0,1} A_k \rho A_k^\dagger \tag{2.37}$$

$$A_0 = \begin{bmatrix} 1 & 0 \\ 0 & \sqrt{1-\gamma} \end{bmatrix} \quad A_1 = \begin{bmatrix} 0 & \sqrt{\gamma} \\ 0 & 0 \end{bmatrix}, \tag{2.38}$$

where the energy eigenstates are chosen to be the computational basis states and $|0\rangle$ is the ground state. The energy exchange with an environment at finite temperature can be modeled by *generalized amplitude damping*, defined as:

$$\mathcal{E}_{GAD}(\rho) = \sqrt{p}\, \mathcal{E}_{AD}(\rho) + \sqrt{1-p}\, \left[\, X\mathcal{E}_{AD}(\rho)X \,\right]. \tag{2.39}$$

where the parameter $p$ is determined by the thermal energy $k_B T$ and the energy difference between the basis states, $\Delta E$, according to $\frac{p}{1-p} = e^{\frac{\Delta E}{k_B T}}$. Amplitude damping is the most important non-unital process known.

## 2.5   Summary

- The fundamental unit of quantum information is a qubit:

  $|\psi\rangle = a|0\rangle + b|1\rangle$ or $\rho = \frac{1}{2}(I + \vec{r} \cdot \vec{\sigma})$.

- The quantum circuit is a model of quantum information processing.

- There are universal sets of quantum logic gates.

  Important universal sets include $\{e^{i\frac{\theta}{2}\vec{\eta}\cdot\vec{\sigma}}, \text{CNOT}\}$ and $\{H, P, \text{CNOT}\}$.

- There are three equivalent approaches to quantum operations:

  1. Subsystem evolution – $\mathcal{E}(\rho) = \text{Tr}_o\left[U(\rho \otimes \rho_a)U^\dagger(I_2 \otimes P_o)\right]$

  2. Operator sum representation – $\mathcal{E}(\rho) = \sum_k A_k \rho A_k^\dagger$.

  3. Axiomatic – $\mathcal{E}$ is (i) linear (ii) trace non-increasing and (iii) completely positive

- Quantum process tomography is a procedure to characterize an unknown quantum operation experimentally.

- Common quantum channels include the phase damping channel, the Pauli channel, and the amplitude damping channel. Unital processes satisfy $\mathcal{E}(I) = I$.

# Part II

# Quantum Error Correction

# Chapter 3

# Theory of quantum error correction

## 3.1 Introduction

Quantum information processing is often described as a series of perfect unitary operations and measurements on some physical system. Imperfections in the operations and interactions with the environment, resulting in gate and storage errors in the system, are inevitable in both classical and quantum information processing. Accumulation of errors will be detrimental in any large scale information processing.

The problem of noise in classical computation and communication is resolved by several remarkable achievements: digitization of data, invention of reliable devices and finally prudent use of redundancy (error correcting codes). At a first glance, generalization to the fragile quantum information is not just technically difficult but fundamentally impossible: It is impossible to clone arbitrary quantum information [112] perfectly. Moreover, quantum states and operators form continuous spaces, and it is impossible to detect infinitesimal errors. The much celebrated discovery of quantum error correcting codes [105, 108] came as a real surprise. The subsequent extension to achieve reliable computation with noisy components set a firm theoretical foundation for quantum information processing [11, 48, 68, 74, 57, 96, 104].

The goal of this chapter is to cover the main ideas in noisy quantum coding, and to develop the language and background useful for the rest of the Dissertation. The development mentioned in this chapter also provides a context in which the original results in this Dissertation may fit. Original contributions from this Dissertation are presented from next chapter onwards.

Elements in classical coding (taken from [20, 83]) will be reviewed in Section 3.2. The theory of quantum error correction and various code constructions is reviewed in Section 3.3. Fault-tolerant quantum computation and the threshold theorem is reviewed in Section 3.4.

## 3.2  Elements of classical Error Correction

### 3.2.1  Fundamental concepts

When information is stored or sent through an unreliable channel, it is possible to represent the information with *redundancy* to improve the probability of recovery. The set of all possible encoded messages (codewords) form an "error correcting code".

We illustrate the basic concepts with a simple example. We consider the *binary symmetric channel*, in which a bit $b$ is flipped to $\bar{b}$ with probability $p$. This channel is symmetric with respect to $b = 0$ and 1.

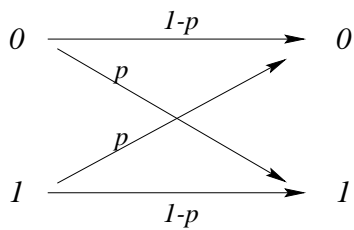

Figure 3.1: The binary symmetric channel

Errors in different uses of the channel are *independent*. Suppose $p$ is small, so that multiple errors are unlikely. One can send $b$ three times and majority-decode to obtain the output. Then, the output is correct unless at least two errors occur, which happens with probability $p' = 3p^2(1-p) + p^3$, which is less than $p$ (an improvement) if $p < 0.5$. This scheme is called the 3-bit repetition code. It illustrates some general concepts in coding:

1. First, one must assume certain noise characteristics of the channel. For example, the binary symmetric channel with independent errors is used above.

2. Given a channel, codewords are chosen from a space *larger* than required for single use, so that the *likely* errors take the original codewords to disjoint sets of received messages to ensure correct decoding. This is best illustrated for the 3-bit repetition code pictorially:

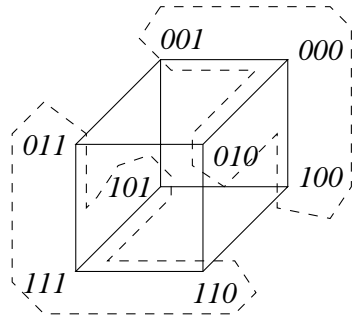

Figure 3.2: The geometry of the 3-bit repetition code. The likely errors take the codewords 000 and 111 to disjoint sets of messages defined by the dashed lines.

3. Coding reduces the error probability at the expense of *more* uses of the channel. If $M$ codewords are encoded in $n$ bits, the *rate* of the code is $(\log_2 M)/n$. For example, the 3-bit repetition code has rate 1/3.

Most of the time, we focus on independent errors with small probability $p$. Coding is designed to handle up to a certain number of errors. For example, the 3-bit repetition code is "1-error correcting". Similarly, a $t$-error correcting code can correct $t$ errors and it takes at least $t + 1$ errors to cause an overall failure. The overall probability of error is improved from $p$ to $\mathcal{O}(p^{t+1})$. Coding entails a trade off between the rate and the reliability of the transmission or storage process, which is captured in a few parameters defined in the next section.

## 3.2.2 The geometry of error correction

Let the symbol set be a field $F$, $|F| = q$, and let messages of length $n$ be vectors in $F^n$. We consider an arbitrary field for most of the time, and restrict to the binary field occasionally to simplify notations.

> *Definition 4*: The Hamming distance between two vectors $a$ and $b$, $d_H(a, b)$, is
> the number of coordinates in which they differ.

Note that the Hamming distance is a metric. We call the set $\{b : d_H(a, b) \leq r\}$ the Hamming
sphere of radius $r$ centered at $a$.

> *Definition 5*: The distance of a code $\mathcal{C}$ is the minimum Hamming distance
> between any two codewords, $d = \min\{d_H(a, b) : a \neq b, a, b \in \mathcal{C}\}$.

Suppose no more than $r$ errors occur. Each codeword is taken to a point within its own
Hamming sphere of radius $r$. These Hamming spheres do not overlap if $2r < d$, therefore
the codeword can be decoded correctly, leading to Theorem 2.

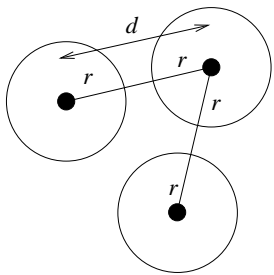

Figure 3.3: Hamming spheres of radius $r$ for $2r < d$.

> *Theorem 2*: A distance $d$ code is $t$-error correcting where $t = \lfloor \frac{d-1}{2} \rfloor$.

> *Definition 6*: An $(n, M, d)$ code is a set of M vectors (codewords) in $F^n$ with
> distance $d$. $n$ is called the *block size* of the code.

Viewing an $(n, M, d)$ code as a set of points in $F^n$, the problem of coding is to pack as many
points as possible while maintaining a certain distance between the points.

The *Hamming Bound* is an important bound relating the distance and the rate of a
code. For a message of length $n$, a Hamming sphere of radius $r$ has $V(r) = \sum_{i=0}^{r} \binom{n}{i}(q-1)^i$
elements. For a $t$-error correcting code, Hamming spheres of radius $t$ around the codewords
do not overlap. Therefore,

*Lemma 1*: (Hamming bound) Let $\mathcal{C}$ be an $(n, M, d)$ code, then

$$n - \log_q M \geq \log_q V(\lfloor \frac{d-1}{2} \rfloor) \qquad (3.1)$$

Codes which saturate the Hamming bound are called *perfect*.

Error correcting codes can be used for error detection. The decoder will check if the received message is a valid codeword. For an $(n, M, d)$ code, it takes $d$ errors to avoid detection. Hence, an $(n, M, d)$ code can detect $d - 1$ errors. In general, an $(n, M, d)$ code can correct $t'$ errors and detect up to $t + t'$ errors for $2t + t' < d$.

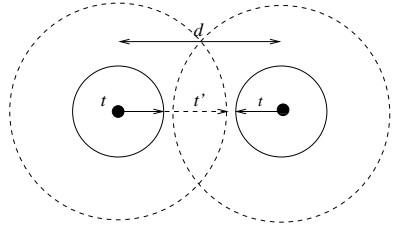

We will describe a special class of codes, the linear codes, with many simplifying properties. There is a beautiful relation between classical linear codes and a class of quantum codes.

### 3.2.3 Classical linear codes

Any error-correcting code $\mathcal{C}$ is a *subset* of $F^n$, the $n$-dimensional vector space over $F$. $\mathcal{C}$ is called *linear* if it is a *subspace* of $F^n$. In other words,

1. $\mathcal{C} \neq \phi$ (the empty set) $\qquad (3.2)$

2. $a, b \in \mathcal{C}, \quad \alpha, \beta \in F \quad \Rightarrow \quad \alpha\, a + \beta\, b \in \mathcal{C} \qquad (3.3)$

If $\dim(\mathcal{C}) = k$, $\mathcal{C}$ is an $(n, q^k, d)$ code, or an $[n, k, d]$ code ($[\cdot, \cdot, \cdot]$ is used for linear codes only). When $q = 2$, $k$ bits are encoded in $n$ bits.

*Definition 7*: The weight of a vector $a$ is the number of non-zero components. It is denoted by $w_H(a)$ and it equals $d_H(a, 0)$ (0 denotes the zero vector).

*Theorem 3*: The distance of a linear code $\mathcal{C}$ is equal to its minimum weight, the minimum of the weights of its non-zero codewords.

A linear error correcting code can be defined by its *generator matrix* or *parity check matrix*. Let $\mathcal{C}$ be an $[n, k, d]$ code. A $k \times n$ matrix $G$ is a *generator matrix* of $\mathcal{C}$ if the rows of $G$ form a basis for $\mathcal{C}$ (codewords are taken as row vectors). An original message $u$ is encoded as $v$ where $v = uG$. An $(n - k) \times n$ matrix $H$ is a *parity check matrix* for $\mathcal{C}$ iff, for any $c \in \mathcal{C}$, $cH^T = 0$. The parity check matrix represents the constraints that define the valid codewords. Alternatively, the rows of $H$ form a basis for the orthogonal complement of $\mathcal{C}$. Note that $GH^T = 0$.

The dual code of $\mathcal{C}$, denoted by $\mathcal{C}^\perp$, is the orthogonal complement of $\mathcal{C}$ in $F^n$. $\mathcal{C}^\perp$ is an $[n, n - k, d^\perp]$ code with generator matrix $H^T$ and parity check matrix $G^T$. In general, there is no simple relation between $d$ and $d^\perp$. Counter-intuitively, a *binary* vector is self-orthogonal if it has even weight, and it is possible for $\mathcal{C}$ and $\mathcal{C}^\perp$ to intersect. If a code is self-dual, $d = d^\perp$. If a code contains its dual, $d^\perp \geq d$.

The Hamming Bound for a linear $[n, k, d]$ code is given by

$$n - k \geq \log_q V(\lfloor \frac{d - 1}{2} \rfloor) \tag{3.4}$$

Another useful concept is that of a "syndrome". Let $\mathcal{C}$ be an $[n, k, d]$ code with parity check matrix $H$. Let $v$ be the encoded message sent and $r$ be the possibly corrupted message received. The syndrome of $r$ is defined to be $s = rH^T$ (an $(n - k)$-vector). If $e$ is the error, $r = v \oplus e$ and

$$s = rH^T = (vH^T) \oplus (eH^T) = eH^T \tag{3.5}$$

Hence, the syndrome only depends on $e$ but *not* on $v$. There is a 1-1 correspondence between errors of weight $\leq (d - 1)/2$ and syndromes, providing a means for *maximum likelihood decoding*.

A coset of a code $\mathcal{C}$ is the set of elements $w \oplus \mathcal{C}$ for $w \in F^n$. An error $e$ takes the original codeword space to the coset $e \oplus \mathcal{C}$. Two cosets $w_1 \oplus \mathcal{C}$ and $w_2 \oplus \mathcal{C}$ are equal iff $w_1, w_2$ are in the same coset. The cosets form a partition of $F^n$. In an $[n, k, d]$ code, there is a 1-1 correspondence between syndromes and cosets. The element of minimum weight in each coset (a unique element of weight $\leq (d-1)/2$) corresponds to the most likely error occurred when the coset is identified. This is best captured by the *standard array* of $\mathcal{C}$, which is a $q^{n-k} \times q^k$ matrix having the following format:

| | $HS(0)$ | $HS(c_1)$ | $HS(c_2)$ | $\cdots$ | $HS(c_{q^k-1})$ |
|---|---|---|---|---|---|
| $\mathcal{C}$ | $0$ | $c_1$ | $c_2$ | $\cdots$ | $c_{q^k-1}$ |
| $e_1 \oplus \mathcal{C}$ | $e_1$ | $e_1 \oplus c_1$ | $e_1 \oplus c_2$ | $\cdots$ | $e_1 \oplus c_{q^k-1}$ |
| $e_2 \oplus \mathcal{C}$ | $e_2$ | $e_2 \oplus c_1$ | $e_2 \oplus c_2$ | $\cdots$ | $e_2 \oplus c_{q^k-1}$ |
| $\cdot$ | $\cdot$ | $\cdot$ | $\cdot$ | $\cdots$ | $\cdot$ |
| $e_l \oplus \mathcal{C}$ | $e_l$ | $e_l \oplus c_1$ | $e_l \oplus c_2$ | $\cdots$ | $e_l \oplus c_{q^k-1}$ |

Table 3.1: The standard array for an $[n, k, d]$ code. $l = q^{n-k} - 1$ in the above.

In the standard array, the rows are the cosets. Each error $e_i$ acts like a mapping from $\mathcal{C}$ to $e_i \oplus \mathcal{C}$, and the $i$-th row is the image of $\mathcal{C}$ under $e_i$. The $j$-th column is like the Hamming sphere around the codeword $c_j$ (here the elements in a column may not form an exact sphere). The first column in each row is the most likely error conditioned on the corresponding syndrome. When $r$ is received, maximum likelihood decoding can be done by reading out the column index. Alternatively, the syndrome can be obtained to find the row index of $r$ and the most likely error $e$. The message sent, $v$, can be recovered by adding $e$ to $r$, *without* knowing $v$. For concreteness, the standard array and the correspondence to the sydrome for a $[5, 2, 3]$ binary code is as follows:

The interpretation of the standard array and the second decoding method that requires no knowledge of $v$ will become important in quantum coding, as discussed next.

## 3.3 Quantum error correction

As discussed earlier in this chapter, generalizing classical error correcting codes to the quantum case appears impossible because of the no-cloning theorem, the impossibility to measure and verify quantum states without disturbing them, and the difficulty in handling a continuum of possible errors. In this section, we will describe the ingenious methods to adapt classical coding theories to construct quantum codes, despite all the apparent difficulties. We assume perfect logic operations, or equivalently the codes are designed to protect against storage or transmission errors but not operational errors. The latter will be discussed in Section 3.4.

| message | 00 | 01 | 10 | 11 | syndrome |
|---------|-------|-------|-------|-------|----------|
| code    | 00000 | 01110 | 10011 | 11101 | 000      |
| coset   | 10000 | 11110 | 00011 | 01101 | 011      |
| coset   | 01000 | 00110 | 11011 | 10101 | 110      |
| coset   | 00100 | 01010 | 10111 | 11001 | 100      |
| coset   | 00010 | 01100 | 10001 | 11111 | 010      |
| coset   | 00001 | 01111 | 10010 | 11100 | 001      |
| coset   | 11000 | 10110 | 01011 | 00101 | 101      |
| coset   | 10100 | 11010 | 00111 | 01001 | 111      |

Table 3.2: The standard array for a $[5, 2, 3]$ code, showing the 1-1 correspondence between the most likely errors, the corresponding cosets and syndromes.

Out of the many beautiful results in quantum error correction, this section will describe the first quantum code due to Shor [105], an elegant necessary and sufficient criteria for quantum error correction due to Nielsen *et al* [89], the CSS codes derived from classical linear codes due to Calderbank, Shor, and Steane [25, 108], and the stabilizer formalism of quantum error correction due to Gottesman [55, 56].

### 3.3.1 The Shor code

How is quantum error correction possible? The answers are best illustrated by Shor's 9-qubit quantum code [105], which is the quantum analog of the classical 3-bit repetition code.

Suppose we encode the states as follows:

$$|0\rangle \rightarrow |0_L\rangle = (|000\rangle + |111\rangle)(|000\rangle + |111\rangle)(|000\rangle + |111\rangle)$$

$$|1\rangle \rightarrow |1_L\rangle = (|000\rangle - |111\rangle)(|000\rangle - |111\rangle)(|000\rangle - |111\rangle) \tag{3.6}$$

where the overall normalization is omitted. The subscript $L$ denotes the encoded or *logical*

states (see Section 2.1). An original superposition $a|0\rangle + b|1\rangle$ is encoded as $a|0_L\rangle + b|1_L\rangle$, delocalizing the encoded information among the 9 qubits.

This quantum code will correct for any one-qubit error for the following reasons. We first show that it is capable of correcting $X$ errors and $Z$ errors on any one qubit. Without loss of generality, suppose the first qubit has an $X$ error. This error can be revealed unambiguously when comparing the pairs of qubits $(1,2)$ and $(2,3)$. In fact, comparing the $i$-th and $j$-th qubits for $(i,j) \in \{(1,2),(2,3),(4,5),(5,6),(7,8),(8,9)\}$ can reveal any single $X$ error. Similarly, a single $Z$ error can be found by comparing the signs in the first and second blocks, and in the second and third blocks. The Shor code can also recover a combined $X$ and $Z$ error on the same qubit, by performing both procedures, which are *independent*. Finally, a qubit suffering an arbitrary error is projected to having $I$, $X$, $Z$, or $XZ$ errors during the "syndrome measurements" of the $X$ and $Z$ errors, and can be corrected.

Many remarkable concepts are illustrated in this code. We provide an informal discussion in the rest of this section, followed by an abstract general framework for quantum error correction in Section 3.3.2.

- The syndrome measurements do not require measuring the qubits. For example, see Fig. 3.4.

  Even more important is that, the syndrome can be found *without* gaining any information on the encoded states. The unitary errors ($X$ and $Z$) can be inverted without knowing the encoded information. This is analogous to the classical decoding method by projecting onto a coset in the standard array and inverting the most likely (minimum weight) error.

- Redundancy is used to embed the codeword space in a larger ambient space, so that correctable errors take the code space to orthogonal subspaces (the quantum analog of the Hamming spheres) which do not overlap. At any time, there is at most one copy of the state.

- $Z$ errors acting on different qubits within the same block have identical effects. It is neither possible nor necessary to distinguish between these errors (interpreting the error as any one of the possibilities will do). A quantum code is *nondegenerate* if all the correctable errors can be identified unambiguously; otherwise, it is degenerate. The Shor code is a degenerate code. A more rigorous definition of degeneracy is given in the next section.

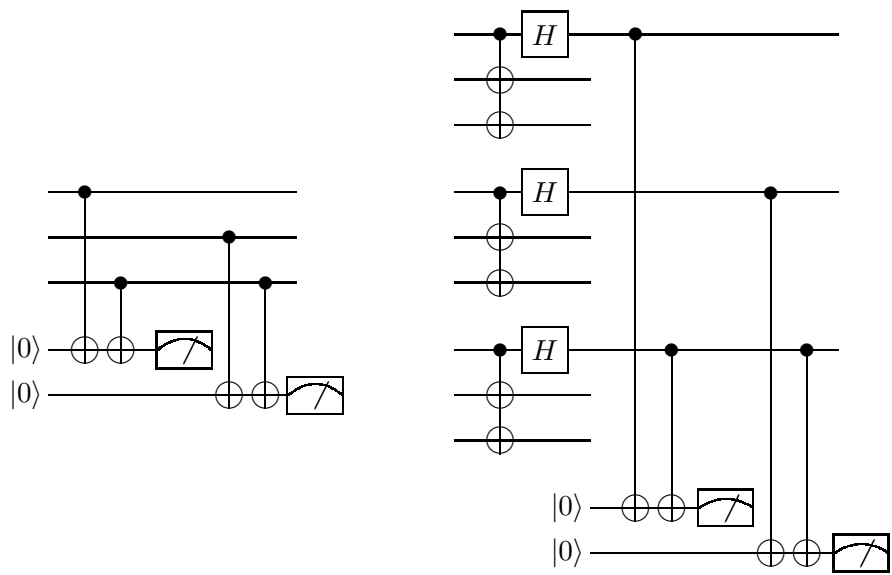

Figure 3.4: Syndrome measurement for the Shor code. The left circuit determines if an $X$ error has occurred to any qubit in a block. The right circuit determines if an $Z$ error has occurred in one of the three blocks.

- The Shor code can correct for an *arbitrary* error that occurs to a single qubit. This is seemingly impossible, since the syndrome space is finite but the possible number of errors are infinite. The crucial observation is that, the continuum of errors can be *discretized*. To see this, suppose an arbitrary error $E$ has occurred to a qubit. Since $E$ can always be written as $c_i I + c_x X + c_y Y + c_z Z$, the erroneous state is the superposition $E|\psi_{in}\rangle = c_i|\psi_{in}\rangle + c_x X|\psi_{in}\rangle + c_y Y|\psi_{in}\rangle + c_z Z|\psi_{in}\rangle$. The syndrome measurements, which identify the $I$, $X$, $Y$ and $Z$ errors, *project* the erroneous state onto one of the four terms which are correctable. Non-unitary errors are correctable for the same reason. Finally, a general noise process can still be discretized using the operator sum representation introduced in Section 2.3. A more rigorous treatment will be given in Section 3.3.2.

- The Shor code is a *concatenated* code – it concatenates two 3-bit repetition codes, correcting the $X$ and the $Z$ errors at the lower and higher level codes respectively.

Having witnessed how the many difficulties in quantum coding can be circumvent in the Shor code, we proceed to discuss the general theory of quantum error correction.

### 3.3.2 General theory of quantum error correction

In this section, we present algebraic conditions for a subspace $\mathcal{C} \le \mathcal{H}$ to be a quantum code that corrects for certain errors acting on $\mathcal{H}$. These conditions [49, 72, 17, 89], often called the "criteria for quantum error correction", are of tremendous value in understanding and constructing quantum codes. They also allow notions in quantum error correction to be made rigorous. We take the approach using *reversible operations* [89].

*Definition 8*: Suppose $\mathcal{E}$ is a quantum operation. We say that $\mathcal{E}$ is reversible on a space $\mathcal{S}$ if there exists a *complete* [1] quantum operation $\mathcal{R}$ such that, $\forall \rho \in \mathcal{S}$

$$\mathcal{R} \circ \mathcal{E}(\rho) = \mathrm{Tr}(\mathcal{E}(\rho))\, \rho\,. \tag{3.7}$$

We require $\mathcal{R}$ to be complete because the reversal should be deterministic. It is important to consider possibly incomplete $\mathcal{E}$. When Eq. (3.7) is satisfied, $\mathrm{Tr}(\mathcal{E}(\rho)) = \mu$ cannot depend on $\rho$ because the left hand side of Eq. (3.7) is linear.

We may connect the notion of reversibility with quantum error correction. Specifically, consider a *complete* quantum operation $\mathcal{E}$ acting on $\mathcal{H}$. For example $\mathcal{E}$ can be independent error processes acting on individual qubits. For a general process $\mathcal{E}$, a code $\mathcal{C}$ may not exist on which $\mathcal{E}$ is exactly reversible. We interpret $\mathcal{E}$ as a stochastic process, with operator sum representation $\mathcal{E}(\rho) = \sum_{n \in \mathcal{K}} A_n \rho A_n^\dagger$. Here, $\mathcal{K}$ is the index set of $\mathcal{A}$, the set of all operation elements $A_n$ appearing in the sum. These $A_n$ have natural interpretation as "errors", occurring with probabilities $\mathrm{Tr}(A_n \rho A_n^\dagger)$. Intuitively, a quantum code $\mathcal{C}$ should correct for the *likely errors*.

To make these notions rigorous, we denote by $\mathcal{A}_{re} \subset \mathcal{A}$ the *reversible subset on $\mathcal{C}$*, and let $\mathcal{K}_{re} = \{n|\ A_n \in \mathcal{A}_{re}\}$ be the index set of $\mathcal{A}_{re}$. The reversible subset is defined so that

---

[1] Complete quantum operations are trace-preserving.

the possibly incomplete process $\mathcal{E}'(\rho) = \sum_{n \in \mathcal{K}_{re}} A_n \rho A_n^\dagger$ is reversible on $\mathcal{C}$ in the sense given by Def. 8. We say that the code $\mathcal{C}$ can correct for $\mathcal{E}'$ or the "errors" $A_n \in \mathcal{A}_{re}$. The error correction criteria, expressed as algebraic conditions for reversibility, are as follow:

*Theorem 4*: $\mathcal{A}_{re}$ is a reversible subset iff

$$P_{\mathcal{C}} A_m^\dagger A_n P_{\mathcal{C}} = g_{mn} P_{\mathcal{C}} \qquad \forall m, n \in \mathcal{K}_{re} \, , \tag{3.8}$$

where $P_{\mathcal{C}}$ is the projector onto $\mathcal{C}$, and $g_{mn}$ are the entries of a positive matrix.

**Remark:** Eq. (3.8) is often written as

$$\langle c_i | A_m^\dagger A_n | c_j \rangle = \delta_{ij} g_{mn} \quad \forall m, n \in \mathcal{K}_{re} \ \forall i, j \tag{3.9}$$

where $|c_i\rangle$ are logical states for the code.

**Proof: [Necessity]** Suppose $\mathcal{E}'$ is reversible on $\mathcal{C}$. Then, there is a complete operation $\mathcal{R}(\cdot) = \sum_l R_l \cdot R_l^\dagger$ and a constant $\mu$ such that

$$\mathcal{R} \circ \mathcal{E}'(\rho) = \mu\rho \qquad \forall \rho \in \mathcal{C} \, . \tag{3.10}$$

For all $\rho \in \mathcal{H}$, $P_{\mathcal{C}} \rho P_{\mathcal{C}} \in \mathcal{C}$, Therefore,

$$\mathcal{R} \circ \mathcal{E}'(P_{\mathcal{C}} \rho P_{\mathcal{C}}) = \sum_l \sum_{n \in \mathcal{A}_{re}} R_l A_n P_{\mathcal{C}} \rho P_{\mathcal{C}} A_n^\dagger R_l^\dagger = \mu P_{\mathcal{C}} \rho P_{\mathcal{C}} \tag{3.11}$$

The last equality is between quantum operations, therefore, $R_l A_n P_{\mathcal{C}}$ and $\sqrt{\mu} P_{\mathcal{C}}$ are related by Theorem 1 in Section 2.3:

$$R_l A_n P_{\mathcal{C}} = v_{ln} \sqrt{\mu} P_{\mathcal{C}} \tag{3.12}$$

where $\sum_{ln} |v_{ln}|^2 = 1$. Using $\sum_l R_l^\dagger R_l = I$ (since $\mathcal{R}$ is complete) and Eq. (3.12), we can write

$$P_{\mathcal{C}} A_m^\dagger A_n P_{\mathcal{C}} = \sum_l P_{\mathcal{C}} A_m^\dagger R_l^\dagger R_l A_n P_{\mathcal{C}} = \mu \sum_l P_{\mathcal{C}} v_{lm}^* v_{ln} P_{\mathcal{C}} = g_{mn} P_{\mathcal{C}} \tag{3.13}$$

where $g_{mn} = \mu \sum_l v_{lm}^* v_{ln}$ are entries of the positive matrix $g = \mu v^\dagger v$.

[**Sufficiency**] Suppose Eq. (3.8) is true with $g$ being a positive matrix. First of all, we can always change to another operator sum representation for $\mathcal{E}'$ to make the corresponding $g$ diagonal: Since $g$ is positive, there is some unitary matrix $u$, and positive diagonal matrix $p$ such that $u^\dagger g u = p$. Therefore, $\sum_{mn} u^\dagger_{km} g_{mn} u_{nl} = p_k \delta_{kl}$ where $p_k = p_{kk}$. Together with Eq. (3.8), we obtain:

$$\sum_{mn} u^\dagger_{km}(P_C A^\dagger_m A_n P_C)u_{nl} = p_k \delta_{kl} P_C \qquad (3.14)$$

$$(P_C \tilde{A}^\dagger_k \tilde{A}_l P_C) = p_k \delta_{kl} P_C \qquad (3.15)$$

where $\tilde{A}_l = \sum_n A_n u_{nl}$. Moreover, by Theorem 1, $\{\tilde{A}_l\}$ and $\{A_n\}$ generate the same quantum operation $\mathcal{E}'$. The desired new operator sum representation $\mathcal{E}'(\rho) = \sum_l \tilde{A}_l \rho \tilde{A}^\dagger_l$ is canonical.

We now construct a recovery operation $\mathcal{R}$ to demonstrate reversibility of $\mathcal{E}'$ on $\mathcal{C}$. Using the diagonal elements of Eq. (3.15), the $\tilde{A}_n$ have polar decompositions

$$\tilde{A}_n P_C = \sqrt{p_n} U_n P_C \qquad \forall n \in \mathcal{K}_{re}, \qquad (3.16)$$

where the $U_n$ are unitary. Using the off-diagonal elements of Eq. (3.15)

$$P_C U^\dagger_n U_m P_C = \delta_{nm} P_C . \qquad (3.17)$$

Equations (3.16) and (3.17) correspond to the well known *non-deformation* and *orthogonality* conditions for quantum error correction. Equation (3.16) means that $\mathcal{C}$ should be chosen so that the correctable errors act like unitary processes on $\mathcal{C}$. Equation (3.17) says that the correctable errors should map $\mathcal{C}$ to orthogonal subspaces. Using the projectors $U_m P_C U^\dagger_m$ onto these subspaces, the corresponding errors $A_m$ can be identified, and be reversed by $U^\dagger_m$.

Mathematically, the recovery operation $\mathcal{R}$ is given by:

$$\mathcal{R}(\rho) = \sum_{k \in \mathcal{K}_{re}} R_k \rho R^\dagger_k + P_E \rho P_E , \qquad (3.18)$$

where $R_k = P_C U_k^\dagger$ is the appropriate reversal process for each error $\tilde{A}_k$ ($k \in \mathcal{K}_{re}$), and $P_E \equiv I - \sum_{k \in \mathcal{K}_{re}} U_k P_C U_k^\dagger$ is required for completeness. We check that, for any $\rho \in \mathcal{C}$

$$
\begin{align}
\mathcal{E}'(\rho) &= \sum_{l \in \mathcal{K}_{re}} \tilde{A}_l P_C \rho P_C \tilde{A}_l^\dagger \tag{3.19} \\
&= \sum_{l \in \mathcal{K}_{re}} p_l U_l P_C \rho P_C U_l^\dagger \tag{3.20} \\
\mathcal{R} \circ \mathcal{E}'(\rho) &= \sum_{kl \in \mathcal{K}_{re}} p_l P_C U_k^\dagger U_l P_C \rho P_C U_l^\dagger U_k P_C \tag{3.21} \\
&= \sum_{l \in \mathcal{K}_{re}} p_l \rho \tag{3.22} \\
&= \mathrm{Tr}(\mathcal{E}'(\rho))\, \rho \tag{3.23}
\end{align}
$$

where we have used Eq. (3.17) to derive Eq. (3.22) and Eq. (3.20) to obtain $\mathrm{Tr}(\mathcal{E}'(\rho)) = \sum_{l \in \mathcal{K}_{re}} p_l$ . This concludes the proof of sufficiency and the theorem itself.

The code is non-degenerate iff $\mathrm{Rank}(g) = |\mathcal{K}_{re}|$. Therefore, a code is degenerate iff $p_n = 0$ for some $n$. The present definition contrasts with the more commonly used definition that the code is non-degenerate iff $g$ is diagonal, which is not satisfactory since it is not invariant under different choices of the operator sum representation. The present definition captures the idea that a code is degenerate if $P_C A_n$ are linearly dependent, and that not all errors $A_n$ are relevant on $\mathcal{C}$.

Note from the proof of the theorem that, if a code $\mathcal{C}$ corrects for the errors $A_n \in A_{re}$, it corrects for any noise process with operation elements which are linear combination of the $A_n$'s. For this reason, a quantum code that corrects for $X$, $Y$, and $Z$ restricted to $t$ qubits can correct for any error on $t$ qubits.

We now define the quantum analog of the probability of successful recovery. We need to quantify how close is one state from another. For a pure input state $|\psi\rangle$ and arbitrary output state $\rho$, we define the *overlap fidelity* between $|\psi\rangle$ and $\rho$ to be

$$
F(|\psi\rangle, \rho) = \mathrm{Tr}(|\psi\rangle\langle\psi|\rho) = \langle\psi|\rho|\psi\rangle \tag{3.24}
$$

which is the probability of projecting $\rho$ onto the span of $|\psi\rangle$. The fidelity for a channel $\mathcal{E}$ quantifies how well an arbitrary input state is being preserved. The *minimum overlap fidelity* for $\mathcal{E}$ is defined as

$$\mathcal{F}_{\mathcal{E}} = \min_{|\psi\rangle} F(|\psi\rangle, \mathcal{E}(|\psi\rangle\langle\psi|)) \qquad (3.25)$$

which is the worse case input-output overlap fidelity.

For a given noise process $\mathcal{E}$ and a code $\mathcal{C}$, we can quantify the effectiveness of $\mathcal{C}$ using the improved fidelity $\mathcal{F}_{\mathcal{R}\circ\mathcal{E}}$. [2] $\mathcal{F}_{\mathcal{R}\circ\mathcal{E}}$ is lower bounded by $P^{det} \equiv \sum_{n\in\mathcal{K}_{re}} p_n$, which is the total detection probability for the reversible subset. This is because

$$\mathcal{F}_{\mathcal{R}\circ\mathcal{E}} = \min_{|\psi\rangle} \left[ F(|\psi\rangle, \mathcal{R}\circ\mathcal{E}'(|\psi\rangle\langle\psi|)) + F(|\psi\rangle, \mathcal{R}\circ\mathcal{E}''(|\psi\rangle\langle\psi|)) \right] \qquad (3.26)$$

$$\geq \min_{|\psi\rangle} F(|\psi\rangle, \mathcal{R}\circ\mathcal{E}'(|\psi\rangle\langle\psi|)) \qquad (3.27)$$

$$= P^{det} \qquad (3.28)$$

where $\mathcal{E}''(\rho) = \sum_{\mathcal{K}-\mathcal{K}_{re}} A_n \rho A_n$ denotes the complementary process which cannot be reversed. To achieve a desired fidelity $\mathcal{F}_d$, we have to include in $\mathcal{A}_{re}$ a sufficient number of highly probable errors $A_n$ so that $\mathcal{F}_{\mathcal{R}\circ\mathcal{E}} \geq P^{det} \geq \mathcal{F}_d$. Thus, $\mathcal{A}_{re}$ is also a high probability subset.

Having covered the general theory of quantum error correction, we turn to the issue of finding codes.

### 3.3.3 Stabilizer codes

The general theory of quantum error correction described in the previous section has provided criteria and a recipe for error recovery. However, it does not provide construction methods. Just like classical coding, code construction relies largely on the ingenuity of their inventors.

We have seen that the Shor code is related to the classical repetition code. The classical and quantum codes also use redundancy in a related manner. A useful connection to make is to adapt known classical codes to construct quantum codes. Steane [108], and independently Shor and Calderbank [25] developed a class of codes (named CSS codes after their inventors) derived from some special linear classical codes over GF(2). [3] Gottesman [55, 56] subsequently developed a group theoretical description of quantum codes which is very useful in constructing and understanding quantum codes. Similar but more general results by Calderbank *et al* [23, 24] connected quantum coding to orthogonal geometry and classical codes over GF(4) and led to the discovery of many codes with important implications. We review the *stabilizer code* formalism due to Gottesman which is more relevant to this Dissertation. It also provides simple explanations of the CSS codes and fault tolerant quantum computation.

---

[2] The enlarged Hilbert space $\mathcal{H}$ is different from the unencoded space. The actual improvement in fidelity should be the difference $\mathcal{F}_{\mathcal{R}\circ\mathcal{E}} - \mathcal{F}_{\mathcal{E}_o}$ where $\mathcal{E}_o$ acts on the unencoded space.

**Stabilizer codes**

Consider the $n$-qubit Hilbert space $\mathcal{H}$ of $2^n$ dimensions, and the "smaller" Pauli group $\mathcal{G}_n$ acting on $\mathcal{H}$ generated by $X^{(i)}$ and $Z^{(i)}$ (see Section 2.2). In this definition, $\sigma_y \notin \mathcal{G}_n$, though $Y = XZ = -i\sigma_y \in \mathcal{G}_n$. Each element $M \in \mathcal{G}_n$ has eigenvalues $\pm 1$ or $\pm i$, and $M^2 = \pm I$ ($M^\dagger = \pm M$). Any two elements in $\mathcal{G}_n$ either commute or anticommute. Let $S$ be an *abelian* subgroup of $\mathcal{G}_n$, with $n - k$ *hermitian* generators. These commuting generators define $2^{n-k}$ simultaneous eigenspaces. The eigenspace $\mathcal{C}$ corresponding to the eigenvalue $+1$ for all generators of $S$ is called the stabilizer code with stabilizer $S$. [4] Any state $|\psi\rangle$ in $\mathcal{C}$ is *stabilized* by any element $M$ in $S$, i.e. $M|\psi\rangle = |\psi\rangle$. $\mathcal{C}$ has $2^k$ dimensions and therefore encodes $k$ qubits.

Let $\{|i_L\rangle\}$ be a basis for $\mathcal{C}$ and $\{E_a\}$ be the set of errors to be corrected. For now, we concentrate on $t$-error correcting codes, therefore $\{E_a\} \subset \mathcal{G}_n$ contains elements in $\mathcal{G}_n$ with

---

[3] Specifically, consider a linear classical code $\mathcal{C}_1$ with subcode $\mathcal{C}_2$ and consider the cosets of $\mathcal{C}_2$ in $\mathcal{C}_1$. We fix some computation basis, and call its Hadamard transform the conjugate basis. We define a quantum code with logical states as follows. Each logical state is an equal superposition of codewords in a coset of $\mathcal{C}_2$ in $\mathcal{C}_1$ in the computation basis. It can be proved that each logical state is some superposition of codewords in $\mathcal{C}_2^\perp$ in the conjugate basis. Hence the quantum code can independently correct for $X$ errors using $\mathcal{C}_1$ in the computation basis and $X$ errors in the conjugate basis due to $\mathcal{C}_2^\perp$. But the $X$ errors in the conjugate basis are just $Z$ errors in the original computation basis. When $\mathcal{C}_1$ and $\mathcal{C}_2^\perp$ are both good codes, a certain number of $X$ and $Z$ errors can be corrected, implying that arbitrary errors in certain number of qubits can be corrected.

[4] The existence of such an eigenspace is guaranteed by the commutivity and hermiticity of its generators. Note that every element $M \in S$ is hermitian with eigenvalues $\pm 1$, and $M^2 = I$. In doing so, we exclude elements of odd number of $Y$'s from $S$.

Hamming weight no greater than $t$.

*Theorem 5*: $\mathcal{C}$ is a code correcting $\{E_a\}$ if $\forall a, b$, $E_a^\dagger E_b$ either (i) anticommutes with some $M \in S$ or (ii) $E_a^\dagger E_b \in S$.

**Proof:** If $E_a^\dagger E_b$ anticommutes with some $M \in S$,

$$\langle i_L | E_a^\dagger E_b | j_L \rangle = \langle i_L | E_a^\dagger E_b M | j_L \rangle = -\langle i_L | M E_a^\dagger E_b | j_L \rangle = -\langle i_L | E_a^\dagger E_b | j_L \rangle = 0 . \quad (3.29)$$

If $E_a^\dagger E_b \in S$,

$$\langle i_L | E_a^\dagger E_b | j_L \rangle = \langle i_L | j_L \rangle = \delta_{ij} . \quad (3.30)$$

Moreover, if $E_a^\dagger E_b \in S$, $(E_a^\dagger E_b)^{-1} = E_b^\dagger E_a \in S$, hence Eq. (3.30) holds when $a$ and $b$ are interchanged. Combining Eqs. (3.29) and (3.30), $\langle i_L | E_a^\dagger E_b | j_L \rangle = \delta_{ij} c_{ab}$, where $c_{ab} = 0$ or 1 is independent of $i, j$, and since $c_{aa} = 1$, $c_{ab}$ form a positive matrix. Therefore $\mathcal{C}$ satisfies the error correction criteria given by Eq. (3.9).

A stabilizer code is non-degenerate if, $\forall a \neq b$, $E_a^\dagger E_b$ anticommutes with some $M \in S$. In this case, $c_{ab}$ is diagonal. A degenerate stabilizer code has some correctable errors $E_a$, $E_b$ such that $E_a^\dagger E_b \in S$, meaning that $E_a$, $E_b$ act identically on $\mathcal{C}$. Therefore, for a stabilizer code, correctable errors are either exactly distinguishable or identical on $\mathcal{C}$.

The distance of a code $d$ is the minimum weight of $E \in N(S) - S$. It follows that a distance $d$ code can correct for $t$ errors, with $d \geq 2t + 1$. The notation $[[n, k, d]]$ is used for a quantum code with distance $d$ that encodes $k$ qubits in $n$ qubits. The double brackets distinguish quantum linear codes from classical linear codes.

We revisit the Shor code, which is a $[[9, 1, 3]]$ stabilizer code with 8 generators:

$$
\begin{aligned}
M_1 &= Z \quad Z \quad I \quad I \quad I \quad I \quad I \quad I \quad I \\
M_2 &= Z \quad I \quad Z \quad I \quad I \quad I \quad I \quad I \quad I \\
M_3 &= I \quad I \quad I \quad Z \quad Z \quad I \quad I \quad I \quad I \\
M_4 &= I \quad I \quad I \quad Z \quad I \quad Z \quad I \quad I \quad I \\
M_5 &= I \quad I \quad I \quad I \quad I \quad I \quad Z \quad Z \quad I \\
M_6 &= I \quad I \quad I \quad I \quad I \quad I \quad Z \quad I \quad Z \\
M_7 &= X \quad X \quad X \quad X \quad X \quad X \quad I \quad I \quad I \\
M_8 &= X \quad X \quad X \quad I \quad I \quad I \quad X \quad X \quad X
\end{aligned}
\tag{3.31}
$$

In Eq. (3.31), rows correspond to generators, and columns correspond to qubits. For example, the first row corresponds to the generator $Z \otimes Z \otimes I \otimes I \otimes I \otimes I \otimes I \otimes I \otimes I$. The fact that the Shor code can correct all single qubit errors can be routinely checked on all Pauli operators with weight no greater than 2. The code is degenerate as all the $Z$ generators are of weight 2. It means some $Z$ errors are not distinguishable from another, as can be verified from Eq. (3.6). The syndrome measurments of comparing qubits or the signs of blocks correspond to measuring the eigenvalues of $M_i$.

The generator matrix of a stabilizer code (such as Eq. (3.31), with rows given by the generators) plays a role similar to the parity check matrix in classical coding. In the stabilizer code, one measures the eigenvalue ($\pm 1$) of each generator on the possibly corrupted received state, which is analogous to a parity check. The measured eigenvalues form the "syndrome" in the quantum code. As each distinguishable error takes $\mathcal{C}$ to a different simultaneous eigenspace of the generators of $S$ corresponding to some different eigenvalues, each syndrome corresponds to a unique error in $\{E_a\}$.

Note that the stabilizer formalism gives a very concise description of the code and the effects of the errors, because it describes operators rather than states. This "Heisenberg" point of view will be even more useful when we discuss fault-tolerant quantum operations.

The Shor code is an example of a CSS code. In the stabilizer language, the CSS codes are stabilizer codes such that each generator can be chosen to be a tensor product of $I$ and either $X$ or $Z$, but not both. The $X$ generators correspond to the parity check matrix of the classical code $\mathcal{C}_2^{\perp}$ which handles the $Z$ errors, and the $Z$ generators correspond to the parity check matrix of $\mathcal{C}_1$ which handles the $X$ errors (using the notation defined in the footnote at the beginning of Section 3.3.3).

A particularly interesting CSS code is the $[[7, 1, 3]]$ Steane code [108] with stabilizer:

$$\begin{matrix} I & I & I & Z & Z & Z & Z \\ I & Z & Z & I & I & Z & Z \\ Z & I & Z & I & Z & I & Z \\ I & I & I & X & X & X & X \\ I & X & X & I & I & X & X \\ X & I & X & I & X & I & X \end{matrix} \qquad (3.32)$$

The code is symmetric with respect to interchanging the $X$ and $Z$ generators ($\mathcal{C}_2 = \mathcal{C}_1^{\perp}$, hence $\mathcal{C}_1$ has to contain its dual.) This code has very attractive fault-tolerant features to be discussed in Section 3.4.

The last example of stabilizer code is the smallest possible code which can encode one qubit and correct for a single qubit error [17, 76]. This $[[5, 1, 3]]$ code is cyclic, and it saturates the quantum Hamming bound.

$$\begin{matrix} X & Z & Z & X & I \\ I & X & Z & Z & X \\ X & I & X & Z & Z \\ Z & X & I & X & Z \end{matrix} \qquad (3.33)$$

It is interesting that no CSS code can encode one qubit in five and correct for any single qubit error. [5]

---

[5] This can be shown by elimination. Suppose such a CSS code with 4 generators exists. The only possibility is having two $X$ generators. They cannot both be $I$ in any coordinate. Therefore, one generator is of weight $\geq 3$. Reordering the qubits, the first generator is of the form $XXX \cdots$. Now, any choice of the second generator commutes with some Pauli operator with two $Z$ in the first 3 coordinates, completing the proof.

## 3.4    Fault-tolerant quantum computation and the threshold theorem

The quantum error correcting codes discussed so far can protect quantum information from storage or transmission errors, but not the errors due to the logic gates. Such errors are our main concern in actual computation. In particular, we have the following questions to address:

1. Quantum states and gates form a continuous space. Inaccuracies in the logic gates will accumulate over a long calculation.

2. It is necessary to perform operations without introducing and spreading errors in a catastrophic manner.

3. The encoding and recovery procedures are complex and not error free. A net reduction of errors is not guaranteed by the existence of quantum codes.

4. Quantum codes can only reduce the error probability. We may still need an infinitesimal physical elementary error rate to achieve arbitrarily reliable and long computation.

5. Finally, even if the feasibility issues are resolved, it is not obvious what are the extra resources required to make the computation reliable.

The above issues were first addressed by Shor [104] and the results were substantially extended and improved by many others [48, 11, 96, 74, 68, 57, 21, 59, 115]. The essence of these remarkable results will be summarized in the rest of this section.

The discussion is based on the the circuit model of quantum computation, and assumes that classical computation is perfect and fast, and that measurements can be performed *during* the computation. We assume independent noise processes on different qubits, and the error rates do not grow rapidly with the system size.

### 3.4.1    Fault-tolerant quantum computation

A device that works effectively even when its elementary components are imperfect is said to be *fault-tolerant*. One can identify some basic rules to ensure that the computation is

fault-tolerant:

1. Encode the states – Protect quantum information using quantum error correcting codes

2. Encode the operations – Compute with encoded qubits without decoding them.

3. Use *fault-tolerant operations* – Use operations such that a single error anywhere can only produce at most one error in any other encoded block. In particular, *transversal operations* which only interact a qubit with the corresponding qubit in another code block or ancilla are fault-tolerant.

4. Discretize elementary operations – Use a discrete universal set of gates.

5. Eliminate potential errors by verification and repetition of measurements.

We devote the rest of this section to discuss some of these elements.

### Discretization

The problem with a continuous set of logic operations is resolved by using a *discrete universal set of gates*. The discrete set can be the *Clifford group* with an additional gate such as the Toffoli gate or $\pi/8$ gate (see Section 2.2).

The goal is to find quantum codes and corresponding sets of operations which are universal and fault-tolerant. When constructing encoded gates, we assume that the usual logic operations (for example, one qubit gates and CNOTs) can be performed on the physical qubits. The gate errors will become errors in the encoded states which can be detected and corrected in the usual manner. We focus on fault-tolerant gates on stabilizer codes, and in particular the CSS codes.

### Encoded Pauli group in stabilizer codes

Recall that the stabilizer $S$ is an abelian subgroup of $\mathcal{G}_n$. Operators which do not commute with all $M \in S$ take $\mathcal{C}$ to its orthogonal complement and are detectable errors. However, there may be operations in $\mathcal{G}_n$ that commute with every $M \in S$ but are not in $S$. They are the "dangerous" errors that change the encoded qubits without being detected. However, these are exactly the legitimate *encoded* operations on the stabilizer code because they evolve codewords to codewords. Mathematically, the *centralizer* of $S$ consists of operations which commute with every $M \in S$. For a stabilizer, the *centralizer* is just the *normalizer* of $S$ in $\mathcal{G}$, denoted by $N(S)$, which is the set of all operations which permute the elements of $S$ by conjugation. [6] Since $M \in S$ acts trivially on $\mathcal{C}$, the *encoded* operations can be taken as elements in $N(S)/S$.

Consider an $[[n, k, d]]$ stabilizer code. $S$ has $n - k$ generators (and $2^{n-k}$ elements). The generator can be extended to a maximal *independent* set of $n$ mutually commuting observables (generating $2^n$ commuting observables) by $k$ extra elements $\bar{Z}_1, \bar{Z}_2, \cdots, \bar{Z}_k$ in $N(S)/S$. Then, the simultaneous eigenstates of $\bar{Z}_i$ in $\mathcal{C}$ can be chosen as the encoded logical states. In particular, the encoded $|c_1, c_2, ..., c_k\rangle$ is the $(-1)^{c_i}$ eigenstate of $\bar{Z}_i$. This turns the $\bar{Z}_i$ into the encoded $\sigma_z$ for the $i$-th logical qubit. The remaining elements in $N(S)/S$ will not commute with all of $\bar{Z}_i$, and each $\bar{X}_i$ can chosen from $N(S)/S$ which commutes with all $\bar{Z}_j$ for $j \neq i$ and anticommutes with $\bar{Z}_i$. $\bar{X}_i$ acts as the encoded $\sigma_x$ for the $i$-th logical qubit.

**Encoded Clifford group operations**

**Stabilizer formalism and Heisenberg representation** [56, 58, 57] The stabilizer formalism describes the transformation of the operators instead of the states. In general, if a state $|\psi\rangle$ is initially stabilized by $M$,

$$UMU^{\dagger}U|\psi\rangle = U|\psi\rangle, \tag{3.34}$$

and $U|\psi\rangle$ is stabilized by $UMU^{\dagger}$. In other words, the stabilizer $M$ is transformed to $UMU^{\dagger}$ when the state is transformed by $U$.

We can use Eq. (3.34) to find encoded operations. Suppose we want to find an implementation of the encoded operation $\bar{U}$ in terms of operations on physical qubits. Any operator $W$ such that $W\bar{X}_i W^{\dagger} = \overline{UX_i U^{\dagger}}$ and $W\bar{Z}_i W^{\dagger} = \overline{UZ_i U^{\dagger}}$ is a valid implementation of $\bar{U}$. We will use this method to find the encoded Clifford group operations. We consider the action of one operator on another *by conjugation.* (Recall $O_1$ takes $O_2$ to $O_3$ means $O_1 O_2 O_1^\dagger = O_3$.)

---

[6] Let $A \in \mathcal{G}_n$ and $M \in S$. $AMA^{\dagger} = M \in S$ if $[M, A] = 0$. $AMA^{\dagger} = -M \notin S$ if $\{M, A\} = 0$. Therefore, $[A, M] \in S \; \forall M \in S$ iff $AMA^{\dagger} \in S \; \forall M \in S$, meaning that $A$ permutes the elements of $S$.

Recall from Section 2.2 that the Clifford group, $N(\mathcal{G})$ can be generated by the Hadamard gate $H$, the phase gate $P = \sqrt{Z} = \mathrm{Diag}(1, i)$, and CNOT. $H$ interchanges $X$ and $Z$, $P$ takes $X$ to $iY$ and $Y$ to $iX$, and CNOT$_{12}$ takes $XI$ to $XX$ and $IZ$ to $ZZ$ and preserves $IX$ and $ZI$. The encoded Clifford group elements are defined similarly: $\bar{H}$ interchanges $\bar{X}$ and $\bar{Z}$, $\bar{P}$ takes $\bar{X}$ to $i\bar{Y}$ and $\bar{Y}$ to $i\bar{X}$ and so on.

**Self-dual CSS codes** We restrict attention to a class of CSS codes which admit simple encoded Clifford group gates. Let $\mathcal{C}$ be an $[n, k, d]$ classical code. Let the generator and parity check matrices be $G$ and $H$ respectively. [7] Suppose $\mathcal{C}$ is self-dual. Then, $G = H$. We have seen that $G$ has $k$ rows while $H$ has $n - k$ rows. Hence, $n = 2k$ and $\mathcal{C}$ is a $[2k, k, d]$ classical code. We represent $\mathcal{C}$ by listing its codewords as rows in a matrix $C$. We order the rows as:

$$C = \begin{bmatrix} \begin{array}{c|c} \begin{matrix} 0 \\ \cdot \\ 0 \end{matrix} & M_0 \\ \hline \begin{matrix} 1 \\ \cdot \\ 1 \end{matrix} & M_1 \end{array} \end{bmatrix} \tag{3.35}$$

The $[2k - 1, k, d']$ punctured code (with $d - 1 \le d' \le d$) is defined by

$$C_p = \begin{bmatrix} M_0 \\ \hline M_1 \end{bmatrix} \tag{3.36}$$

Let the parity check matrix for the punctured code be $H_p$. We claim that $H_p$ generates

---

[7]The symbol $H$ is used for both the Hadamard gate and the parity check matrix. The usage should be clear from the context.

$M_0$. Let $v_p \in M_0$. From Eq. (3.35), $v_p$ is orthogonal to every row in $M_0$ or $M_1$ since $C$ is self-dual. Hence, $v_p$ can be generated by $H_p$. Conversely, if $v_p$ can be generated by $H_p$, it is orthogonal to every row in $C_p$. Define the vector $v$ to be that with 0 prepended to $v_p$. $v$ is orthogonal to all rows in $C$ and is therefore in $C$. Since the first coordinate of $v$ is 0, $v_p \in M_0$. Hence $H_p$ generates $M_0$.

Note that $M_0$ is closed as a subspace (since $C$ is linear) and $M_1$ is a coset of $M_0$. Constructing a quantum CSS code using cosets of $M_0$ in $C_p$, we get a $[[2k-1, 1, d-1]]$ quantum code, with both $X$ and $Z$ matrices in the generator of the stabilizer equal to $H_p$.

**Encoded Clifford group elements in CSS codes**  These CSS codes have remarkably simple encoded operations. First of all, the $X$ and $Z$ matrices of the generator are both $H_p \subset M_0$. As $M_0$ is self-dual, every generator has even weight. Therefore, $\bar{X}$ and $\bar{Z}$ can simply be chosen to be the bitwise [8] $X$ and $Z$ which commute with the stabilizer and anticommute with each other. $\bar{H}$ which interchanges $\bar{X}$ and $\bar{Z}$ can be chosen to be the bitwise $H$. A $\overline{\text{CNOT}}$ between two code blocks is just the bitwise CNOT between the corresponding qubits in the code blocks. If the classical code $\mathcal{C}$ is chosen to be *doubly even* (i.e. weight of every word is a multiple of 4), the weight of every generator in the stabilizer is a multiple of 4, and bitwise $P$ and $P^\dagger$ are in $N(S)$. $\bar{P}$ is bitwise $P$ or $P^\dagger$ depending on whether $k$ is odd or even. Bitwise operations are automatically transversal.

The Clifford group elements can be performed on a general stabilizer code, though they are more complicated. These can be found in [56, 57].

**Fault tolerant syndrome measurements**

So far, we have discussed unitary operations only. We now consider measurements, which are essential in error recovery and preparation of standard states. We consider measurements of bitwise operators which have eigenvalues $\pm 1$. These include (but are not restricted to) Pauli operators. We first consider the scheme detailed in [104, 48], and then derive another scheme with a slightly different interpretation.

---

[8] If a procedure of applying an operation to every qubit in a code block implements the encoded version of the operation, it is bitwise. We also call an operation bitwise if it is a tensor product of operations each acting on a single qubit of a code block.

Consider $n$ qubits, and a subset, $\mathcal{K}$, of $k$ qubits. Suppose we want to measure the parity of the qubits in $\mathcal{K}$. This corresponds to measuring the operator $Z_{\mathcal{K}} = \prod_{i \in \mathcal{K}} Z^{(i)}$. $Z_{\mathcal{K}}$ can be measured by performing a CNOT from each qubit in $\mathcal{K}$ to a fixed ancilla $|0\rangle$, which is flipped iff the parity is odd (see Fig. 3.5).

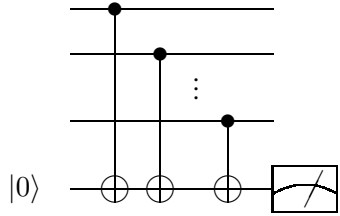

Figure 3.5: Circuit measuring the parity of a subset $\mathcal{K}$ of qubits. Only qubits in $\mathcal{K}$ are shown.

However, this method is *not* fault-tolerant, as a phase error in the ancilla can propagate to all the qubits it has interacted with. The solution in [104] is to prepare an ancilla in the "cat-state" $|c_k\rangle = |0\rangle^{\otimes k} + |1\rangle^{\otimes k}$, and apply bitwise $H$ to obtain an equal superposition of all even parity states. Now, one can apply a CNOT from each qubit in $\mathcal{K}$ to the corresponding qubit in the ancilla, and finally apply direct measurements to the ancilla to determine the parity. Note that no extra information on the encoded state, besides the parity of $\mathcal{K}$, can be obtained, and an error in the ancilla can propagate to at most one qubit in the code block.

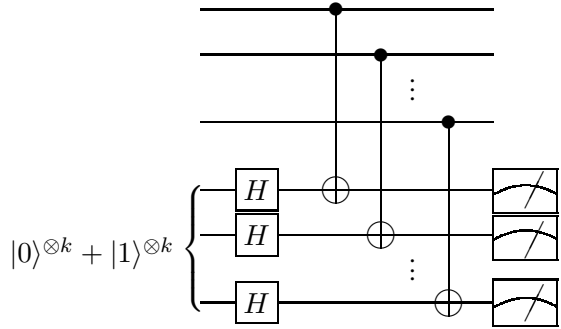

Figure 3.6: Fault-tolerant circuit measuring the parity of $\mathcal{K}$.

In [48], this method was extended to any bitwise operator $M = M_1 \otimes \cdots \otimes M_n$ with eigenvalues $\pm 1$ as follows. Each $M_i \neq I$ can be diagonalized to $Z$ by some $U_i$. Hence, $M = U Z_{\mathcal{K}} U^\dagger$ where $U = U_1 \otimes \cdots \otimes U_n$ is bitwise and $\mathcal{K} = \{i | M_i \neq I\}$. One can measure $M$ by applying $U^\dagger$ to the state, measuring $Z_{\mathcal{K}}$ and applying $U$.

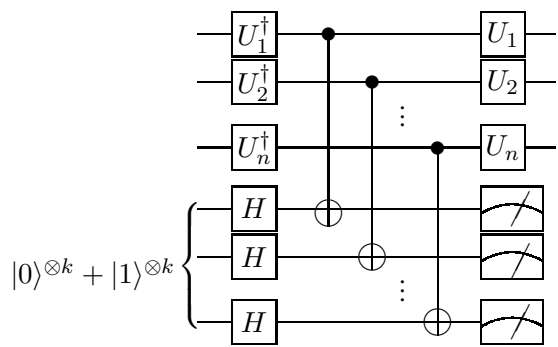

Figure 3.7: Fault-tolerant circuit measuring $M_1 \otimes \cdots \otimes M_n$

Using $X = HZH$, the above figure can be transformed to

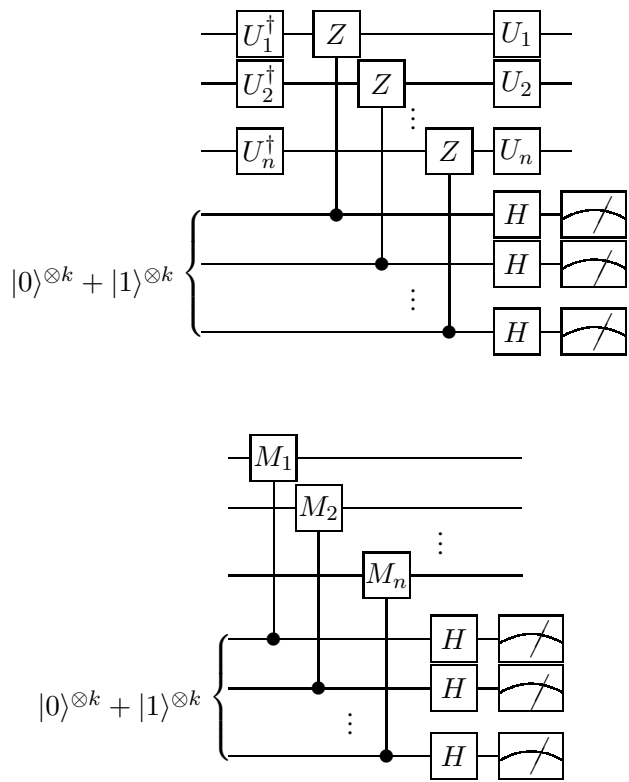

Figure 3.8: Fault-tolerant circuit measuring $M_1 \otimes \cdots \otimes M_n$

The procedure is to apply a "cat-controlled-M" bitwise from the cat-state to the code block, which results in a phase flip to the cat-state iff the measurement outcome is $-1$. This phase flip can be measured in the conjugate basis defined by the Hadamard transform. It is necessary to prepare and verify the cat-state to high fidelity, and to repeat the measurement to avoid measurement errors, but these steps can be done.

**Measurement induced logical operations**

It is possible to use measurements to effect an evolution or to prepare certain states [56, 57]. We first consider what happens to the stabilizer $S$ and the encoded Pauli operators when a measurement of $K \in \mathcal{G}$ is made. If $K \in S$, the measurement is trivial. If $K \in N(S)/S$, measuring $K$ corresponds to a measurement of the encoded logical state. Therefore, we assume $K \notin N(S)$. In this case, $K$ anticommutes with some element $M_1$ in $S$, and we may choose the first generator of $S$ to be $M_1$, and all other generators to commute with $K$. Measuring $K$ projects the state to some $\pm 1$ eigenstate of $K$, leaving $M_i$ unchanged in $S$ for $i \geq 2$. We can "put" $K$ into the stabilizer in the following way. Nothing needs to be done if the post-measurement state $|\psi\rangle$ is a $+1$ eigenstate of $K$. If $|\psi\rangle$ is a $-1$ eigenstate of $K$, then $A|\psi\rangle$ is stabilized by $K$ for any $A$ that anticommutes with $K$ because $A|\psi\rangle = A(-K)|\psi\rangle = K(A|\psi\rangle)$. If furthermore, $A$ commutes with $M_i$ for $i \geq 2$, $A|\psi\rangle$ has stabilizer generated by $A$ and $\{M_i\}_{i\geq 2}$. (Such $A$ always exists, since $A = M_1$ is one possibility.) Therefore, by measuring $K$ and "fixing-up" if necessary, one can replace $M_1$ by $K$ in $S$. The transformation in the (encoded) state is given by the changes in $N(S)/N$. Note that if $N$ is a generator for $N(S)/S$, $MN$ is still a generator $\forall M \in S$. Therefore, the effect of measuring $K$ is to replace $N$ by $NM_1$ if (and only if) $N$ anticommutes with $K$.

In summary, when we measure an operator $K$, we perform the following procedure on the stabilizer and the encoded operations $\bar{X}$ and $\bar{Z}$:

1. Identify an element $M_1 \in S$ that anticommutes with $K$.

2. Rewrite $\bar{X}$, $\bar{Z}$ and the remaining generators of $S$ to commute with $K$ by multiplying by $M_1$ if necessary. These rewritten operators are *equivalent* to the old ones.

3. Replace $M_1$ by $K$ to obtain the new $S$ and $N(S)/S$.

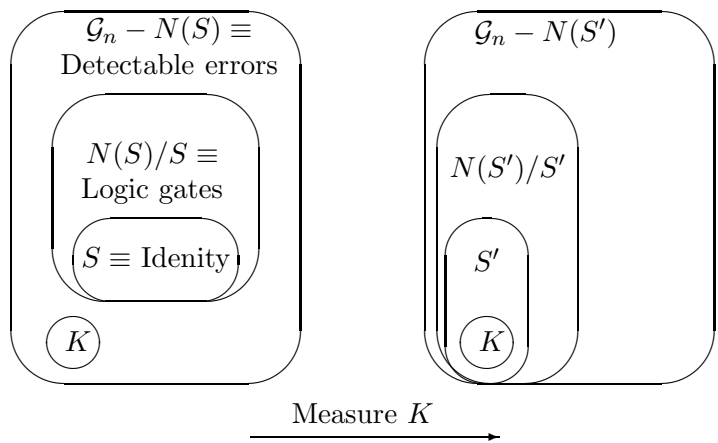

Figure 3.9: Using measurement as logic operation.

This technique to transform the stabilizer is very useful for preparing known states, as will be used in Chapter 5.

**Universality**

As mentioned before, the Clifford group is not universal. However, the Clifford group together with an extra gate such as the Toffoli gate or the $\pi/8$ gate is universal. The construction of these gates is made possible by special ancilla preparation and measurements which can be made fault-tolerant. A detailed description of this is the subject of Chapter 5.

### 3.4.2   The Threshold Theorem and Overhead

We have described how quantum error correcting codes can reduce the error probability if the logic operations are perfect and how fault tolerant methods can be applied to avoid the propagation of errors. However, the above results do not guarantee a net reduction of errors. The coding procedure increases the complexity of the circuit and therefore can increase the net error probability, especially if the elementary error rate is too high. Moreover, such a procedure may require a lot of space and time resources. The *threshold theorem for quantum computation* provides a reassuring answer to the above concerns. It can be stated as follows:

Provided the noise in individual quantum gates (or storage) is below a certain constant threshold, it is possible to efficiently perform an arbitrarily long quantum computation reliably.

The threshold theorem is made possible by concatenating well chosen quantum error correcting codes to obtain super-exponential reduction in net error probability while increasing the space-time requirement only exponentially. To be specific, suppose a computation of $T$ steps is to be performed. The error per step should be about $p_{aim} = 1/T$. Suppose the elementary error probability is $p$, and a $t$-error correcting code of block size $n$ is used, reducing the error from $p$ to $cp^{t+1}$ for some constant $c$. If we further use $n$ of these first-level-encoded qubits to encode one second-level-encoded qubit, and if the structure of the circuit is preserved (*self-similar*) the error will be reduced to $c(cp^{t+1})^{t+1}$. If $L$ levels of concatenation are used, the error will be reduced to $c^{(1+(t+1)+\cdots+(t+1)^{L-1})}p^{(t+1)^L} = c^{-1/t}(c^{1/t}p)^{(t+1)^L}$. It is clear that concatenation is useful iff $p \leq p_{th} = c^{-1/t}$. To achieve the desired level of precision, we set $p_{aim} \approx c^{-1/t}(c^{1/t}p)^{(t+1)^L}$, in which case $(t+1)^{L+1} = \frac{\log(p_{aim}/p_{th})}{\log(p/p_{th})}$. The space requirement increases by a factor of $n^L = \left[\frac{\log(p_{aim}/p_{th})}{\log(p/p_{th})}\right]^{(\log_{t+1} n)}$, and similarly for the time requirement. The requirements are only polylog in the desired accuracy or the length of the computation, which is efficient enough in most applications.

## 3.5 Summary and preview

In this chapter, we have summarized the major result in the development of quantum error correction. We have reviewed classical coding theory. We have discussed the Shor code, the criteria for quantum error correction, the stabilizer formalism and the stabilizer codes, fault-tolerant quantum computation and the threshold theorem.

This Dissertation contains various results related to quantum error correction, and they are organized as follows.

- Chapter 4 consists of a series of related results on the direct construction of quantum codes without classical analogs. This approach is complementary to the one described in this chapter and is useful for exploiting knowledge of the noise process to improve the efficiency of the codes. Codes with rates unmatched by general codes will be presented. Surprisingly, these codes violate the quantum error correction criteria, and the understanding of such violation leads to a relaxed criteria for quantum error correction.

- Chapter 5 describes a systematic method to construct gates outside the Clifford group fault-tolerantly. These gates are needed in addition to the Clifford group operations to form a universal set of gates. The existing constructions are difficult to understand and generalize. A systematic construction which explains and generalizes most known constructions will be given.

- Chapter 8 describes an experimental test of quantum error correction in NMR. It details the modifications required in NMR to perform the coding scheme, and the analysis that conclusively demonstrates the net reduction of error due to coding. A systematic study of deviations from the ideal behavior is presented.

# Chapter 4

# Amplitude damping codes & approximate QEC

In this chapter, we will describe direct construction of quantum codes without classical analogs. We describe a class of bosonic codes and a 4-qubit binary code for amplitude damping, which exploit knowledge of the noise process to outperform the rate of the general codes. We also obtain a relaxed criteria for quantum error correction. These original results are reported in [31, 78].

## 4.1   Bosonic codes for amplitude damping

We have seen in Chapter 3 how quantum error correction is possible theoretically, and how it can be useful for reliable computation even when the coding operations are imperfect. Most known codes assume the Pauli error basis ($I$, $X$, $Y$, $Z$), and coding is performed to allow correction of arbitrary unknown errors on a number of qubits. The smallest general 1-error correcting code requires 5 qubits to encode one qubit. However, in a given physical system, the dominant decoherence process is of a specific nature which may admit a simpler description. An important question therefore arises: given a particular decoherence process, what is the optimal quantum error correction scheme?

While a general solution to the above question remains to be found, we describe progress towards such a solution. In Sections 4.1.1-4.1.8, we demonstrate a new class of quantum error correcting codes which correct only one particular noise process known as amplitude damping. In contrast to other previous work, we consider bosonic systems which occupy the Hilbert space $|0\rangle \cdots |N\rangle$. Moreover, our codes are constructed directly from the quantum error correction criteria without classical origins. Some of these *bosonic codes* achieve better rates than any general code. In particular, we found a code with corrects for one amplitude damping error using effectively $n = 4.6$ qubits to encode one qubit. We present necessary and sufficient conditions for the codes, and describe construction algorithms, and performance bounds.

### 4.1.1   Amplitude damping model

Amplitude damping [82, 52], first introduced in Section 2.4 for the qubit case, describes the energy loss from the system to a zero temperature environment. This is a good approximation to many real life systems. In this section, we extend the discussion to higher dimensions. Amplitude damping can be studied by modeling the system as a simple harmonic oscillator. The energy exchange between the system and the environment is given by the interaction Hamiltonian:

$$H_I = \chi(a^\dagger b + b^\dagger a) \tag{4.1}$$

where $a$, $b$ are the annihilation operators of the system and the environment respectively and $\chi$ is a coupling constant. As mentioned in Section 2.4, complicated interactions can often be described by much simpler models. In this case, a single harmonic oscillator for the environment is sufficient to model the dynamics of interest.

We denote the amplitude damping process between the times $t$ and $t + \Delta t$ by $\mathcal{E}$. We can derive an operator sum representation of $\mathcal{E}$ by assuming that the environment is initially in the ground state $|0\rangle$ and by taking the partial trace along the number eigenstates $|k\rangle$:

$$\mathcal{E}(\rho) = \text{Tr}_e\left[e^{-iH_I\Delta t}(\rho \otimes |0\rangle\langle 0|)e^{iH_I\Delta t}\right] \tag{4.2}$$

$$= \sum_k A_k \rho A_k^\dagger \tag{4.3}$$

$$\text{where} \quad A_k = {}_b\langle k|e^{-i\chi\Delta t(a^\dagger b + b^\dagger a)}|0\rangle_b \,, \tag{4.4}$$

and $b$ denotes the environment model state. $A_k$ describes the event of losing $k$ quanta from the system to the environment. Operator algebra techniques[82] can be used to explicitly evaluate the inner product, giving

$$A_k = \sum_n \sqrt{\binom{n}{k}} \sqrt{(1-\gamma)^{n-k}\gamma^k} \, |n-k\rangle\langle n| \, . \tag{4.5}$$

In Eq. (4.5), $\gamma = 1 - \cos^2(\chi\Delta t)$ is the probability of losing a single quantum from the system during time $\Delta t$. [1] Note that $A_0 \neq I$ – even when no quantum is lost to the environment, the state of the system is still changed.

For a pure initial state $|\psi\rangle$,

$$\mathcal{E}(|\psi\rangle\langle\psi|) = \sum_{k=0}^{N} A_k|\psi\rangle\langle\psi|A_k^\dagger \tag{4.6}$$

is a mixture of unnormalized pure states. In this case, we may use a shorthand

$$[\psi'] = \bigoplus_{k=0}^{N} A_k|\psi\rangle \, , \tag{4.7}$$

where the symbols $[\psi']$ and "$\oplus$" are reserved for statistical mixtures of pure states. Identical states in the mixed sum can be combined using the rule

$$a|\phi\rangle \oplus b|\phi\rangle = \sqrt{|a|^2 + |b|^2}|\phi\rangle \, . \tag{4.8}$$

The normalization of each pure state component gives its probability of occurrence.

So far, we have described the effect of amplitude damping on a single register. Consider now a system with $m$ registers undergoing independent amplitude damping. An initial pure state

$$|\psi_{in}\rangle = |n_1 \dots n_m\rangle \tag{4.9}$$

---

[1] This time behavior in our simple model differs from a more detailed model, such as the one in Appendix A.1. Such distinction is irrelevant for the present purpose.

becomes the mixed state

$$[\psi_{out}\rangle = \left[\bigoplus_{k=0}^{N} A_k|n_0\rangle\right] \otimes \cdots \otimes \left[\bigoplus_{k=0}^{N} A_k|n_m\rangle\right],\tag{4.10}$$

with $(N+1)^m$ possible final states. It is convenient to use the shorthand

$$A_{\tilde{k}} = A_{k_0} \otimes \cdots \otimes A_{k_m},\tag{4.11}$$

where $k_j$ is the $j$-th digit of the number $\tilde{k}$ written in base N+1, so that we may rewrite Eq. (4.10) as

$$[\psi_{out}\rangle = \bigoplus_{\tilde{k}=0}^{(N+1)^m-1} A_{\tilde{k}}|\psi_{in}\rangle.\tag{4.12}$$

As an example, consider the amplitude damping of the state

$$|\psi_{in}\rangle = a|01\rangle + b|10\rangle.\tag{4.13}$$

Using

$$A_0 = |0\rangle\langle 0| + \sqrt{1-\gamma}|1\rangle\langle 1|\tag{4.14}$$

$$A_1 = \sqrt{\gamma}|0\rangle\langle 1|,\tag{4.15}$$

the output state can be found to be

$$[\psi_{out}\rangle = A_{00}|\psi_{in}\rangle \oplus A_{01}|\psi_{in}\rangle \oplus A_{10}|\psi_{in}\rangle \oplus A_{11}|\psi_{in}\rangle\tag{4.16}$$

$$= \sqrt{1-\gamma}|\psi_{in}\rangle \oplus \sqrt{\gamma}|00\rangle.\tag{4.17}$$

This result can be understood intuitively: The original state only contains a single quantum, thus, whenever it is lost, the final state must be the vacuum. This example indicates that the state of Eq. (4.13) is useful for detection of a single quantum loss. However, it cannot recover the input state upon detection of an error, so cannot be used for error *correction*.

### 4.1.2   Example

Let us motivate our code construction by considering the following example: We *encode* the logical zero and one states of a single qubit as

$$|0_L\rangle = \left[\frac{|40\rangle + |04\rangle}{\sqrt{2}}\right] \quad |1_L\rangle = |22\rangle \,, \tag{4.18}$$

such that the initial state is the arbitrary logical qubit

$$|\psi_{in}\rangle = a|0_L\rangle + b|1_L\rangle \,. \tag{4.19}$$

The possible outcomes after amplitude damping may be written as

$$[\psi_{out}\rangle = \bigoplus_{\tilde{k}} |\phi_{\tilde{k}}\rangle = \bigoplus_{\tilde{k}} A_{\tilde{k}}|\psi_{in}\rangle \,, \tag{4.20}$$

where we express $\tilde{k}$ as a base 5 numeral, and $|\phi_{\tilde{k}}\rangle$ is an unnormalized pure state (the norm of which gives its probability to occur in the mixture). For small loss probability $\gamma$, the most likely final state will be

$$|\phi_{00}\rangle = (1-\gamma)^2|\psi_{in}\rangle \,, \tag{4.21}$$

corresponding to no quanta being lost to the bath. The next most likely states result from the loss of a single quantum:

$$|\phi_{01}\rangle = \sqrt{2\gamma}(1-\gamma)^{3/2}\left[a|03\rangle + b|21\rangle\right] \tag{4.22}$$

$$|\phi_{10}\rangle = \sqrt{2\gamma}(1-\gamma)^{3/2}\left[a|30\rangle + b|12\rangle\right] \,. \tag{4.23}$$

States resulting from the loss of more than one quantum occur with probabilities of order $\gamma^2$. Therefore, we aim at correcting up to losing one quantum. Each such error $E_i$ takes $|0_L\rangle$ and $|1_L\rangle$ to states $|0_L\rangle_i$ and $|1_L\rangle_i$ respectively. The key is that $|0_L\rangle$, $|1_L\rangle$, $|0_L\rangle_i$ and $|1_L\rangle_i$ $\forall i$ are mutually orthogonal, and so are $|\phi_{00}\rangle$, $|\phi_{01}\rangle$, and $|\phi_{10}\rangle$. In principle, a ("quantum non-demolition") measurement scheme can detect all error syndromes. Furthermore, for each $i$, the norms of $|0_L\rangle_i$ and $|1_L\rangle_i$ are equal. After detecting an error syndrome $i$, one can apply an appropriate unitary transformation converting $|0_L\rangle_i$ and $|1_L\rangle_i$, to $|0_L\rangle$ and $|1_L\rangle$ respectively. This makes possible the correction:

$$a|0_L\rangle_i + b|1_L\rangle_i \rightarrow \alpha \left[ a|0_L\rangle + b|1_L\rangle \right] , \qquad (4.24)$$

where $\alpha$ is independent of $a$, $b$. Note, this is done without any information about $a$, $b$, and without diminishing the amplitude of the erroneous state. For this particular code, the output state has fidelity [102, 72] (see also Eq. (4.60)) $\mathcal{F} = 1 - 6\gamma^2$ with respect to the input.

As a comparison, if $|0_L\rangle = |11\rangle$, $|1_L\rangle = |22\rangle$, the most probable state is $|\phi_{00}\rangle = a(1 - \gamma)|11\rangle + b(1 - \gamma)^2|22\rangle$. No unitary transformation will bring it back to $a|11\rangle + b|22\rangle$ unless $a$, $b$ is known. Alternatively, a fixed non-unitary transformation can revert the change, but it will reduce the fidelity of the correction process to $1 - \mathcal{O}(\gamma)$.

In the next section, we turn to the criteria for a scheme in which $k$ qubits may be encoded so that losses up to $t$ quanta may be corrected.

### 4.1.3   Code Criteria

We consider a non-degenerate code which will correct up to $t$ losses. The code encodes $l_o + 1$ logical states in $m$ bosonic registers each having a maximum of $N$ quanta and $N + 1$ dimensions. We define $\mathcal{K}(s)$ to be the set of all $m$-digit base-$(N + 1)$ numbers whose digits sum to $s$ (corresponding to the errors having exactly $s$ losses). The logical states must satisfy the criteria for quantum error correction, Eq. (3.9),

$$\langle c_{l_1}|A_{\tilde{k}}^{\dagger}A_{\tilde{k}'}|c_{l_2}\rangle = 0 \qquad \text{for } l_1 \neq l_2 \text{ or } \tilde{k} \neq \tilde{k}' \qquad (4.25)$$

$$\langle c_l|A_{\tilde{k}}^{\dagger}A_{\tilde{k}}|c_l\rangle = g_{\tilde{k}} \qquad \forall l \qquad (4.26)$$

for all $\tilde{k}, \tilde{k}' \in \bigcup_{s \leq t} \mathcal{K}(s)$. In Eq. (4.26), $g_{\tilde{k}}$ is some constant which depends only on $\tilde{k}$. Equation (4.25) requires that all erroneous states be orthogonal, and Eq. (4.26) requires that the encoded Hilbert space not be deformed.

We now present an explicit statement of these two conditions as algebraic conditions on the code construction. We first consider a logical state $|c_l\rangle$ which is an equal superposition of $N_l$ energy eigenstates, $|n_1 \ldots n_m\rangle$. We shall borrow from [55] the name *quasi-classical states*, "QCS" for short, because these states resemble the classical codewords. When all the QCS are equally weighted, we call the code "balanced". Otherwise, the code is referred to as "unbalanced". Each logical state can be represented by a matrix with $m$ columns and $N_l$ rows, each row being one of the QCS in the codeword. For instance, if

$$|c_l\rangle = \frac{1}{\sqrt{N_l}} \left[ |n_{11} \cdots n_{1m}\rangle + \cdots + |n_{N_l 1} \cdots n_{N_l m}\rangle \right] , \qquad (4.27)$$

then the corresponding matrix $\mathcal{M}_l$ is:

$$\begin{bmatrix} n_{11} & n_{12} & \ldots & n_{1m} \\ n_{21} & n_{22} & \ldots & n_{2m} \\ \ldots & \ldots & \ldots\ldots & \\ n_{N_l 1} & n_{N_l 2} & \ldots & n_{N_l m} \end{bmatrix} . \qquad (4.28)$$

**Non-deformation condition**

For $t = 0$ errors, we have $\mathcal{K}(0) = \{0\}$, and $A_0 |n_{i1} \ldots n_{im}\rangle = (1 - \gamma)^{RS_i/2} |n_{i1} \ldots n_{im}\rangle$ where the row sum $RS_i = \sum_{j=1}^m n_{ij}$. Criteria given by Eq. (4.26) require the norm squares of $A_0 |c_l\rangle$ be the same for all $|c_l\rangle$, that is:

$$\frac{1}{N_l} \sum_{i=1}^{N_l} (1 - \gamma)^{RS_i} \qquad (4.29)$$

be the same for all $|c_l\rangle$. A *sufficient* condition for Eq. (4.29) is that $RS_i$ be independent of $i$ and $l$. [2] In other words, all QCS in all logical states have a total of $N_T$ quanta. Physically, this requirement stems from the fact that a state with higher number of quanta decays faster. To preserve the a posteriori probability of each logical state $|c_l\rangle$, we must encode them in a subspace in which the decay probabilities are equal for all of them. Denote the set of all QCS with $N_T$ quanta in $m$ registers as $\mathcal{Q}(N_T, m)$. The non-deformation constraint

---

[2] This is not a necessary condition. The Shor code of Eq. (3.6) satisfies Eq. (4.29) without equal row sums.

Eq. (4.26) is satisfied for $t = 0$ if we construct all the $|c_l\rangle$ from states in $\mathcal{Q}(N_T, m)$. Similarly, for $t = 1$ error, we have $\mathcal{K}(1) = \{0 \cdots 01, 0 \cdots 10, \ldots, 1 \cdots 00\}$, and, for example,

$$A_{0 \cdots 1}|c_l\rangle = A_{0 \cdots 1} \frac{1}{\sqrt{N_l}} \sum_{i=1}^{N_l} |n_{i1} \ldots n_{im}\rangle \tag{4.30}$$

$$= \sum_{i=1}^{N_l} \sqrt{\frac{n_{im}\gamma(1-\gamma)^{N_T-1}}{N_l}} |n_{i1} \ldots n_{im} - 1\rangle, \tag{4.31}$$

where the criteria of equal row sums for $t = 0$ is assumed. Taking the norm square of Eq. (4.31), we obtain

$$\langle c_l | A_{0 \cdots 1}^{\dagger} A_{0 \cdots 1} | c_l \rangle = \frac{\gamma(1-\gamma)^{N_T-1}}{N_l} \sum_{i=1}^{N_l} n_{im}. \tag{4.32}$$

The non-deformation criteria require the above sum to be the same for all $|c_l\rangle$. Equivalently, the column sum of the $m$-th column of each $\mathcal{M}_l$ divided by $N_l$ has to be independent of $l$. Similar expressions for all the $A_{\tilde{k}}$ give rise to a full set of criteria:

*Lemma 2*: Let each logical state $|c_l\rangle$ be expressed as an $m$ column, $N_l$ row matrix with elements $n_{ij}$. If $|c_l\rangle$ satisfy the equal row sum criteria for $t = 0$ losses, and $\sum_i n_{ij}/N_l = y_j \ \forall \ |c_l\rangle$, then $\langle c_l | A_{\tilde{k}}^{\dagger} A_{\tilde{k}} | c_l \rangle = g_{\tilde{k}}, \ \forall \tilde{k} \in \mathcal{K}(1)$.

These criteria correspond to certain symmetry requirements in the $|c_l\rangle$. A similar result can be derived for $t = 2$:

*Lemma 3*: Same setting as in Lemma 2. Let us choose $|c_l\rangle$ which satisfy the non-deformation criteria for $t = 1$, and such that $\sum_i n_{ij_1} n_{ij_2}/N_l = y_{j_1,j_2}$ for all $|c_l\rangle$, where $(j_1, j_2) \in [1, N_l] \times [1, N_l]$. Then $\langle c_l | A_{\tilde{k}}^{\dagger} A_{\tilde{k}} | c_l \rangle = g_{\tilde{k}}, \ \forall \tilde{k} \in \mathcal{K}(2)$.

**Proof:** We need to work out $A_{\tilde{k}}|c_l\rangle$ for each $\tilde{k}$ and apply the criteria for $t = 0, 1, 2$. For $t = 2$, $\mathcal{K}(2) = \{0 \cdots 02, 0 \cdots 20, \ldots, 2 \cdots 00, 0 \cdots 11, 0 \cdots 101, \ldots, 11 \cdots 0\}$. For instance,

$$A_{0 \cdots 02}|c_l\rangle = A_{0 \cdots 02} \frac{1}{\sqrt{N_l}} \sum_{i=1}^{m} |n_{i1} \ldots n_{im}\rangle \tag{4.33}$$

$$= \sum_{i=1}^{m} \sqrt{\frac{\binom{n_{im}}{2}\gamma^2(1-\gamma)^{N_T-2}}{N_l}} |n_{i1}, \ldots, (n_{im} - 2)\rangle, \tag{4.34}$$

where $\binom{\cdot}{\cdot}$ is the usual binomial coefficient. The norm square of this state is

$$\langle c_l | A_{0\cdots02}^{\dagger} A_{0\cdots02} | c_l \rangle = \frac{\gamma^2(1-\gamma)^{N_T-2}}{2N_l} \sum_{i=1}^{m} n_{im}(n_{im}-1) \,. \qquad (4.35)$$

The $\sum_{i=1}^{m} n_{im}$ term is independent of $l$ by the criteria for $t=1$; hence,

$$\frac{1}{N_l} \sum_{i=1}^{m} n_{im}^2 \qquad (4.36)$$

has to be independent of $l$ to satisfy the non-deformation criteria. Other $A_{\tilde{k}}$ with $\tilde{k} = 0\cdots02, \ldots, 2\cdots00$ impose the above requirement on other columns. Similarly, $\tilde{k} = 0\cdots11$ changes $|c_l\rangle$ to:

$$\sum_{i=1}^{m} \sqrt{\frac{n_{im-1}n_{im}\gamma^2(1-\gamma)^{N_T-2}}{N_l}} |n_{i1}\ldots(n_{im-1}-1)(n_{im}-1)\rangle \,. \qquad (4.37)$$

which has norm square

$$\frac{\gamma^2(1-\gamma)^{N_T-2}}{N_l} \sum_{i=1}^{m} n_{im-1}n_{im} \,. \qquad (4.38)$$

Eq. (4.26) requires the following to be independent of $l$:

$$\frac{1}{N_l} \sum_{i=1}^{m} n_{im-1}n_{im} \,. \qquad (4.39)$$

Similar results can be obtained for other $\tilde{k}$ with 1's at any two registers $j_1$ and $j_2$. When we allow $j_1 = j_2$, we include the previous result for two losses at the same register. This completes the proof of the lemma.

Generalization to arbitrary $t$ is as follows:

*Theorem 6*: Let each $|c_l\rangle$ be expressed as an $m$ column, $N_l$ row matrix with elements $n_{ij}$. If we choose $|c_l\rangle$ such that:

$$\sum_{i} n_{ij_1}n_{ij_2}\cdots n_{ij_s}/N_l = y_{j_1,j_2,\ldots,j_s} \qquad (4.40)$$

are independent of $|c_l\rangle$ $\forall l$, $\forall (j_1, j_2, \ldots, j_s) \in [1, N_l]^s$ and $\forall s \in [1, t]$, then

$$\langle c_l | A_{\tilde{k}}^\dagger A_{\tilde{k}} | c_l \rangle = g_{\tilde{k}}, \; \forall \tilde{k} \in \bigcup_{s \le t} \mathcal{K}(s), \; \forall l.$$

**Proof:** For an arbitrary $s$, the non-deformation constraints for $s$ losses are equations involving up to $s$ powers of $n_{ij}$. Using the constraints for fewer than $s$ losses, we reduce the new constraint to involving exactly $s$ powers of $n_{ij}$. By mathematical induction, the result for an arbitrary $s \le t$ can be obtained. This completes the proof of the theorem.

The above theorem can be generalized to unbalanced codes in which logical states are unequally weighted superpositions of QCS. If the amplitudes of the QCS in $|c_l\rangle$ are $(\sqrt{\mu_1}, \sqrt{\mu_2}, \cdots, \sqrt{\mu_{N_l}})$, we replace the sum $\sum_i n_{ij_1} n_{ij_2} \cdots n_{ij_t}/N_l$ by $\sum_i \mu_i n_{ij_1} n_{ij_2} \cdots n_{ij_t}$, i.e., we replace the equal weights $\frac{1}{N_l}$ by the $\mu_i$'s.

As $t$ increases, the non-deformation criteria become very restrictive. We have found unbalanced codes by numerical search correcting up to $t \le 4$ (Section 4.1.7) which have no analogs in the balanced codes. On the other hand, for $t \le 2$, we found simple construction algorithms for balanced codes with no apparent counterparts for the unbalanced codes.

**Orthogonality condition**

The other criteria, the orthogonality constraints given by Eq. (4.25) can be satisfied as follows. Let $|u\rangle = |u_1 \ldots u_m\rangle$ and $|v\rangle = |v_1 \ldots v_m\rangle$ be two states in $\mathcal{Q}(N_T, m)$. We define a *distance* between $u$ and $v$ as

$$\mathcal{D}(u, v) = \frac{1}{2} \sum_i |u_i - v_i|. \tag{4.41}$$

Clearly, $0 \le \mathcal{D} \le N_T$. Moreover, $\mathcal{D}(u, v) = \mathcal{D}(v, u)$, $\mathcal{D}(u, u) = 0$, and

$$\mathcal{D}(u, v) + \mathcal{D}(v, w) = \frac{1}{2} \sum_i |u_i - v_i| + |v_i - w_i| \tag{4.42}$$

$$\ge \frac{1}{2} \sum_i |u_i - w_i| \tag{4.43}$$

$$= \mathcal{D}(u, w). \tag{4.44}$$

Thus $\mathcal{D}$ is a metric on the discrete space $\mathcal{Q}(N_T, m)$. (For binary states, $\mathcal{D}$ is half of the Hamming distance.) Define the distance between any two logical states $|c_{l_1}\rangle$ and $|c_{l_2}\rangle$ to be the minimum of $\mathcal{D}(u_1, u_2)$ over all QCS $|u_1\rangle$, $|u_2\rangle$ in $|c_{l_1}\rangle$ and $|c_{l_2}\rangle$ respectively. Two logical states with non-negative amplitudes of the constituent QCS are orthogonal *iff* their distance is non-zero. We therefore have the following:

> *Theorem 7*: Let $|c_{l_1}\rangle$ and $|c_{l_2}\rangle$ be formed from the states in $\mathcal{Q}_1$ and $\mathcal{Q}_2$ respectively, where $\mathcal{Q}_1, \mathcal{Q}_2 \subset \mathcal{Q}(N_T, m)$ and $\mathcal{D}(u_1, u_2) > t \;\forall u_1 \in \mathcal{Q}_1, u_2 \in \mathcal{Q}_2$. Then $\langle c_{l_1} | A_{\tilde{k}}^\dagger A_{\tilde{k}'} | c_{l_2}\rangle = 0$, $\forall \tilde{k}, \tilde{k}' \in \bigcup_{s \le t} \mathcal{K}(s)$.
>
> **Proof:** Let $A_{\tilde{k}'} |c_{l_1}\rangle = |d_{l_1}\rangle$, $A_{\tilde{k}} |c_{l_2}\rangle = |d_{l_2}\rangle$, and let $|v_1\rangle$, $|v_2\rangle$ be QCS in $|d_{l_1}\rangle$, $|d_{l_2}\rangle$ respectively *s.t.* $\mathcal{D}(d_{l_1}, d_{l_2}) = \mathcal{D}(v_1, v_2)$. Let $|u_1\rangle$, $|u_2\rangle$ be the QCS in $|c_{l_1}\rangle$, $|c_{l_2}\rangle$ that are mapped to $|v_1\rangle$, $|v_2\rangle$ by $A_{\tilde{k}'}$, $A_{\tilde{k}}$. Then, $\mathcal{D}(u_1, v_1) = \mathcal{D}(u_2, v_2) \le t/2$, and $\mathcal{D}(v_1, v_2) + \mathcal{D}(u_1, v_1) + \mathcal{D}(u_2, v_2) \ge \mathcal{D}(u_1, u_2) > t$. Hence, $\mathcal{D}(v_1, v_2) > 0$ and $\mathcal{D}(d_{l_1}, d_{l_2}) > 0$. Therefore, $|d_{l_1}\rangle$ and $|d_{l_2}\rangle$ are orthogonal states. $\square$

In other words, by constructing logical states using QCS which are sufficiently far apart, the orthogonality conditions of the erroneous states can be maintained.

**Example revisited**

In view of the above non-deformation and orthogonality conditions, it is obvious why the example in Section 4.1.2 can correct for one loss. The logical states are represented by

$$\mathcal{M}_0 = \begin{bmatrix} 4 & 0 \\ 0 & 4 \end{bmatrix} \qquad \mathcal{M}_1 = \begin{bmatrix} 2 & 2 \end{bmatrix} \tag{4.45}$$

with all QCS taken from $\mathcal{Q}(4, 2)$ and $\mathcal{D}(d_0, d_1) = 2$. We now generalize this example to a systematic construction.

## 4.1.4 Construction Algorithm For $t \le 2$ Balanced Codes

In this section, an explicit procedure to obtain a class of balanced codes to correct for $t = 1$ and $t = 2$ errors is presented.

To correct for $t = 1$ error, consider ordered $m$-tuples $(x_1, x_2, \ldots, x_m)$ such that $x_1 + x_2 + \ldots + x_m = n$. We will use the same symbol $\mathcal{Q}(n, m)$ for the space of all such $m$-tuples as well as the space of all QCS $\{|x_1 x_2 \ldots x_m\rangle\}$. Let $\mathcal{R}$ be the cyclic permutation on $\mathcal{Q}(n, m)$, i.e. $\mathcal{R}((x_1, x_2, \ldots, x_m)) = (x_2, \ldots, x_m, x_1)$. Define the order of an $m$-tuple to be the size of its orbit under $\mathcal{R}$. Since the order $p$ must divide m, let $m = pq$. An element of order $p$ looks like $(x_1, \ldots, x_p, x_1, \ldots, x_p, \ldots, x_1, \ldots, x_p)$ with the string $(x_1, \ldots, x_p)$ repeated $q$ times. The orbit looks like:

$$
\begin{aligned}
&(x_1, x_2, \ldots, x_p, \ldots \ldots, x_1, x_2, \ldots, x_p) \\
&(x_2, x_3, \ldots, x_1, \ldots \ldots, x_2, x_3, \ldots, x_1) \\
&\cdots \\
&(x_p, x_1, \ldots, x_{p-1}, \ldots \ldots, x_p, x_1, \ldots, x_{p-1}) \,.
\end{aligned}
\tag{4.46}
$$

Each logical state is taken to be an equal superposition of QCS in some orbit:

$$
\begin{aligned}
|c\rangle = \tfrac{1}{\sqrt{p}} ( \quad &|x_1 x_2 \cdots x_p \cdots \cdots x_1 x_2 \cdots x_p\rangle \\
+ \quad &|x_2 x_3 \cdots x_1 \cdots \cdots x_2 x_3 \cdots x_1\rangle \\
+ \quad &\cdots \\
+ \quad &|x_p x_1 \cdots x_{p-1} \cdots \cdots x_p x_1 \cdots x_{p-1}\rangle ) \,.
\end{aligned}
\tag{4.47}
$$

States formed by distinct orbits are orthogonal, as the orbits partition $\mathcal{Q}(n, m)$. Furthermore, we multiply each number in the QCS by $d$. The minimal distance between distinct QCS is at least $d$, since distances come as multiples of $d$ only. Hence, all the erroneous states will remain orthogonal after $t$ losses if $t < d$. Logical states are now in the form:

$$
\begin{aligned}
|c\rangle = \tfrac{1}{\sqrt{p}} ( \quad &|dx_1 dx_2 \cdots dx_p \cdots dx_1 dx_2 \cdots dx_p\rangle \\
+ \quad &|dx_2 dx_3 \cdots dx_1 \cdots dx_2 dx_3 \cdots dx_1\rangle \\
+ \quad &\cdots \\
+ \quad &|dx_p dx_1 \cdots dx_{p-1} \cdots dx_p dx_1 \cdots dx_{p-1}\rangle ) \,.
\end{aligned}
\tag{4.48}
$$

For the non-deformation criteria, the row sum is $nd = N_T$ by construction. The column sum divided by $N_l$ $(= p)$ is:

$$\frac{dx_1 + \cdots + dx_p}{p} = \frac{dn}{m} \tag{4.49}$$

in any logical state, independent of the order of the constituent QCS. Codes in examples (1)-(3) in Section 4.1.7 are constructed in this way.

To correct for $t = 2$ errors, the $t = 1$ criteria have to be satisfied as well. We take a subset of the $t = 1$ logical states which satisfy the extra non-deformation criteria for $t = 2$. We also replace $d = 2$ by $d = 3$. For $m > 2$, pairs of logical states in the form

$$|0_L\rangle \;=\; \frac{1}{\sqrt{m}}\left[\; |dx_1 dx_2 \cdots dx_m\rangle + |dx_2 dx_3 \cdots dx_1\rangle + \cdots + |dx_m dx_1 \cdots dx_{m-1}\rangle\;\right] \tag{4.50}$$

$$|1_L\rangle \;=\; \frac{1}{\sqrt{m}}\left[\; |dx_m \cdots dx_2 dx_1\rangle + |dx_{m-1} \cdots dx_1 dx_m\rangle + \cdots + |dx_1 dx_m \cdots dx_2\rangle\;\right] \tag{4.51}$$

always satisfy the non-deformation criteria for $t = 2$ (proof omitted). Example (4) in Section 4.1.7 is constructed in this way. These codes encode one qubit.

For $t \geq 3$, we perform a numerical search for special QCS in which the system of linear equations for the weights is linearly dependent. In the best case, the number of linear equations to be solved can be much reduced. Therefore we can find codewords involving fewer QCS, fewer number of registers and fewer number of quanta. Although encoding is certainly possible with a much smaller Hilbert space, we have not found a systematic way to generate such QCS. Codes correcting $t \leq 4$ errors are exhibited in Section 4.1.7.

### 4.1.5   Existence of codes

How large must $N$, $N_T$, $m$, and $N_l$ be to satisfy both the non-deformation constraint, Eq. (4.26), and the orthogonality constraint, Eq. (4.25)? We now show that an unbalanced code exists for arbitrarily large $t$ if $N_T$ is allowed to be arbitrarily large, and give an upper bound for the required $N_T$.

Let $|c_0\rangle$, $|c_1\rangle$, …, $|c_l\rangle$, …, $|c_{l_o}\rangle$, be $l_o + 1$ logical states, each being an unequally-weighted superposition of $N_l$ QCS in $\mathcal{Q}(n, m)$. For convenience, define

$$\mathcal{P}(n, m) = \binom{n + m - 1}{m - 1} \equiv \frac{(n + m - 1)!}{n!(m - 1)!} \tag{4.52}$$

as the number of all possible partitions of $n$ into $m$ non-negative parts, i.e., the number of ways to write $x_1 + x_2 + \ldots + x_m = n$ [63]. Suppose we choose $N_T = nd$ such that

$$\mathcal{P}\left(\frac{N_T}{t + 1}, m\right) = \mathcal{P}(n, m) \geq N_0 + N_1 + N_2 + \cdots + N_{l_o} = N_{QCS}, \tag{4.53}$$

where $N_{QCS}$ is the total number of QCS in all the logical states. By *Theorem 2*, all the QCS involved can be chosen to be distinct, and multiplication of the number states by $d = t + 1$ allows the orthogonality condition to be satisfied.

On the other hand, the non-deformation condition involves satisfying a certain number of constraint equations, given by the total number of possible errors times $l_o$. The possible number of ways to lose $s$ quanta from $m$ registers is just the number of partitions of $s$ into $m$ parts, $\mathcal{P}(s, m)$. Take the QCS to be arbitrary, and solve the non-deformation constraint equations (of *Theorem 1*, generalized to include unbalanced codes) as linear equations for the weights of the QCS. As long as the number of variables ($N_{QCS}$) are no fewer than the number of equations, solutions always exist. We may also augment the system of equations by $l_o + 1$ equations to ensure the correct normalization of each logical state. Hence, for $N_T$ satisfying

$$1 + l_o + l_o \sum_{s=0}^{t} \mathcal{P}(s, m) \leq N_{QCS} \leq \mathcal{P}\left(\frac{N_T}{t + 1}, m\right), \tag{4.54}$$

codes with $m$ registers correcting $t$ errors exist.

We simplify Eq. (4.54) by writing explicitly the expression for $\mathcal{P}(N_T/(t + 1), m)$ and $\mathcal{P}(s, m)$ and performing the summation (by writing the summand as a telescopic sum). We obtain

$$m! \, (1 + l_o) + l_o \left[\frac{(t + m)!}{t!}\right] \leq m \left[\frac{(\frac{N_T}{t + 1} + m - 1)!}{\frac{N_T}{t + 1}!}\right]. \tag{4.55}$$

For example, when $m = 2$, Eq. (4.55) becomes

$$\frac{l_o(t+2)(t+1)}{2} + l_0 + 1 \leq \frac{N_T}{t+1} + 1 \,, \tag{4.56}$$

which gives a scaling law $N_T \approx t^3 l_o / 2$. The scaling of $N_T$ as a function of $t$ for arbitrary but fixed $m$ can be obtained by approximating the factorials involving $N_T$ and $t$ in Eq. (4.55) using the Stirling approximation. We found that $N_T \approx \left(\frac{l_o}{em}\right)^{\frac{1}{m-1}} t^{\frac{2m-1}{m-1}}$. We have also assumed $\frac{N_T}{t+1}$ to be large in obtaining the scaling law, and this is a consistent assumption. Note that this upper bound is generally much larger than necessary, as can be seen in the examples for $t = 3$ and $t = 4$. Much more efficient codes may be obtained, because the QCS may be chosen to give redundant constraint equations. This may be accomplished either systematically (Section 4.1.4), or by numerical search (Section 4.1.7).

### 4.1.6   Rates and Fidelities

The performance of these bosonic quantum codes can be characterized by the *rate* – number of qubits communicated per qubit transmitted, and by the *fidelity* – the worst-case overlap between the input and the decoded and corrected output. We now discuss these two measures.

The rate $r$ is given by the ratio of the number of encoded qubits to the equivalent number of qubits in our ambient Hilbert space:

$$r = \frac{k}{m \log_2(N+1)} \,, \tag{4.57}$$

where $2^k$ = number of logical states, and $(N+1)^m$ is the size of the Hilbert space in our code. The exact number of possible logical states depends on the choice of $N$ (maximum number of excitations in any single register) and $m$ (the number of registers). For $t = 1$, we have worked out a counting scheme, but omit the details here. However, the majority of the QCS have order $m$. Hence, to a good approximation, the number of logical states obtained is:

$$2^k = \frac{\mathcal{P}(n,m)}{m} \,. \tag{4.58}$$

Thus, the asymptotic rate of our codes for large $n$ is $\frac{m-1}{m}$. For small $n$, logical states involving fewer than $m$ QCS allow slightly more qubits to be encoded compared with Eq. (4.58) (see examples (1)-(3) in Section 4.1.7). This small gain can be important in certain applications. For arbitrary but fixed $t$, recall from Section 4.1.5 that $N_T \approx \left(\frac{l_o}{em}\right)^{\frac{1}{m-1}} t^{\frac{2m-1}{m-1}}$ is large enough to guarantee the existence of a code with $l_o + 1$ logical states. Together with the fact $N \leq N_T$, a loose asymptotic lower bound for the achievable rate $r = \frac{m-1}{m}$ can be obtained for our code.

We now turn to the code fidelity $\mathcal{F}$, which we desire to know as a function of the parameters $N_T$, $m$ and $t$. Recall from Section 3.3.2 that the minimum overlap fidelity is defined as:

$$\mathcal{F} = \min_{\psi_{in}} \langle \psi_{in} | \rho_f | \psi_{in} \rangle,\qquad(4.59)$$

where $\rho_f$ is the final output state after correction. When the correction criteria are satisfied, recovery procedures exist for the correctable errors which recover the input states. Therefore, the fidelity is at least the total detection probability of the correctable errors:

$$\mathcal{F} = \min_{\psi_{in}} \sum_{\tilde{k} \in \bigcup_{s \leq t} \mathcal{K}(s)} \langle \psi_{in} | A_{\tilde{k}}^{\dagger} A_{\tilde{k}} | \psi_{in} \rangle.\qquad(4.60)$$

Let the input state be $|\psi_{in}\rangle = \sum_l \alpha_l |c_l\rangle$. Using the orthogonality and non-deformation conditions,

$$\langle \psi_{in} | A_{\tilde{k}}^{\dagger} A_{\tilde{k}} | \psi_{in} \rangle = |\langle c_l | A_{\tilde{k}}^{\dagger} A_{\tilde{k}} | c_l \rangle|,\qquad(4.61)$$

for any $|c_l\rangle$. Expressing $|c_l\rangle$ as

$$
\begin{aligned}
|c_l\rangle \;=\;& \sqrt{\mu_1} |n_{11} n_{12} \cdots n_{1m}\rangle && (4.62)\\
+\;& \sqrt{\mu_2} |n_{21} n_{22} \cdots n_{2m}\rangle && (4.63)\\
+\;& \cdots && (4.64)\\
+\;& \sqrt{\mu_{N_l}} |n_{N_l 1} n_{N_l 2} \cdots n_{N_l m}\rangle, && (4.65)
\end{aligned}
$$

it follows that for $\tilde{k} = (k_1, k_2, \ldots, k_m) \in \mathcal{K}(s)$:

$$|\langle c_l | A_{\tilde{k}}^\dagger A_{\tilde{k}} | c_l \rangle| = (1 - \gamma)^{N_T - s} \gamma^s \sum_{i=1}^{N_l} \mu_i \binom{n_{ij}}{k_j}, \tag{4.66}$$

and using the following relation for binomial coefficients:

$$\binom{N_T}{s} = \sum_{\tilde{k} \in \mathcal{K}(s)} \sum_{i=1}^{N_l} \mu_i \binom{n_{i1}}{k_1} \binom{n_{i2}}{k_2} \cdots \binom{n_{im}}{k_m}, \tag{4.67}$$

we find that the fidelity is

$$\mathcal{F} = \sum_{s=1}^{t} (1 - \gamma)^{N_T - s} \gamma^s \binom{N_T}{s} \tag{4.68}$$

$$= 1 - \binom{N_T}{t+1} \gamma^{t+1} + \mathcal{O}(\gamma^{t+2}). \tag{4.69}$$

This expression holds for balanced codes as well as unbalanced codes. The amazing feature is that given a code which satisfies the orthogonality and non-deformation constraints, $\mathcal{F}$ is independent of $m$; it is determined only by $N_T$ and $t$.

One should note that although codes can be constructed to correct for an arbitrary number $t$ of losses, $N_T$ increases with $t$, which in turns implies a higher loss probability for the system as a whole. These two effects compete against each other to give an upper bound on the fidelity, which can be estimated as follows. Let $N_T$ be the required total number of quanta, and $t$ be the total number of losses to be corrected. As previously discussed, the quantum error correction criteria reduce the two parameters to one degree of freedom, in terms of which we may estimate the optimal achievable fidelity. In terms of $t$, the optimum fidelity for fixed $\gamma$ is obtained by setting

$$\frac{d}{dt} \ln(1 - \mathcal{F}) = 0. \tag{4.70}$$

From Eq. (4.69), this gives to first order in $\gamma$

$$\frac{1}{\binom{N_T}{t+1}} \frac{\partial \binom{N_T}{t+1}}{\partial N_T} \frac{dN_T}{dt} + \frac{1}{\binom{N_T}{t+1}} \frac{d \binom{N_T}{t+1}}{dt} + \ln \gamma = 0. \tag{4.71}$$

Using the Stirling approximation for the factorials in $\binom{N_T}{t+1}$, we obtain

$$\ln\left(\frac{N_T}{N_T-t-1}\right)\frac{dN_T}{dt} + \ln\left(\frac{N_T-t-1}{t+1}\right) + \ln\gamma = 0\,. \tag{4.72}$$

In general, $N_T$ is much larger than $t$, which allows further simplification of Eq. (4.72):

$$\frac{dN_T}{dt}\frac{t}{N_T} - \ln\left(\frac{t}{N_T}\right) + \ln\gamma = 0\,. \tag{4.73}$$

The exact dependence of $N_T$ on $t$ is generally very complicated. This point can be appreciated from the explicit code examples following this section. In particular, the minimum $N_T$ depends on the existence of "good solutions" to the criteria, therefore, the optimal fidelity is analytically intractable. We only have a bound on the fidelity of a bosonic code with arbitrary QCS. There is no theoretical bound on the number of correctable errors.

For concreteness, we will illustrate the above assuming that $N_T$ asymptotically follows a power scaling law in $t$. As illustrated in the previous section, $N_T$ is *bounded* by such polynomials in $t$. Therefore the following gives a *loose* lower bound of the upper bound of the fidelity. Suppose $N_T \approx f l_o t^\alpha$ where the prefactor $f$ and exponent $\alpha$ are approximately constant. Eq. (4.73) can be solved for the optimum $t$:

$$t_{opt} \approx \left(e^{-\alpha}/\gamma f l_0\right)^{1/(\alpha-1)}\,. \tag{4.74}$$

Plugging back into the Eq. (4.69) would give an estimate for the optimal achievable fidelity. However, these gross estimates are not expected to be meaningful in actual applications, because $N_T(t)$ will be the determining factor, and as previously mentioned, is analytically unobtainable.

### 4.1.7   Explicit Codes

Some explicit codes resulting from our work are presented here. Normalization factors are omitted whenever they are common to all logical states. Codes are specified as $((N_T, m, l_o + 1, d))$, where $N_T$ is the total number of excitations in the QCS, $m$ is the number of registers for each QCS, $l_o + 1$ is the number of logical states and $d$ is the minimal distance between the logical states. The fidelity of all the codes are given by $\mathcal{F} \approx 1 - \binom{N_T}{t+1}\gamma^{t+1}$.

Example (1) – $((4, 2, 2, 2))$, $n = 2$, $t = 1$, fidelity $\mathcal{F} \approx 1 - 6\gamma^2$:

$$|0_L\rangle = \frac{1}{\sqrt{2}}[|40\rangle + |04\rangle] \tag{4.75}$$

$$|1_L\rangle = |22\rangle \tag{4.76}$$

Example (2) – $((12, 3, 10, 2))$, $n = 6$, $t = 1$, fidelity $\mathcal{F} \approx 1 - 66\gamma^2$, labels given in hexadecimal ($c = 12$, $a = 10$):

$$|c_0\rangle = \frac{1}{\sqrt{3}}[|00c\rangle + |c00\rangle + |0c0\rangle] \tag{4.77}$$

$$|c_1\rangle = \frac{1}{\sqrt{3}}[|02a\rangle + |a02\rangle + |2a0\rangle] \tag{4.78}$$

$$|c_2\rangle = \frac{1}{\sqrt{3}}[|048\rangle + |804\rangle + |480\rangle] \tag{4.79}$$

$$|c_3\rangle = \frac{1}{\sqrt{3}}[|066\rangle + |606\rangle + |660\rangle] \tag{4.80}$$

$$|c_4\rangle = \frac{1}{\sqrt{3}}[|084\rangle + |408\rangle + |840\rangle] \tag{4.81}$$

$$|c_5\rangle = \frac{1}{\sqrt{3}}[|0a2\rangle + |20a\rangle + |a20\rangle] \tag{4.82}$$

$$|c_6\rangle = \frac{1}{\sqrt{3}}[|228\rangle + |822\rangle + |282\rangle] \tag{4.83}$$

$$|c_7\rangle = \frac{1}{\sqrt{3}}[|246\rangle + |624\rangle + |462\rangle] \tag{4.84}$$

$$|c_8\rangle = \frac{1}{\sqrt{3}}[|264\rangle + |642\rangle + |264\rangle] \tag{4.85}$$

$$|c_9\rangle = |444\rangle \tag{4.86}$$

Example (3) – $((6, 3, 4, 2))$, $n = 3$, $t = 1$, fidelity $\mathcal{F} \approx 1 - 15\gamma^2$:

$$|c_0\rangle = |600\rangle + |060\rangle + |006\rangle \tag{4.87}$$

$$|c_1\rangle = |420\rangle + |204\rangle + |042\rangle \tag{4.88}$$

$$|c_2\rangle = |240\rangle + |402\rangle + |024\rangle \tag{4.89}$$

$$|c_3\rangle = |222\rangle \,. \tag{4.90}$$

Example (4) – $((9, 3, 2, 3))$, $n = 3$, $t = 2$, fidelity $\mathcal{F} \approx 1 - 84\gamma^3$: Note this code differs from the previous one from having $d = 3$ instead of $d = 2$. We take only $|c_1\rangle$ and $|c_2\rangle$ as codewords.

$$|0_L\rangle = |306\rangle + |063\rangle + |630\rangle \tag{4.91}$$

$$|1_L\rangle = |036\rangle + |360\rangle + |603\rangle \tag{4.92}$$

Example (5) – $((6, 4, 2, 2))$, $n = 6 = 0+1+2+3$, fidelity $\mathcal{F} \approx 1 - 15\gamma^2$: The minimal distance between QCS is $d = 2$. However, the QCS are not generated by multiplying each number by $d = 2$.

$$|0_L\rangle = |0321\rangle + |1032\rangle + |2103\rangle + |3210\rangle \tag{4.93}$$

$$|1_L\rangle = |0123\rangle + |1230\rangle + |2301\rangle + |3012\rangle. \tag{4.94}$$

Example (6) – $((7, 2, 2, 2))$, fidelity $\mathcal{F} \approx 1 - 21\gamma^2$: The logical states are not formed by cyclic permutations of the QCS. Note that column one and two have different column sums.

$$|0_L\rangle = |70\rangle + |16\rangle \tag{4.95}$$

$$|1_L\rangle = |52\rangle + |34\rangle \tag{4.96}$$

Example (7) – $((9, 2, 2, 3))$, fidelity $\mathcal{F} \approx 1 - 84\gamma^3$: Unbalanced code that will tolerate $t = 2$ errors. Note that one codeword is formed from the other by reversing the order of the registers. (This symmetry between the two registers is a sufficient condition for balanced codes with $t = 2$, $m \geq 3$.)

$$|0_L\rangle = \frac{1}{2}|90\rangle + \frac{\sqrt{3}}{2}|36\rangle \tag{4.97}$$

$$|1_L\rangle = \frac{1}{2}|09\rangle + \frac{\sqrt{3}}{2}|63\rangle \tag{4.98}$$

Example (8) – $((9, 3, 2, 3))$, fidelity $\mathcal{F} \approx 1 - 84\gamma^3$: Unbalanced code that will tolerate $t = 2$

errors, showing that the symmetry is not a necessary condition for correcting $t = 2$ errors.

$$|0_L\rangle = \frac{1}{\sqrt{3}}[|036\rangle + |306\rangle + |360\rangle] \tag{4.99}$$

$$|1_L\rangle = \frac{1}{3}[\sqrt{6}|333\rangle + \sqrt{2}|009\rangle + |090\rangle] \tag{4.100}$$

Example (9) – $((16, 2, 2, 4))$, fidelity $\mathcal{F} \approx 1 - 1820\gamma^4$: Unbalanced code that will tolerate $t = 3$ errors. Labels are given in base 17. $c$ and $g$ denote 12 and 16 respectively.

$$|0_L\rangle = \frac{1}{\sqrt{8}}[|0g\rangle + |g0\rangle + \sqrt{6}|88\rangle] \tag{4.101}$$

$$|1_L\rangle = \frac{1}{\sqrt{2}}[|4c\rangle + |c4\rangle] \tag{4.102}$$

Example (10) – $((20, 3, 2, 4))$, fidelity $\mathcal{F} \approx 1 - 4845\gamma^4$: Another unbalanced code that will tolerate $t = 3$ errors. Labels are given in base 21. $c$, $g$ and $k$ denote 12, 16, and 20 respectively.

$$|0_L\rangle = \frac{1}{5}[|04g\rangle + 2|40g\rangle + 2\sqrt{5}|0k0\rangle] \tag{4.103}$$

$$|1_L\rangle = \frac{1}{\sqrt{5}}[\sqrt{2}|44c\rangle + \sqrt{3}|488\rangle] \tag{4.104}$$

Example (11) – $((50, 2, 2, 5))$, fidelity $\mathcal{F} \approx 1 - 2118760\gamma^5$: Note the rapid growth in the numerical factor in the second term. To correct for large number of errors, we need to encode a qubit in a large Hilbert space, but emission probabilities are large for high number states. This puts a limit of performance in our codes. The actual code involves numbers five times the numbers shown below. $a$ denotes 10.

$$|0_L\rangle = \sqrt{\frac{1}{18}}|0a\rangle + \sqrt{\frac{5}{9}}|46\rangle + \sqrt{\frac{1}{3}}|82\rangle + \sqrt{\frac{2}{45}}|91\rangle \tag{4.105}$$

$$|1_L\rangle = \sqrt{\frac{1}{18}}|19\rangle + \sqrt{\frac{1}{6}}|28\rangle + \sqrt{\frac{33}{90}}|55\rangle + \sqrt{\frac{1}{3}}|73\rangle + \sqrt{\frac{7}{90}}|a0\rangle. \tag{4.106}$$

### 4.1.8  Discussion

Our treatment of amplitude damping errors contrasts from the usual standpoint of quantum error correction, which deals with $X$ and $Z$ errors. The relationship can be understood by expressing the $A_0$ and $A_1$ operators as coherent superpositions of such errors; from Eqs. (4.14) and (4.15),

$$A_0 = \frac{1}{2}\left[(1+\sqrt{1-\gamma})I + (1-\sqrt{1-\gamma})Z\right] \qquad (4.107)$$

$$A_1 = \frac{\sqrt{\gamma}}{2}\left[X - Y\right]. \qquad (4.108)$$

With probabilities up to $\mathcal{O}(\gamma)$, a binary code with $m$ qubits will either project $A_0^{\otimes m}|c_l\rangle$ onto a state with no errors, or project $A_0^{\otimes m-1}A_1|c_l\rangle$ onto a state with one $X$ or $Y$ error. Hence, a binary code correcting for any one qubit error *will* indeed correct all amplitude damping errors up to losing one quantum, although *not* to all orders in $\gamma$. One reason we have studied bosonic codes is to exploit the possibilities for achieving higher efficiencies or easier physical implementation, though the study is theoretically interesting on its own.

It is important to realize that amplitude damping errors are not independent qubit errors, since the decay factor of each QCS depends on the total number of excitations in it.

Rates from our bosonic codes contrast with those achievable by the usual binary codes. For the code in Example (1), $N = N_T = nd = 4$, $m = 2$ and $k = 1$, so $r = 0.22$. This is slightly better than $r = 0.20$ for the five qubit binary perfect code described in Eq. (3.33), and much better than the eight qubit code of Plenio *et. al.* [93].

The code fidelities may also be compared. Our $((4,2,2,2))$ code achieves $\mathcal{F} \approx 1 - 6\gamma^2$. In comparison, the five-qubit binary code achieves fidelity $\approx 1 - 1.75\gamma^2$, while the eight-qubit code achieves only $\approx 1 - 6\gamma^2$! This agreement with the bosonic code is not accidental; it stems from the use of the same total excitation number. However, it is worthwhile to point out that despite the effort to balance the codewords, the five-bit code still has better performance on average, due to the smaller number of excitations involved in the system.

In conclusion, we have given general criteria for quantum error correction of amplitude damping in bosonic states and have constructed codes based on the amplitude damping operation elements. This generalizes the binary error correcting codes for Pauli errors, and specializes to address the dominant decoherence process of many systems such as photons transmitted through optical fibers and trapped ions. We classify our errors according to the number of excitations lost, instead of the more common classification of the number of qubits or registers corrupted. We have shown that specialization to correct amplitude damping can improve the rate in some cases. However, direct construction of codes can be difficult. In Sections 4.2-4.3, we will describe an improved method which leads to even better codes.

## 4.2 Approximate quantum error correction can lead to better codes

We have seen that codes constructed for specific noise processes can have better efficiency than general codes based on classical codes and the Pauli error basis. Unfortunately, without the specific structure in the Pauli error basis, there is no general method to construct quantum codes. There is no apparent method to adapt classical coding techniques without the Pauli-error basis. The approach to solve the code criteria directly can be very difficult, as can be seen in the bosonic code construction. Consequently, few such codes are known [33, 22, 34, 93, 31].

As the quantum error correction criteria are both necessary and sufficient, code construction has always aimed at satisfying the criteria exactly. One important result of this Dissertation is the observation that the criteria is *not* necessary. We developed a new approach to quantum error correction based on *approximate* satisfaction of the original quantum error correction criteria. These new criteria are important in two ways: they admit more codes and they can be much simpler algebraically. Therefore, they are very useful in seeking more efficient codes for specific noise processes and in direct code constructions. [3]

The usefulness of this approach is immediate. Using the approximate criteria, we have discovered a four-qubit binary code which corrects for single qubit amplitude damping errors. Incidentally, this code was not found even after a substantial amount of effort by the community, because it violates the exact criteria. Moreover, such a short *non-degenerate* code is impossible using the Pauli basis. The reason can be understood by examining the operation elements of amplitude damping:

$$A_0 = \frac{1}{2}\left[(1 + \sqrt{1-\gamma})I + (1 - \sqrt{1-\gamma})Z\right] \tag{4.109}$$

$$A_1 = \frac{\sqrt{\gamma}}{2}\left[X - Y\right]. \tag{4.110}$$

---

[3] In the general code construction based on the Pauli error basis, the criteria are naturally satisfied exactly.

To first order in the error probability $\gamma$, $n+1$ possible errors may happen to an $n$-qubit code using the $A_0$, $A_1$ error basis, so it follows that $n \geq 3$ is required. In contrast, in the Pauli basis, any $X$ or $Y$ error must be corrected by the code, so that $2n+1$ possible errors must be dealt with. A non-degenerate code has to map the codeword space to orthogonal spaces if the syndrome is to be detected unambiguously. Hence, the minimum allowable size for the encoding space is the product of the dimension of the codeword space and the number of operation elements to be corrected. It follows that $n \geq 5$ qubits are required for a non-degenerate Pauli basis code, in contrast to our four-qubit code.

The lessons are that (1) better codes may be found for specific error processes, and (2) approximate error correction simplifies code construction and admits more codes. Approximate error correction is a property with no analog in digital classical error correction, because it makes use of the slight non-orthogonality possible only between quantum states. We describe our approach to this problem by first exhibiting our four-qubit example code in detail. We then generalize our results to provide new, relaxed error correction criteria and specific procedures for decoding and recovery. We conclude by discussing possible extensions to our work.

### 4.2.1   Four Bit Amplitude Damping Code

Recall that single qubit amplitude damping is defined by

$$\mathcal{E}(\rho) = \sum_{k=0,1} A_k \rho A_k^\dagger \tag{4.111}$$

$$A_0 = \begin{bmatrix} 1 & 0 \\ 0 & \sqrt{1-\gamma} \end{bmatrix} \qquad A_1 = \begin{bmatrix} 0 & \sqrt{\gamma} \\ 0 & 0 \end{bmatrix}. \tag{4.112}$$

The probability of losing a photon, $\gamma$, is assumed to be small. To correct errors induced by this process, we encode one qubit using four, with the logical states

$$|0_L\rangle = \frac{1}{\sqrt{2}} \Big[ |0000\rangle + |1111\rangle \Big] \tag{4.113}$$

$$|1_L\rangle = \frac{1}{\sqrt{2}} \Big[ |0011\rangle + |1100\rangle \Big]. \tag{4.114}$$

A circuit for encoding the logical state is shown in Fig. 4.1.

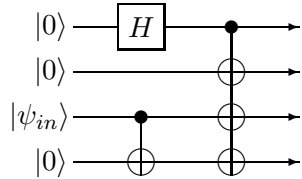

Figure 4.1: Circuit for encoding a qubit. The third register contains the input qubit.

Using the notation defined in Section 4.1.1, the possible outcomes after amplitude damping may be written as

$$[\psi_{out}\rangle = \bigoplus_{\tilde{k}} |\phi_{\tilde{k}}\rangle \equiv \bigoplus_{\tilde{k}} A_{\tilde{k}} |\psi_{in}\rangle, \tag{4.115}$$

where $\tilde{k}$ are binary strings labeling the errors (for example $A_{010\cdots} = A_0 \otimes A_1 \otimes A_0 \cdots$). For the input state

$$|\psi_{in}\rangle = a|0_L\rangle + b|1_L\rangle, \tag{4.116}$$

*all* possible final states occurring with probabilities $\mathcal{O}(\gamma)$ or above are

$$|\phi_{0000}\rangle = a\left[ \frac{|0000\rangle + (1-\gamma)^2|1111\rangle}{\sqrt{2}} \right] + b\left[ \frac{(1-\gamma)[|0011\rangle + |1100\rangle]}{\sqrt{2}} \right]$$

$$|\phi_{1000}\rangle = \sqrt{\frac{\gamma(1-\gamma)}{2}} \Big[ a(1-\gamma)|0111\rangle + b|0100\rangle \Big]$$

$$|\phi_{0100}\rangle = \sqrt{\frac{\gamma(1-\gamma)}{2}} \Big[ a(1-\gamma)|1011\rangle + b|1000\rangle \Big] \tag{4.117}$$

$$\begin{aligned}
|\phi_{0010}\rangle &= \sqrt{\frac{\gamma(1-\gamma)}{2}}\left[a(1-\gamma)|1101\rangle + b|0001\rangle\right] \\
|\phi_{0001}\rangle &= \sqrt{\frac{\gamma(1-\gamma)}{2}}\left[a(1-\gamma)|1110\rangle + b|0010\rangle\right].
\end{aligned}$$

The usual quantum error correction criteria Eq. (3.9) require $\langle 0_L|A_{\tilde{k}}^{\dagger}A_{\tilde{k}}|0_L\rangle = \langle 1_L|A_{\tilde{k}}^{\dagger}A_{\tilde{k}}|1_L\rangle$, but

$$\begin{aligned}
\langle 0_L|A_{0000}^{\dagger}A_{0000}|0_L\rangle &= 1 - 2\gamma + 3\gamma^2 + \mathcal{O}(\gamma^3) & (4.118) \\
\langle 1_L|A_{0000}^{\dagger}A_{0000}|1_L\rangle &= 1 - 2\gamma + \gamma^2. & (4.119)
\end{aligned}$$

So the code we have constructed does not satisfy the usual criteria. We will demonstrate that the code satisfies new approximate error correction conditions later on, and revisit the recovery procedure afterwards. First, we exhibit how the code works.

### Decoding and Recovery Circuit

Let us denote each of the four qubits by $n_1, \ldots, n_4$. Error correction is performed by distinguishing the five possible outcomes of Eq. (4.117), and then applying the appropriate correction procedure. The first step is *syndrome calculation*, which may be done using the circuit shown in Fig. 4.2A. There are three possible measurement results from the two meters: $(M_2, M_4) = (0,0), (1,0)$ and $(0,1)$. Conditioned on $(M_2, M_4)$, recovery processes $W_k$ implemented by the other three circuits of Fig. 4.2 can be applied to the output in $n_1$ and $n_3$.

If $(M_2, M_4) = (0,0)$, then $n_1$ and $n_3$ are in the state:

$$a\left[\frac{|00\rangle + (1-\gamma)^2|11\rangle}{\sqrt{2}}\right] + b\left[\frac{(1-\gamma)(|01\rangle + |10\rangle)}{\sqrt{2}}\right]. \qquad (4.120)$$

To regenerate the original qubit, the circuit of Fig. 4.2B is used: a CNOT is applied using $n_3$ as control, giving

$$a|0\rangle\left[\frac{|0\rangle + (1-\gamma)^2|1\rangle}{\sqrt{2}}\right] + b|1\rangle\left[\frac{(1-\gamma)(|1\rangle + |0\rangle)}{\sqrt{2}}\right]. \qquad (4.121)$$

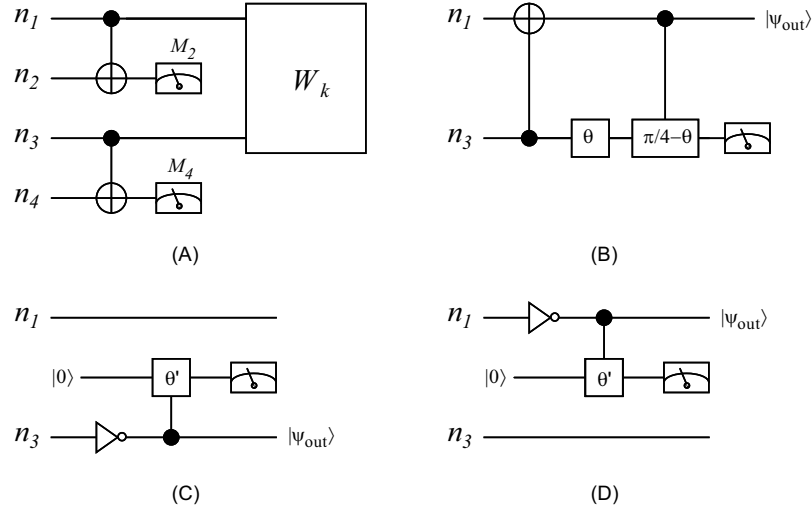

Figure 4.2: (A) Circuit for error syndrome detection. The measurement result is used to select $W_k$ out of three actions. If the result $(M_2, M_4)$ is 00, 10, or 01, circuits (B), (C), or (D) are applied, respectively, to recover the state. The NOT gate, $X$, is represented by the usual classical symbol. The angles $\theta$, $\theta'$ are given by $\tan\theta = (1-\gamma)^2$ and $\cos\theta' = 1-\gamma$. The rotation gate and controlled-rotation gate specified by an angle $\tilde{\theta}$ perform the functions $\exp(i\tilde{\theta}\sigma_y)$ and $\Lambda_1(\exp(i\tilde{\theta}\sigma_y))$ respectively in the notation of [13].

$n_1$ can now be used as a control to rotate $n_3$ to be parallel to $|0\rangle$. We obtain as the final output in $n_1$ and $n_3$:

$$\left[ a\sqrt{\frac{(1-\gamma)^4 + 1}{2}}|0\rangle + b(1-\gamma)|1\rangle \right]|0\rangle \qquad (4.122)$$

$$= \left[ (1-\gamma)(a|0\rangle + b|1\rangle) + \mathcal{O}(\gamma^2)|0\rangle \right]|0\rangle , \qquad (4.123)$$

with the corrected and decoded qubit left in $n_1$ as desired.

If $(M_2, M_4) = (1,0)$, the inferred state before syndrome measurement is $\phi_{1000}$ or $\phi_{0100}$. In either case, $n_1$ and $n_3$ are in a product state and $n_3$ is in the distorted state:

$$\sqrt{\frac{(1-\gamma)\gamma}{2}} \left[ a(1-\gamma)|1\rangle + b|0\rangle \right] . \qquad (4.124)$$

To undo the distortion, we apply the *non-unitary* transformation in Fig. 4.2C. The combined operation on $n_3$ due to the NOT gate, the controlled-rotation gate and the measurement of the ancilla bit can be expressed in the operator sum representation: $\mathcal{N}(\rho) = N_0 \rho N_0^\dagger + N_1 \rho N_1^\dagger$, where

$$N_0 = \begin{bmatrix} 0 & 1 \\ 1-\gamma & 0 \end{bmatrix}, \quad N_1 = \sqrt{\gamma(2-\gamma)} \begin{bmatrix} 0 & 0 \\ 1 & 0 \end{bmatrix}. \qquad (4.125)$$

The $N_0$ and $N_1$ operators correspond to measuring the ancilla to be in the $|0\rangle$ and $|1\rangle$ states respectively. If the ancilla state is $|0\rangle$, we obtain the state:

$$|\psi_{out}\rangle = \sqrt{\frac{(1-\gamma)^3 \gamma}{2}} \left[ a|0\rangle + b|1\rangle \right], \qquad (4.126)$$

in the third register because $N_0$ preferentially damps out the $b|0\rangle$ component in Eq. (4.124). We get an error message if the ancilla is in the $|1\rangle$ state. Finally, if $(M_2, M_4) = (0,1)$ the same procedure can be applied as in the $(M_2, M_4) = (1,0)$ case, with $n_1$ and $n_3$ swapped.

The fidelity, defined as the worst (over all input states) possible overlap between the original qubit and the recovered qubit is

$$\mathcal{F} = (1-\gamma)^2 + 4 \left[ \frac{(1-\gamma)^3 \gamma}{2} \right] = 1 - 5\gamma^2 + \mathcal{O}(\gamma^3), \qquad (4.127)$$

Note that the final state Eq. (4.123) is slightly distorted. This occurs because the recovery operation is not exact, due to the failure to satisfy the code criteria exactly. Furthermore, the circuits in Fig. 4.2C and 4.2D have a finite probability for failure. However, these are second order problems, and do not detract from the desired fidelity order.

## 4.2.2   Approximate Sufficient Conditions

We now explain why our code works despite its violation of the usual error correction criteria. The reason is simple: small deviations from the criteria are allowed as long as they do not detract from the desired *fidelity order*. To make this idea mathematically concrete, we present a simple generalization of the usual error correction criteria described in Section 3.3.2. These *approximate error correction criteria* are *sufficient* conditions for approximate error correction.

We first summarize the exact criteria described in Section 3.3.2, retaining all previously defined notations. We consider a quantum code $\mathcal{C}$ which is a subspace in a larger Hilbert space $\mathcal{H}$, acted on by a noise process $\mathcal{E}(\rho) = \sum_{n \in \mathcal{K}} A_n \rho A_n^\dagger$. The criteria for $\mathcal{C}$ to correct for the errors $A_n \in \mathcal{A}_{re}$, or equivalently, for $\mathcal{E}'(\rho) = \sum_{n \in \mathcal{K}_{re}} A_n \rho A_n^\dagger$ to be reversible on $\mathcal{C}$, is given by

$$P_C A_m^\dagger A_n P_C = g_{mn} P_C \qquad \forall m, n \in \mathcal{K}_{re} \,, \tag{4.128}$$

where $P_C$ is the projector onto $\mathcal{C}$, and $g_{mn}$ are entries of a positive matrix. It is always possible to rewrite $\mathcal{E}'(\rho) = \sum_{n \in \tilde{\mathcal{K}}_{re}} \tilde{A}_n \rho \tilde{A}_n^\dagger$ such that

$$P_C \tilde{A}_m^\dagger \tilde{A}_n P_C = p_n \delta_{mn} P_C \qquad \forall m, n \in \tilde{\mathcal{K}}_{re} \,, \tag{4.129}$$

where $p_n$ are non-negative c-numbers. Without loss of generality, we will use Eq. (4.129) and omit the tilde to simplify notations.

The sufficiency of Eq. (4.129) can be established as follows. When Eq. (4.129) is satisfied, the $A_n$ operators have polar decompositions

$$A_n P_C = \sqrt{p_n} U_n P_C \qquad \forall n \in \mathcal{K}_{re} \,, \tag{4.130}$$

where the $U_n$'s are unitary and $P_C U_n^\dagger U_m P_C = \delta_{nm} P_C$. The recovery operation $\mathcal{R}$ is defined as

$$\mathcal{R}(\rho) = \sum_{k \in \mathcal{K}_{re}} R_k \rho R_k^\dagger + P_E \rho P_E \,, \tag{4.131}$$

where $R_k = P_C U_k^\dagger$ is the appropriate reversal process for each $A_k \in \mathcal{A}_{re}$. When we apply $\mathcal{R}$ to $\mathcal{E}(\rho)$, the first term in Eq. (4.131) is given by $P^{det}\rho$ where $P^{det} = \sum_{n \in \mathcal{K}_{re}} p_n$ is the total detection probability for the reversible subset. $P^{det}$ is a lower bound for the fidelity $\mathcal{F}$ for pure input states. [4]

Now we generalize Eq. (4.129)-(4.130) based on the following assumption: the error is parametrized by certain small quantities with physical origins such as the strength and duration of the coupling between the system and the environment. For simplicity, we consider only one-parameter processes, and let $\epsilon$ be the small parameter. For example, $\epsilon$ can be the single qubit error probability. Suppose the aim is to find a code for a known $\mathcal{E}$ with fidelity:

$$\mathcal{F} \geq 1 - \mathcal{O}(\epsilon^{t+1}) \,. \tag{4.132}$$

---

[4] The fidelity can be generalized to mixed input states $\mathcal{F} = \text{Tr}(\rho_{out}^{1/2} \rho_{in} \rho_{out}^{1/2})$, which is minimized at pure input states. Therefore, $P^{det}$ lower bounds the fidelity for all input states.

In the new criteria, it is still necessary that $P^{det} \geq \mathcal{F}$, that is, $\mathcal{A}_{re}$ has to include all $A_n$ with large detection probability $\max_{|\psi_{in}\rangle \in \mathcal{C}} \text{Tr}(|\psi_{in}\rangle \langle \psi_{in}| A_n^\dagger A_n) \approx \mathcal{O}(\epsilon^s)$, $s \leq t$. However, it is *not* necessary to recover the exact input state; only a good overlap between the input and output states is needed. In terms of the condition on the codeword space, it suffices for the $A_n$ to be *approximately* unitary and mutually orthogonal on $\mathcal{C}$. These observations can be expressed as relaxed *sufficient* conditions for error correction. Suppose

$$A_n P_C = U_n \sqrt{P_C A_n^\dagger A_n P_C} \,, \tag{4.133}$$

is a polar decomposition for $A_n$. We define c-numbers $p_n$ and $\lambda_n$ so that $p_n$ and $p_n \lambda_n$ are the largest and the smallest eigenvalues of $P_C A_n^\dagger A_n P_C$, considered as an operator on $\mathcal{C}$. The relaxed conditions for error correction are that:

$$p_n(1 - \lambda_n) \;\; \leq \;\; \mathcal{O}(\epsilon^{t+1}) \quad\quad \forall n \in \mathcal{K}_{re} \tag{4.134}$$

$$P_C U_m^\dagger U_n P_C \;\; = \;\; \delta_{mn} P_C \,. \tag{4.135}$$

Note that when $\lambda_n = 1$, Eqs. (4.133)-(4.135) reduce to the exact criteria. In the approximate case, $P^{det} = \sum_{n \in \mathcal{K}_{re}} \text{Tr}(|\psi_{in}\rangle \langle \psi_{in}| A_n^\dagger A_n)$ is not a constant, but depends on the input state $|\psi_{in}\rangle$. Since $\mathcal{A}_{re}$ includes enough errors so that $P^{det} \geq 1 - \mathcal{O}(\epsilon^{t+1})$, when Eq. (4.134) is satisfied, we also have $\sum_{n \in \mathcal{K}_{re}} p_n \geq 1 - \mathcal{O}(\epsilon^{t+1})$ and $\sum_{n \in \mathcal{K}_{re}} p_n \lambda_n \geq 1 - \mathcal{O}(\epsilon^{t+1})$.

We now prove that $\sum_{n \in \mathcal{K}_{re}} p_n \lambda_n$ is a lower bound on the fidelity. Defining the *residue operator*

$$\pi_n = \sqrt{P_C A_n^\dagger A_n P_C} - \sqrt{p_n \lambda_n} P_C \,, \tag{4.136}$$

we find, for the operator norm of $\pi_n$,

$$0 \leq |\pi_n| \leq \sqrt{p_n} - \sqrt{p_n \lambda_n} \,, \tag{4.137}$$

and

$$A_n P_C = U_n(\sqrt{p_n \lambda_n} I + \pi_n) P_C \,. \tag{4.138}$$

The sufficiency of our conditions to obtain the desired fidelity may be proved as follows. Though Eq. (4.130) is not satisfied, as long as Eq. (4.135) is true, we can still define the *approximate* recovery operation

$$\mathcal{R}(\rho) = \sum_{k \in \mathcal{K}_{re}} R_k \rho R_k^\dagger + P_E \rho P_E \tag{4.139}$$

with $R_k = P_C U_k^\dagger$ being the *approximate* recovery operation for $A_k$, and $P_E$ is as defined in the case of exact error correction. For a pure input state $|\psi_{in}\rangle\langle\psi_{in}|$, applying $\mathcal{R}$ on $\mathcal{E}(|\psi_{in}\rangle\langle\psi_{in}|)$, and ignoring the last term which is positive definite produces an output with fidelity

$$
\begin{aligned}
\mathcal{F} &\equiv \min_{|\psi_{in}\rangle \in \mathcal{C}} \mathrm{Tr}\Big[\, |\psi_{in}\rangle\langle\psi_{in}|\ \mathcal{R}(\mathcal{E}(|\psi_{in}\rangle\langle\psi_{in}|))\,\Big] \\
&\geq \min_{|\psi_{in}\rangle \in \mathcal{C}} \sum_{k,n \in \mathcal{K}_{re}} |\langle\psi_{in}|U_k^\dagger A_n|\psi_{in}\rangle|^2 \,.
\end{aligned}
\tag{4.140}
$$

Omitting all terms for which $k \neq n$ and applying Eq. (4.138) gives

$$\mathcal{F} \geq \min_{|\psi_{in}\rangle \in \mathcal{C}} \sum_{n \in \mathcal{K}_{re}} |\langle\psi_{in}|\sqrt{p_n \lambda_n} + \pi_n|\psi_{in}\rangle|^2 \geq \sum_{n \in \mathcal{K}_{re}} p_n \lambda_n \,, \tag{4.141}$$

where in the last step, we have used Eq. (4.137). Hence, the fidelity is at least $\sum_{n \in \mathcal{K}_{re}} p_n \lambda_n \geq 1 - \mathcal{O}(\epsilon^{t+1})$ and the desired fidelity order is achieved as claimed.

An explicit procedure for performing this recovery is as follows. First, a measurement of the projectors $P_k \equiv U_k P_C U_k^\dagger$ is performed. Conditioned on the measurement result, $k$, the unitary operator $U_k$ is applied to complete the recovery.

### 4.2.3 4-bit Code Revisited

In terms of the approximate quantum error correction criteria Eqs. (4.134)-(4.135), we may understand why our four-bit amplitude damping quantum code works as follows. We present matrices with respect to the orthonormal basis ordered as:

$$|0000\rangle, |0011\rangle, |1100\rangle, |1111\rangle, |0111\rangle, |0100\rangle, \ldots \tag{4.142}$$

The projection operator $|0_L\rangle\langle 0_L| + |1_L\rangle\langle 1_L|$ onto the codeword space $\mathcal{C}$ is

$$P_C = \frac{1}{2} \begin{bmatrix} 1 & 0 & 0 & 1 \\ 0 & 1 & 1 & 0 \\ 0 & 1 & 1 & 0 \\ 1 & 0 & 0 & 1 \end{bmatrix}, \tag{4.143}$$

where the irrelevant null space is omitted. We are interested in the restriction of $A_k$ to $\mathcal{C}$, therefore we exhibit rows and columns in $A_k$ that have nontrivial contributions to $A_k P_C$. The operation element corresponding to no loss to the environment is

$$A_{0000} = \begin{bmatrix} 1 & 0 & 0 & 0 \\ 0 & 1-\gamma & 0 & 0 \\ 0 & 0 & 1-\gamma & 0 \\ 0 & 0 & 0 & (1-\gamma)^2 \end{bmatrix}. \tag{4.144}$$

The eigenvalues of $P_C A_{0000}^\dagger A_{0000} P_C$ are $(1-\gamma)^2$ and $\frac{1}{2}(1+(1-\gamma)^4)$. Interested readers can check for themselves that

$$A_{0000} P_C = U_{0000} \left[ (1-\gamma)I + (\gamma^2 + \mathcal{O}(\gamma^4))\tilde{\pi}_{0000} \right] P_C \tag{4.145}$$

(the order of $\gamma$ in $\pi_{0000}$ is factored out of $\tilde{\pi}_{0000}$) with the choice:

$$U_{0000} = \begin{bmatrix} \cos(\theta - \frac{\pi}{4}) & 0 & 0 & -\sin(\theta - \frac{\pi}{4}) \\ 0 & 1 & 0 & 0 \\ 0 & 0 & 1 & 0 \\ \sin(\theta - \frac{\pi}{4}) & 0 & 0 & \cos(\theta - \frac{\pi}{4}) \end{bmatrix} \tag{4.146}$$

$$\tilde{\pi}_{0000} \;=\; \begin{bmatrix} 1 & 0 & 0 & 0 \\ 0 & 0 & 0 & 0 \\ 0 & 0 & 0 & 0 \\ 0 & 0 & 0 & 1 \end{bmatrix} \tag{4.147}$$

where $\tan\theta = (1-\gamma)^2$, and only the nontrivial restriction to $\mathcal{C}$ is exhibited. The exact quantum error correction criteria are not satisfied, as $P_C A_{0000}^\dagger A_{0000} P_C$ has different eigenvalues. However, the difference is of order $\mathcal{O}(\gamma^2)$ and thus the relaxed condition Eq. (4.134) is satisfied.

For the error which describes losing a quantum from $n_1$, we have

$$A_{1000} = (1-\gamma)^{\frac{1}{2}}\sqrt{\gamma} \begin{bmatrix} 0 & 0 & 0 & 0 & 0 & 0 \\ 0 & 0 & 0 & 0 & 0 & 0 \\ 0 & 0 & 0 & 0 & 0 & 0 \\ 0 & 0 & 0 & 0 & 0 & 0 \\ 0 & 0 & 0 & 1-\gamma & 0 & 0 \\ 0 & 0 & 1 & 0 & 0 & 0 \end{bmatrix}. \tag{4.148}$$

The eigenvalues of $P_C A_{1000}^\dagger A_{1000} P_C$ are $\gamma(1-\gamma)$ and $\gamma(1-\gamma)^3$. The difference is $(2\gamma^2 - \gamma^3)(1-\gamma)$. We have the decomposition

$$A_{1000}P_C = \sqrt{\frac{(1-\gamma)\gamma}{2}} U_{1000} \left[ (1-\gamma)I + \gamma\tilde{\pi}_{1000} \right] P_C \tag{4.149}$$

$$U_{1000} \;=\; \begin{bmatrix} 1 & 0 & 0 & 0 & 0 & 0 \\ 0 & 1 & 0 & 0 & 0 & 0 \\ 0 & 0 & 0 & 0 & 0 & 1 \\ 0 & 0 & 0 & 0 & 1 & 0 \\ 0 & 0 & 0 & 1 & 0 & 0 \\ 0 & 0 & 1 & 0 & 0 & 0 \end{bmatrix} \begin{bmatrix} \frac{1}{\sqrt{2}} & 0 & 0 & -\frac{1}{\sqrt{2}} & 0 & 0 \\ 0 & \frac{1}{\sqrt{2}} & -\frac{1}{\sqrt{2}} & 0 & 0 & 0 \\ 0 & \frac{1}{\sqrt{2}} & \frac{1}{\sqrt{2}} & 0 & 0 & 0 \\ \frac{1}{\sqrt{2}} & 0 & 0 & \frac{1}{\sqrt{2}} & 0 & 0 \\ 0 & 0 & 0 & 0 & 1 & 0 \\ 0 & 0 & 0 & 0 & 0 & 1 \end{bmatrix}$$

$$\tilde{\pi}_{1000} \;=\; \begin{bmatrix} 0 & 0 & 0 & 0 & 0 & 0 \\ 0 & 0 & \frac{1}{\sqrt{2}} & 0 & 0 & 0 \\ 0 & 0 & \frac{1}{\sqrt{2}} & 0 & 0 & 0 \\ 0 & 0 & 0 & 0 & 0 & 0 \\ 0 & 0 & 0 & 0 & 0 & 0 \\ 0 & 0 & 0 & 0 & 0 & 1 \end{bmatrix}. \tag{4.150}$$

Other one loss cases are similar. For different $k$, the $U_k P_C$ matrices have non-zero entries in different rows, and are orthogonal to each other. Hence, all the approximate code criteria are satisfied. Using these explicit matrices, we obtain

$$\begin{aligned} \mathcal{R}(\mathcal{E}(|\psi_{in}\rangle\langle\psi_{in}|)) \;&\approx\; \sum_{k\in\mathcal{K}_{re}} P_C U_k^\dagger A_k |\psi_{in}\rangle\langle\psi_{in}| A_k^\dagger U_k P_C \\ &=\; (1-3\gamma^2)|\psi_{in}\rangle\langle\psi_{in}| + \dots. \end{aligned} \tag{4.151}$$

The fidelity is thus at least $1-3\gamma^2$, and is of the desired order.

The recovery procedure suggested in Eq. (4.151) contrasts with the decoding and recovery circuits in Section 4.2.1. It is an interesting exercise to check that the composition of the operations in Fig. 4.2A and 4.2B, followed by re-encoding the recovered qubit has the same effect on $\mathcal{C}$ as applying $U_{0000}^\dagger$ for recovery. For the case in which an emission occurs in the first qubit, the composition of operations in Fig. 4.2A and 4.2C, followed by re-encoding the recovered qubit has the same effect on $\mathcal{C}$ as preferentially damp out the $|n_3\rangle = |0\rangle$ component followed by applying $U_{1000}^\dagger$ for recovery. Note that it costs $2\gamma^2$ in the fidelity to remove the distortion.

### 4.2.4  Applications to other codes

Our approximate criteria may also be used to simplify code construction using a non-Pauli error basis. For example, consider the bosonic codes in Sections 4.1.2-4.1.8. For logical states $|c_1\rangle, |c_2\rangle, \dots$ of the form:

$$|c_l\rangle \;=\; \sqrt{\mu_1}|n_{11}n_{12}\dots n_{1m}\rangle$$

$$+ \quad \sqrt{\mu_2}|n_{21}n_{22}\ldots n_{2m}\rangle$$

$$+ \quad \cdots$$

$$+ \quad \sqrt{\mu_{N_l}}|n_{N_l 1}n_{N_l 2}\ldots n_{N_l m}\rangle , \qquad (4.152)$$

the original non-deformation conditions for correcting up to one loss of quantum require the following to be constant for all logical states:

$$\langle c_l|A_0^\dagger A_0|c_l\rangle \;=\; \sum_{i=1}^{N_l}(1-\gamma)^{RS_i}\mu_i \qquad (4.153)$$

$$\langle c_l|A_{0\cdots 1\cdots 0}^\dagger A_{0\cdots 1\cdots 0}|c_l\rangle \;=\; \sum_{i=1}^{N_l}(1-\gamma)^{RS_i-1}\gamma\mu_i n_{ij} . \qquad (4.154)$$

In the above, $RS_i = \sum_{j=1}^m n_{ij}$ is the total number of quanta in the $i$th QCS in $|c_l\rangle$. We have seen that it is difficult to find a solution for Eq. (4.154) when $RS_i$ is not constant for all $i$.

With the new criteria, it suffices for the following to be constant for all logical states:

$$\sum_{i=1}^{N_l}\gamma\mu_i n_{ij} \qquad \forall j . \qquad (4.155)$$

That is, the equality of the excitation number in all QCS in all logical states is relaxed to the equality of the *average* number of excitations over the QCS in each codeword. This provides an alternative explanation of why the five-bit code

$$|0_L\rangle \;=\; |00000\rangle + |11000\rangle - |10011\rangle - |01111\rangle$$

$$+ \quad |11010\rangle + |00110\rangle + |01101\rangle + |10101\rangle$$

$$|1_L\rangle \;=\; |11111\rangle - |00011\rangle + |01100\rangle - |10000\rangle$$

$$- \quad |00101\rangle + |11010\rangle + |10010\rangle - |01010\rangle \qquad (4.156)$$

can correct for one amplitude damping error (as described in Eq. (4.112)): although the code does not satisfy the exact non-deformation criteria Eq. (4.154) for the non-Pauli error basis, it satisfies the *approximate* ones leading to Eq. (4.155).

### 4.2.5    Stabilizer description for approximate error correction

Gottesman has subsequently given an explanation of approximate error correction of amplitude damping using the stabilizer formalism [56]. The amplitude damping operators $A_0$ and $A_1$ can be expressed in other error bases:

$$A_0 = I + \frac{1 - \sqrt{1-\gamma}}{2}(I - Z) = I + \frac{1 - \sqrt{1-\gamma}}{2}B \qquad (4.157)$$

$$A_1 = \frac{\sqrt{\gamma}}{2}X(1 - Z) = \frac{\sqrt{\gamma}}{2}A \qquad (4.158)$$

where $B \equiv (I-Z)$ and $A \equiv X(I-Z)$. The elements of the reversible set are tensor products of $I$, $A$, and $B$ with low weights. To correct for $t$ amplitude damping errors, terms up to order $\gamma^t$ are relevant. Therefore, we need to consider $E^\dagger F$ for all $E, F$ in the reversible set that contain $r$ factors of $A$ and $s$ factors of $B$ where $\frac{r}{2} + s \leq t$.

It is sufficient to use CSS codes with the $X$ generators capable of detecting $t$ $Z$ errors and the $Z$ generators correcting $t$ $X$ errors. If $E^\dagger F$ contains at least one $A$, it anticommutes with some $Z$ generators (the $Z$ factor in $A$ commutes with the $Z$ generator and does not affect the argument). If $E^\dagger F$ has no $A$ at all, it can have up to $t$ $B$ and therefore up to $t$ $Z$ which anticommutes with some $X$ generator. Comparing with correcting $t$ general errors, the requirement of correcting $t$ $Z$ errors is reduced to detecting $t$ errors only. As an example, the following code

$$\begin{array}{ccccccc}
X & X & X & X & X & X & X \\
Z & Z & Z & Z & I & I & I \\
Z & Z & I & I & Z & Z & I \\
Z & I & Z & I & Z & I & Z
\end{array} \qquad (4.159)$$

can correct for one amplitude damping error and encode 3 qubits.

It is interesting to note that the above construction which takes into account the approximate criteria and only detects $Z$ errors still does not admit a 4 bit code. This can be proved by elimination. First of all, no such 3 bit codes exist, as the minimal set of generators is $\{XXX, ZZI, IZZ\}$. To eliminate any 4 bit code, it suffices to show that at least 3 $Z$ generators are required for correcting up to one $X$ error in four. The minimum weight of the $Z$ stabilizers can either be 1 or 2. If the minimum weight is 1, without loss of generality, $ZIII$ is a generator. It takes at least 2 other $Z$ generators to handle $X$ errors in the last three qubits. If the minimum weight is 2, and $ZZII$ is a generator, it is obvious that no single $Z$ generator anticommutes with all of $IIXI$, $IIIX$ and $IIXX$. The lesson is, this construction using specific error model and approximate error correction is still not optimal, since the error description using $X$ and $Z$ does not exploit special properties of the $A_0, A_1$ operators.

Consider the four-bit code again. It is in fact the 4 bit CSS code we have ruled out:

$$
\begin{aligned}
M_1 &= X \quad X \quad X \quad X \\
M_2 &= Z \quad Z \quad I \quad I \\
M_3 &= I \quad I \quad Z \quad Z \\
\overline{X} &= X \quad X \quad I \quad I \\
\overline{Z} &= Z \quad I \quad Z \quad I
\end{aligned}
\tag{4.160}
$$

Recall that the difficulty is that no stabilizer anticommutes with $XXII$ and $IIXX$, so that the error correction criteria cannot be established for $A^\dagger AII$ and $IIA^\dagger A$. However, anticommutivity is not required to establish the criteria. In fact, $A^\dagger AII \times ZZII = -A^\dagger AII$, therefore, $P_C A^\dagger AII P_C = -P_C A^\dagger AII P_C = 0$ establishing the error correction criteria without anticommutivity. More precisely, $E^\dagger F$ can be negated by multiplication by a stabilizer (instead of by conjugation). This is unique to a non-Pauli basis, since no stabilizer can negate $E^\dagger F$ in the Pauli basis by multiplication. This can be generalized, and an immediate application is the delightful result that the Shor code can correct for two amplitude damping errors. It is trivial that the two $X$ generators can detect up to any two $Z$ errors. It remains to consider $E^\dagger F$ with up to 4 factors of $A$. Almost all of them anticommute with some $Z$ generators except for $AAA^\dagger IIIIII$ (and others with exactly one or two $A^\dagger$, and similar ones in the other two blocks). This is negated by the stabilizer $IZZIIIIII$. Similar arguments apply to all other cases.

This result is very remarkable: such a binary code correcting for 2 losses is very difficult to find in the first place and is also very lengthy to verify directly without the stabilizer description, and is impossible with the Pauli error basis.

## 4.3   Summary

We have constructed amplitude damping codes in bosonic and qubit systems which have rates better than any general codes:

- $|0_L\rangle = \dfrac{1}{\sqrt{2}}\Big[|04\rangle + |40\rangle\Big]$        $|1_L\rangle = |22\rangle$

- $|0_L\rangle = \dfrac{1}{\sqrt{2}}\Big[|0000\rangle + |1111\rangle\Big]$     $|1_L\rangle = \dfrac{1}{\sqrt{2}}\Big[|0011\rangle + |1100\rangle\Big]$

These examples demonstrate that choosing an appropriate error basis can potentially reduce the requirements in coding schemes. We also suggest an approximate method to enable code construction without the Pauli basis to be done more easily. Approximate error correction is particularly interesting because it is a property with no analog in digital classical error correction, as it makes use of slight non-orthogonality possible only between quantum states. It also extends the current scope of quantum error correction, which is closely related to digital classical error correction.

# Chapter 5

# Fault-tolerant logic gates

## 5.1  Motivation

We have seen in Section 3.4 that, to robustly perform quantum computation in the presence of noise, one needs to perform quantum gates and measurements directly on encoded states in a fault-tolerant manner. These fault-tolerant quantum gates and measurements must prevent a single error from propagating to more than one error in any code block, so that small correctable errors will not grow to exceed the correction capability of the code. This requirement greatly restricts the types of unitary operations which can be performed on the encoded qubits. We have seen how the Clifford group operations can be performed *bitwise* on CSS codes, and have seen how fault-tolerant measurements of certain class of operators can be performed. (These can also be performed on any stabilizer codes.) However, to obtain a universal set of gates, at least one additional gate has to be constructed, such as the Toffoli gate or the $\pi/8$ gate. The first construction of such gates was reported in [104], and was followed by many others [11, 73, 68, 57, 21]. Unfortunately, these constructions are given, but not systematically derived. Thus they cannot be easily generalized. The construction can also be complicated. This point can be better appreciated by considering the circuits for performing the Toffoli gate and the $\pi/8$ gate (denoted by $T$) in Fig. 5.1.

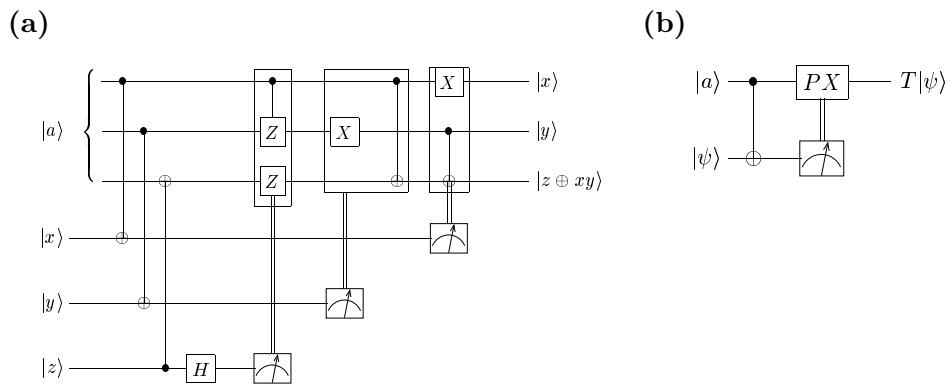

Figure 5.1: Fault-tolerant implementation of the Toffoli gate and the $\pi/8$ gate. $|a\rangle$ denotes the corresponding special ancilla states required. $|a\rangle = (|000\rangle + |010\rangle + |100\rangle + |111\rangle)/2$ for the Toffoli gate and $|a\rangle = \frac{1}{\sqrt{2}}(|0\rangle + e^{i\pi/4}|1\rangle)$ for the $\pi/8$ gate.

While the validity of these circuits can be directly verified, it is not apparent how they are constructed. Furthermore, all known constructions [1] share some intriguing similarities which are not understood. They all involve some special ancilla states, and special operations conditioned on some measurement outcomes.

It was first pointed out by Shor [104] that the use of operations conditioned on classical measurement outcomes is reminiscent to teleportation [16]. In teleportation, quantum states are sent using pre-shared EPR states, measurements made by the sender, classical communication performed, and operations made by the receiver conditioned on the measurement outcomes. Such connection between teleportation and fault tolerant gates were not understood, until teleportation is used as a basic primitive to systematically construct many fault tolerant gates in [59]. However, the resulting circuits are not as simple as prior *ad-hoc* constructions.

In this chapter, we uses a simpler primitive dubbed "one-bit-teleportation" to construct fault tolerant gates. This provides a systematic and unified construction of a large class of important fault-tolerant gates which parallel the simplest existing schemes and improve on many others.

We describe our basic primitive, one-bit teleportation, in Section 5.2. Its application to fault-tolerant gate construction is presented in Section 5.3, which is followed in Section 5.4 with specific circuits for the $\pi/8$, controlled-phase, and Toffoli gates. We summarize our results in Section 5.7.

---

[1] Except for the polynomial codes [11] and toric codes [68].

## 5.2 One-bit teleportation

We consider a class of "one-bit-teleportation" circuits, each of which is essentially the following swap circuit for two qubits:

$$(5.1)$$

Throughout this section, the first and second qubits refer to the registers with respective initial states $|0\rangle$ and $|\psi\rangle$. Since $X = HZH$, Eq. (5.1) is equal to:

$$(5.2)$$

Since measurement commutes with a controlled-quantum-gate when the control qubit is being measured [60], we have

$$(5.3)$$

where all notations are as defined in Section 2.2. Operations which are performed conditioned on classical measurement results are called classically-controlled operations. In Eq. (5.2), the two qubits are disentangled before the second Hadamard gate. Therefore, the second qubit can be measured without affecting the unknown state of the first qubit. Using

Eq. (5.3), Eq. (5.2) is equal to

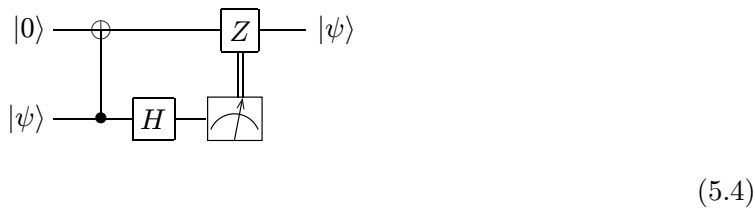

$$(5.4)$$

Applying Eq. (5.4) to $U^\dagger|\psi\rangle$, we obtain the left-hand-side of the following:

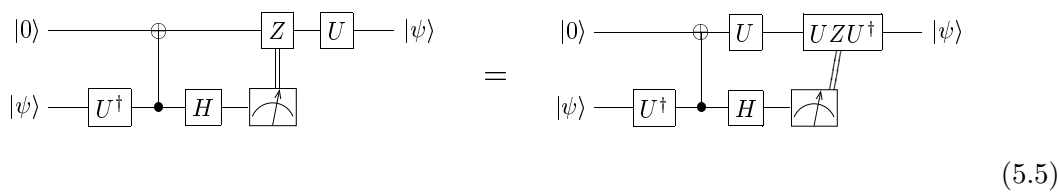

$$(5.5)$$

In particular, when $U = H$, the right-hand-side of Eq. (5.5) is equal to

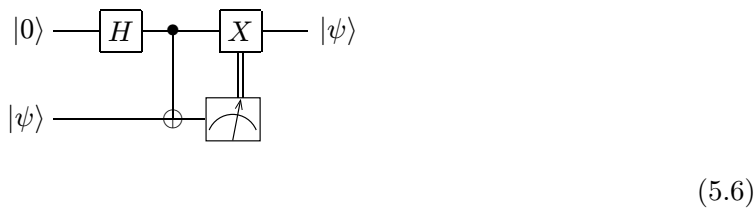

$$(5.6)$$

The circuits in Eqs. (5.4) and (5.6) are referred to as "$Z$-teleportation" and "$X$-teleportation" after the classically-controlled-operation involved. $X$ and $Z$-teleportation circuits can both be represented using the same general structure:

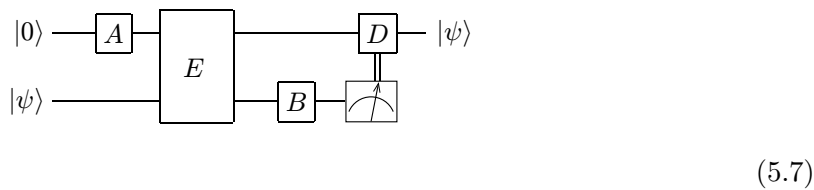

$$(5.7)$$

where for $Z$-teleportation, $A = I, B = H, D = Z$, and $E$ is a single CNOT with the first qubit as its target. For $X$-teleportation, $A = H, B = I, D = X$, and $E$ is a single CNOT with the first qubit as its control.

## 5.3 Fault-tolerant gate constructions using one-bit teleportation

In this section, we develop a general method for fault-tolerant gate construction using one-bit teleportation as a basic primitive. We will confine our attention to the self-dual doubly-even CSS codes [23, 108], although the results can be extended to any stabilizer code [59].

### 5.3.1 Fault-tolerant gate hierarchy

We first summarize the fault-tolerant gate hierarchy introduced in [59]. Let $C_1$ denote the Pauli group. $C_2$, the Clifford group, is the set of gates which map Pauli operators into Pauli operators under conjugation. We can recursively define an infinite class of quantum gates as

$$C_k \equiv \{U | U C_1 U^\dagger \subseteq C_{k-1}\}, \tag{5.8}$$

for $k \geq 2$. For every $k$, $C_{k-1} \subset C_k$ is strictly increasing.

It is now easy to understand what is involved in all existing schemes to complete the universality requirement. In all cases, one gate in the set difference, $C_3 - C_2$, is added to $C_2$ to form a universal set. These include the $\pi/8$ gate $T$ [21], ($T|x\rangle = e^{i\pi x/4}|x\rangle$ for $x \in \{0, 1\}$), the controlled-phase gate CP [69] (CP$|xy\rangle = i^{x \cdot y}|xy\rangle$ for $x, y = \{0, 1\}$) and the Toffoli gate [104]. We will see in the next few sections that there is a reduction which allows the $C_k$ gates to be recursively constructed from performing and measuring $C_{k-1}$ gates. This reduction is what makes it possible to complete the universality requirement out of the available $C_2$ primitives.

### 5.3.2 $C_3$ gate construction using one-bit teleportation

We now consider a general method to perform $C_3$ gates fault-tolerantly using the one-bit teleportation scheme as a primitive. Let $U \in C_3$ be an $n$-qubit gate, to be applied to $|\psi\rangle$, an encoded quantum state with $n$ *logical* qubits. One can first teleport $|\psi\rangle$ and then apply $U$ to the teleported version of $|\psi\rangle$ (why this is useful will be explained shortly). Each logical qubit can be teleported using either $Z$-teleportation or $X$-teleportation given by Eq. (5.7). We call the gates in Eq. (5.7) $A_i$, $B_i$, $E_i$ and $D_i$ for the teleportation of the $i$-th qubit. We relabel the tensor product $A_1 \otimes \cdots \otimes A_n$ as $A$, and similarly for $B$, $D$ and $E$. This is summarized in Eq. (5.9):

$$(5.9)$$

In Eq. (5.9), the symbol "$/^n$" represents a bundle of $n$ logical qubits. The measurement box outputs an $n$-bit classical outcome represented by the double line. The $i$-th classical bit controls whether $D_i$ is performed on the $i$-th logical state in the first register. For simplicity, we draw a box with label $D$ where the double line ends. If $Z$-teleportation is applied to the $i$-th logical qubit, $A_i = I, B_i = H, D_i = Z$, and $E_i$ is a CNOT with the $i$-th qubit in the first register as target; if $X$-teleportation is applied instead, $A_i = H, B_i = I, D_i = X$, and $E_i$ is a CNOT with the $i$-th qubit in the first register as control.

The usefulness of the extra teleportation of $|\psi\rangle$ before applying $U$ is the following. We can commute $U$ backwards in time, so that $U$ acts on the known ancilla state $A|0\rangle^{\otimes n}$. Commuting $U$ with each classically-controlled operation $D_i$ changes $D_i$ to $UD_iU^\dagger \in C_2$ as $D_i \in C_1$. For simplicity, we represent the combined controlled-operation as $UDU^\dagger$ in Eq. (5.10). Likewise, commuting $U$ with $E$ changes $E$ to $UEU^\dagger$. [2] To ensure $UEU^\dagger$ is in $C_2$ for an arbitrary $U \in C_3$, we consider $U$ and $E$ such that $[U, E] = 0$, in which case, $UEU^\dagger = E \in C_2$. Then Eq. (5.9) becomes

$$(5.10)$$

All the circuit elements outside the dotted box can be implemented fault-tolerantly. Inside the dotted box, we need not apply $U$ to $A|0\rangle^{\otimes n}$; instead, we can create the logical state $UA|0\rangle^{\otimes n}$ directly. As $|0\rangle^{\otimes n}$ has stabilizers $Z_i$ ($i = 1, \ldots, n$), where $Z_i$ is the encode $\overline{Z}$ acting on the $i$-th encoded qubit, the stabilizers of $UA|0\rangle^{\otimes n}$ are $UAZ_iA^\dagger U^\dagger = UA_iZ_iA_i^\dagger U^\dagger = UD_iU^\dagger$ (following the discussion after Eq. (5.9)). As $UD_iU^\dagger \in C_2$ (which can be performed bitwise in CSS codes) and have eigenvalues $\pm 1$, these can be fault-tolerantly measured using the method described in Section 3.4. Measuring $UD_iU^\dagger \; \forall i$ allows $UA|0\rangle^{\otimes n}$ to be prepared from any convenient state. Therefore, a beautiful reduction is obtained, from performing $U \in C_3$ to performing and measuring $UD_iU^\dagger$ which is in $C_2$. This completes the discussion on how to perform $U$ fault-tolerantly on any encoded state, provided $U \in C_3$ and teleportation circuits can be found to satisfy $[U, E] = 0$.

---

[2] We write $UEU^\dagger$ instead of $(U \otimes I^{\otimes n})E(U^\dagger \otimes I^{\otimes n})$ for simplicity. Unimportant identity operators are likewise suppressed throughout the discussion.

The above construction can be used to systematically construct interesting gates in $C_3 - C_2$. We now exhibit the examples of the $\pi/8$ gate, the controlled-phase gate and the Toffoli gate. Any one of these gates, together with $C_2$, form a universal set of gates.

## 5.4 Examples

### 5.4.1 The $\pi/8$ gate

The $\pi/8$ gate, $T$, has the matrix representation

$$T = \begin{bmatrix} 1 & 0 \\ 0 & e^{i\pi/4} \end{bmatrix} . \tag{5.11}$$

Note that $T$ is diagonal and commutes with the CNOT in the $X$-teleportation circuit. Therefore, we choose to apply $X$ teleportation to $|\psi\rangle$ and apply $T$ to the teleported $|\psi\rangle$:

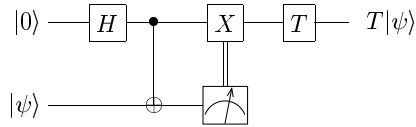

We commute $T$ backwards to obtain

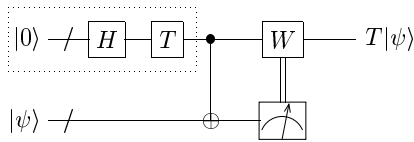

where $W = TXT^\dagger = \sqrt{i}PX$. It remains to prepare the ancilla state,

$$|\phi_+\rangle = TH|0\rangle = \frac{|0\rangle + e^{i\pi/4}|1\rangle}{\sqrt{2}} \tag{5.12}$$

with stabilizer $W$. We can prepare any convenient state, such as the logical $|0\rangle$, by measuring $Z$ fault-tolerantly. We replace $Z$ by $W$ in the stabilizer by measuring $W$ and "fixing" by applying $Z$ if we obtain a $-1$ eigenstate of $W$ (as $\{Z,W\} = 0$) (see Section 3.4). This completes the scheme to perform the encoded version of $T$ fault-tolerantly. We remark that we obtain the same scheme as in [21].

### 5.4.2   The controlled-phase gate

The controlled-phase gate, CP (defined in Section 5.3.1), acts on basis states according to $\mathrm{CP}|xy\rangle = i^{x \cdot y}|xy\rangle$. $\mathrm{CP} \in C_3$, and together with $H$ and CNOT, form a universal set of gates [69, 88]. We use the following circuit symbol for CP:

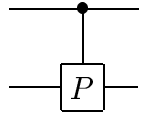

$$\tag{5.13}$$

CP commutes with $Z_i$ and it acts on $X_i$ ($i = 1, 2$) as follows:

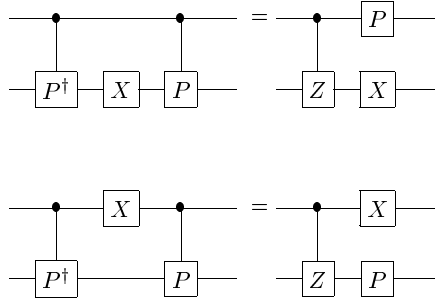

$$\tag{5.14}$$

To perform CP, we teleport $|\psi\rangle$ and apply CP. We apply $X$-teleportation to both qubits such that the CNOTs in the circuit commute with CP. As the circuit acts coherently on the input, it suffices to consider an arbitrary basis state $|xy\rangle$.

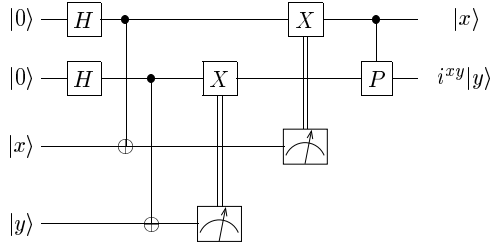

(5.15)

Commuting CP backwards using the commutation rules in Eq. (5.14), we obtain a circuit to perform CP

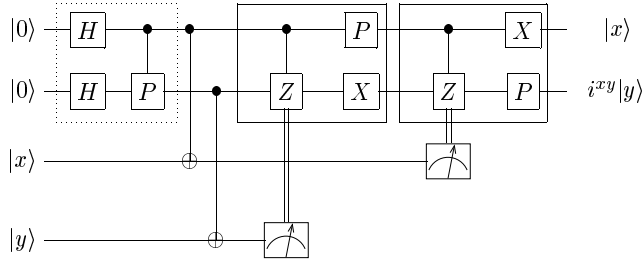

(5.16)

where the double lines control all the operations in the corresponding boxes. It remains to prepare the special ancilla state $|a\rangle$:

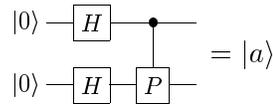

with stabilizers $W_i = \text{CP} X_i \text{CP}^\dagger$ for $i = 1, 2$. We can conveniently prepare the state $|a'\rangle$:

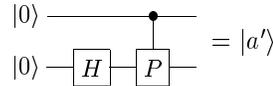

The stabilizer of $|a'\rangle$ can be generated by $Z_1$ and $X_2$ since CP has no effect when the first bit is $|0\rangle$. Therefore, $|a'\rangle$ can easily be prepared by measuring $Z_1$ and $X_2$. On the other hand, the stabilizer of $|a'\rangle$ is generated by $Z_1 = \text{CP} Z_1 \text{CP}^\dagger$ and $W_2 = \text{CP} X_2 \text{CP}^\dagger$. Therefore,

we only need to replace $Z_1$ by $W_1$, by measuring $W_1$ to obtain $|a\rangle$. This is possible because $\{W_1, Z_1\} = 0$ and $[W_1, W_2] = 0$ by construction. This ancilla completes the fault-tolerant construction of the controlled-phase gate.

### 5.4.3   The Toffoli gate

To construct the Toffoli gate (controlled-controlled-NOT), we begin with some useful commutation rules:

$$
\tag{5.17}
$$

$$
\tag{5.18}
$$

Since the Toffoli gate is diagonalized by a Hadamard transform on the target bit, using $X$ ($Z$) teleportation for the control (target) bit ensures the three CNOTs to commute with the Toffoli gate.

$$
\tag{5.19}
$$

Commuting the Toffoli gate backwards using Eqs. (5.17)-(5.18), Eq. (5.19) is equivalent to

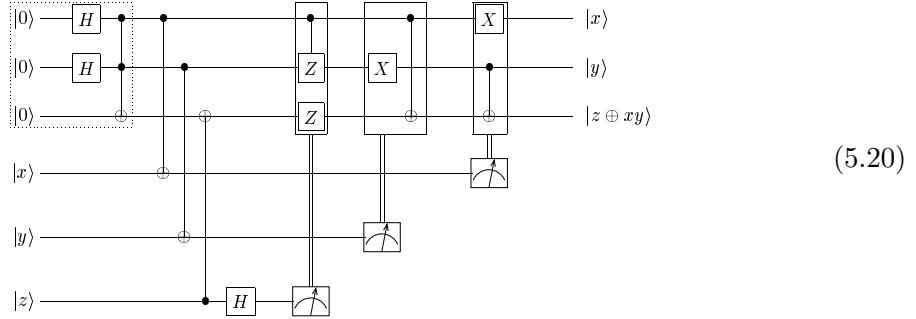

$$(5.20)$$

It remains to prepare the ancilla $|a\rangle$:

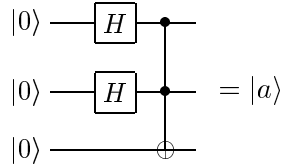

with stabilizers $W_1 = X_1 \otimes \text{CNOT}_{23}$, $W_2 = X_2 \otimes \text{CNOT}_{13}$ and $W_3 = Z_3 \otimes \text{CZ}_{12}$, where CZ denotes a controlled-$Z$, and the *ordered* subscripts for CNOT and CZ specify the control and target bits. We start with the state $|a'\rangle$:

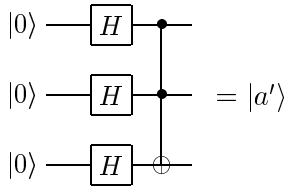

on which the Toffoli gate has no effect. Therefore $|a'\rangle$ has stabilizers generated by $X_i$ for $i = 1, 2, 3$ and can easily be prepared. At the same time, the generators of the stabilizer can be given by $W_1$, $W_2$ and $X_3$. Again, $X_3$ and $W_3$ anticommute due to the addition of $H$ in the third qubit, but the $W_i$ are mutually commuting. Hence, $|a\rangle$ can be prepared by replacing $X_3$ by $W_3$ in the stabilizer by measurement of $W_3$. This completes the ancilla preparation and therefore the gate construction.

Note that the ancilla and the quantum circuit derived are the same as those in Shor's original construction [104]. The one-bit teleportation scheme elucidates the choice of the ancilla state and the procedure in [104].

## 5.5  Recursive construction

In this section, we discuss what gates can be constructed with one-bit teleportation as a primitive. We extend our discussion to gates in $C_k$ and characterize a class of gates which can be recursively constructed.

We will prove by induction that the diagonal subset of $C_k$, defined by $F_k = \{U \in C_k$ and $U$ is diagonal$\}$, can be recursively constructed. First of all, for $U \in F_k$, we apply $X$-teleportation to each logical qubit to ensure commutivity with the CNOTs. Second, for $U \in F_k$, and $D \in C_1$, $UDU^\dagger = \tilde{U}D$ where $\tilde{U} \in F_{k-1}$ [54]. If the gates in $F_{k-1}$ can be performed, the classically-controlled operator $UDU^\dagger$ for $U \in F_k$ can also be performed. Third, it can be proved by induction that $UA|0\rangle^{\otimes n}$ can be prepared fault-tolerantly [59]. [3] Finally, the gates in $F_2 \subset C_2$ have transversal implementation. By induction, all the gates in $F_k$ can be performed fault-tolerantly by a recursive construction.

The sets $F_k$ contain many interesting gates, such as the single qubit $\pi/2^k$ rotations $V^k = \mathrm{Diag}(1, e^{i\pi/2^k})$ and the controlled-$V^{k-1} = \mathrm{Diag}(1, 1, 1, e^{i\pi/2^{k-1}})$, which are used in the quantum Fourier transform circuit [103, 39] essential to Shor's factoring algorithm [103]. $F_k$ also includes the multiple-qubit gates $\Lambda_n(V^l)$ for $n + l \leq k$ [54], where $\Lambda_n(V^l)$ applies $V^l$ to the $(n+1)$-th qubit if and only if the first $n$ qubits are all in the state $|1\rangle$. By the closure property of $F_k$ [54], all products of $\Lambda_n(V^l)$ for $n + l \leq k$ are in $F_k$. For small $k$, recursive construction can be more efficient than approximating these gates to an equal accuracy using a universal set of fault tolerant quantum logic gates.

The gates in $F_k$ are not the only ones which can be constructed using the one-bit teleportation scheme. If $U \in C_3$ is related to an element in $F_3$ by conjugation by Hadamard gates in the $i_1$-th, ..., $i_l$-th qubits, one can apply $X$ teleportation to those qubits and $Z$ teleportation to the rest to ensure $[U, E] = 0$. The Toffoli gate is an example. More generally, $C_3$ gates in the form $U = G_b V G_a$ for $V \in F_3$ and $G_a, G_b \in C_2$ can be performed simply by performing $G_a, G_b$ directly before and after the scheme for $V$. The controlled-Hadamard gate is an example.

---

[3] Since states with stabilizers in $F_k$ can be prepared if cat-state-controlled-$F_k$ operations can be performed, it suffices to prove the latter. For the induction step, suppose the cat-state-controlled-$F_{k-1}$ operations can be performed, one can also prepare states with stabilizers in $F_{k-1}$. Together, they can be used to perform any cat-state-controlled-$F_k$ operation, completing the induction step. The base case $F_2$ is true.

Note that so far, we only give sufficient conditions for a gate to be implemented by one-bit teleportation. The necessary conditions are intriguing, but due to our present lack of knowledge about the $C_k$ gates, they are far from being understood.

## 5.6    Extensions

We briefly mention two other applications of the one-bit teleportation circuits beyond fault-tolerant gate constructions.

### Hybrid circuits

The one-bit teleportation circuits can be used to convert data types. For example, the data type can be unencoded qubits, qubits encoded with various codes or some cat-states. Consider Eqs. (5.4) and (5.6) again. They can convert the input data type to the data type of the ancilla, if the format of the operations are chosen properly. One immediate application is to encode qubits. Another application is to change some encoded data from a code $\mathcal{C}_1$ to another $\mathcal{C}_2$ without ever decoding the data.

### Distributed computation

The current construction is also well suited to explain distributed computation, in which certain non-local operations are not allowed, and have to be performed using pre-shared entangled states and classical communications. Teleportation is one such protocol. More can be found in [115].

## 5.7    Conclusion

We have presented a systematic technique to construct a variety of quantum operations, by using a primitive one-bit teleportation scheme to reduce difficult gate constructions to measurements and ancilla state preparations. We applied this technique to fault-tolerant quantum computation, and have demonstrated simple derivations of the $\pi/8$, controlled-phase, and Toffoli gates, and have gained better understanding of how $C_3$ gates are made possible. These constructions are easily generalized to realize an infinite family of gates. The possibility to directly construct gates outside $C_2$ without going through the universality argument can bring large reductions in the resources required for quantum computation. Clearly, this means that one-bit teleportation may be useful for designing and optimizing computation and communication protocols [35, 37]. Even more intriguing, perhaps, is that this result gives us a first glimpse at what might someday be a standard architecture for a quantum computer: a simple assembly of one-bit teleportation primitives, capable of universal quantum computation on quantum data, given the assistance of standard quantum states which are obtained as commercial resources. Definition of such an architecture could be pivotal in the development of this field, much as the von Neumann or Harvard architectures [65] were important in classical computation.

# Part III

# NMR Quantum Computation

# Chapter 6

# Theory of NMR Quantum Computation

## 6.1  Introduction

In this part of the Dissertation, we turn to the more difficult issue of realizing quantum information processing. The obstacle to building such devices is more than just decoherence, which can at least be overcome theoretically. The problem is the many simultaneous requirements of quantum information processing which contradict each other. It is necessary that macroscopically controlled measurements and manipulations can be made on microscopic quantum degrees of freedom, to accomplish certain basic tasks: (1) prepare the system in a fiducial initial state (a pure state such as the ground state), (2) perform a universal set of logic gates, (3) implement projective measurement to read out the computational results and (4) maintain long coherence times. Many candidate implementations have been proposed, but very few have advanced sufficiently to experimentally demonstrate a multi-qubit logic gate. Most experimental problems are believed to be technological, and will be resolved in the future. However, it is highly desirable to have an immediately accessible system to put our theories to test and to motivate efforts in potentially useful problems. Solution NMR quantum computation at room temperature was proposed in late 1996 as such a possibility (by Gershenfeld and Chuang [53] and independently by Cory *et al* [40]).

Solution NMR is a limited computation model, yet it allows a tradeoff between resources and computation power. Implementation is relatively straightforward for a system of up to 10-20 qubits, and systems of several qubits are immediately available. Moreover, problems in NMR are well characterized and therefore present well-defined and meaningful challenges to the community. (The problems in most other implementations are not yet understood or identified, due to the meager experimental data. Ironically, NMR quantum computation has received more skepticism because the problems are better understood.)

In this part of the Dissertation, some of the theoretical problems in solution NMR quantum computation are discussed and resolved. It also describes some interesting experiments. This chapter describes the original proposal in [53] as well as contributions from this Dissertation that are extensions of the original proposal. Chapters 7 and 8 present two primary results of this Dissertation related to the subject.

## 6.2   Primitives in NMR quantum computation

Nuclear spin systems are good candidates for quantum computers for many reasons. Nuclear spins can have long coherence times. Manipulation by complex sequences of RF pulses can be carried out easily using modern spectrometers. Coupled operations are built in as coupling of spins within molecules. However, the signal from a single spin is too weak to be detected using current technology. Bulk NMR quantum computation addresses the detection problem by using a bulk sample of identical and independent spin systems, such as molecules in solution. These identical systems (quantum computers) run the same macroscopically defined quantum algorithm in classical parallelism. However, projective measurement of each individual system is impossible. Moreover, a pure initial state cannot be easily prepared. The breakthrough in [53, 40] is to realize that computation can be performed at room temperature starting with thermal initial states by utilizing the signal of the small *excess* ground state population resembling that of a pure state. Moreover, existing quantum algorithms can be modified to use ensemble measurements. These observations simplify the implementation to the extent that it becomes immediately realizable. The rest of this section describes logic gates, measurements and initial states in NMR. How the small excess population evolves is described in Section 6.3. Adaptations to enable quantum process tomography are described in Section 6.4. How to manipulate the small excess population to compute properly is described in Section 6.5. A short review in signal strength and scaling is given in Section 6.6. Finally, the possibility to directly compute on thermal inputs will be discussed in Section 6.7. We aim to develop the background for Chapters 7 and 8 as well as to elucidate limitations in NMR and methods to resolve them.

### 6.2.1 The quantum system (hardware)

We shall consider a physical system which consists of a solution of identical molecules. Each molecule has $n$ magnetically inequivalent nuclear spins which serve as qubits. A static magnetic field is applied externally along the $+\hat{z}$ direction. This magnetic field splits the energy levels of the spin states aligned with and against the external field. Let $|0\rangle$ and $|1\rangle$ be the ground and excited states. The Hamiltonian for the Zeeman splitting is given in the energy eigenbasis as:

$$\mathcal{H}_{\text{Z}} = -\frac{1}{2} \sum_i \omega_i Z^{(i)}, \qquad (6.1)$$

where $i$ is the spin index, $\omega_i/2\pi$ is the *Zeeman frequency* for the $i$-th spin, and $Z^{(i)}$ is the usual notation for $\sigma_z$ acting on the $i$-th spin. The convention $\hbar = 1$ is used. We assume the spins have very different Zeeman frequencies, a situation loosely termed as "heteronuclear".

Nuclear spins can interact via the dipolar coupling or the indirect coupling mediated by electrons [10, 107]. If the molecules tumble fast and isotropically in the solution, dipolar coupling and the tensor part of the indirect coupling will be averaged away. In any case, in the presence of a strong external magnetic field, only the *secular* part (the energy conserving terms which commute with $\mathcal{H}_{\text{Z}}$) is important [10, 107]. For a heteronuclear system, the resulting coupling (known as the $J$ coupling) takes the form

$$\mathcal{H}_{\text{c}} = \sum_{i<j} g_{ij} Z^{(i)} \otimes Z^{(j)}, \qquad (6.2)$$

which is independent of the exact original coupling. In Eq.(6.2), $g_{ij}$ denotes the coupling constant between the $i$-th and the $j$-th spins. We also write $g_{ij} = \frac{\pi J_{ij}}{2}$, because the spectral lines of the $i$-th and the $j$-th spins are split by $J_{ij}$ Hz.

Combining the Zeeman and coupling terms, the *reduced* Hamiltonian for our system is well approximated by [10, 107]

$$\mathcal{H} = \mathcal{H}_{\mathrm{Z}} + \mathcal{H}_{\mathrm{c}} + \mathcal{H}_{\mathrm{env}} \tag{6.3}$$

where $\mathcal{H}_{env}$ represents coupling to the reservoir, such as interactions with other nuclei, and higher order terms in the spin-spin coupling.

As an example, the energy diagram for the two-qubit case is shown in Fig. 6.1.

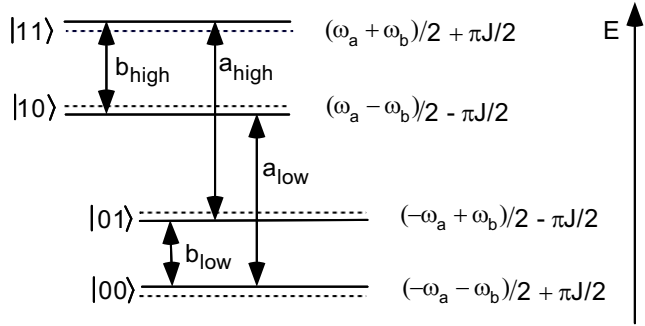

Figure 6.1: Energy diagram for the two-spin nuclear system. $a$ and $b$ are spin indices, $J = J_{ab}$ is the unique coupling constant. The transitions labeled $a_{low}$, $a_{high}$, $b_{low}$ and $b_{high}$ give rise to 4 spectral lines, and represent transitions $(|0\rangle \leftrightarrow |1\rangle)|0\rangle$, $(|0\rangle \leftrightarrow |1\rangle)|1\rangle$, $|0\rangle(|0\rangle \leftrightarrow |1\rangle)$ and $|1\rangle(|0\rangle \leftrightarrow |1\rangle)$ respectively.

### 6.2.2 Universal set of quantum logic gates

We consider the universal set of any coupled two-qubit operation together with the set of all single qubit transformations [47, 12, 45, 13] (also Section 2.2). Both requirements are satisfied in NMR as follows.

**Single qubit operations**

Single qubit operations can be induced by pulsed radio frequency (RF) magnetic fields, oriented in the $\hat{x}\hat{y}$-plane perpendicular to the static field. An RF pulse can selectively address spin $i$ by oscillating at angular frequency $\omega_i$. An RF pulse along the axis $\hat{\eta}$ induces the rotation operator $e^{-i\frac{\theta}{2}\vec{\sigma}\cdot\hat{n}}$, where $\theta$ is proportional to the pulse duration and amplitude. The rotation operator transforms the Bloch vector given by Eq. (2.29) according to the right-hand-rule.

For example, the Pauli matrix $X = ie^{-i\frac{\pi}{2}X}$ can be performed in a single $\pi$ pulse about $\hat{x}$. The Hadamard gate $H = ie^{-i\frac{\pi}{2}\sigma_x}e^{-i\frac{\pi}{4}\sigma_y}$ can be implemented by a $\pi/2$ pulse about $\hat{y}$ followed by a $\pi$ pulse about $\hat{x}$. This is an example of a *pulse sequence*. (Both $X^{(i)}$, $Z^{(i)}$ or the full notation $\sigma_j^{(i)}$ may be used in this part of the Dissertation.)

**Coupled operations**

Coupled logic gates can be naturally performed by the time evolution of the system. However, the Hamiltonian in Eq. (6.3) does not couple specific pairs of qubits. Rather, all couplings occur simultaneously along with the intended one. The fundamental tasks of turning on and off specific coupling terms are called "recoupling" and "decoupling". In Chapter 7, we show how to recouple and decouple efficiently. For now, we simply assume that individual coupling term can be selectively turned on. To perform a coupled operation between spins $i$ and $j$, we turn on the coupling term $g_{ij}Z^{(i)} \otimes Z^{(j)}$ for time $t$, leading to the evolution or logic gate $e^{-ig_{ij}tZ^{(i)}\otimes Z^{(j)}}$. Entanglement can be created because the evolution depends on the state of *both* spins. A frequently used coupled "operation" is $Z\!\!\!Z_{ij} \equiv e^{-i\frac{\pi}{4}Z^{(i)}\otimes Z^{(j)}}$, which corresponds to an evolution time of $t = \frac{\pi}{4g_{ij}} = \frac{1}{2J_{ij}}$. Appending $Z\!\!\!Z_{ij}$ with the *single qubit rotations* $e^{i\frac{\pi}{4}Z^{(i)}}$ and $e^{i\frac{\pi}{4}Z^{(j)}}$ leads to the unitary operation

$$\chi_{ij} = e^{i\pi/4}\begin{bmatrix} 1 & 0 & 0 & 0 \\ 0 & 1 & 0 & 0 \\ 0 & 0 & 1 & 0 \\ 0 & 0 & 0 & -1 \end{bmatrix}, \tag{6.4}$$

which is the controlled-z operation acting on spins $i$ and $j$. Together with the set of all single qubit transformations, the $\chi_{ij}$'s complete the requirement for universality. For instance, CNOT$_{ij} = H^{(j)} \chi_{ij} H^{(j)}$ can be implemented by concatenating the sequences for each constituent operation. It is crucial in decoupling and recoupling that the free unitary evolution can be reversed.

### 6.2.3  Measurement

The measured quantity in NMR experiments is the time varying voltage induced in a pick-up coil in the $\hat{x}\hat{y}$-plane:

$$V(t) = -V_0 \, \text{Tr} \left[ e^{-i\mathcal{H}t} \rho(0) e^{i\mathcal{H}t} \times \sum_i ( \, i\sigma_x^{(i)} + \sigma_y^{(i)} \, ) \right]. \tag{6.5}$$

The signal $V(t)$, known as the *free induction decay* (FID), is recorded with a phase-sensitive detector. In Eq.(6.5), the onset of acquisition of the FID is taken to be $t = 0$.

In the following, we consider a two-qubit system, with spin labels $a$ and $b$ and $J_{ab} = J$. The generalization to $n$ qubits is straightforward. Suppose the density matrix $\rho(0)$ has Pauli decomposition $\rho(0) = \sum_{i,j=0}^{3} c_{ij}\sigma_i \otimes \sigma_j$ (recall $\sigma_{0,1,2,3} = I, \sigma_{x,y,z}$). One can calculate $V(t)$ as a function of $t$ and the coefficients $c_{ij}$ using Eq. (6.5). Furthermore, if one takes the Fourier transform of $V(t)$, one obtains in the spectrum four peaks at frequencies $\frac{\omega_a}{2\pi} + \frac{J}{2}$, $\frac{\omega_a}{2\pi} - \frac{J}{2}$, $\frac{\omega_b}{2\pi} + \frac{J}{2}$, $\frac{\omega_b}{2\pi} - \frac{J}{2}$, with corresponding integrated areas ("peak integrals")

$$I_{a_{high}} = -\Big[ i(c_{10} - c_{13}) + c_{20} - c_{23} \Big] \tag{6.6}$$

$$I_{a_{low}} = -\Big[ i(c_{10} + c_{13}) + c_{20} + c_{23} \Big] \tag{6.7}$$

$$I_{b_{high}} = -\Big[ i(c_{01} - c_{31}) + c_{02} - c_{32} \Big] \tag{6.8}$$

$$I_{b_{low}} = -\Big[ i(c_{01} + c_{31}) + c_{02} + c_{32} \Big]. \tag{6.9}$$

Note that the expression $c_{10} - c_{13}$ occurring in the *high* frequency line of spin $a$ is the coefficient of $\sigma_x \otimes |1\rangle\langle 1|$ in $\rho(0)$; $c_{10} + c_{13}$ in the *low* frequency line of spin $a$ is the coefficient of $\sigma_x \otimes |0\rangle\langle 0|$. Likewise, $c_{20} - c_{23}$ is the coefficient of $\sigma_y \otimes |1\rangle\langle 1|$ and $c_{20} + c_{23}$ is the coefficient of $\sigma_y \otimes |0\rangle\langle 0|$. These quantities signify the transitions $|0\rangle \leftrightarrow |1\rangle$ for spin $a$ conditioned on spin $b$ being in $|1\rangle$ or $|0\rangle$. Similar observations hold for the high and low frequency lines of spin $b$ (see Fig. 6.1).

**Quantum state tomography**

From Eqs. (6.6)-(6.9), we can determine the coefficients $c_{10}, c_{13}, c_{20}, c_{23}, c_{01}, c_{31}, c_{02}, c_{32}$. It is also possible to determine other $c_{ij}$ in the following manner.

Suppose a $\pi/2$ $\hat{x}$-pulse is applied to the first spin right before acquisition, inducing the transformation $\rho(0) = \sum_{ij} c_{ij}\sigma_i \otimes \sigma_j \to \rho'(0) = \sum_{ij} c'_{ij}\sigma_i \otimes \sigma_j$. It is easily checked that $c'_{3j} = c_{2j}$ and $c'_{2j} = -c_{3j}$. Therefore, the acquisition of $\rho'(0)$ yields $c'_{10}, c'_{13}, c'_{20}, c'_{23}, c'_{01}, c'_{31}, c'_{02}, c'_{32}$ or $c_{10}, c_{13}, -c_{30}, -c_{33}, c_{01}, c_{21}, c_{02}, c_{22}$. Note that $c_{30}, c_{33}, c_{21}, c_{22}$ add to what can be determined in Eqs. (6.6)-(6.9) without the extra pulse before acquisition. Applying other *readout pulses* before acquisition allows different sets of readout coefficients to be obtained.

To perform quantum state tomography – that is, to determine $\rho(0)$ – one can prepare $\rho(0)$ nine times and apply no pulses or $\pi/2$ pulses about the $\hat{x}$ or $\hat{y}$ directions on the two spins independently to find all $c_{ij}$ $\forall (i,j) \neq (0,0)$.

The method for state tomography can be generalized to $n$ qubits provided the coupling between any two spins is non-negligible. In general the resources required are exponential in the number of qubits, since the number of parameters to be determined in the density matrix is exponential in $n$. [1]

### 6.2.4 Thermal Initial States

In bulk NMR quantum computation at room temperature, a pure initial state is not available due to large thermal fluctuations ($\hbar\omega_i \ll k_B T$ the thermal energy). Instead, a convenient class of initial states are the states at thermal equilibrium (thermal states). The thermal state density matrix $\rho_{th}$ is diagonal in the energy eigenbasis, with diagonal entries proportional to the Boltzmann factors:

$$\rho_{th} = \frac{1}{\mathcal{Z}} e^{-\frac{\mathcal{H}}{k_B T}} \tag{6.10}$$

---

[1] For a system of $n$ qubits, there are $n2^{n-1}$ spectral lines if all pairwise couplings are non-negligible. However, there are $2^{2n} - 1$ free parameters in the density matrix to be determined.

where $\mathcal{Z}$ is the partition function normalization factor. At room temperature, $\langle\frac{\mathcal{H}}{k_B T}\rangle \approx 10^{-4}$, $\mathcal{Z} \approx \text{Dim}(\rho_{th}) = 2^n$ and $\rho_{th} = \frac{1}{2^n}(I - \frac{\mathcal{H}}{k_B T})$ to first order. Moreover, since $\langle\mathcal{H}_c\rangle \ll \langle\mathcal{H}_Z\rangle$,

$$\rho_{th} \approx \frac{1}{2^n}\left(I - \frac{\mathcal{H}_Z}{k_B T}\right) \tag{6.11}$$

### 6.2.5 Example

As an example of the above theories, consider applying the pulse sequence in Fig. 6.2 to the thermal state of the 2-spin system considered in previous sections.

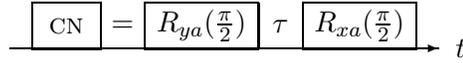

Figure 6.2: Pulse sequence for CN. Time runs from left to right. $R_{ij}(\theta)$ stands for a rotation of angle $\theta$ about the $i$-axis for spin $j$. $\tau$ stands for a time evolution of duration $\frac{1}{2J}$ which results in the operation $\mathbb{Z}_{ab}$.

In a heteronuclear system, the pulses are short compared to other relevant time scales. Therefore, other changes of the system during the pulses are ignored. The unitary operation implemented by the above sequence is given by ($a, b$ are the first and second qubits)

$$\begin{aligned} \text{CN} &= e^{-i\frac{\pi}{4}\sigma_x \otimes I} e^{-i\frac{\pi}{4}\sigma_z \otimes \sigma_z} e^{-i\frac{\pi}{4}\sigma_y \otimes I} \tag{6.12} \\ &= \frac{1}{\sqrt{2}}\begin{bmatrix} 1-i & 0 & 0 & 0 \\ 0 & 0 & 0 & -1-i \\ 0 & 0 & 1+i & 0 \\ 0 & 1-i & 0 & 0 \end{bmatrix}, \tag{6.13} \end{aligned}$$

similar to CNOT$_{ba}$ up to phase factors on each computational basis state.

The thermal initial state is given by Eq. (6.11). Since the identity component is invariant under unitary transformations and gives no signal, we only consider the term proportional to $-\mathcal{H}_Z$, and omit the constant of proportionality:

$$\begin{aligned} \rho_{th} &\sim \frac{\omega_a}{2}\sigma_z \otimes I + \frac{\omega_b}{2} I \otimes \sigma_z \tag{6.14} \\ &= \frac{1}{2}\text{Diag}(\omega_a + \omega_b, \ \omega_a - \omega_b, \ -\omega_a + \omega_b, \ -\omega_a - \omega_b). \tag{6.15} \end{aligned}$$

The sequence transforms $\rho_{th}$ to

$$
\begin{aligned}
\rho_{cn} &= \text{CN}\rho_{th}\text{CN}^\dagger &&(6.16)\\
&= \frac{1}{2}\,\text{Diag}(\omega_a + \omega_b,\ -\omega_a - \omega_b,\ -\omega_a + \omega_b,\ \omega_a - \omega_b) &&(6.17)\\
&= \frac{\omega_a}{2}\,\sigma_z \otimes \sigma_z + \frac{\omega_b}{2}\,I \otimes \sigma_z\,, &&(6.18)
\end{aligned}
$$

in which the populations of $|01\rangle$ and $|11\rangle$ are interchanged. Note that CN and $\text{CNOT}_{ba}$ effect the same evolution *on the thermal state*, because the extra phases in CN are irrelevant for a state diagonal in the computational basis. This fact can be used to simplify the pulse sequences for the initialization procedures to be discussed in Section 6.5.

By inspection of Eqs. (6.14) and (6.18), it can be seen that $\rho_{th}$ and $\rho_{cn}$ have zero peak integrals given by Eqs. (6.6)-(6.9). To obtain information about the states, a readout pulse $R_{xa}(\frac{\pi}{2})$ can be applied to transform the two states to

$$
\begin{aligned}
\rho'_{th} &= -\frac{\omega_a}{2}\,\sigma_y \otimes I + \frac{\omega_b}{2}\,I \otimes \sigma_z &&(6.19)\\
\rho'_{cn} &= -\frac{\omega_a}{2}\,\sigma_y \otimes \sigma_z + \frac{\omega_b}{2}\,I \otimes \sigma_z\,. &&(6.20)
\end{aligned}
$$

In $\rho'_{th}$ is a term $\sigma_y \otimes I$ with coefficient $c'_{20} = -\frac{\omega_a}{2}$ which contributes to two spectral lines at $\frac{\omega_a}{2\pi} \pm \frac{J}{2}$ with equal and positive, real peak integrals. The readout pulse transforms the unobservable coefficient $c_{30}$ in $\rho_{th}$ to the observable $-c'_{20}$ in $\rho'_{th}$, yielding information on *the state before the readout pulse*. Similarly, $\rho'_{cn}$ has a $\sigma_y \otimes \sigma_z$ term with coefficient $c'_{23} = -\frac{\omega_a}{2}$ which gives rise to two spectral lines with real and opposite peak integrals (Fig. 6.3). Outputs in NMR experiments are peak integrals of this type carrying information on the states.

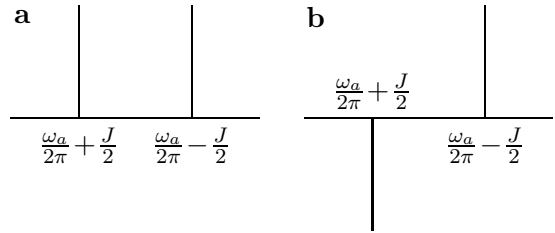

Figure 6.3: The spectra of spin $a$ after a readout pulse (a) on the thermal state $\rho_{th}$ and (b) on $\rho_{cn}$.

## 6.3   Deviation density matrix and effective evolutions

The thermal initial state in Eq. (6.11) decomposes into the identity term $\frac{I}{2^n}$ and a small traceless term $-\frac{1}{2^n}\frac{\mathcal{H}_Z}{k_B T}$ called the *deviation density matrix*. As the identity does not contribute to any signal, it is the *induced effective evolution* on the deviation which is observed experimentally, rather than the original quantum operation. In this section, we discuss the relation between the observed induced evolution and the original quantum operation, due to Nielsen [86].

The problem to infer the original quantum operation from the observed effective evolution in NMR will be addressed in Section 6.4, as original contribution from this Dissertation [27]. We apply the theory in effective evolution to analyze quantum error correction in NMR in Chapter 8.

As it turns out, the distinction between the original quantum operation and the observed induced evolution can be neglected in most applications. The identity does not contribute to any signal, and is invariant under the large important class of *unital* processes (see Section 2.4), in which case the identity can be neglected [53, 29]. In NMR, the discrepancy from unitality is caused by thermalization (generalized amplitude damping defined by Eq. (2.39) in Section 2.4). In most applications such as quantum algorithms, the relevant time scales are much shorter than the thermalization time scale, and unitality is a good approximation. Such operation regime will be considered in Sections 6.5-6.7.

### 6.3.1   Effective evolution

For a non-unital but trace-preserving process $\mathcal{E}$, the observable evolution of the deviation can be understood as follows. Rewriting

$$\rho = vI + \rho_\Delta \,, \tag{6.21}$$

where $\rho_\Delta = \rho - vI$ is the traceless deviation from the identity and $v = 1/\mathrm{Dim}(\rho)$,

$$\begin{aligned}
\mathcal{E}(\rho) &= v\mathcal{E}(I) + \mathcal{E}(\rho_\Delta) \tag{6.22} \\
&= vI + v(\mathcal{E}(I) - I) + \mathcal{E}(\rho_\Delta) \tag{6.23}
\end{aligned}$$

$$= \upsilon I + \mathcal{E}_\Delta(\rho_\Delta) \qquad (6.24)$$

where

$$\mathcal{E}_\Delta(\rho_\Delta) = \upsilon(\mathcal{E}(I) - I) + \mathcal{E}(\rho_\Delta), \qquad (6.25)$$

and $\rho_\Delta \to \mathcal{E}_\Delta(\rho_\Delta)$ describes the observed evolution of the deviation. $\mathcal{E}_\Delta = \mathcal{E}$ iff $\mathcal{E}$ is unital. This is because $\mathcal{E}_\Delta(0) = \upsilon(\mathcal{E}(I) - I) \neq 0$ cannot be linear for any non-unital $\mathcal{E}$, and cannot be equal to $\mathcal{E}$.

## 6.4 Quantum process tomography in NMR

Recall from Section 2.3.2 that quantum process tomography (QPT) is an experimental procedure to determine the unknown quantum process of an open quantum system. We consider using Method I for QPT described in Section 2.3.2 to test quantum gates and understand decoherence in NMR. The procedure relies upon the ability to prepare a complete set of basis states $\rho_{in}$ as input to the unknown process $\mathcal{E}$, and the ability to measure the output density matrices $\mathcal{E}(\rho_{in})$. In NMR, there are two practical problems associated with tomography. First, unitary operations only manipulate the deviation, preventing a complete set of basis states from being prepared by unitary actions alone. Second, one can only measure traceless observables, and thus one cannot obtain the entire output density matrix of a process.

In this section, we show how these hurdles can be overcome theoretically. We first review the original QPT recipe of interest. Then we explain our extensions to the basic procedure which enable QPT with NMR. Experiments using QPT to investigate the fidelity of the CNOT gate and the validity of the independent error model in NMR are underway [27].

### 6.4.1 Review of QPT

Let the unknown quantum process be given by the operator-sum representation $\mathcal{E}(\rho) = \sum_k A_k \rho A_k^\dagger$. In this context, $\mathcal{E}$ can always be taken as trace-preserving. $\mathcal{E}(\rho)$ can be re-expressed as $\mathcal{E}(\rho) = \sum_{m,n} \chi_{mn} B_m \rho B_n^\dagger$, where $B_m$ are a fixed basis for operators on $\rho$, and $\chi_{mn}$ are entries of a positive Hermitian matrix. In the new representation, the information about the process is represented by the coefficients $\chi_{mn}$ instead of the operation elements $A_k$.

Let the $N^2$ matrices $\rho_j$ be a basis for density matrices. The result of applying $\mathcal{E}$ to $\rho_j$ can in turns be expressed in terms of the basis:

$$\mathcal{E}(\rho_j) = \sum_k \lambda_{jk}\rho_k\,. \tag{6.26}$$

The coefficients $\lambda_{jk}$, which fully specify $\mathcal{E}$, can be determined experimentally using quantum state tomography (see Sections 2.3.2 and 6.2.3). To determine $\chi$ from $\lambda$, let $B_m\rho_j B_n^\dagger = \sum_k \beta_{jk}^{mn}\rho_k$ such that $\sum_{mn} \beta_{jk}^{mn}\chi_{mn} = \lambda_{jk}$. Taking $\beta$ as a matrix and $\lambda, \chi$ as vectors, with composite column and row indices $mn$ and $jk$, we have $\beta\vec{\chi} = \vec{\lambda}$, which can be inverted to obtain $\vec{\chi}$.

## 6.4.2   QPT in NMR

Consider Eq. (6.26) again:

$$\mathcal{E}(\rho_j) = \sum_k \lambda_{jk}\rho_k\,. \tag{6.27}$$

The original goal is to prepare a complete set of basis states $\rho_j$ and to measure $\mathcal{E}(\rho_j)$ to obtain full information of $\lambda_{jk}$. We describe the complications in NMR and possible resolutions, first in an abstract setting, followed by concrete methods in NMR.

Equation (6.27) can be thought of as a linear representation of $\mathcal{E}$, in which $\rho_j$ form a basis and $\lambda_{jk}$ are entries of a matrix. In real life, one prepares independent $\rho_j$ and obtains $\mathcal{E}(\rho_j)$ where $\rho_j$ and $\mathcal{E}(\rho_j)$ are physical states. However, in this abstract setting, we change basis to an orthogonal one, with respect to the inner product $\langle A, B \rangle = \text{Tr}(A^\dagger B)$. Expressing $\rho$ and $\mathcal{E}(\rho)$ as vectors, $\mathcal{E}(\rho) = \lambda\rho$. We choose $\rho_1 = I$. By the orthogonality of $\rho_j$, $\text{Tr}(\rho_j) = 0 \ \forall \ j \geq 2$. Since $\mathcal{E}$ is trace-preserving, $\text{Tr}(\mathcal{E}(\rho_j)) = 0 \ \forall \ j \geq 2$, and have vanishing

$\rho_1$ component. Therefore, the matrix representation of $\lambda$ has the form

$$
\lambda = \begin{bmatrix} 1 & 0 & \cdots & 0 \\ \hline & & & \\ R & & M & \\ & & & \end{bmatrix}
\tag{6.28}
$$

Experimentally, only the traceless components of the output can be measured, with an amplification factor depending on many experimental details such as the amplifier gains, the RF coils, and the sample size. In other words, at most $dR$, $dM$ are obtainable, where $d$ is the unknown amplification factor. Moreover, if we can only manipulate the traceless deviation of the physical input unitarily, we only obtain $d(M + R')$ where $R'$ is a square matrix with every column equal to $R/2^n$. The first problem can be resolved by calibrating our apparatus with some known process to obtain $d$. The second problem can be resolved by perform an extra *non-unitary* procedure to prepare the input $\frac{I}{2^n}$ to obtain $dR$. How these adaptions translate to the NMR case is described next.

In NMR, the thermal initial state for a system of $n$ spins is given by

$$
\rho_{th} = \frac{1}{2^n} \left( I + \frac{\hbar}{2k_B T} \sum_j \omega_j Z^{(j)} \right).
\tag{6.29}
$$

By performing unitary operations on the system, we may rotate the $\sum_j \omega_j Z^{(j)}$ deviation to prepare $2^n - 1$ linearly independent inputs. We augment the basis set with the identity matrix, prepared by applying a uniformly spatially varying RF pulse to the sample. The pulse rotates different parts of the sample by different amounts, so that effectively, the pulse acts as random $X$ errors which depolarize the thermal state to the maximally mixed state. One can also calibrate the apparatus using the known initial state of the system resulting from thermalization. By state tomography, we can measure the traceless part of the thermal state given by Eq. (6.29), and obtain $\frac{d}{2^n} \frac{\hbar}{2k_B T} \sum_j \omega_j Z^{(j)}$. Comparing to the known deviation $\frac{1}{2^n} \frac{\hbar}{2k_B T} \sum_j \omega_j Z^{(j)}$, $d$ can be determined.

## 6.5   Effective Pure States and State Labeling

In this section, we outline one possible method to compute on the highly mixed thermal input states. We consider states which decompose into an "active" computing component and a "quiet" remainder. For example, when the density matrix decomposes into an identity and a pure-state-like deviation, only the latter evolves and contributes to the output signal under *unital* processes, precisely simulating the computation on a pure state. The possibility to perform quantum information processing task directly on the mixed state input is discussed in Section 6.7.

### 6.5.1   Effective Pure States

An effective pure state is a state that behaves for all computational purposes as a pure state. A computation (possibly with initial state preparation and measurement procedures) is generally a trace-preserving quantum operation $\mathcal{C}$. The density matrix $\rho_\epsilon$ is an effective pure state for a computation $\mathcal{C}$ corresponding to a pure state $|\psi\rangle\langle\psi|$, if, for all meaningful observables $O_i$, $\mathcal{C}$ induces another computation $\mathcal{C}'$ and observable $O_i'$ such that

$$\mathrm{Tr}(\mathcal{C}'(\rho_\epsilon)O_i') = \alpha\,\mathrm{Tr}(\mathcal{C}(|\psi\rangle\langle\psi|)O_i) \tag{6.30}$$

for some fixed known constant $\alpha$. In other words, proportional outcomes are obtained whether $\mathcal{C}$ is run on $|\psi\rangle$ or $\mathcal{C}'$ is run on $\rho_\epsilon$ for all meaningful measurements.

Physically, the standard measurement is to project onto the $|0\rangle$ or $|1\rangle$ states of the measured qubits, therefore $Z$ is the measured operator. In NMR, $X$ and $Y$ are the measured operators. Therefore, we restrict to traceless observables $O_i$.

For example, the following state is an effective pure state for any computation $\mathcal{C}$ which is unital:

$$\rho_\epsilon = \frac{1-\alpha}{2^N}I + \alpha|\psi\rangle\langle\psi| \tag{6.31}$$

This is because

$$\mathcal{C}(\rho_\epsilon) = \frac{1-\alpha}{2^N}I + \alpha\mathcal{C}(|\psi\rangle\langle\psi|) \quad \text{and} \tag{6.32}$$

$$\text{Tr}\left[\, \mathcal{C}(\rho_\epsilon)O_i \,\right] = \alpha\text{Tr}\left[\, \mathcal{C}(|\psi\rangle\langle\psi|)O_i \,\right] \tag{6.33}$$

This effective pure state requires no modification of the original computation ($\mathcal{C} = \mathcal{C}'$).

Our eventual goal is to find modifications to the computation so that the input of interest (such as the thermal state) is an effective pure state. This can be achieved, for example, by finding a preparation procedure $\mathcal{P}$ which transforms the initial state to the form in Eq. (6.31). Alternatively, we may also consider other effective pure states which require non-trivial readout procedures $\mathcal{R}$ and other modifications to $\mathcal{C}$. We devote the rest of this section to the preparation of effective pure states.

### 6.5.2 State Labeling

We consider preparing an effective pure state from an arbitrary initial state $\rho$. There are three major techniques: logical labeling, temporal labeling and spatial labeling. Specialization to thermal initial states in NMR will be discussed. A hybrid method, as original work in this Dissertation (unpublished), will be described. Such method was independently reported in [71].

**Logical Labeling**

In this method, extra qubits are used to label the subspaces of the input states. One prepends the computation $\mathcal{C}$ with an initial preparation step $\mathcal{P}$ and appends $\mathcal{C}$ with a readout step $\mathcal{R}$, such that

$$\mathcal{C}'(\rho) = \mathcal{R} \circ (\mathcal{I}_{label} \otimes \mathcal{C}) \circ \mathcal{P}(\rho)\,. \tag{6.34}$$

The purpose of $\mathcal{P}$ is to modify $\rho$ by concentrating some of its randomness into the label states. In particular, we may choose $\mathcal{P}$ such that

$$\mathcal{P}(\rho) = \sum_k |k\rangle\langle k| \otimes \rho_k\,, \tag{6.35}$$

where $\rho_0 = \alpha|\psi\rangle\langle\psi|$ is the desired initial state and $\rho_k$ for $k \geq 1$ are some "garbage" states. The $|k\rangle$ are labeling degrees of freedom. The computation $\mathcal{I}_{label} \otimes \mathcal{C}$ is to operate only on the Hilbert space of $\rho_k$, leaving the label states invariant. We have

$$\rho' = (\mathcal{I}_{label} \otimes \mathcal{C}) \circ \mathcal{P}(\rho) = \sum_k |k\rangle\langle k| \otimes \mathcal{C}(\rho_k) \qquad (6.36)$$

For a measurement of $O_i$ on the computation degrees of freedom, the readout preparation $\mathcal{R}$ is chosen as

$$\mathcal{R}(\rho') = \frac{1}{2}\left[\,\rho' + U_i\rho'U_i^\dagger\,\right] \qquad (6.37)$$

where $U_i = |0\rangle\langle 0| \otimes I + \sum_{k \geq 1} |k\rangle\langle k| \otimes A_i$ and $A_i$ anticommutes with $O_i$. Then,

$$\text{Tr}\left[\,\mathcal{C}'(\rho)(I \otimes O_i)\,\right] = \text{Tr}\left[\,\mathcal{R}(\rho') \times (I \otimes O_i)\,\right] \qquad (6.38)$$

$$= \text{Tr}\left[\,\rho' \times \frac{1}{2}\left[\,I \otimes O_i + |0\rangle\langle 0| \otimes O_i + \sum_{k \geq 1}|k\rangle\langle k| \otimes (A_i^\dagger O_i A_i)\,\right]\,\right] \qquad (6.39)$$

$$= \text{Tr}\left[\,\rho' \times (|0\rangle\langle 0| \otimes O_i)\,\right] \qquad (6.40)$$

$$= \text{Tr}\left[\,\mathcal{C}(\rho_0)\,O_i\,\right] \qquad (6.41)$$

where Eq. (6.39) is obtained from Eq. (6.37) and cyclic permutation of the operators inside the trace. It can be viewed as applying the dual of $\mathcal{R}$ to the observable, an extension of the Heisenberg's picture to non-unitary quantum operations.

Note that in logical labeling, one only needs to prepare a state of the form given in Eq. (6.31) *in the subspace* labeled by $|0\rangle\langle 0|$. This comes at a price – extra qubits and special readout procedures may be required.

For NMR, the initial state $\rho$ is the thermal state. When extra spins are used as the labeling degrees of freedom, the operation $\mathcal{R}$ may be omitted because the spectral lines are separated for different $|k\rangle$ in the labeling states (see Section 6.2.3). Alternatively, $\mathcal{R}$ can be implemented in two separate runs of the experiment, each with the identity or $U_i$ in place of $\mathcal{R}$. The measurement outcomes are then added to obtain the desired result. This method to add up the outcomes to simulate a quantum operation is closely related to temporal labeling as described next.

**Temporal labeling**

This method requires no extra qubits nor special readout procedure, but it requires a number of repetitions of the experiment. The idea is to add up the results of a series of experiments that begin with different preparation operations $P_k$ before the same intended computation $\mathcal{C}$. By linearity,

$$\sum_k \mathcal{C}(P_k \rho_{th} P_k^\dagger) = \mathcal{C}\left(\sum_k P_k \rho_{th} P_k^\dagger\right) = \mathcal{C}(\rho_\epsilon) \tag{6.42}$$

$$\sum_k \mathrm{Tr}\left[\, O_i \mathcal{C}(P_k \rho_{th} P_k^\dagger)\,\right] = \mathrm{Tr}\left[\, O_i \mathcal{C}\left(\sum_k P_k \rho_{th} P_k^\dagger\right)\,\right] = \mathrm{Tr}\left[\, O_i \mathcal{C}(\rho_\epsilon)\,\right] \tag{6.43}$$

In other words, summing over the experimental results (on the left side of Eq. (6.43)) is equivalent to performing the experiment with the unnormalized initial state $\rho_\epsilon = \sum_k P_k \rho_{th} P_k^\dagger$ (on the right side). The aim is to choose $P_k$ such that $\rho_\epsilon$ is the effective state in Eq. (6.31).

In the case of diagonal initial states, such as the thermal state, $P_k$ can be chosen to be cyclic permutations of the states $|l\rangle$ for $l \geq 1$, which fix $|0\rangle$. For example,

$$\rho = P_1 \rho P_1^\dagger = \begin{bmatrix} a_0 & 0 & 0 & 0 & 0 \\ 0 & a_1 & 0 & \cdot & \cdot \\ 0 & 0 & a_2 & \cdot & \cdot \\ 0 & \cdot & \cdot & \cdot & \cdot \\ 0 & \cdot & \cdot & \cdot & a_L \end{bmatrix}, \;\; P_2 \rho P_2^\dagger = \begin{bmatrix} a_0 & 0 & 0 & 0 & 0 \\ 0 & a_2 & 0 & \cdot & \cdot \\ 0 & 0 & a_3 & \cdot & \cdot \\ 0 & \cdot & \cdot & \cdot & \cdot \\ 0 & \cdot & \cdot & \cdot & a_1 \end{bmatrix}, \tag{6.44}$$

$$\cdots, \;\; P_L \rho P_L^\dagger = \begin{bmatrix} a_0 & 0 & 0 & 0 & 0 \\ 0 & a_L & 0 & \cdot & \cdot \\ 0 & 0 & a_1 & \cdot & \cdot \\ 0 & \cdot & \cdot & \cdot & \cdot \\ 0 & \cdot & \cdot & \cdot & a_{L-1} \end{bmatrix} \tag{6.45}$$

Then,

$$\sum_k P_k \rho P_k^\dagger = \sum_{l=1}^{L} a_l \begin{bmatrix} 1 & 0 & 0 & 0 & 0 \\ 0 & 1 & 0 & \cdot & \cdot \\ 0 & 0 & 1 & \cdot & \cdot \\ 0 & \cdot & \cdot & \cdot & \cdot \\ 0 & \cdot & \cdot & \cdot & 1 \end{bmatrix} + (La_0 - \sum_{l=1}^{L} a_l) \begin{bmatrix} 1 & 0 & 0 & 0 & 0 \\ 0 & 0 & 0 & \cdot & \cdot \\ 0 & 0 & 0 & \cdot & \cdot \\ 0 & \cdot & \cdot & \cdot & \cdot \\ 0 & \cdot & \cdot & \cdot & 0 \end{bmatrix} \tag{6.46}$$

which is an effective pure state of the form of Eq. (6.31). Such permutation works well for a small number of qubits, but requires resources exponential in the number of qubits. In [71], more efficient methods are presented.

### Spatial labeling

Effective pure states have been created by Cory *et al* [42] by applying different unitary operations $P_k$ which vary continuously with a spatial degree of freedom $k$. This is experimentally implemented using a static gradient magnetic field before the computation, which rotates each spin by an amount determined by the location of the molecule in the physical apparatus such that the integral over $k$ is an effective pure state. No extra qubits or repetitions are required.

### Hybrid method

This method is a special procedure for thermal initial states, using both logical and temporal labeling. For a system of $n$ qubits, it creates an effective pure state of $n-1$ qubits using 2 temporal labeling experiments. The number of gates required for the preparation is *linear* in $n$.

Recall that $\rho_{th} = c(I + \gamma \sum_i \omega_i Z^{(i)})$, where $c = \frac{1}{2^n}$ and $\gamma = \frac{\hbar}{2k_B T}$. The preparation procedure is defined by

$$\mathcal{P}(\rho) = \frac{1}{2} U_2(\rho + U_1 \rho U_1^\dagger) U_2^\dagger \tag{6.47}$$

where $U_1 = \prod_{k=2}^n \text{CNOT}_{1k}$, $U_2$ applies $X$ to the first qubit if all other qubits are in $|1\rangle$:

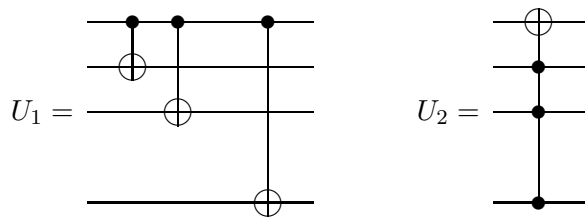

Intuitively, $\mathcal{P}$ works as follows. The density matrix of the thermal state looks like:

$$\frac{1}{c}\rho_{th} = \left[\begin{array}{c|c} I & 0 \\ \hline 0 & I \end{array}\right] + \gamma \left[\begin{array}{c|c} \omega_1 I & 0 \\ \hline 0 & -\omega_1 I \end{array}\right] + \gamma \left[\begin{array}{c|c} M & 0 \\ \hline 0 & M \end{array}\right] \qquad (6.48)$$

where $2^{n-1} \times 2^{n-1}$ block matrices are shown and

$$M = \left[\begin{array}{cccc} \omega_2 + \omega_3 + \cdots + \omega_n & 0 & 0 \cdots & \\ 0 & \omega_2 + \omega_3 + \cdots - \omega_n & 0 \cdots & \\ \cdots & & & \\ 0 & -\omega_2 - \omega_3 - \cdots + \omega_n & 0 & \\ 0 & -\omega_2 - \omega_3 - \cdots - \omega_n & & \end{array}\right] \qquad (6.49)$$

is the restriction of $\sum_{i \leq 2} \omega_i Z^{(i)}$ to the space of the last $n-1$ spins. When the first spin is in state $|1\rangle$, $U_1$ applies $\prod_{i=2}^{n} X^{(i)}$ which negates $M$ by conjugation. Therefore,

$$\frac{1}{c}U_1\rho_{th}U_1^\dagger = \left[\begin{array}{c|c} I & 0 \\ \hline 0 & I \end{array}\right] + \gamma \left[\begin{array}{c|c} \omega_1 I & 0 \\ \hline 0 & -\omega_1 I \end{array}\right] + \gamma \left[\begin{array}{c|c} M & 0 \\ \hline 0 & -M \end{array}\right] \qquad (6.50)$$

Hence,

$$\frac{1}{2c}\left[\rho_{th} + U_1\rho_{th}U_1^\dagger\right] = \left[\begin{array}{c|c} I & 0 \\ \hline 0 & I \end{array}\right] + \gamma \left[\begin{array}{c|c} \omega_1 I & 0 \\ \hline 0 & -\omega_1 I \end{array}\right] + \gamma \left[\begin{array}{c|c} M & 0 \\ \hline 0 & 0 \end{array}\right] \qquad (6.51)$$

The final step $U_2$ exchanges the last diagonal elements of the upper and lower diagonal block matrices,

$$\frac{1}{2c}U_2\left[\rho_{th} + U_1\rho_{th}U_1^\dagger\right]U_2^\dagger = \left[\begin{array}{c|c} (1+\gamma\omega_1)I & 0 \\ \hline 0 & (1-\gamma\omega_1)I \end{array}\right] + \gamma \left[\begin{array}{c|c} M_1 & 0 \\ \hline 0 & M_2 \end{array}\right] \qquad (6.52)$$

where

$$M_2 = \begin{bmatrix} 0 & 0 & 0 & 0 & 0 \\ 0 & 0 & & & \\ 0 & \cdot & 0 & & \\ 0 & \cdot & \cdot & 0 & \\ 0 & \cdot & 2\omega_1 - \omega_2 - \cdots - \omega_n & & \end{bmatrix} \tag{6.53}$$

represents a pure-state deviation in the last $n-1$ spins. Therefore, conditioned on the first spin being $|1\rangle$, we obtain an effective pure state for the other $n-1$ spins.

**Required Resources**  For the preparation $\mathcal{P}$, $U_1$ takes $n-1$ elementary gates. For $U_2$, recall that extra phases are irrelevant on diagonal density matrices. Therefore, we can replace the conditional $X^{(1)}$ by a conditional $iX^{(1)} \in \mathrm{SU}(2)$. The resulting operation takes $\mathcal{O}(n)$ elementary gates [13]. For the readout operation, provided each spin couples to the first spin, the spectral lines conditioned on the first spin being $|0\rangle$ or $|1\rangle$ are split. No extra readout procedure is required and we need only two temporal labeling experiments. Otherwise, four temporal labeling experiments are required. We obtain an effective pure state of $n-1$ spins out of $n$ spins.

The hybrid method illustrates that labeling can be performed using very simple techniques.

## Remark

State labeling is not just a way to create effective pure states; it is also a method to create *robust* quantum computation procedures. Equation (6.30) can be understood as transformation from a given quantum computation $\mathcal{C}$ (which nominally operates on pure state inputs) into another one, $\mathcal{C}'$, which is robust in the sense that it can operate on a class of mixed state inputs. This notion of robust quantum computation significantly expands the physical systems suitable for quantum computation. This idea is also related to the algorithms described in Section 6.7 which are robust against errors in the initial states. The fact that logical labeling can be viewed as an error detection procedure is discussion in Section 8.6.

## 6.6 Signal Strength and Scaling

Even though all effective pure states are mathematically equivalent, physically, the signal has to be above the detection noise (in other words, $\alpha$ in Eq. (6.30) has to be large enough). We discuss the limitations of the signal strength when using the labeling techniques discussed in the previous section.

For a sample of $N$ molecules each with $n$ spins in a static field B, the transverse magnetization of the $i$-th spin is given by

$$M_{xi} = N\gamma_i\hbar\text{Tr}(\rho X^{(i)}),$$ (6.54)

where $\gamma_i$ is the gyromagnetic ratio of the $i$-th spin. [2] $\text{Tr}(\rho X^{(i)})$ is upper bounded by the largest eigenvalue of the deviation density matrix of $\rho$. The labeling methods described so far are unital processes, which cannot increase the largest eigenvalue [87]. Therefore, the upper bound can be rewritten as

$$|M_{xi}| \leq \frac{N\gamma_i\hbar}{2^{n+1}k_BT}\sum_i\omega_i = \frac{nN\gamma_i\bar{\gamma}B\hbar^2}{2^{n+1}k_BT}$$ (6.55)

where $\omega_i = \gamma_i B$ and $\bar{\gamma}$ is the mean gyromagnetic ratio of the $n$ spins. For protons at room temperature in an 11.8 Tesla field, $\gamma\hbar B/k_BT \approx 10^{-4}$.

Usually, NMR signals are detected inductively in some pick-up coil with $K$ turns and area A, in a resonant tank with a quality factor $Q$. The time-varying magnetization leads to a changing flux $\Phi$ in the coil which produces a peak-to-peak voltage

$$V = QK\frac{d\Phi}{dt} = QK\frac{d}{dt}\mu_0 M_{xi}A \quad .$$ (6.56)

In the lab frame the readout magnetization oscillates at the Larmor frequency $\gamma_i B$, therefore the amplitude of the oscillating voltage in the pick-up coil will be

$$V = QK(\gamma_i B)\mu_0\left[\frac{nN\gamma_i\bar{\gamma}B\hbar^2}{2^{n+1}k_BT}\right]A \, .$$ (6.57)

---

[2] The energy difference of the spin states is given by $\hbar\omega = \hbar\gamma B = \mu B$ where $\gamma$ and $\mu$ are the gyromagnetic ratio and magnetic dipole moment of the spin.

Major improvements can be brought by using electron spins to cool down and read out nuclear spins. The electron gyromagnetic ratio is $\sim 10^3$ times larger than those of the nuclei. Transferring the electron thermal polarization to the nuclear spins can increase the initial polarization by a factor of $10^3$. Transferring the output from the nuclear spins to electron spins gives another factor of $10^3$ improvement. For quantum computation, the bandwidth of interest is known a-priori and high-Q resonators can be used instead. The quality factor $Q$ can be made at least $10^2$ times better. Since $2^{10} \approx 10^3$, each three orders of magnitude increase in the signal can accommodate about 10 more qubits. Therefore, $30 - 40$ qubits are possible without large scale modifications of the experimental set up. Further scaling up will necessitate high spin polarizations, and may require optical pumping, phase transitions in ordered systems or cooling down to millikelvin temperatures. More sophisticated initial state preparation technique using algorithmic cooling [100] has been developed which yields $\mathcal{O}(n)$ spins in the ground state starting with molecules of $n$ spins.

## 6.7  Hot qubit algorithms

We have seen how the stringent limitations on the initial states can be circumvent using effective pure states. Likewise, problems in bulk measurements can be resolved by determinizing quantum algorithms such that computation results do not average away [53]. However, such methods come at a cost: either an exponential reduction in signal strength or a linear reduction [100] in the number of usable qubits or extra computation steps, with all other resources held constant. It is an intriguing open question how limited the *bulk quantum computation* model at high temperature is compared to the *standard quantum computation* model. In this section, we exhibit non-trivial overlap between the two computation models: they are polynomially equivalent for a class of quantum algorithms including the well known Deutsch-Jozsa (DJ) algorithm [46]. These "hot qubit algorithms" tolerate independent bit flip errors in the input states and remain informative when ensemble measurements are used. Our result does not resolve the general question of equivalence between bulk computation at high temperature and standard quantum computation; rather, it provides new insight into the possibility to trade for important simplifications at the price of a restricted computation model while retaining polynomial equivalence to the standard computation model. The discussion is motivated by bulk NMR quantum computation, though it applies to any physical system with similar abstract descriptions.

### 6.7.1 Computation models

We first define our computation models:

1. Standard quantum computation (SQC), using a single quantum system with pure initial state (taken to be $|0\rangle^{\otimes n}$) and projective measurements.

2. Bulk quantum computation (BQC), using a large number of identical and independent single quantum systems, and ensemble measurements. Depending on the available initial states, the BQC model can be subdivided:

   (a) BQC$_P$ with pure initial state $\rho^{BQC_P} = (|0\rangle\langle 0|)^{\otimes n}$ (for example, see [113]).

   (b) BQC$_T$ with thermal initial state $\rho^{BQC_T}$, in which the $i$-th qubit is in a statistical mixture of $|0\rangle$ and $|1\rangle$ with probabilities $p_i$ and $(1 - p_i)$, independent of other qubits.

   (c) BQC$_E$ with effective pure state as input. The effective pure state with parameter $\alpha$ is given by $\rho^{BQC_E} = \alpha(|0\rangle\langle 0|)^{\otimes n} + \frac{1}{2^n}(1 - \alpha)I$.

In SQC, projective measurement of the $i$-th spin (onto $|0\rangle$ or $|1\rangle$) is a measurement of $Z^{(i)}$. Note that $Z|x\rangle = (1 - 2x)|x\rangle$ for $x \in \{0, 1\}$. In BQC, the corresponding output from the final state $\rho$ is $E_i = \text{Tr}(\rho Z^{(i)})$.

### 6.7.2 Noise sources in BQC

**Algorithmic uncertainty**

An algorithm which outputs a superposition of eigenstates of the measured operator is intrinsically random. Using a bulk sample, the measurement outcome has variance inversely proportional to the sample size. The algorithmic uncertainty can therefore be made negligible using a sufficiently large sample (independence of problem size), a situation which we assume for the rest of the discussion.  Furthermore, we assume the maximum allowed sample size is used.  We call a single run using such a sample a "single shot experiment".

## Channel noise

A realistic measurement apparatus has imperfections which can be modeled as some noisy channel.  The situation is like sending the *correct* experimental outcomes from the system to the experimentalist through a noisy channel (the measuring apparatus).  Therefore, the random noise due to the apparatus is called "channel noise".

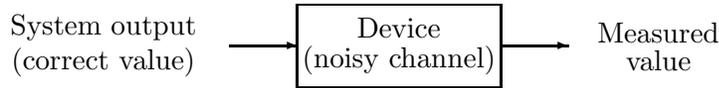

Figure 6.4: Measurement errors caused by an imperfect device can be modeled as sending the correct result through a noisy channel.

We model channel noise as a random variable $N$ with zero mean and variance $\Delta_{CN}^2$. Channel noise restricts the distinguishability of different outcomes.  Suppose $S$ is the correct outcome, then the measured outcome is $S' = S + N$, with mean $S$ and variance $\Delta_{CN}^2$. In most applications, $S$ is digitized by identifying $S$ with the closest value in a discrete (ordered) set $\{d_1, d_2, \cdots\}$.  Usually, $d = \frac{1}{2} \min_i |d_i - d_{i+1}|$ sets the required accuracy level. The probability to infer $d_i$ incorrectly from $S'$ depends on the ratio $d : \Delta_{CN}$.  As this ratio drops below unity, the error probability becomes large.  In this case, $K \sim (\frac{\Delta_{CN}}{d})^2$ repetitions of the experiment should be run to achieve a certain probability of success.  $\Delta_{CN}$ is generally independent on the algorithm and the problem size $n$, while $d$ can depend on both.  For a fixed algorithm, if $d = 1/f(n)$ then $K \sim f(n)^2$.

## Initial state noise

One can unify the various BQC models by viewing $\mathrm{BQC}_E$ and $\mathrm{BQC}_T$ as computation on $\mathrm{BQC}_P$ *prepended* with noise processes $\mathcal{E}_\mu(\rho^{BQC_P}) = \rho^{BQC_\mu}$ for $\mu = T, E$.

Let the noiseless channel on a single qubit be denoted by $\mathcal{I}$, and the bit flip channel with flip probability $1-p$ be denoted by $\mathcal{X}_p(\rho) = p\rho + (1-p)X\rho X$.  Then, $\mathcal{E}_\mu$ can be represented

as

$$\mathcal{E}_T = \mathcal{X}_{p_1} \otimes \cdots \otimes \mathcal{X}_{p_n} \tag{6.58}$$

$$\mathcal{E}_E = \alpha \mathcal{I}^{\otimes n} + (1 - \alpha) \mathcal{X}_{\frac{1}{2}}^{\otimes n} . \tag{6.59}$$

The initial state noise propagates in an algorithm in the following manner. Let $\mathcal{C}$ represent a computation in $\text{BQC}_P$, which outputs $\rho_o^{BQC_P}$ to give the outcomes

$$E_j^P = \text{Tr}(M_j \rho_o^{BQC_P}) , \tag{6.60}$$

where $M_j$ is a measured operator on the $j$-th qubit. Then, running $\mathcal{C}$ in $\text{BQC}_\mu$ is equivalent to running $\mathcal{C} \circ \mathcal{E}_\mu$ in $\text{BQC}_P$. Furthermore, if $\mathcal{F}_\mu$ exists such that [3]

$$\mathcal{C} \circ \mathcal{E}_\mu = \mathcal{F}_\mu \circ \mathcal{C} , \tag{6.61}$$

then, the mixed state computation outputs the state

$$\rho_o^{BQC_\mu} = \mathcal{F}_\mu(\rho_o^{BQC_P}) \tag{6.62}$$

In other words, mixed state computation is like pure state computation *appended* with $\mathcal{F}_\mu$. $\mathcal{C}$ *propagates* $\mathcal{E}_\mu$ to $\mathcal{F}_\mu$. Furthermore, the measurement outcomes from $\text{BQC}_\mu$ are

$$E_j^\mu = \text{Tr}(M_j \rho_o^{BQC_\mu}) = \text{Tr}(M_j \mathcal{F}_\mu(\rho_o^{BQC_P})) = \text{Tr}(\mathcal{F}_\mu^\dagger(M_j) \rho_o^{BQC_P}) . \tag{6.63}$$

Comparing with Eq. (6.60), computations in $\text{BQC}_\mu$ is equivalent to computations in $\text{BQC}_P$ with modified measured operators.

### 6.7.3 Hot Qubit Algorithms

In this section, we introduce the hot qubit algorithms (HQA) for which $\text{BQC}_T$, $\text{BQC}_P$ and SQC are polynomially equivalent. We first define HQA, and provide an important example afterwards. Other examples can be found in [114].

---

[3] If $\mathcal{C}$ is unitary, $\mathcal{F}_\mu = \mathcal{C} \circ \mathcal{E}_\mu \circ \mathcal{C}^\dagger$ always exists and is unique.

**Definition**

We consider ensemble measurements $E_i(\rho) = \text{Tr}(\rho Z^{(i)})$ and unitary computations $\mathcal{C}(\rho) = U\rho U^\dagger$. Let $\mathcal{F}_T$ be as defined in the previous section. That is,

$$\mathcal{C} \circ \mathcal{E}_T = \mathcal{F}_T \circ \mathcal{C} \,. \tag{6.64}$$

Then, the quantum algorithm $\mathcal{C}$ is an HQA if it satisfies the following two conditions:

1. $\text{BQC}_P$ is polynomially equivalent to SQC when implementing $\mathcal{C}$.

2. $E_i(\mathcal{F}_T(\rho)) = \gamma_i E_i(\rho)$ where $\gamma_i$ is independent of the input size.

From Eq. (6.62) and condition 2, we conclude that for HQA, $\text{BQC}_T$ and $\text{BQC}_P$ are equivalent. In other words, $\mathcal{C}$ propagates $\mathcal{E}_T$ to $\mathcal{F}_T$ which only causes a constant signal reduction. Together with condition 1, $\text{BQC}_T$ and SQC are polynomially related in implementing HQA.

**Example: The Deutsch-Jozsa algorithm**

We now illustrate the meaning of hot qubit algorithms with an example, the Deutsch-Jozsa algorithm [46, 36]. The goal is to solve the Deutsch-Jozsa problem [35]:

$f : \{0,1\}^n \rightarrow \{0,1\}$ is a binary function on $n$ bits which is either constant or balanced (balanced means that $|\{x : f(x) = 0\}| = |\{x : f(x) = 1\}|$). The question is to find out whether $f$ is constant or balanced, using "oracle calls" which output $|x\rangle|f(x) \oplus w\rangle$ upon the input of $|x\rangle|w\rangle$.

$2^{n-1}+1$ classical queries are required in the worst case to solve the problem deterministically. However, there is a single-query algorithm under the SQC model: 1. Starting from the state $|0\rangle^{\otimes n}|1\rangle$, apply $H^{\otimes(n+1)}$ to prepare a query $\sum_x |x\rangle(|0\rangle-|1\rangle)$. 2. Call the oracle $U_f$, obtaining $\sum_x |x\rangle(|0\rangle - |1\rangle) \rightarrow \sum_x (-1)^{f(x)}|x\rangle(|0\rangle - |1\rangle)$. 3. Apply $H^{\otimes(n+1)}$ to the state. These three steps can be represented as:

$$|0\rangle^{\otimes n}|1\rangle \quad \xrightarrow{1.H^{\otimes n}\otimes H} \quad \frac{1}{\sqrt{2^{n+1}}} \sum_x |x\rangle(|0\rangle - |1\rangle) \tag{6.65}$$

$$\overset{2.U_f}{\longrightarrow} \quad \frac{1}{\sqrt{2^{n+1}}} \sum_x (-1)^{f(x)} |x\rangle (|0\rangle - |1\rangle) \tag{6.66}$$

$$\overset{3.H^{\otimes n} \otimes H}{\longrightarrow} \quad \frac{1}{2^n} \sum_x \sum_y (-1)^{x\cdot y \oplus f(x)} |y\rangle |1\rangle \,, \tag{6.67}$$

where all summations are over $\{0,1\}^n$. The final state of the input register (the first $n$ qubits) is therefore

$$|\phi_f^{SQC}\rangle = \frac{1}{2^n} \sum_y \sum_x (-1)^{f(x) \oplus x\cdot y} |y\rangle = \sum_y g(y) |y\rangle \,, \tag{6.68}$$

with

$$g(y) = \frac{1}{2^n} \sum_x (-1)^{f(x) \oplus x\cdot y} \,. \tag{6.69}$$

If $f$ is constant, $|g(0)| = 1$ and if $f$ is balanced, $|g(0)| = 0$. Projecting the input register along the computational basis gives a definite value of $y$. The cases $y = 0$ and $y \neq 0$ correspond to the deterministic answers $f$ being constant and balanced respectively.

We now consider running the algorithm in $\text{BQC}_P$. Recall that we measure $Z^{(i)}$ (note that $\langle y \rangle \neq (\langle Z^{(1)} \rangle, \cdots, \langle Z^{(n)} \rangle)$). As $g(y)$ is a complicated function of $f$, it is not obvious if the measurements are still informative. Surprisingly, an affirmative answer can be obtained by the following simple argument. For SQC, the output is $|y = 0\rangle$ for $f$ constant and $|y \geq 1\rangle$ for $f$ balanced. They are eigenvectors of the coarse grained operator $Z_T = \sum_i Z^{(i)}$ with eigenvalues $\lambda = n$ and $\lambda \leq n - 2$. More importantly, $E(Z_T) = \sum_i E_i$ and one needs to distinguish between $E(Z_T) = n$ and $E(Z_T) \leq n - 2$. The digitization therefore requires an accuracy level of $d = 1$. If the channel noise causes a standard deviation of $\Delta_{CN}$ in each measured $E_i$, the measured $E(Z_T)$ has standard deviation $\sqrt{n}\Delta_{CN}$. When $n$ is large and $d : \Delta_{CN} \ll 1$, $\mathcal{O}(n)$ repetitions are sufficient to amplify the probability of success. Thus, $\text{BQC}_P$ can solve the DJ problem with $\mathcal{O}(n)$ queries. The Deutsch-Jozsa algorithm satisfies condition 1 for being an HQA.

For condition 2, we first show that the noise does *not* propagate. In other words, $\mathcal{E}_T \circ \mathcal{C} = \mathcal{C} \circ \mathcal{E}_T$ and $\mathcal{E}_T = \mathcal{F}_T$. We then show that $\mathcal{E}$ affects all measurements only by a factor independent of $n$.

First of all, the $(n+1)$-th qubit (or the work bit) is unaffected by the algorithm, and can be omitted from the discussion. We first assume that the work bit is pure and take into account a possibly mixed state at the conclusion. Let $\tilde{U}_f$ and $\tilde{U}$ denote the restriction of $U_f$ (the oracle) and $U$ (the computation) to the input register: $\tilde{U}_f|x\rangle = (-1)^{f(x)}|x\rangle$ and $\tilde{U} = H^{\otimes n}\tilde{U}_f H^{\otimes n}$ following Eqs. (6.65)-(6.67). For each $i$, $[X^{(i)}, \tilde{U}] = [X^{(i)}, H^{\otimes n}\tilde{U}_f H^{\otimes n}] = H^{\otimes n}[Z^{(i)}, \tilde{U}_f]H^{\otimes n} = 0$. Recall that $\mathcal{E}_T = \otimes_i \mathcal{X}_{p_i}$ and $\mathcal{X}_p(\rho) = p\rho + (1-p)X\rho X$. Therefore, each operation element of $\mathcal{E}_T$ commutes with $\tilde{U}$; thus, the operations $\mathcal{E}_T$ and $\mathcal{C}$ commute. Hence, the signal from the $i$-th qubit is given by

$$
\begin{aligned}
E_i^{BQC_T} &= \mathrm{Tr}(Z^{(i)}\mathcal{E}_T(\rho_o^{BQC_P})) && (6.70) \\
&= \mathrm{Tr}(\mathcal{E}_T^\dagger(Z^{(i)})\rho_o^{BQC_P}) && (6.71) \\
&= (2p_i - 1)\mathrm{Tr}(Z^{(i)}\rho_o^{BQC_P}) && (6.72) \\
&= (2p_i - 1)E_i^{BQC_P} && (6.73)
\end{aligned}
$$

The second to last line holds because $\mathcal{E}_T$ is an independent qubit process, and $\mathcal{X}_p(Z) = (2p-1)Z$. We conclude that the outputs of $\mathrm{BQC}_T$ and $\mathrm{BQC}_P$ are identical up to the scaling factors $(2p_i - 1)$ and $\mathrm{BQC}_T$ and $\mathrm{BQC}_P$ are equivalent. Condition 2 for being an HQA is satisfied by the Deutsch-Jozsa algorithm.

For simplicity, suppose $p_i = p$ $\forall i$. When the ancilla starts in a thermal state, the probability is $p$ for it to be $|1\rangle$ as in Eq. (6.65) and $1 - p$ for it to be $|0\rangle$. In the latter, the oracle does nothing to the input register $(|x\rangle \rightarrow |x\rangle)$ and it outputs the signal of a constant function. We need to distinguish between $\sum_i E_i = (2p-1)n$ and $\sum_i E_i \leq (2p-1)(p(n-2)+(1-p)n) = (2p-1)(n-2p)$. The smaller accuracy threshold $d = p(2p-1)$ will only increase the number of repetitions by a constant factor.

### Discussion on HQA

We have established that $\mathrm{BQC}_T$ and SQC are polynomially equivalent in solving the DJ problem; however, we have not been able to show that $\mathrm{BQC}_T$ outperforms its classical counterpart. The reason is, $\mathrm{BQC}_T$ and SQC are equivalent in a probabilistic setting, while SQC is exponentially better than classical computation only in the deterministic case. $\mathrm{BQC}_T$

loses its advantage over classical computation when small probability of error is allowed, since probabilistic classical computation can solve the DJ problem with only a constant number of queries.

The difficulty in finding HQA versions of other important algorithms lies in having to satisfy both conditions 1 and 2 simultaneously. For example, Simon's algorithm [106], unmodified, satisfies condition 2, so that $BQC_T$ and $BQC_P$ are polynomially equivalent. However, the ensemble measurement in $BQC_P$ erases all the useful information in the answer for most inputs. Therefore, $BQC_P$ fails to solve Simon's problem with the original algorithm. While Simon's algorithm can be modified to function in $BQC_P$, the modified algorithm no longer satisfies condition 2. Thus, we have not been able to construct an HQA for Simon's problem.

### 6.7.4 Discussion on bulk computation models

To summarize,

- $BQC_P$ is polynomially equivalent to SQC in running all the known quantum algorithms, which can be determinized with polynomial overhead.

- $BQC_E$ with large $\alpha$ can efficiently simulate $BQC_P$.

- $BQC_T$ can simulate $BQC_E$ (with parameter $\alpha$) by preparing effective pure states. However, $\alpha \leq \frac{n\epsilon}{2^n}$ where $(1+\epsilon)/2 = \max_i p_i$ and $n$ is the problem size. Therefore, $BQC_T$ can efficiently simulate $BQC_P$ and SQC only for sufficiently large $\frac{n\epsilon}{2^n}$.

- For HQA, $BQC_T$ and SQC are polynomially equivalent. In contrast, transforming $BQC_T$ to $BQC_E$ to implement HQA requires an exponential number of repetitions.

In the rest of this section, we discuss the difference between $BQC_T$ and $BQC_E$. We can consider the distance or the fidelity (using for instance, the definitions in [88]) between $\rho^{BQC_T}$, $\rho^{BQC_E}$ and $|0\rangle^{\otimes n}$, $I/2^n$.

First of all, for $\rho^{BQC_T}$, the entropy $nH(p)$ never approaches $n$ and the distance between $\rho^{BQC_T}$ and the "useless" state $I/2^n$ is greater than $2p - 1$. In contrast, converting $\rho^{BQC_T}$ to $\rho^{BQC_E}$ increases the entropy from $nH(p)$ to more than $(1-\alpha)n$ where $\alpha \xrightarrow{n\to\infty} 0$. The distance between $\rho^{BQC_E}$ and $I/2^n$ is vanishing. A more interesting fact is that both $\rho^{BQC_T}$ and $\rho^{BQC_E}$ have vanishing overlap with $|0\rangle^{\otimes n}$, yet there is an exponential gap in their computation capacity. The implication is that, states other than $|0\rangle^{\otimes n}$ in $\rho^{BQC_T}$ must be contributing in an HQA, but they are only arranged to be "quiet" in $\rho^{BQC_E}$.

Finally, $\rho^{BQC_T}$ is a product state but $\rho^{BQC_E}$ cannot be written as $\rho_1 \otimes \cdots \otimes \rho_n$. This in turns means that $\mathcal{E}_E$ cannot be written as a product channel, or else $\rho^{BQC_E} = \mathcal{E}_E(|0\rangle^{\otimes n})$ will be a product state. Rather, $\mathcal{E}_E$ represents a classically correlated noise and propagates to diminish the output signal.

Even though it remains inconclusive what $BQC_T$ can accomplish, the lesson is that the poor scaling in $BQC_E$ is due to the inefficient conversion of the thermal state to an effective pure state for a large number of spins. As the thermal state is the stationary state of a natural decoherence process, and is easily obtained, it is important to find efficient simulations of SQC using $BQC_T$, perhaps via yet another computation model which tolerates more noise than $BQC_T$ but can be more efficiently simulated by $BQC_T$.

## 6.8   Summary and preview

In this chapter, we have seen how various techniques have made it possible to perform quantum computation on bulk NMR systems with thermal initial states. We have discussed logic operations, measurements, state tomography and process tomography in NMR. We have seen how the problems associated with mixed state inputs can be resolved for small scale computations by "state labeling". The possibility to perform some quantum algorithms directly on a thermal state was also demonstrated.

In the next chapter, we will describe an efficient method to selectively perform coupled logic operations in NMR. In Chapter 8 we will describe an experiment on quantum error correction. The following facts or notations will be frequently used for the next two chapters.

- Hamiltonians of an NMR system:

  Zeeman energy: $\mathcal{H}_Z = -\frac{1}{2} \sum_i \omega_i Z^{(i)}$

  J-coupling: $\mathcal{H}_c = \sum_{i<j} g_{ij} Z^{(i)} \otimes Z^{(j)}$ where $g_{ij} = \pi J_{ij}/2$

- Time evolution: $\tau$

  Primitive coupled operation: $Z\!\!\!Z_{ij} \equiv e^{-i\frac{\pi}{4}Z^{(i)}\otimes Z^{(j)}}$

- Thermal initial state: $\rho_{th} = \frac{1}{\mathcal{Z}}e^{-\frac{\mathcal{H}}{k_B T}} \approx \frac{1}{2^n}\left(I - \frac{\mathcal{H}_Z}{k_B T}\right)$.

# Chapter 7

# Logic gates in NMR - decoupling and recoupling

## 7.1 Introduction

Quantum computation requires the ability to perform coupled logic operations, which can only originate from the natural couplings in the quantum systems involved. In most circumstances, the natural interactions do not couple specific pairs of qubits as desired in most applications of quantum computation. Rather, many couplings occur simultaneously along with the intended one. Moreover, the problem of simultaneous and undesired coupling generally becomes worse with larger systems and stronger couplings, which are essential for quantum computation to be useful. The fundamental task to turn off spurious evolution is so difficult that, coercing a complex system to *do nothing* [67] – ceasing all evolution – can be just as difficult as making it do something computationally useful.

In this chapter, we address a simpler problem: to stop the spurious coupling and to perform specific coupled logic gates in NMR quantum computation. The task of turning off all couplings is known in the art of NMR as *decoupling*; doing this for all but a select subset of couplings is known as *selective recoupling*. The basic idea is to interrupt the free evolution by carefully chosen pulses. These pulses are single qubit operations that transform the Hamiltonian in the time between the pulses in such a manner that unwanted couplings in consecutive evolutions cancel out each other. Ingenious schemes have been found [50, 107, 84] but they do not address the problems relevant to quantum computation. The primary interest in these schemes is to reveal complex structures in the spectra rather than to achieve precise quantum evolutions. Quantum computation brings new requirements, and initial efforts [80] have been made to develop pulse sequences to satisfy these needs; however, these schemes have necessitated resources (such as the total number of pulses applied) exponential in the number of spins being controlled. Schemes for selective recoupling are generally difficult to find for a large system. Each pulse simultaneously affects many coupling terms in the Hamiltonian. To turn off all but one of the coupling terms, these pulses have to satisfy many simultaneous requirements.

In this chapter, we present *efficient* schemes for decoupling and selective recoupling. For an $n$-spin system, in which any pair of spins can be coupled, our schemes concatenate $cn$ time intervals and use fewer than $cn^2$ pulses, where $c \approx 1$ for most $n$ with strict upper bound $c \leq 2$. Our method exploits simplifications in the couplings in heteronuclear spin systems. In this case, we show that the conditions for decoupling and selective recoupling are special orthogonality conditions, with solutions given by a class of well-known matrices called *Hadamard matrices*. These are generalizations of the well known Hadamard transformation in quantum computation. The efficiency of the scheme originates from the existence of general Hadamard matrices in many dimensions.

This chapter is structured as follows. Section 7.2 contains the precise statement of the problem. In Section 7.3, we first motivate the construction of the decoupling scheme with examples, and then derive conditions for decoupling and describe the general construction related to Hadamard matrices. Important properties of Hadamard matrices are summarized. Modifications of the decoupling scheme to perform selective recoupling are described. We conclude with various properties and limitations of the scheme in Section 7.4.

## 7.2   The statement of the problem

We shall consider a heteronuclear $n$-qubit system. Recall from Section 6.2.1 that the Hamiltonian of such a system is given by

$$\mathcal{H} = \mathcal{H}_{\mathrm{Z}} + \mathcal{H}_{\mathrm{c}} + \mathcal{H}_{\mathrm{env}} \tag{7.1}$$

where

$$\mathcal{H}_{\mathrm{Z}} = -\frac{1}{2} \sum_i \hbar \omega_i Z^{(i)} \, , \tag{7.2}$$

denotes the Zeeman terms, and

$$\mathcal{H}_{\mathrm{c}} = \sum_{i<j} g_{ij} Z^{(i)} \otimes Z^{(j)} \, , \tag{7.3}$$

denotes the coupling terms.

A single qubit operation is performed by applying a *pulsed* radio frequency (RF) magnetic field along some direction $\hat{\eta}$ perpendicular to the static field. To address the $i$-th spin, the frequency of the RF field is chosen to approximate $\omega_i/2\pi$. When the $\omega_i$'s are very different, a very short pulse can be used, so that during the pulse, all other evolutions are negligible except for the rotation operator $e^{-i\frac{\theta}{2}\vec{\sigma}^{(i)}\cdot\hat{\eta}}$ where $\theta$ is proportional to the pulse duration and the power. The Lie group of all single qubit operations can be generated by rotations about $\hat{x}$ and $\hat{y}$.

Coupled operations such as controlled-z or CNOT acting on the $i$-th and the $j$-th spins can be performed given the primitive

$$Z\!\!Z_{ij} = e^{-i\frac{\pi}{4} Z^{(i)} \otimes Z^{(j)}} \, . \tag{7.4}$$

The ultimate goal is to be able to efficiently realize arbitrary quantum operations on an $n$-spin system with arbitrary couplings. In this chapter, we consider a more limited objective, which can now be stated precisely, using the definitions of Eqs. (7.2)-(7.4):

Given a heteronuclear system of $n$ spins with free evolution $e^{-i(\mathcal{H}_Z + \mathcal{H}_c)t}$, controlled using typical RF pulses, how can $Z\!Z_{ij}$ be implemented efficiently?

Following NMR tradition, we refer to this task as "recoupling".

## 7.3    Construction of the schemes

A problem closely related to recoupling is the following:

Given a heteronuclear system of $n$ spins with free evolution $e^{-i\mathcal{H}_c t}$, controlled using typical RF pulses, how can the identity, $I$, be implemented efficiently?

We refer to this task as "decoupling". It is conceptually easier to first construct a decoupling scheme. The scheme is derived from Hadamard matrices, which will be reviewed. Modifications to implement selective recoupling will be described afterwards.

### 7.3.1    Decoupling scheme for two qubits

To motivate the general construction, we analyze the simplest example of decoupling two spins. From Eq. (7.3), the evolution operator for an arbitrary duration $t$ is given by $\tau = e^{-ig_{12}tZ^{(1)} \otimes Z^{(2)}}$. Recall from Section 6.2.2 that $X$ is a rotation of $\theta = \pi$ along $\hat{x}$ up to an irrelevant overall phase. $X^{(i)}$ is physically performed by an RF pulse at frequency $\omega_i$. The important observation is,

$$X^{(2)}(Z^{(1)} \otimes Z^{(2)})X^{(2)} = -Z^{(1)} \otimes Z^{(2)} \tag{7.5}$$

and therefore

$$
\begin{aligned}
X^{(2)}\tau X^{(2)} &= X^{(2)}e^{-ig_{12}tZ^{(1)} \otimes Z^{(2)}}X^{(2)} &\tag{7.6}\\
&= e^{-ig_{12}tZ^{(1)} \otimes (X^{(2)}Z^{(2)}X^{(2)})} &\tag{7.7}\\
&= e^{-ig_{12}tZ^{(1)} \otimes (-Z^{(2)})} &\tag{7.8}\\
&= \tau^{-1} &\tag{7.9}
\end{aligned}
$$

where Eq. (7.7) is obtained using a Taylor series expansion of the matrix exponent and the fact $(X^{(2)})^2 = I$. This observation implies that adding the gate $X^{(2)}$ before and after the evolution $\tau$ results in $\tau^{-1}$, so that the sequence of events $X^{(2)}\tau X^{(2)}\tau = I$ has no net coupling although the spins are actually coupled all the time. This is called refocusing in NMR, and clearly illustrate how single qubit operations can transform the Hamiltonian so that unwanted couplings in consecutive evolutions cancel out each other.

We now extract the essential features of the above decoupling scheme by rewriting the sequence $X^{(2)}\tau X^{(2)}\tau$ as

$$e^{-ig_{12}t\ (+)Z^{(1)}\otimes(-)Z^{(2)}} \times e^{-ig_{12}t\ (+)Z^{(1)}\otimes(+)Z^{(2)}}, \qquad (7.10)$$

and referring to $\tau$ and $X^{(2)}\tau X^{(2)}$ as time intervals. We note the following facts:

**1.** Since the matrix exponents commute, negating the coupling for exactly half of the total time is sufficient to cancel out the coupling.

**2.** Since the coupling is bilinear in $Z^{(1)}$ and $Z^{(2)}$, it is unchanged (negated) when the signs of $Z^{(1)}$ and $Z^{(2)}$ agree (disagree).

**3.** The sign of $Z^{(i)}$ is $(-)$ or $(+)$ depending on whether $X^{(i)}$ gates are applied before and after the interval. In other words, the sign of $Z$ for each spin in each time interval is controlled by inserting $X$ gates for that spin before and after that interval.

In summary, the most crucial point leading to decoupling is that, the signs of the $\sigma_z$ matrices of the coupled spins, controlled by the $X$ gates, disagree for half of the time.

### 7.3.2 Sign matrix and decoupling criteria

In general, we consider schemes which concatenate a certain number of equal-time intervals and use $X$ gates to control the signs of $Z$ for each spin. The essential information on the signs can be represented by a "sign matrix" defined as follows. The "sign matrix" of a pulse scheme for $n$-spins with $m$ time intervals is the $n \times m$ matrix with the $(i, a)$ entry being the *sign* of $Z^{(i)}$ in the $a$-th time interval. We denote any sign matrix for $n$ spins by $S_n$. For example, the sequence in Eq. (7.10) can be represented by the sign matrix

$$S_2 = \begin{bmatrix} + & + \\ + & - \end{bmatrix}.$$  (7.11)

Each row represents a sequence of $m$ intervals for each spin and each interval given by $-$ is preceded and followed by $X$ gates for that spin. Therefore, each sign matrix corresponds to a sequence of events for the whole system. Following the discussion in Sec. 7.3.1, decoupling is achieved whenever any two rows in the sign matrix disagree in exactly half of the entries (all couplings are negated for exactly half of the time). The general construction of decoupling scheme is now reduced to finding sign matrices satisfying the above criteria.

As an illustration, we construct a decoupling scheme for four spins. We first find a correct sign matrix and then derive the corresponding pulse sequence. For example, a possible sign matrix is given by

$$S_4 = \begin{bmatrix} + & + & + & + \\ + & + & - & - \\ + & - & - & + \\ + & - & + & - \end{bmatrix},$$  (7.12)

in which any two rows disagree in exactly two entries. The sequence corresponding to $S_4$ can be obtained by converting each column to a time interval before and after which $X$ pulses are applied to spins (rows) given by $-$'s. No pulses are applied to spins (rows) with $+$'s. The resulting sequence,

$$\tau(X^{(3)}X^{(4)}\tau X^{(3)}X^{(4)})(X^{(2)}X^{(3)}\tau X^{(2)}X^{(3)})(X^{(2)}X^{(4)}\tau X^{(2)}X^{(4)}),$$  (7.13)

is the identity by construction and this can also be verified directly. Note that $\mathcal{H}_c$ in $\tau = e^{-i\mathcal{H}_c t}$ now denotes the sum of six possible coupling terms for four spins. Note also Eq. (7.13) is written in such a way that it corresponds visually to the sign matrix, though the evolutions are actually in reverse time order relative to $S_4$. However, such ordering is irrelevant for commuting evolutions. Since $X^{(i)}X^{(i)} = I$, Eq. (7.13) can be simplified to

$$\tau(X^{(3)}X^{(4)}\tau X^{(4)})(X^{(2)}\tau X^{(3)})(X^{(4)}\tau X^{(2)}X^{(4)})\,. \tag{7.14}$$

This simplified pulse sequence can also be obtained directly from Eq. (7.12) by converting columns to time intervals and inserting $X^{(i)}$ between intervals whenever the $i$-th row changes sign or whenever a $-$ sign reaches either end of the row. The relation between the sequences in Eq. (7.13) and Eq. (7.14) and $S_4$ is illustrated in Fig. 7.1.

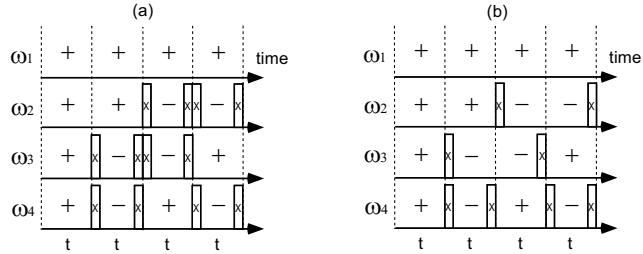

Figure 7.1: (a) Pulse sequence corresponding to Eq. (7.13). From $S_4$, each "$-$" sign in the $i$-th row and $a$-th column translates to two $X$ pulses at $\omega_i$ before and after the $a$-th time interval. (b) Pulse sequence obtained from simplifying (a). This corresponds to Eq. (7.14), and can be constructed directly from $S_4$ by translating each change of sign in the $i$-th row to an $X$ pulse at $\omega_i$. A "$-$" sign at the end of the row also gives rise to an $X$ pulse at end of the last time interval.

The above scheme can be generalized to decouple $n$ spins with $m$ time intervals as follows:

Construct the $n \times m$ sign matrix $S_n$, with entries $+$ or $-$, such that *any* two rows disagree in exactly half of the entries. For each $-$ sign in the $i$-th row and the $a$-th column, apply $X^{(i)}$ before and after the $a$-th time interval.

Because of the pulses, in each time interval, each $Z^{(i)}$ has a sign as given by the sign matrix. The $\sigma_z$ matrices of any two spins therefore have opposite signs for half of the time, during which their coupling is negated, and the evolution is always canceled.

For $n$ spins, $n \times m$ sign matrices which correspond to decoupling schemes do not necessarily exist for arbitrary $m$, but they always exist for large and special values of $m$. A

possible structure is:

$$S_n = \begin{bmatrix} + & \cdots & + & + & \cdots & + & + & \cdots & + & + & \cdots & + \\ + & \cdots & + & + & \cdots & + & - & \cdots & - & - & \cdots & - \\ & \cdots & & & \cdots & & & \cdots & & & \cdots & \\ + & \cdots & + & - & \cdots & - & + & \cdots & + & - & \cdots & - \\ & \cdots & & & \cdots & & & \cdots & & & \cdots & \\ + & \cdots & - & + & \cdots & - & + & \cdots & - & + & \cdots & - \end{bmatrix},$$

in which intervals are bifurcated when rows (spins) are added. Such bifurcation takes place whenever it is impossible to add an extra row that is orthogonal to all the existing ones ("depletion"). If such depletion occurs frequently, the sign matrix will have exponential number of columns, and decoupling will take exponential number of steps as $n$ increases. The challenge is to find correct sign matrices with subexponential number of columns.

### 7.3.3   Equivalent decoupling criteria

The criteria for a sign matrix $S_n$ to represent a valid decoupling scheme is that any two rows disagree in exactly half of the entries. It is useful to rephrase the complicated criteria concisely. Suppose $\pm$ is replaced by $\pm 1$ in $S_n$. If $S_n$ satisfies the decoupling criteria, any two rows have zero inner product and therefore $S_n S_n^T = nI$. Conversely, any $n \times m$ matrix $M$ with entries $\pm 1$ satisfying $MM^T = nI$ is a valid sign matrix giving a decoupling scheme that requires $m$ time intervals. We now present very efficient solutions to this simple criteria, namely the Hadamard matrices.

### 7.3.4   Hadamard matrices

Hadamard matrices have applications in many areas such as the construction of designs, error correcting codes and Hadamard transformations [38, 1, 111, 83].

A Hadamard matrix of order $n$, denoted by $H(n)$, is an $n \times n$ matrix with entries $\pm 1$, such that

$$H(n)H(n)^T = nI \ . \tag{7.15}$$

The rows are pairwise orthogonal, therefore any two rows agree in exactly half of the entries. Likewise columns are pairwise orthogonal. We abbreviate "±1" as "±". $S_2$ and $S_4$ in Eqs. (7.11) and (7.12) are simple examples of $H(2)$ and $H(4)$. An example of $H(12)$ is given by

$$H(12) = \begin{bmatrix}
+ & + & + & + & + & + & - & + & + & + & + & + \\
+ & + & + & - & - & + & + & - & + & - & - & + \\
+ & + & + & + & - & - & + & + & - & + & - & - \\
+ & - & + & + & + & - & + & - & + & - & + & - \\
+ & - & - & + & + & + & + & - & - & + & - & + \\
+ & + & - & - & + & + & + & + & - & - & + & - \\
- & + & + & + & + & + & - & - & - & - & - & - \\
+ & - & + & - & - & + & - & - & - & + & + & - \\
+ & + & - & + & - & - & - & - & - & - & + & + \\
+ & - & + & - & + & - & - & + & - & - & - & + \\
+ & - & - & + & - & + & - & + & + & - & - & - \\
+ & + & - & - & + & - & - & - & + & + & - & -
\end{bmatrix} .$$  (7.16)

The following is a list of useful facts about Hadamard matrices (details and proofs omitted):

1. *Equivalence*  Permutations or negations of rows or columns of Hadamard matrices leave the orthogonality condition invariant. Two Hadamard matrices are equivalent if one can be transformed to the other by a series of such operations. Each Hadamard matrix is equivalent to a *normalized* one, which has only +'s in the first row and column. For instance, $H(12)$ in Eq. (7.16) can be *normalized* by negating the 7-th row and column.

2. *Necessary conditions*  $H(n)$ exists only for $n = 1$, $n = 2$ or $n \equiv 0 \bmod 4$. This is obvious if the matrix is normalized, and the columns are permuted so that the first

three rows become:

$$\begin{bmatrix} + & \cdots & + & + & \cdots & + & + & \cdots & + & + & \cdots & + \\ + & \cdots & + & + & \cdots & + & - & \cdots & - & - & \cdots & - \\ + & \cdots & + & - & \cdots & - & + & \cdots & + & - & \cdots & - \\ \cdot & \cdot & \cdot & \cdot & \cdot & \cdot & \cdot & \cdot & \cdot & \cdot & \cdot & \cdot \end{bmatrix}.$$

3. *Hadamard's conjecture* [62]  $H(n)$ exists for every $n \equiv 0 \bmod 4$. This famous conjecture is verified for all $n < 428$.

4. *Sylvester's construction* [109]  If $H(n)$ and $H(m)$ exist, then $H(nm)$ can be constructed as $H(n) \otimes H(m)$. In particular, $H(2^r)$ can be constructed as $H(2)^{\otimes r}$, which is proportional to the matrix representation of the Hadamard transformation for $r$ qubits.

5. *Paley's construction* [90]  Let $q$ be an odd prime power. If $q \equiv 3 \bmod 4$, then $H(q+1)$ exists; if $q \equiv 1 \bmod 4$, then $H(2(q+1))$ exists.

6. *Numerical facts* [38]    For an arbitrary integer $n$, let $\underline{n}$ and $\overline{n}$ be the largest and smallest integers that satisfy $\underline{n} < n \leq \overline{n}$ with *known* $H(\underline{n})$ and $H(\overline{n})$. We define the "gap", $\delta_n$, to be $\overline{n} - \underline{n}$ (see Fig. 7.2). For $n \leq 1000$, $H(n)$ is known for every possible order except for 6 cases, and the maximum gap is 8. For $n \leq 10000$, $H(n)$ is unknown for 192 possible orders and the maximum gap is 32.

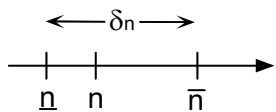

Figure 7.2: The gap $\delta_n$ between two existing orders of Hadamard matrices.

The importance of the full connection to Hadamard matrices will become clear after we construct the scheme for an arbitrary number of spins in the next section.

### 7.3.5   Hadamard matrices and decoupling scheme

It is immediate from previous discussions that each $H(n)$ is a valid sign matrix giving a decoupling scheme for $n$ spins using only $n$ time intervals.

For example, $S_2$ and $S_4$ in Eqs. (7.11) and (7.12) are $H(2)$ and $H(4)$. Whenever $H(n)$ exists, there is a decoupling scheme for $n$ spins concatenating only $n$ time intervals. However, $H(n)$ may or may not exist for a given $n$. For an arbitrary integer $n$, let $\overline{n}$ be the smallest integer that satisfies $n \leq \overline{n}$ with *known* $H(\overline{n})$. To construct a decoupling scheme for $n$ spins when $H(n)$ does not necessarily exist, we start with $H(\overline{n})$ and take $S_n$ to be any $n \times \overline{n}$ submatrix of $H(\overline{n})$. In other words, $S_n$ is formed by choosing $n$ rows from $H(\overline{n})$, which still achieves decoupling because subsets of rows of $H(\overline{n})$ are still pairwise orthogonal. The resulting decoupling scheme for $n$ spins requires $\overline{n}$ time intervals.

As an example, $S_9$ can be chosen to be the first nine rows of $H(12)$ in Eq. (7.16):

$$
S_9 = \begin{bmatrix}
+ & + & + & + & + & + & - & + & + & + & + & + \\
+ & + & + & - & - & + & + & - & + & - & - & + \\
+ & + & + & + & - & - & + & + & - & + & - & - \\
+ & - & + & + & + & - & + & - & + & - & + & - \\
+ & - & - & + & + & + & + & - & - & + & - & + \\
+ & + & - & - & + & + & + & + & - & - & + & - \\
- & + & + & + & + & + & - & - & - & - & - & - \\
+ & - & + & - & - & + & - & - & - & + & + & - \\
+ & + & - & + & - & - & - & - & - & - & + & +
\end{bmatrix} .
\tag{7.17}
$$

Note that the scheme is efficient if $\overline{n} - n \ll n$. A detailed analysis of the efficiency will be given after we present the recoupling scheme.

### 7.3.6   Recoupling Scheme

We first construct a scheme which removes both $\mathcal{H}_Z$ and $\mathcal{H}_c$. To remove both $\mathcal{H}_Z$ and $\mathcal{H}_c$, note that the Zeeman term for the $i$-th spin is linear in $Z^{(i)}$, and negating $Z^{(i)}$ for half of the time results in no net Zeeman evolution for the $i$-th spin. Therefore, Zeeman evolution for all spins can be removed if the sign matrix has identically zero row sum. Such a sign matrix can be constructed by starting with a *normalized* $H(\overline{n})$ and excluding the first row of $H(\overline{n})$ in the sign matrix. Since a normalized $H(\overline{n})$ has only +'s in the first row, all other rows have zero row sums by orthogonality. Such construction is possible unless $n = \overline{n}$, in which case construction should start with $H(\overline{n+1})$. For instance, the last nine rows of the normalized $H(12)$ is a valid $S_9$.

To implement selective recoupling between the $i$-th and the $j$-th spins, the sign matrix should have equal $i$-th and $j$-th rows but any other two rows should be orthogonal. The coupling term $g_{ij} Z^{(i)} \otimes Z^{(j)}$ never changes sign and is implemented selectively, while all other couplings are removed. The sign matrix can be obtained from a *normalized* $H(\overline{n})$ by first excluding the 1-st row and taking the 2-nd row of $H(\overline{n})$ to be the $i$-th and the $j$-th rows of $S_n$. The other $n - 2$ rows of $S_n$ can be chosen from the remaining $\overline{n} - 2$ rows of $H(\overline{n})$. This scheme also removes $\mathcal{H}_Z$ and requires no more than $\overline{n}$ time intervals. To implement $Z\!\!\!Z_{ij}$, the duration of each interval $t$ is chosen to satisfy $g_{ij}\overline{n}t = \pi/4$. Note that the total time used to implement $Z\!\!\!Z_{ij}$ is the shortest possible, since the coupling is always "on".

For example, starting from the normalized $H(12)$, $S_9$ performing $Z\!\!\!Z_{34}$ can be chosen as

$$
S_9 = \begin{bmatrix}
+ & + & + & + & - & - & - & + & - & + & - & - \\
+ & - & + & + & + & - & - & - & + & - & + & - \\
+ & + & + & - & - & + & - & - & + & - & - & + \\
+ & + & + & - & - & + & - & - & + & - & - & + \\
+ & - & - & + & + & + & - & - & - & + & - & + \\
+ & + & - & - & + & + & - & + & - & - & + & - \\
+ & - & - & - & - & - & - & + & + & + & + & + \\
+ & - & + & - & - & + & + & - & - & + & + & - \\
+ & + & - & + & - & - & + & - & - & - & + & +
\end{bmatrix} .
\qquad (7.18)
$$

### 7.3.7 Efficiency

The decoupling and recoupling schemes require $\overline{n}$ time intervals. They require at most $n\overline{n}$ pulses, since $XX = I$ and the $X$ pulses are only used in pairs. The remaining question is: how does $\overline{n}$ depend on $n$? If Hadamard matrices exist and can be constructed for all orders, $\overline{n} = n$. However, some Hadamard matrices are missing, either because no construction methods are known or they simply cannot exist. Therefore, $\overline{n} = cn$ where $c \geq 1$. We will use the facts about the existence of Hadamard matrices described in Section 7.3.4. As $H(n)$ exists for every $n \equiv 0 \bmod 4$ for $n < 428$, therefore, $\overline{n} - n \leq 3 \ \forall n < 428$. We argue for *arbitrary n* that the schemes are still very efficient. First of all, we prove that $c < 2$. For each $n$, there exists $r$ such that $2^{r-1} \leq n < 2^r$. Since $H(2^r)$ exists by Sylvester's construction, $cn = \overline{n} \leq 2^r < 2n$. We now show that $c$ is close to the ideal value 1 in most cases, due to the existence of Hadamard matrices of orders other than powers of 2. This is why the full connection to Hadamard matrices is useful. First of all, $\overline{n} - n \leq 31 \ \forall n \leq 10000$. In Fig. 7.3, $c$ as a function of $n$ is plotted for $n \leq 10000$. Within this technologically relevant range of $n$, $c$ deviates significantly from 1 only for a few exceptional values of *n when n is small*. For completeness, we present arguments for $c \approx 1$ for *arbitrarily large n* in Appendix C.1. This is based on Paley's construction and the prime number theorem. Finally, if Hadamard's conjecture is proven, $\overline{n} - n \leq 3 \ \forall n$.

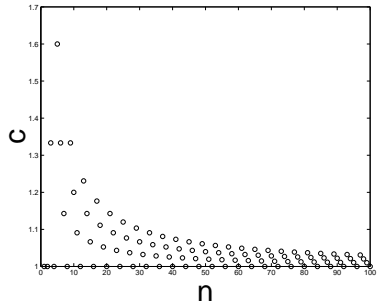

Figure 7.3: Plots of $c$ vs $n$, where $cn = \overline{n}$ is the minimum number of time intervals required to perform decoupling or selective recoupling for an *n*-spin system. $c$ for $n \leq 100$ and $101 \leq n \leq 10000$ are plotted separately.

## 7.4    Conclusion

We reduce the problem of decoupling and selective recoupling in heteronuclear spin systems to finding sign matrices which is further reduced to finding Hadamard matrices. While the most difficult task of constructing Hadamard matrices is not discussed, solutions already exist in the literature. Even more important is that the connection to Hadamard matrices results in very efficient schemes.

Some properties of the scheme are as follows. First of all, the scheme is optimal in the following sense. The rows of Hadamard matrices and their negations form the codewords of the first order Reed-Muller codes, which are *perfect codes* [111, 83]. It follows that, for each Hadamard matrix, it is impossible to add an extra row which is orthogonal to all the existing ones. Therefore, for a given $n$, $\overline{n}$ is in fact the minimum number of time intervals necessary for decoupling or recoupling, if one restricts to the class of schemes considered. Second, the scheme applies for arbitrary duration of the time intervals. This is a consequence of the commutivity of all the terms in the Hamiltonian, which in turn comes from the large separations of the Zeeman frequencies compared to the coupling constants. Spin systems can be chosen to satisfy this condition. Finally, disjoint pairs of spins can be coupled in parallel.

We outline possible simplifications of the scheme for systems with restricted range of coupling. For example, a linear spin system with $n$ spins but only $k$-nearest neighbor coupling can be decoupled by a scheme for $k$ spins only. The $i$-th row of the $n \times \overline{k}$ sign matrix can be chosen to be the $r$-th row of $H(\overline{k})$, where $i \equiv r \bmod k$. Selective recoupling can be implemented using a decoupling scheme for $k+1$ spins. The sign matrix is constructed as in decoupling using $H(\overline{k+1})$ but the rows for the spins to be coupled are chosen to be the $(k+1)$-th row different from all existing rows [2]. This method involving periodic boundary conditions generalizes to other spatial structures. The size of the scheme depends on $k$ and the exact spatial structure but not on $n$.

Our scheme has several limitations. First of all, it only applies to systems in which spins can be individually addressed by short pulses and couplings have the simplified form given by Eq. (7.3). These conditions are essential to the simplicity of the scheme. They can all be satisfied if the Zeeman frequencies have large separations. Second, generalizations to include couplings of higher order than bilinear remain to be developed. Furthermore, in practice, RF pulses are inexact and have finite durations, leading to imperfect transformations and residual errors.

The present discussion is only one example of a more general issue, that the naturally occurring Hamiltonian in a system does not directly give rise to convenient quantum logic gates or other computations such as simulation of quantum systems [110]. Efficient conversion of the given system Hamiltonian to a useful form is necessary and is an important challenge for future research.

# Chapter 8

# Quantum error correction in NMR

## 8.1 Introduction

We have described how quantum error correction can be used to protect information. We have also discussed the theories that enable small scale quantum information processing in NMR. In this chapter, we describe an experimental implementation of a simple phase error detection scheme [30] that *encodes one qubit in two* and detects a single phase error in either one of the two qubits. The output state is rejected if an error is detected so that the probability to accept an erroneous state is reduced to the smaller probability of having multiple errors. Our aim is to study the effectiveness of quantum error correction in a real experimental system, focusing on effects arising from imperfections of the logic gates. Therefore, the experiment is designed to eliminate potential artificial origins of bias in the following ways. First, we compare output states stored with and without coding (the latter is unprotected but less affected by gate imperfections). Second, by ensuring that all qubits used in the code decohere at nearly the same rate, we eliminate apparent improvements brought by having an ancilla with lifetime much longer than the original unencoded qubit. Third, our experiment utilizes only *naturally occurring* error processes. Finally, the main error processes and their relative importance to the experiment are thoroughly studied and simulated to substantiate any conclusions. In these aspects, our study differs significantly from previous work [41] demonstrating quantum error correction working only in principle.

We performed two sets of experiments using NMR: (1) the "coding experiments" in which input states were encoded, stored and decoded, and (2) the "control experiments" in which encoding and decoding were omitted. Comparing the output states obtained from the coding and the control experiments, both error correction by coding and extra errors caused by the imperfect coding operations were taken into account when evaluating the actual advantage of coding. In our experiments, coding reduced the net error probabilities to second order as predicted, at the cost of small additional errors which decreased with the original error probabilities. We identified the major imperfection in the logic gates to be the inhomogeneity in the radio frequency (RF) field used for single spin rotations. Simulation results including both phase damping and RF field inhomogeneity confirmed that the additional errors were mostly caused by RF field inhomogeneity. The causes and effects of other deviations from theory were also studied.

The rest of this chapter is structured into five sections: Section 8.2 reviews the phase damping model, the two-bit coding scheme, and aspects of bulk NMR quantum computation useful for the present discussion. These are reviewed in Sections 8.2.1, 8.2.2 and 8.2.3. Section 8.3 describes the methods to implement the two-bit coding scheme in NMR, and the fidelity measures to evaluate the scheme. Section 8.4 presents the experimental details. Section 8.5 consists of the experimental results together with a thorough analysis. The effects of coding, gate imperfections and the causes and effects of other discrepancies are studied in detail. In Section 8.6, we conclude with a summary of our results. We also discuss syndrome measurement in bulk NMR, the equivalence between logical labeling and coding, the applicability of the two-bit detecting code as a correcting code exploiting classical redundancy in the bulk sample and the issue of signal strength in error correction in bulk NMR.

## 8.2   Theory

### 8.2.1   Phase damping

We first introduced phase damping in Section 2.4. We provide a more physically motivated explanation in the following. Phase damping can be caused by random phase shifts $e^{-i\theta\sigma_z/2}$ of the system due to its interaction with the environment, causing the state change:

$$\rho = \begin{bmatrix} a & b^* \\ b & c \end{bmatrix} \rightarrow \rho' = \begin{bmatrix} a & b^* e^{-i\theta} \\ b e^{i\theta} & c \end{bmatrix} . \tag{8.1}$$

We model phase randomization as a stochastic Markov process with $\theta$ drawn from a normal distribution. The density matrix resulting from averaging over $\theta$ is

$$\langle \rho' \rangle_\theta = \int \frac{1}{\sqrt{2\pi} s} e^{-\frac{\theta^2}{2s^2}} \rho' d\theta = \begin{bmatrix} a & b^* e^{-\frac{s^2}{2}} \\ b e^{-\frac{s^2}{2}} & c \end{bmatrix} , \tag{8.2}$$

where $s^2$ is the variance of the distribution of $\theta$. By the Markov assumption, the total phase shift during a time period $t$ is a random walk process with variance proportional to $t$. Therefore, we replace $s^2/2$ by $\lambda t$ in Eq. (8.2) when the time elapsed is $t$. The effect of phase damping is:

$$\begin{bmatrix} a & b^* \\ b & c \end{bmatrix} \quad \rightarrow \quad \begin{bmatrix} a & e^{-\lambda t} b^* \\ e^{-\lambda t} b & c \end{bmatrix} . \tag{8.3}$$

Since the diagonal and the off-diagonal elements represent the populations of the basis states and the quantum coherence between them, the exponential decay of the off-diagonal elements caused by phase damping signifies the loss of coherence without net change of quanta.

Phase damping describes the axisymmetric exponential decay of the $\hat{x}$ and $\hat{y}$ components of any Bloch vector, $\rho = \frac{1}{2}(I + x\sigma_x + y\sigma_y + z\sigma_z)$, as depicted in Fig. 8.1.

Recall that phase damping can also be described as a discrete process. Eq. (8.3) describing phase damping can be rewritten as

$$\mathcal{E}(\rho) = (1 - p) \, I \rho I^\dagger + \, p \, \sigma_z \rho \sigma_z^\dagger , \tag{8.4}$$

where $p = (1 - e^{-\lambda t})/2$. In Eq. (8.4), the output $\mathcal{E}(\rho)$ can be considered as a $(1\text{-}p)$:$p$ mixture of $\rho$ and $\sigma_z \rho \sigma_z^\dagger$; in other words, $\mathcal{E}(\rho)$ is a mixture of the states after the event "no jump" ($I$) or "a phase error" ($\sigma_z$) has occurred. The weights $1 - p$ and $p$ are the probabilities of the two possible events.

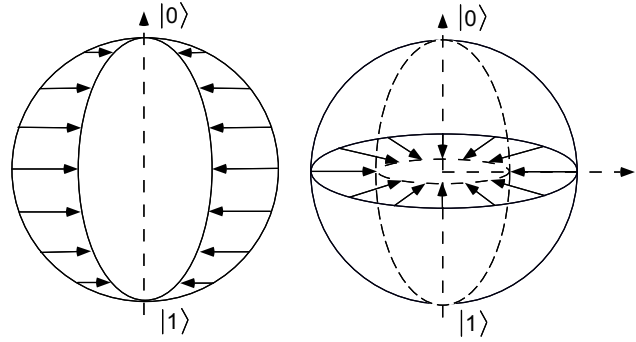

Figure 8.1: Trajectories of different points on the Bloch sphere under the effect of phase damping. Points move along perpendiculars to the $\hat{z}$-axis at rates proportional to the distances to the $\hat{z}$-axis. As a result, the Bloch sphere turns into an ellipsoid.

We emphasize that Eqs. (8.3) and (8.4) describe the same physical process. Equation (8.4) provides a discrete interpretation of phase damping, with the continuously changing parameter $e^{-\lambda t}$ embedded in the probabilities of the possible events.

For a system of multiple qubits, we *assume* independent decoherence on each qubit. For example, for two qubits $a$ and $b$ with error probabilities $p_a$ and $p_b$, the joint process is given by

$$
\begin{aligned}
\mathcal{E}(\rho) \;=\; & (1-p_a)(1-p_b) \; (I \otimes I) \; \rho \; (I \otimes I) \\
+ \; & (1-p_a) \; p_b \quad (I \otimes \sigma_z) \; \rho \; (I \otimes \sigma_z) \\
+ \; & p_a \; (1-p_b) \quad (\sigma_z \otimes I) \; \rho \; (\sigma_z \otimes I) \\
+ \; & p_a \; p_b \quad\quad (\sigma_z \otimes \sigma_z) \; \rho \; (\sigma_z \otimes \sigma_z) \,,
\end{aligned} \tag{8.5}
$$

where $\rho$ denotes the $4 \times 4$ density matrix for the two qubits. The events $\sigma_z \otimes I$ and $I \otimes \sigma_z$ are first order errors, while $\sigma_z \otimes \sigma_z$ is second order. First and second order events occur with probabilities linear and quadratic in the small error probabilities.

Having described the noise process, we now proceed to describe a coding scheme that will correct for it.

## 8.2.2 The two-bit phase damping detecting code

For a code to detect errors, it suffices to choose the codeword space $\mathcal{C}$ such that all errors to be detected map $\mathcal{C}$ to its orthogonal complement. In this way, detection can be done unambiguously by a projection onto $\mathcal{C}$ *without* distinguishing individual codewords; hence without disturbing the encoded information. To make this concrete, consider the code [30]

$$|0_L\rangle = \frac{1}{\sqrt{2}}(|00\rangle + |11\rangle) \tag{8.6}$$

$$|1_L\rangle = \frac{1}{\sqrt{2}}(|01\rangle + |10\rangle). \tag{8.7}$$

An arbitrary encoded qubit is given by

$$|\psi\rangle = \alpha|0_L\rangle + \beta|1_L\rangle \tag{8.8}$$

$$= \frac{1}{\sqrt{2}}\left[\alpha(|00\rangle + |11\rangle) + \beta(|01\rangle + |10\rangle)\right]. \tag{8.9}$$

After the four possible errors in Eq. (8.5), the possible outcomes are

$$|\psi_{II}\rangle = I \otimes I\,|\psi\rangle = \alpha\,\frac{|00\rangle + |11\rangle}{\sqrt{2}} + \beta\,\frac{|01\rangle + |10\rangle}{\sqrt{2}} \tag{8.10}$$

$$|\psi_{ZI}\rangle = \sigma_z \otimes I\,|\psi\rangle = \alpha\,\frac{|00\rangle - |11\rangle}{\sqrt{2}} + \beta\,\frac{|01\rangle - |10\rangle}{\sqrt{2}} \tag{8.11}$$

$$|\psi_{IZ}\rangle = I \otimes \sigma_z|\psi\rangle = \alpha\,\frac{|00\rangle - |11\rangle}{\sqrt{2}} + \beta\,\frac{-|01\rangle + |10\rangle}{\sqrt{2}} \tag{8.12}$$

$$|\psi_{ZZ}\rangle = \sigma_z \otimes \sigma_z|\psi\rangle = \alpha\,\frac{|00\rangle + |11\rangle}{\sqrt{2}} + \beta\,\frac{-|01\rangle - |10\rangle}{\sqrt{2}}, \tag{8.13}$$

with the *first order* erroneous states $|\psi_{ZI}\rangle$ and $|\psi_{IZ}\rangle$ orthogonal to the correct state $|\psi_{II}\rangle$. Therefore, it is possible to distinguish (8.10) from (8.11) and (8.12) by a projective measurement during decoding, which is described next.

The encoding and decoding can be performed as follows. We start with an arbitrary input state and a ground state ancilla, represented as qubits $a$ and $b$ in the circuit in Fig. 8.2.

To encode the input qubit, $H$ is applied to the ancilla, followed by CNOT$_{ba}$. Let spins $a$

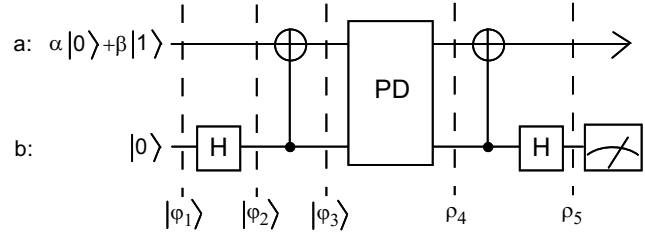

Figure 8.2: Circuit for encoding and decoding. Qubit $a$ is the input qubit. $|\psi_{1-3}\rangle$ are given by Eqs. (8.14)-(8.16). $\rho_4$, $\rho_5$ are mixtures of the states in Eqs. (8.10)-(8.13) and in Eqs. (8.18)-(8.21). A phase error in either one of the qubits will be revealed by qubit $b$ being in $|1\rangle$ after decoding, and in that case, qubit $a$ will be rejected.

and $b$ be the first and second register. Then, the qubits transform according to Fig. 8.2 as

$$|\psi_1\rangle = (\alpha|0\rangle + \beta|1\rangle)|0\rangle \tag{8.14}$$

$$\xrightarrow{I \otimes H} |\psi_2\rangle = \frac{1}{\sqrt{2}}(\alpha|0\rangle + \beta|1\rangle)(|0\rangle + |1\rangle) \tag{8.15}$$

$$\xrightarrow{\text{CNOT}_{ba}} |\psi_3\rangle = \frac{1}{\sqrt{2}}\Big[(\alpha|0\rangle + \beta|1\rangle)|0\rangle + (\alpha|1\rangle + \beta|0\rangle)|1\rangle\Big] \tag{8.16}$$

$$= \frac{1}{\sqrt{2}}\Big[\alpha(|00\rangle + |11\rangle) + \beta(|01\rangle + |10\rangle)\Big] , \tag{8.17}$$

where Eq. (8.17) is the desired encoded state.

The decoding operation is the inverse of the encoding operation (see Fig. 8.2) so as to recover the input $(\alpha|0\rangle + \beta|1\rangle)|0\rangle$ in the absence of errors. Phase errors lead to other decoded outputs. The possible decoded states are given by:

$$|\psi_{II}\rangle \xrightarrow{decode} (\alpha|0\rangle + \beta|1\rangle)|0\rangle \tag{8.18}$$

$$|\psi_{ZI}\rangle \implies (\alpha|0\rangle - \beta|1\rangle)|1\rangle \tag{8.19}$$

$$|\psi_{IZ}\rangle \implies (\alpha|0\rangle + \beta|1\rangle)|1\rangle \tag{8.20}$$

$$|\psi_{ZZ}\rangle \implies (\alpha|0\rangle - \beta|1\rangle)|0\rangle . \tag{8.21}$$

Note that the ancilla decodes to $|1\rangle$ if and only if a *single* phase error has occurred. Moreover, qubits $a$ and $b$ are in product states but they are classically correlated. Therefore, the syndrome can be read out by a projective measurement on $b$ without measuring the encoded state. The decoding operation transforms the codeword space and its orthogonal complement to the subspaces spanned by $|0\rangle$ and $|1\rangle$ in qubit $b$, while all the encoded information, either with or without errors, goes to qubit $a$.

We quantify the error correcting effect of coding using the discrete interpretation of the noise process, leaving a full discussion of the fidelity to Section 8.3. Recall from Eq. (8.5) that the errors $I \otimes I$, $I \otimes \sigma_z$, $\sigma_z \otimes I$, and $\sigma_z \otimes \sigma_z$ occur with probabilities $(1-p_a)(1-p_b)$, $(1-p_a)p_b$, $p_a(1-p_b)$ and $p_a p_b$ respectively, and only in the first and the last cases will the output state be accepted. The probability of accepting the output state is $(1-p_a)(1-p_b) + p_a p_b$ whereas the probability of accepting the correct state is $(1-p_a)(1-p_b)$. The *conditional* probability of a correct, accepted state is therefore

$$\frac{(1-p_a)(1-p_b)}{(1-p_a)(1-p_b) + p_a p_b} \approx 1 - p_a p_b \tag{8.22}$$

for small $p_a$, $p_b$. The code improves the conditional error probability to second order, as a result of screening out the first order erroneous states.

We conclude this section with a discussion of some properties of the two-bit code. First, the code also applies to mixed input states since the code preserves all constituent pure states in the mixed input. Second, we show here that two qubits are the minimum required to encode one qubit and to detect any phase errors. Let $\mathcal{C}$ be the 2-dimensional codeword space and $E$ be a non-trivial error to be detected. For phase damping, $E$ is unitary and therefore $E\mathcal{C}$ is also 2-dimensional. Moreover, $\mathcal{C}$ and $E\mathcal{C}$ must be orthogonal if $E$ is to be detected. Therefore the minimum dimension of the system is 4, which requires two qubits. However, using only two qubits implies other intrinsic limitations. First, the code can detect but cannot distinguish errors. Therefore, it cannot correct errors. This affects the absolute fidelity (the overall probability of successful recovery) but not the conditional fidelity (the probability of successful recovery if the state is accepted). Second, the error $\sigma_z \otimes \sigma_z$ cannot be detected. This affects both fidelities but only in second order. To understand why these limitations are intrinsic, let $\{E_k\}$ be the set of non-trivial errors to be detected. Since $E_k\mathcal{C}$ has to be orthogonal to $\mathcal{C}$ for all $k$, and since $\mathcal{C}$ has a unique orthogonal complement of dimension 2 in a two-bit code, it follows that all $E_k\mathcal{C}$ are equal, and it is impossible to distinguish (and correct) the different errors. By the same token, for any distinct errors $E_{k'}$ and $E_k$, $E_{k'}E_k\mathcal{C} = \mathcal{C}$ because they are both orthogonal to $E_k\mathcal{C}$, which has a unique 2-dimensional orthogonal complement. Therefore, a two-bit code that detects single phase errors can never detect double errors. Finally, since a detecting code cannot correct errors, it can only improve the conditional fidelity of the *accepted* states but not the absolute fidelity. We remark that the conditional fidelity is a better measure in our experiments due to the bulk nature of the system used to implement the two-bit code. A discussion of fidelity measures in our experiments and quantum error correction in bulk systems will be postponed until Sections 8.3 and 8.6. The system in which the two-bit code is implemented will be described next.

### 8.2.3 Bulk NMR Quantum Computation

We use a two qubit NMR system, with *reduced* Hamiltonian (see also Fig. 8.3)

$$\mathcal{H} = -\frac{\omega_a}{2}\sigma_z \otimes I - \frac{\omega_b}{2}I \otimes \sigma_z + \frac{\pi J}{2}\sigma_z \otimes \sigma_z + \mathcal{H}_{env}\,, \qquad (8.23)$$

where the symbols have their usual meaning defined in Section 6.2 and $g_{ab} = \pi J/2$.

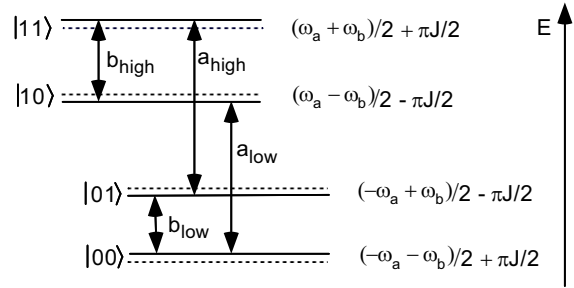

Figure 8.3: Energy diagram for the two-spin nuclear system. The transitions labeled $a_{low}$, $a_{high}$, $b_{low}$ and $b_{high}$ refer to transitions $(|0\rangle \leftrightarrow |1\rangle)|0\rangle$, $(|0\rangle \leftrightarrow |1\rangle)|1\rangle$, $|0\rangle(|0\rangle \leftrightarrow |1\rangle)$ and $|1\rangle(|0\rangle \leftrightarrow |1\rangle)$ respectively.

We define some frequently used gates based on the discussion in Sections 2.2 and 6.2. The single qubit rotation $e^{-i\frac{\theta}{2}\vec{\sigma}\cdot\hat{\eta}}$ is denoted by $R_\eta(\theta)$. We denote rotations of $\pi/2$ along the $\hat{x}$ and $\hat{y}$ axes for spins $a$ and $b$ by $R_{xa}$, $R_{ya}$, $R_{xb}$, and $R_{yb}$ with respective matrix representations $e^{-i\frac{\pi}{4}\sigma_x \otimes I}$, $e^{-i\frac{\pi}{4}\sigma_y \otimes I}$, $e^{-i\frac{\pi}{4}I \otimes \sigma_x}$, and $e^{-i\frac{\pi}{4}I \otimes \sigma_y}$. The rotations in the reverse directions are denoted by an additional "bar" above the symbols of the original rotations, such as $\overline{R}_{xa}$.

In the respective rotating frames of the spins (tracing the free precession of the uncoupled spins), only the $J$-coupling term, $e^{-i\frac{\pi Jt}{2}\sigma_z \otimes \sigma_z}$, is relevant in the time evolution. In this chapter, an evolution of duration $t = \frac{1}{2J}$ is denoted by $\tau$. It corresponds to the evolution $e^{-i\frac{\pi}{4}\sigma_z \otimes \sigma_z}$ ($Z\!Z_{ab}$ in the previous chapter).

Recall from Section 6.2.3 that if the density matrix is $\sum_{i,j=0}^{3} c_{ij}\sigma_i \otimes \sigma_j$ at the onset of the measurement, four spectral lines at frequencies $\frac{\omega_a}{2\pi} + \frac{J}{2}, \frac{\omega_a}{2\pi} - \frac{J}{2}, \frac{\omega_b}{2\pi} + \frac{J}{2}, \frac{\omega_b}{2\pi} - \frac{J}{2}$, can be obtained, with corresponding peak integrals:

$$I_{a_{high}} = -\Big[i(c_{10} - c_{13}) + c_{20} - c_{23}\Big] \tag{8.24}$$

$$I_{a_{low}} = -\Big[i(c_{10} + c_{13}) + c_{20} + c_{23}\Big] \tag{8.25}$$

$$I_{b_{high}} = -\Big[i(c_{01} - c_{31}) + c_{02} - c_{32}\Big] \tag{8.26}$$

$$I_{b_{low}} = -\Big[i(c_{01} + c_{31}) + c_{02} + c_{32}\Big]. \tag{8.27}$$

Note that the expression $c_{10} - c_{13}$ occurring in the *high* frequency line of spin $a$ is the coefficient of $\sigma_x \otimes |1\rangle\langle 1|$ in $\rho(0)$; $c_{10} + c_{13}$ in the *low* frequency line of spin $a$ is the coefficient of $\sigma_x \otimes |0\rangle\langle 0|$. Likewise, $c_{20} - c_{23}$ is the coefficient of $\sigma_y \otimes |1\rangle\langle 1|$ and $c_{20} + c_{23}$ is the coefficient of $\sigma_y \otimes |0\rangle\langle 0|$. These quantities signify the transitions $|0\rangle \leftrightarrow |1\rangle$ for spin $a$ conditioned on spin $b$ being in $|1\rangle$ or $|0\rangle$. Similar observations hold for the high and low lines of spin $b$ (see Fig. 6.1).

Generalized amplitude damping is much slower than phase damping in our system so that unitality is a good approximation. When the evolution of the density matrix is given by $\mathcal{E}$, the effective evolution of the deviation $\rho_\Delta$ is given by (see Section 6.3.1)

$$\rho_\Delta \rightarrow \upsilon(\mathcal{E}(I) - I) + \mathcal{E}(\rho_\Delta). \tag{8.28}$$

The second term in Eq. (8.28) represents the result of applying $\mathcal{E}$ to the deviation $\rho_\Delta$ when neglecting the identity, and the first term is the correction due to non-unitality. In our experiment, $\mathcal{E}(I) - I$ is small compared to other effects and can be treated as a small extra distortion of the state when making the unitality assumption.

## 8.3   Two-bit code in NMR

We now describe how the two-bit code experiment can be implemented in an ensemble of two-spin systems. Modifications of the standard theories in Section 8.2.2 are needed. These include methods for state preparation, designing encoding and decoding pulse sequences, methods to store the qubit with controllable phase damping, and finally methods to read out the decoded qubit. Fidelity measures for deviation density matrices are also defined.

Spins $a$ and $b$ are designated to be the input and the ancilla qubits respectively. The output states of spin $a$ are reconstructed from the peak integrals at frequencies $\omega_a/2\pi \pm J/2$. Fig. 8.4 schematically summarizes the major steps in the experiments, with details given in the text.

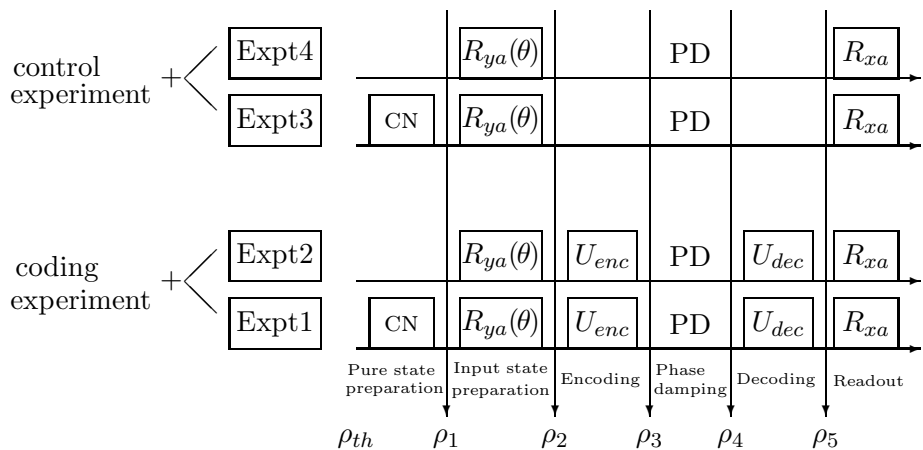

Figure 8.4: Schematic diagram for the two-bit code experiment. CN is used to prepare the initial state. $R_{ya}(\theta)$ is a variable angle rotation applied to prepare an arbitrary input state, which is then subject to phase damping (PD). In the coding experiment, encoding and decoding operations, $U_{enc}$ and $U_{dec}$, are performed before and after phase damping, whereas in the control experiment, these operations are omitted. $R_{xa}$ is used as a readout pulse on spin $a$ to determine the output state $\rho_5$ in spin $a$. $\rho_i$ corresponds to $|\psi_i\rangle$ or $\rho_i$ in Fig. 8.2. Details are described in the text.

Some notation is defined as follows. Initial states and input states refer to $\rho_0$ and $\rho_1$ in Fig. 8.4. The phrase "ideal case" refers to the scenario of having perfect logic operations throughout the experiments and pure phase damping during storage.

**Initial state preparation**

It is necessary to initialize spin $b$ to $|0\rangle$ before the experiment. This can be done with temporal labeling. The idea is to add up the results of a series of experiments that begin with different preparation operations $P_k$ before the intended computation. The result is equivalent to performing the computation on the initial state $\sum_k P_k \rho_{th} P_k^\dagger$. To prepare spin $b$ in an effective pure state, only two experiments are needed: the first experiment starts with no additional pulses; the second experiment starts with CN (Fig. 6.2) which acts as CNOT$_{ba}$ on the thermal state. The equivalent initial state is $\rho_{th} + \rho_{cn}$ (symbols are as defined in Section 6.2.5):

$$\begin{bmatrix} \omega_a + \omega_b & 0 & 0 & 0 \\ 0 & -\omega_b & 0 & 0 \\ 0 & 0 & -\omega_a + \omega_b & 0 \\ 0 & 0 & 0 & -\omega_b \end{bmatrix} = \omega_a \; \sigma_z \otimes |0\rangle\langle 0| + \omega_b \; I \otimes \sigma_z \,. \tag{8.29}$$

The first term in Eq. (8.29) is the desired initial state. The second term cannot affect the observable of interest, the spectral lines at $\omega_a/2\pi$, because of the following. The identity in $a$ is invariant under the preparation pulse $R_{ya}(\theta)$. The input state is thus the identity, which has no coherence to start with. Therefore, the output state after phase damping in both the control and the coding experiment is still the identity. This is non-trivial in the coding experiment. However, inspection of Eqs. (8.18)-(8.21) shows that spin $a$ is changed at most by a phase in the coding experiment. While Eqs. (8.18)-(8.21) apply only to the case when $b$ starts in $|0\rangle$, the result can be generalized to any arbitrary diagonal density matrix in $b$ (proof omitted). It follows from Eqs. (8.24)-(8.27) that the second term is not observable in the output spectral lines of $a$.

In contrast, the input state in spin $a$ can be a mixed state as given by the first term in Eq. (8.29), since the phase damping code is still applicable. Different input states can be prepared by rotations about the $\hat{y}$-axis of different angles $\theta \in [0, \pi]$ to span a semi-circle in the Bloch sphere in the $\hat{x}\hat{z}$-plane. Due to the axisymmetry of phase damping (Eq. (8.2)), these states suffice to represent all the states to test the code.

We conclude with an alternative interpretation of the initial state preparation. Let the fractional populations of $|00\rangle$, $|01\rangle$, $|10\rangle$, and $|11\rangle$ be $p_{00}$, $p_{01}$, $p_{10}$, and $p_{11}$ in the thermal state. Then, the initial state after temporal labeling is

$$(p_{01} + p_{11})\ I \otimes |1\rangle\langle 1| + 2\, p_{10}\ I \otimes |0\rangle\langle 0| +\ 2\ (p_{00} - p_{10})\ |00\rangle\langle 00|\,, \qquad (8.30)$$

where the identity term is not omitted, unlike Eq. (8.29). Temporal labeling serves to randomize spin $a$ in the first term in Eq. (8.30) when $b$ is $|1\rangle$. We have shown previously that the identity input state of $a$ is preserved throughout both the coding and the control experiments in the ideal case. Consequently, only the last term in Eq. (8.30) contributes to any detectable signal in all the experiments, and we can consider the last term as the initial state.

Having justified both pictures to identify the first term in Eq. (8.29) and the last term in Eq. (8.30) as the initial state, both pictures will be used throughout the discussion.

**Encoding and decoding**

The original encoding and decoding operations are composed of the Hadamard transformation and CNOT$_{ba}$, as defined in Section 6.2. The actual sequences can be simplified and are shown in Fig. 8.5.

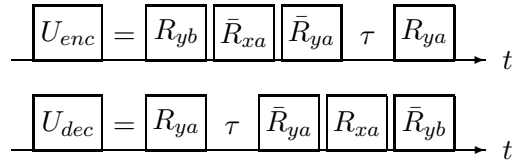

Figure 8.5: Pulse sequences to implement the encoder $U_{enc}$ and the decoder $U_{dec}$. Time runs from left to right.

The operator $U_{enc}$ can be found by multiplying the component operators in Fig. 8.5,

giving

$$U_{enc} = \frac{1}{\sqrt{2}} \begin{bmatrix} 1 & -1 & 0 & 0 \\ 0 & 0 & i & i \\ 0 & 0 & 1 & -1 \\ i & i & 0 & 0 \end{bmatrix}. \tag{8.31}$$

The encoded states are slightly different from those in section 8.2.2:

$$|0_L\rangle = \frac{1}{\sqrt{2}}(|00\rangle + i|11\rangle) \tag{8.32}$$

$$|1_L\rangle = \frac{1}{\sqrt{2}}(i|01\rangle + |10\rangle), \tag{8.33}$$

but the scheme is nonetheless equivalent to the original one. The decoding operation $U_{dec}$ is given by

$$U_{dec} = \frac{1}{\sqrt{2}} \begin{bmatrix} 1 & 0 & 0 & -i \\ -1 & 0 & 0 & -i \\ 0 & -i & 1 & 0 \\ 0 & -i & -1 & 0 \end{bmatrix} = U_{enc}^{\dagger}. \tag{8.34}$$

The possible decoded outputs are the same as in Section 8.2.2 except for an overall sign in the single error cases.

**Storage**

The time delay between encoding and decoding corresponds to storage time of the quantum state. During this delay time, phase damping, amplitude damping and $J$-coupling occur simultaneously. How to single out the effects of phase damping during storage is explained as follows.

First of all, the time constants of amplitude damping, $T_1$'s, are much longer than those of phase damping, $T_2$'s. Storage times $t_d$ are chosen to satisfy $t_d \leq T_2 \ll T_1$. This ensures that the effects of amplitude damping are small.

The remaining two processes, phase damping and $J$-coupling, can be considered as independent and commuting processes in between any two pulses since all the phase damping operators commute with the $J$-coupling evolution $\exp(-i\,\sigma_z \otimes \sigma_z \,\pi J t_d/2)$. We choose $J t_d$ to be even integers to approximate the identity evolution. As $J$ is known with limited accuracy, we add refocusing $\pi$-pulses [50] to spin $b$ (about the $\hat{y}$-axis) in the middle and at the end of the phase damping period to ensure trivial evolution under $J$-coupling. These pulses flip the $\hat{z}$ axis for $b$ during the second half of the storage time so that evolution in the first half is always reversed by that in the second half. In this way, controllable amount of phase damping is achieved to good approximation.

**Control Experiment**

For each storage time $t_d$, input state, and temporal labeling experiment, a control experiment is performed with the coding and decoding operations omitted. Since phase damping and $J$-coupling can be considered as independent processes, and $J$-coupling is arranged to act trivially, the resulting states illustrate phase damping of spin $a$ without coding.

**Output and readout**

For an input state prepared with $R_{ya}(\theta)$, the state after encoding, dephasing and decoding ($\rho_5$ in Fig. 8.4) is derived in Appendix C.2 and is given by Eq. (C.7) (from now on, $\omega_a$ is omitted):

$$
\begin{aligned}
\rho_5^{coded} &= \Big[ \cos\theta \ (1 - p_a - p_b + 2p_a p_b) \ \sigma_z + \sin\theta \ (1 - p_a - p_b) \ \sigma_x \Big] \otimes \ (I + \sigma_z)/2 \\
&+ \Big[ \cos\theta \ (p_a + p_b - 2p_a p_b) \ \sigma_z + \sin\theta \ (-p_a + p_b) \ \sigma_x \Big] \otimes \ (I - \sigma_z)/2 \,. \quad (8.35)
\end{aligned}
$$

In the control experiment, the corresponding output state is given by Eq. (C.2):

$$
\rho_5^{control} = \Big[ \cos\theta \ \sigma_z + (1 - 2p_a)\sin\theta \ \sigma_x \Big] \otimes (I + \sigma_z)/2 \,. \quad (8.36)
$$

The initial state used in the derivation of Eq. (8.36) is the first term in Eq. (8.29), and the encoding and decoding operations are as given by Eqs. (8.31) and (8.34).

In the ideal case, the output state can be read out in a single spectrum. Recall that the coefficients of $-\sigma_y \otimes (I \pm \sigma_z)$ and $-\sigma_x \otimes (I \pm \sigma_z)$ are the real and imaginary parts of the low and the high frequency lines of spin $a$. Therefore, the coefficients of $-\sigma_z \otimes (I \pm \sigma_z)$ and $-\sigma_x \otimes (I \pm \sigma_z)$ in $\rho_5^{coded}$ and $\rho_5^{control}$ can be read out as the real and imaginary parts of the low and the high frequency lines of spin $a$, if $R_{xa}$ is applied before acquisition. This pulse transforms the $z$-component of spin $a$ to the $y$-component leaving the $x$-component unchanged, as described in Section 6.2. Note that only states with spin $b$ being $|0\rangle$ ($|1\rangle$) contribute to the low (high) frequency line. Therefore, in the coding experiments, the accepted (rejected) states of $a$ can be read out separately in the low (high) frequency line. There are no rejected states in the control experiments.

The rest of the chapter makes use of the following notation. "Output states" or "accepted states" refer to the reduced density matrices of $b$ *before* the readout pulse, and are denoted by $\rho_a^{coded} \equiv {}_b\langle 0|\rho_5^{coded}|0\rangle_b$ and $\rho_a^{control} \equiv {}_b\langle 0|\rho_5^{control}|0\rangle_b$. Rejected states refer to ${}_b\langle 1|\rho_5^{coded}|1\rangle_b$ from the coding experiments.

The accepted and rejected states for a given input as calculated from Eq. (8.35) and Eq. (8.36) are summarized in Table 8.1.

|  | $z$-component | $x$-component |
|---|---|---|
| input state | $\cos\theta$ | $\sin\theta$ |
| coding expt.: accepted state rejected state | $(1 - p_a - p_b + 2p_ap_b)\cos\theta$ <br> $(p_a + p_b - 2p_ap_b)\cos\theta$ | $(1 - p_a - p_b)\sin\theta$ <br> $(-p_a + p_b)\sin\theta$ |
| control expt.: accepted state rejected state | $\cos\theta$ <br> $0$ | $(1 - 2p_a)\sin\theta$ <br> $0$ |

Table 8.1: Input and output states of spin $a$ in the coding and the control experiments.

The output states $\rho_a^{coded}$ and $\rho_a^{control}$, as predicted by Table 8.1, are plotted in Fig. 8.6 in the $\hat{x}\hat{z}$-plane of the Bloch sphere of spin $a$. The north and south poles represent the Bloch vectors $\pm\hat{z}$ ($|0\rangle$ and $|1\rangle$). The time trajectories of various initial states are indicated by the arrows. The Bloch sphere is distorted to an ellipsoid after each storage time. We concentrate on the cross-section in one half of the $\hat{x}\hat{z}$-plane, and call the curve an "ellipse" for convenience. The storage times plotted have equal spacing and correspond to $p_a = 0, 0.071, 0.133, 0.185, 0.230, 0.269$. For each ellipse, $p_b$ is chosen to be the same as $p_a$. The main experimental results will comprise of information of this type.

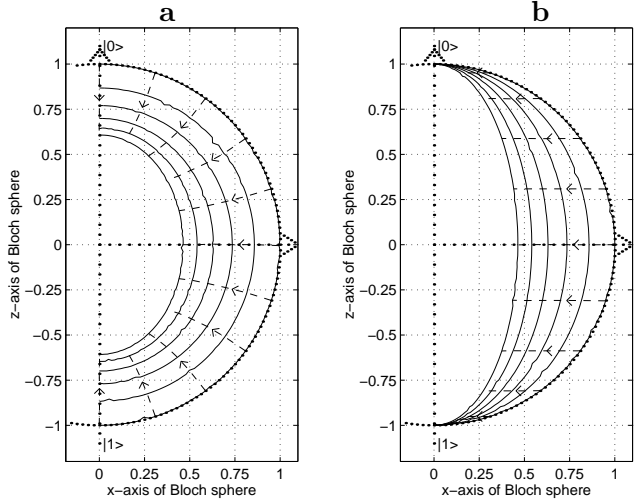

Figure 8.6: Predicted output states (a) with or (b) without coding. The arrows indicate the direction of time and the ellipses represent snapshots of the original surface of the Bloch sphere.

**Fidelity**

One can quantify how well the input states are preserved using various fidelity measures. In classical communication, the fidelity can be defined as the probability of successful recovery of the input bit string in the worse case. In quantum information processing, when the input $\rho_{in}$ is *pure*, the above definition generalizes to the *minimum overlap fidelity* (see also Eq. (3.25)),

$$\mathcal{F} = \min_{\rho_{in}} \mathrm{Tr}(\rho_{out}\rho_{in}).\tag{8.37}$$

We emphasize that Eq. (8.37) applies to *pure* input states only. The reason why Eq. (8.37) is sufficient for our purpose will become clear later.

When $\rho_{in}$ and $\rho_{out}$ are qubit states of unit trace with respective Bloch vectors $\hat{r}_{in}$ and $\vec{r}_{out}$, Eq. (8.37) can be rewritten as

$$\mathcal{F} = \min_{\hat{r}_{in}} \frac{1}{2}(1 + \hat{r}_{in} \cdot \vec{r}_{out}).\tag{8.38}$$

From Eq. (8.3) for phase damping, when $\hat{r}_{in} = (r_x, r_y, r_z)$, $\vec{r}_{out} = (e^{-\lambda t} r_x, e^{-\lambda t} r_y, r_z)$. Therefore,

$$
\begin{aligned}
\hat{r}_{in} \cdot \vec{r}_{out} &= e^{-\lambda t}(r_x^2 + r_y^2) + r_z^2 & (8.39) \\
&= -2p(r_x^2 + r_y^2) + 1 \,, & (8.40)
\end{aligned}
$$

where we have used the fact $|\hat{r}_{in}|^2 = 1$ for pure states and $p = (1 - e^{-\lambda t})/2$. The minimum in Eq. (8.38) is attained for input states on the equatorial plane with $r_x^2 + r_y^2 = 1$. Therefore

$$
\mathcal{F} = 1 - p = \frac{1}{2}(1 + e^{-\lambda t}) \,. \tag{8.41}
$$

With coding, the accepted state is (see Eqs. (8.18)-(8.21))

$$
\rho_a^{coded} = (1 - p_a)(1 - p_b)\rho_{in} + p_a p_b \sigma_z \rho_{in} \sigma_z \,. \tag{8.42}
$$

If one considers the conditional fidelity in the accepted state, $\rho_{out}$ in Eq. (8.37) should be taken as the post measurement density matrix,

$$
\begin{aligned}
\rho_{out} &= \frac{\rho_a^{coded}}{\mathrm{Tr}(\rho_a^{coded})} = \frac{\rho_a^{coded}}{(1 - p_a)(1 - p_b) + p_a p_b} & (8.43) \\
&\approx (1 - p_a p_b)\rho_{in} + p_a p_b \sigma_z \rho_{in} \sigma_z \,. & (8.44)
\end{aligned}
$$

Note that the above expression is identical to the expression for single qubit phase damping but with error probability $p = p_a p_b$. Therefore, coding changes the conditional error probability to second order, and the conditional fidelity is improved to $\mathcal{F}_C = 1 - p_a p_b$.

The amount of distortion can also be summarized by the ellipticities of the "ellipses" that result from phase damping. The ellipticity $\epsilon$ is defined to be the ratio of the major axis to the minor axis. Without coding, the major axis remains unchanged under phase damping, and the minor axis shrinks by a factor of $e^{-\lambda t}$, therefore $\epsilon = e^{\lambda t}$. Using Eq. (8.41),

$$
\mathcal{F} = \frac{1}{2}(1 + \frac{1}{\epsilon}) \,. \tag{8.45}
$$

With coding, $\mathcal{F}_C$ is given by the same expression on the right hand side of Eq. (8.45). In the ideal case, the overlap fidelity and the ellipticity have a one-to-one correspondence. In the presence of imperfections, the overlap fidelity and the ellipticity, one being the minimum of the input-output overlap and the other being an average parameter of distortion, are more effective in reflecting different types of distortion.

We now generalize to new definitions of fidelity for deviation density matrices for the two-bit code. In NMR, quantum information is encoded in the small deviation of the state from a completely random mixture. The problem with the usual definitions of fidelity is that they do not change significantly even when the small deviation changes completely. This is true whether the fidelities are defined for pure or mixed input states. To overcome this problem, we introduce the strategy of identifying the initial excess population in $|00\rangle$ as the pure initial state so that usual definitions of fidelity for pure input states are applicable. This improves the sensitivity of the fidelity measures and provides a closer connection to the pure state picture.

The initial state in Eq. (8.30) can be rewritten as

$$\rho = \alpha \rho^{pure} + (1 - \alpha) \rho^{quiet}, \tag{8.46}$$

where $\alpha = 2(p_{00} - p_{10}) = \hbar \omega_a / 2 k_B T$, and

$$\rho^{pure} = |00\rangle\langle00| \tag{8.47}$$

$$\rho^{quiet} \approx \frac{1}{1 - \alpha} \Big[ (p_{01} + p_{11}) \ I \otimes |1\rangle\langle1| + 2 \ p_{10} \ I \otimes |0\rangle\langle0| \Big]. \tag{8.48}$$

It has already been shown that $\rho^{quiet}$ is irrelevant to the evolution and the measurement of $\rho^{pure}$ when all processes are unital. Therefore $\rho^{quiet}$ is neglected and the small signal resulting from the slow non-unital processes will be treated as extra distortion to the observable component. The input state prepared by $R_{ya}(\theta)$ can be written as

$$\rho_{in} = \alpha \rho_{in}^{pure} + (1 - \alpha) \rho^{quiet}. \tag{8.49}$$

For the state change $\rho_{in} \rightarrow \mathcal{E}(\rho_{in})$, we consider the overlap between $\rho_{in}^{pure}$ and $\mathcal{E}(\rho_{in}^{pure})$ in place of the overlap between $\rho_{in}$ and $\mathcal{E}(\rho_{in})$. This defines a new overlap fidelity $\mathcal{F}_\Delta =$ $\min_{\rho_{in}^{pure}} \text{Tr}(\rho_{in}^{pure} \mathcal{E}(\rho_{in}^{pure})) = \min_{\hat{r}_{in}} \frac{1}{2}(1 + \hat{r}_{in} \cdot \vec{r}_{out})$ similar to the pure state case.

$\mathcal{F}_\Delta$ can be calculated from the experimental results in the following manner. The measured Bloch vector of $a$, $\vec{r}_m$, is proportional to that defined by $_b\langle 0|\mathcal{E}(\rho_{in}^{pure})|0\rangle_b$. Due to limitations in the measurement process, this proportionality constant $\tilde{\alpha}$ is not known a priori. However, when $\theta = 0$ in the control experiment, $\mathcal{E}(\rho_{in}^{pure}) = \rho_{in}^{pure}$ and $\vec{r}_m = \tilde{\alpha}\hat{r}_{in}$. Therefore, $\tilde{\alpha} = |\vec{r}_m|_{\theta=0}$ can be determined. In other words, $|\vec{r}_m|_{\theta=0}$ is used to normalize all other measured output states before using the expression for $\mathcal{F}_\Delta$.

The expression for $\mathcal{F}_\Delta$ can also be used for the conditional fidelity in the coding experiment if the post-measurement accepted output state is known. This requires $\text{Tr}(\rho_a^{coded}) = (1 - p_a - p_b + 2p_a p_b)$ to be determined for each storage time. The correct normalization is again given by the output at $\theta = 0$, which equals $\vec{r}_m = \text{Tr}(\rho_a^{coded})\tilde{\alpha}\hat{r}_{in}$.

In summary, each ellipse obtained in the coding and the control experiment is normalized by the amplitude at $\theta = 0$:

$$\mathcal{F}_\Delta = \min_{\hat{r}_{in}} \frac{1}{2} \left[ 1 + \frac{\hat{r}_{in} \cdot \vec{r}_m}{|\vec{r}_m(\theta = 0)|} \right] . \tag{8.50}$$

It is interesting to note that in contrast to the fidelity measure, the ellipticity measure naturally performs an equivalent normalization, and thus can be used for deviations without modifications.

We now turn to the experimental results, beginning with a description of our apparatus.

## 8.4 Apparatus and experimental parameters

We performed our experiments on carbon-13 labeled sodium formate $(CHOO^-Na^+)$ (Fig. 8.7) at $15°C$. The nuclear spins of proton and carbon were used as input and ancilla respectively. Note that the system was heteronuclear. The sodium formate sample was a 0.6 milliliter 1.26 molar solution (8:1 molar ratio with anhydrous calcium chloride) in deuterated water [3]. The sample was degassed and flame sealed in a thin walled, 5mm NMR sample tube.

The time constants of phase damping and amplitude damping are shown in Table 8.2. The fact $T_2 \ll T_1$ ensures that the effect of amplitude damping is small compared to phase damping. The experimental conditions are chosen so that proton and carbon have almost equal $T_2$'s. This eliminates potential bias caused by having a long-lived ancilla when evaluating the effectiveness of coding. This also realizes a common assumption in coding theory that identical quantum systems are available for coding. Subsidiary experiments with qubits having very different $T_2$'s are described in Appendix C.3.

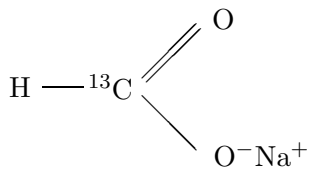

Figure 8.7: $^{13}$C-labeled formate. The nuclear spins of the neighboring proton and carbon represent qubits $a$ and $b$.

|            | $T_1$   | $T_2$   |
|------------|---------|---------|
| $^1$H      | 9 s     | 0.65 s  |
| $^{13}$C   | 13.5 s  | 0.75 s  |

Table 8.2: $T_1$'s and $T_2$'s for CaCl$_2$-doped formate at 15°C, measured using standard inversion recovery and Carr-Purcell-Meiboom-Gill pulse sequences respectively.

Phase damping arises from constant or low-frequency non-uniformities of the "static" magnetic field which randomize the phase evolution of the spins in the ensemble. Several processes contribute to this inhomogeneity on microscopic or macroscopic scales. Which process dominates phase damping varies from system to system [10]. For instance, intermolecular magnetic dipole-dipole interaction dominates phase damping in a solution of small molecules, whereas the modulation of direct electron-nuclear dipole-dipole interactions becomes more important if paramagnetic impurities are present in the solution. For molecules with quadrupolar nuclei (spin > 1/2), modulation of the quadrupolar coupling dominates phase damping. Other mechanisms such as chemical shift anisotropy can also dominate phase damping in other circumstances. These microscopic field inhomogeneities have no net effects on the static field when averaged over time, but they result in irreversible phase randomization with parameters intrinsic to the sample. Another origin of inhomogeneity comes from the macroscopic applied static magnetic field. In contrast to the intrinsic processes, phase randomization due to this inhomogeneity can be reversed by applying refocusing pulses as long as diffusion of molecules is insignificant.

Phase damping caused by the intrinsic irreversible processes alone has a time constant denoted by $T_2$, while the combined process has a shorter time constant denoted by $T_2^*$. $T_2$ is measured by the Carr-Purcell-Meiboom-Gill [50] experiment using multiple refocusing pulses. $T_2^*$ can be estimated from the line-width of the NMR spectral lines: during acquisition, the signal decays exponentially due to phase damping, resulting in Lorentzian spectral lines with line-width $1/\pi T_2^*$.

In our experiment, $T_2^*$'s for proton and carbon were estimated to within 0.05 s to be $\approx 0.35$ and 0.50 s. The storage times $t_d$ were approximately 0, 62, 123, 185, 246, 308 ms ($n/J$ for $n = 0, 12, 24, 36, 48, 60$). The maximum storage time was 120 $\tau$, long compared to the clock cycle and was comparable to $T_2^*$. The decay constant $\lambda$, defined in Section 8.2.1, was given by $\lambda = 1/T_2^*$. The error probabilities after a storage time of $t_d$ were $p_i = (1 - \exp(-t_d/T_{2i}^*))/2$ for spins $i = a, b$. To reconstruct the ellipse for each storage time, 11 experiments were run with input states spanning a semi-circle in the $\hat{x}\hat{z}$-plane. Each input state was prepared by a $R_{ya}(\theta)$ pulse with $\theta = n\pi/10$ for $n = 0, 1, \cdots, 10$.

All experiments were performed on an Oxford Instruments superconducting magnet of 11.7 Tesla, giving precession frequencies of $\omega_a/2\pi \approx 500$ MHz for proton and $\omega_b/2\pi \approx 125$ MHz for carbon. A Varian $^{\mathsf{UNITY}}$*Inova* spectrometer with a triple-resonance probe was used to send the pulsed RF fields to the sample and to measure the FID's. The RF pulses selectively rotated a particular spin by oscillating on resonance with it. The $\pi/2$ pulse durations were calibrated, and they were typically 8 to 14 $\mu$s. To perform logical operations in the respective rotating frames of the spins, reference oscillators were used to keep track of the free precession of both spins, leaving only the $J$-coupling term of 195.0 Hz in the time evolution. Each FID was recorded for $\approx 6.8$ s (until the signal had faded completely). The thermal state was obtained after a relaxation time of 80 s ($\gg T_1$'s) before each pulse sequence.

Using the above apparatus and procedures, we performed the experiments as outlined in Section 8.3. The experimental results are described in the next section.

## 8.5   Results and discussion

### 8.5.1   Decoded Bloch spheres

The output states, $\rho_a^{coded}$ and $\rho_a^{control}$, obtained as described in Sections 8.3 and 8.4, and the analysis that confirms the correction effects of coding, are presented in this section.

Figure 8.8 shows the accepted states in the $\hat{x}\hat{z}$-plane of the Bloch sphere of spin $a$. $\rho_5^{coded}$ and $\rho_5^{control}$ are plotted in Figs. 8.8 a and b. The ellipse for each storage time is obtained by a least-squares fit described later.

**Main result**

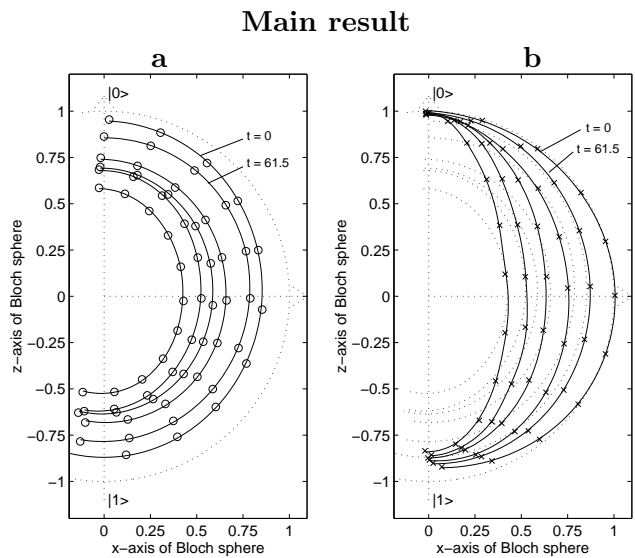

Figure 8.8:    (a) Experimental data (circles) showing the output states from the coding experiment. Each ellipse (solid line) corresponds to one storage time and is obtained by a least-squares fit (Eq. (8.56)) to the data. The storage times are $n\times61.5$ms for $n = 0, 1, \cdots, 5$, and shrinking ellipses correspond to increasing $n$. (b) Experimental data (crosses) and fitted ellipses (solid lines) for the control experiment. A replica of figure (a) is plotted in dotted lines for comparison. In both figures, uncertainties in the data are much smaller than the circles and crosses.

The most important feature in Fig. 8.8 is the reduction of the ellipticities of the ellipses due to coding, which represents partial removal of the distortion caused by phase damping - the signature of error correction. Coding is effective throughout the range of storage times tested.

We quantify the correction effects due to coding using the ellipticities. When deviations from the ideal case such as offsets of the angular positions of the points along the ellipses and attenuation of signal strength with increasing $\theta$ exist, the minimum overlap fidelities and the ellipticities are no longer related by Eq. (8.45). Since the ellipticity is an average measure of distortion which is less susceptible to scattering of individual data points, we first study the ellipticities. Moreover, since the deviations from the ideal case are small, we can still *infer* the fidelities from the ellipticities using Eq. (8.45). A discussion of the discrepancies and the exact overlap fidelities will given later.

**Ellipticities** In the ideal case, the ellipticity for each ellipse can be obtained experimentally as

$$\epsilon = \sqrt{\frac{\mathcal{I}(\theta = 0)}{\mathcal{I}(\theta = \frac{\pi}{2})}}\,,\tag{8.51}$$

where $\mathcal{I}$ denotes the intensity (amplitude square) of the peak integral. $\mathcal{I}$ is given by the $\hat{x}$ and $\hat{z}$-components of the output states as

$$\mathcal{I} = r_x^2 + r_z^2\,.\tag{8.52}$$

In the ideal case, $\mathcal{I}(\theta)$ can be found from Table 8.1:

$$\mathcal{I}_{control}(\theta) = 1 - 4(p_a - p_a^2)\sin^2\theta \tag{8.53}$$

$$\mathcal{I}_{coded}(\theta) = (1 - p_a - p_b + 2p_ap_b)^2 - 4p_ap_b(1 - p_a - p_b + p_ap_b)\sin^2\theta\,,\tag{8.54}$$

and both are of the functional form

$$\mathcal{I}_{ideal}(\theta) = A + B\sin^2\theta\,.\tag{8.55}$$

Experimentally, the output Bloch vectors do not form perfect ellipses. We modify Eq. (8.55) to include signal strength attenuation with increasing $\theta$ and constant offsets in the angular positions:

$$\mathcal{I}_{exp}(\theta) = (A + B\sin^2(\theta + D))(1 - C(\theta + D))\,,\tag{8.56}$$

and perform non-linear least-squares fits of the experimental data to determine $A, B, C$, and $D$. The fitted ellipses plotted in Fig. 8.8 follow from Eq. (8.56) and the fitted parameters. The ellipticities $\epsilon$ are found using Eq. (8.51) by interpolating the intensities at $\theta = 0$ and $\theta = \frac{\pi}{2}$. The ellipticities are plotted in Fig. 8.9 a. The uncertainties of the fitted parameters originate from the uncertainties in the data, which are estimated to be $\approx 1\%$ for the amplitude and 1.5 degrees for the phase in the measured peak integrals. These uncertainties are propagated numerically to the ellipticities as plotted in Fig. 8.9 a. Ideal case predictions and simulation results are also plotted in Fig. 8.9 a. The simulation takes into account the major imperfection in the pulses and will be described later.

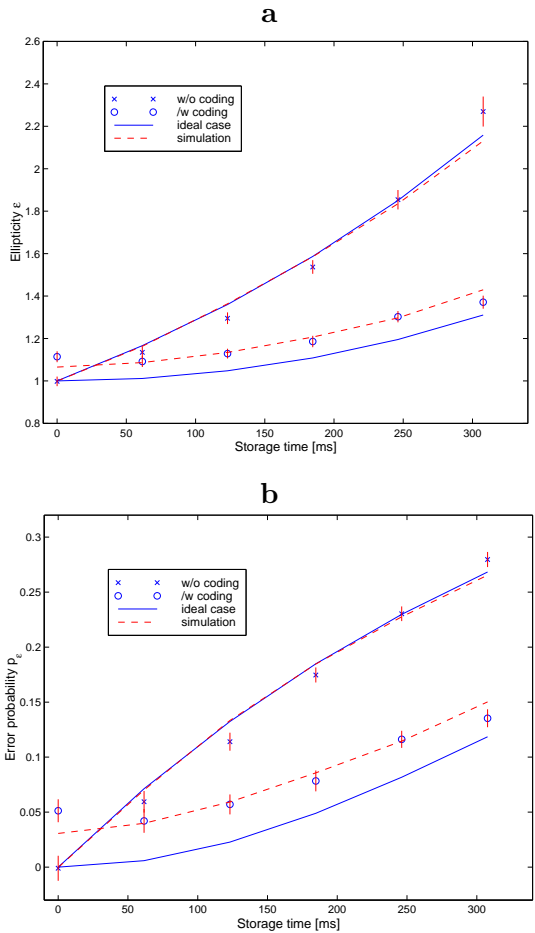

Figure 8.9: (a) Ellipticity and (b) inferred fidelity as a function of the storage time in the coding and the control experiments. Error bars represent 95% confidence level.

**Error correction** The effectiveness of coding to correct errors is evident when comparing the ellipticities from the coding and the control experiments (Fig. 8.9 a). Without coding, the ellipticity grows exponentially as $e^{t_d/T_{2a}^*}$ (for $T_{2a}^*$ fitted to be $\approx 0.4s$). With coding, the growth is slowed down, with almost zero growth for small $t_d$. The suppression of linear growth of the ellipticity can be further quantified by weighted quadratic fits $\epsilon = c_0 + c_1 t_d + c_2 t_d^2$ to the ellipticities. For the control experiments, $c_0 = 1.00 \pm 0.01$, $c_1 = 1.31 \pm 0.21$ and $c_3 = 8.8 \pm 0.8$ whereas for the coding experiments, $c_0 = 1.10 \pm 0.02$, $c_1 = -0.24 \pm 0.29$ and $c_3 = 3.8 \pm 0.9$ ($t_d$ in seconds). Therefore, the linear term "vanishes" due to coding. The small negative coefficient for the linear term originates from the scattering of the data point at zero storage time.

To quantify the "cost of the noisy gates" caused by the imperfect pulses, we compare the ellipticities from the coding experiments and from the ideal case, the quadratic fits of which are respectively $\epsilon_{expt}^{coded} = 1.10 - 0.24 t_d + 3.80 t_d^2$ and $\epsilon_{ideal}^{coded} = 1.00 + 0.15 t_d + 2.50 t_d^2$. The imperfections cause the ellipticity to increase by 0.1 at $t_d = 0$ and this extra distortion *decreases* with $t_d$. We take advantage of the fact that the simulation results are close to the data points but are not as scattered to have a better estimate of this "cost of the noisy gates". The simulation data can be fitted by $\epsilon_{sim}^{coded} = 1.06 + 0.32 t_d + 2.47 t_d^2$. Compared to the ideal case, the coding operations increase the ellipticity by $\approx 0.06$ at $t_d = 0$, and this extra distortion remains almost constant for all $t_d$.

The error probabilities as inferred from the ellipticities $p_\epsilon = 1 - \mathcal{F}_\epsilon = \frac{1}{2}(1 - \frac{1}{\epsilon})$ are plotted in Fig. 8.9 b as a function of storage time.

Error correction is also manifest by expressing $p_\epsilon$ in the coding experiment as a function of the original $p_\epsilon$ in the control experiment, as plotted in Fig. 8.10. The quadratic fit to the experimental results gives $p_{exp}^{coded} = c_0 + c_1 p + c_2 p^2$ where $p$ stands for $p_\epsilon$ in the control experiment, $c_0 = 0.047 \pm 0.008$, $c_1 = -0.05 \pm 0.12$ and $c_2 = 1.38 \pm 0.40$. Therefore, the expected improvement $p \to p_a p_b$ is confirmed. Experimentally, the error probabilities are larger than in the ideal case by at most 4.7% and these extra errors decrease with $p$. The quadratic fit to the simulation results (which is a good approximation of the experimental data) gives $p_{sim}^{coded} = 0.032 - 0.032 p + 1.783 p^2$ and differs from the ideal case by a constant amount of $\approx 0.033 \pm 0.003$ for all $p$, which represents the cost of the noisy gates. In conclusion, coding with noisy gates is still effective in our experiments, even though the noisy gates add a constant amount of distortion.

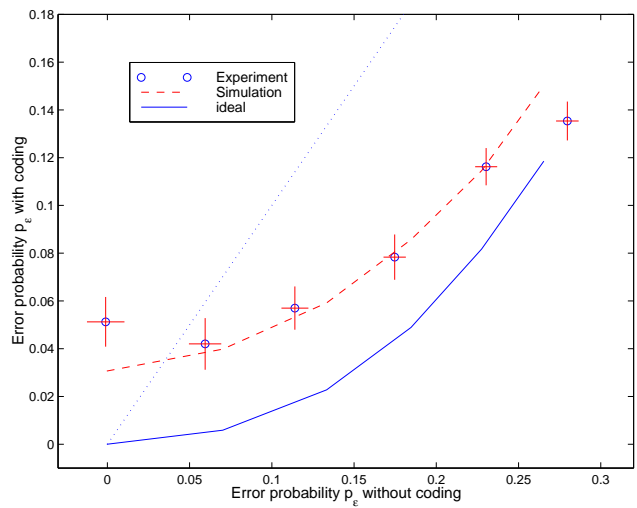

Figure 8.10: Error probabilities in the coding experiments vs the corresponding values in the control experiments. Error bars represent 95% confidence level. The 45° line is plotted as a dotted line.

### 8.5.2   Discrepancies

While the data exhibit a clear correction effect, there are notable deviations from the ideal case. First, the ellipses with coding are smaller than their counterparts without coding. This is most obvious when the storage time is zero, in which case the coding and the control experiments should produce equal outputs. Second, the signal strength is attenuated with increasing $\theta$ relative to ideal ellipses. Third, although the data points are well fitted by ellipses, their angular positions are not exactly as expected ("$\theta$-offsets"). Finally, the spacings between the ellipses deviate from expectation. The causes of these discrepancies and their implications on error correction are discussed next.

**Gate imperfections: RF field inhomogeneity**   The major cause of experimental errors is RF field inhomogeneity, which causes gate imperfections. This was determined by a series of experiments (details of which are not given here), and a thorough numerical simulation, as described below. The physical origin of the problem is as follows. The coil windings produce inhomogeneous RF fields that randomize the angles of rotation among molecules. For a single rotation, the signal averaged over the ensemble decreases exponentially with the pulse duration to good approximation. A measure of the RF field inhomogeneity is the signal strength after a $\pi/2$ pulse. They are measured to be $\approx 0.96$ and $0.92$ for proton and carbon respectively. In other words, a single $\pi/2$ pulse has an error of $\approx$ 4-8%.

RF field inhomogeneity affects our experiments in many ways. First, it attenuates the signal in both the coding and the control experiments, but the effects are much more severe in the coding experiments which have eight extra pulses. For instance, when the storage time is zero, the two experiments should have identical outputs, but the ellipse in the coding experiment is actually 5-15 % smaller. Second, for each ellipse, attenuation increases with $\theta$ as the preparation pulse $R_{ya}(\theta)$ becomes longer.

The effects of the RF field inhomogeneity are complicated, because the errors from different RF pulses are correlated, and the correlation depends on the temporal separation between the pulses and the diffusion rate of the molecules. The correlation time of the RF field inhomogeneity is comparable to the experimental time scales. For this reason, predictions of the effects of RF field inhomogeneity are analytically intractable.

Numerical simulations were performed to model the dominant effects of RF field inhomogeneity. We followed the evolution of the states assuming random RF field strengths drawn from Lorentzian distributions (also known as Cauchy distributions) with means and standard deviations matching pulse calibration and attenuation for the $\pi/2$ pulses. All parameters in the simulation, including $T_2^*$'s, were determined experimentally without introducing any free parameters. As the exact time correlation function for the errors was unknown, except for numerical evidence of a long correlation time, we *assumed* perfect correlation in the errors. The simulated ensemble output signal was obtained by Gaussian integration with numerical errors bounded to below 1.5%. The results were shown in Fig. 8.11.

Besides phase damping and error correction effects in the data, the simulations also reproduce extra signal attenuation in the coding experiments. The ellipticities obtained from simulations (see Fig. 8.9) approximate the experimental values very well. Such agreement to experimental results is surprising in the absence of free parameters in the model. Simulation results allow the discrepancy between the observed and the ideal ellipticities to be explained in terms of the RF field inhomogeneity and allow the "cost of coding" to be better estimated to be the constant 6 % increase in ellipticity or the $\approx 3$ % increase in error probabilities.

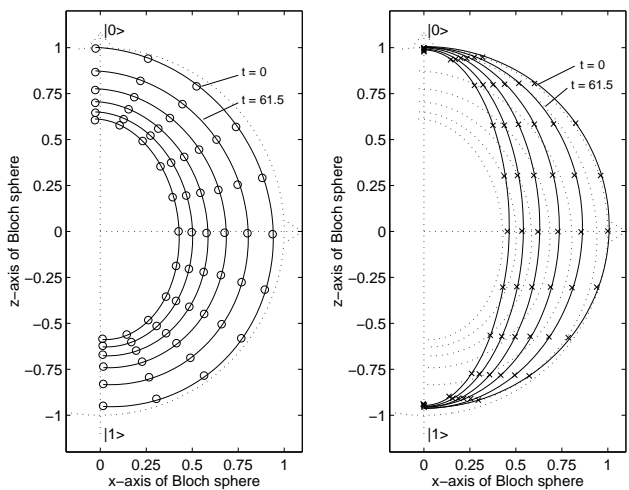

Figure 8.11: Simulated output states, plotted similarly as in Fig. 8.8. The simulation results are fitted by ellipses similar to those of the experimental data.

The simulation results also predict increasing attenuation with $\theta$. From the fitted parameter $C$ (see Eq. (8.56)), the amplitudes at $\theta = \pi$ are $\approx 4\%$ *weaker* than the corresponding values at $\theta = 0$ in the simulations. Experimentally, this attenuation increases from $\approx 8$ to 15% (as storage time increases from 0 to 308 ms) in the control experiments, and remains $\approx 8\%$ in the coding experiments. Therefore, RF field inhomogeneity contributes to the attenuation but only partially.

We conclude that RF field inhomogeneity as we have modeled explains the diminished signal strength in the coding experiments. The simulation quantifies the "cost of the noisy gates". RF field inhomogeneity also explains part of the attenuation with increasing $\theta$. We can also conclude that other discrepancies not predicted by the simulations are *not* caused by RF field inhomogeneity and these discrepancies are described next.

**Other discrepancies** The simulation results show that RF field inhomogeneity does *not* explain why the attenuation at large $\theta$ increases with storage time without coding, and it does not explain the $\theta$-offsets along the ellipses and the unexpected spacings between them.

The increased attenuation with storage time at large $\theta$ can be caused by amplitude damping. A precise description [4] of amplitude damping during storage is out of scope and we consider only the dominant effect predicted by a simple picture. *Phenomenologically*, the loss of energy to the lattice is described in the NMR literature by

$$z(t) = z(\infty) + (z(0) - z(\infty)) \; e^{-t/T_1} \,, \tag{8.57}$$

where $z(\infty) = 1$ is the thermal equilibrium magnetization. As $z(0) = \pm 1$ at $\theta = 0$ and $\pi$, we expect no changes at $\theta = 0$, but expect $|z|$ at $\theta = \pi$ to decrease by $0 - 7$ % for $t_d \approx 0 - 308$ ms and $T_1 \approx 9$ s for proton. Note that refocusing does not affect spin $a$ in the control experiments [5] but it swaps $|0_L\rangle$ and $|1_L\rangle$ halfway during storage, symmetrizing the amplitude damping effects in the coding experiments. Therefore, we expect increased attenuation with storage time in the control experiments only. This matches our observations that the attenuation of $\mathcal{I}(\theta = \pi)$ with respect to $\mathcal{I}(\theta = 0)$ increases from 8 to 16 % in the control experiments, and remains 8% in the coding experiments. Moreover, earlier data taken without refocusing (not presented) have the same trend of increased attenuation with storage time in both the coding and the control experiments. These are all in accord with the hypothesis that amplitude damping is causing the observed effect.

The second unexplained discrepancy is that the output states span more than a semi-ellipse in the coding experiment but slightly less than a semi-ellipse in the control experiment. We are not aware of any quantum process that can be a cause of it. It is notable that the output states and the fitted ellipticities can be used to infer the initial values of $\theta$, and they are roughly proportional to the expected values for each ellipse. The proportionality constants are $5 - 8$ % higher than unity in the coding experiment, and $0 - 1.6$ % lower in the control experiments. Moreover, similar effects are observed in many other experimental runs. Therefore, this is likely to be a systematic error.

We have no convincing explanation for the anomalous spacings between the ellipses in the experiment. However, from the fact that all the data points belonging to the same ellipse are well fitted by it, the anomalous spacings are unlikely to be caused by random fluctuations on the time scales of each ellipse-experiment. The effect of the anomalous spacings is reflected in the scattering of the ellipticities of the data, and the large uncertainties in the quadratic fits.

While it is impossible to eliminate or to fully explain these imperfections, it is possible to show that the deviations cannot affect the conclusion that error correction is effective.

**Effects of the discrepancies**   We now consider the effects of the discrepancies on the ellipticities and the inferred fidelities in the experiments. First of all, radial attenuation of the signal due to RF field inhomogeneity does not affect the ellipticities nor the inferred fidelities (taken as conditional fidelities). Second, different expressions for the "ellipticity" are not equivalent when the output states do not form perfect ellipses. However, they differ by no more than 7 and 3 % in the control and the coding experiments. $\theta$-offsets along the ellipses are not reflected in the ellipticities, resulting in overestimated inferred fidelities. This is bounded by 3%. The scattering of the ellipticities due to anomalous spacings between the ellipses is averaged out with curve-fits to the data. The most crucial point is, none of these effects have a dependence on the storage time that can be mistaken as error correction. Therefore, the effects of error correction can still be confirmed in the presence of all these small discrepancies.

### 8.5.3   Overlap fidelity

The two previous subsections dealing with the ellipticities provide an analysis of the global performance of the code. A stricter analysis is provided in this section using the overlap fidelities given by Eq. (8.50) in Section 8.3. The minimal overlap reflects all defects and deviations hidden in the ellipticities as well as other distortions such as that caused by amplitude damping.

All measurements are normalized using the amplitudes at $\theta = 0$ as described in Section 8.3. In the control experiments, the output states at $\theta = 0$ are least affected by amplitude damping and RF field inhomogeneity. Therefore, the normalization can be done accurately. In the coding experiment, the signal attenuation at $\theta = 0$ due to RF field inhomogeneity can lead to overestimated fidelities. We determine the uncertainties due to RF field inhomogeneity by the following method. For each storage time, the amplitudes of the accepted and the rejected states at $\theta = 0$ are summed. The sum is compared with the corresponding amplitude at $\theta = 0$ in the control experiment to estimate the attenuation due to RF field inhomogeneity. The effects on the overlap fidelities are bounded to below 2%. The errors in the measured peak integrals are propagated to the fidelities which result in standard deviations no more than 0.7%. We apply similar procedures to the simulation results. The net error probabilities, given by $1 - \mathcal{F}$ for the control and $1 - \mathcal{F}_C$ for the coding experiments, are plotted in Fig. 8.12.

The large difference in the rates of growth of error probabilities confirms the effectiveness of coding even when a stricter measure of fidelity is used.

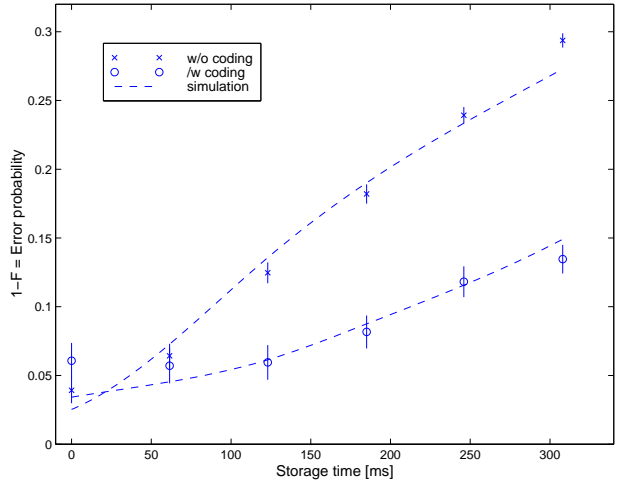

Figure 8.12:  Overlap fidelity as defined in Eq. (8.50). Points indicate experimental data and dashed lines indicate simulation results. Error bars represent 95% confidence level.

## 8.6   Conclusion

We have demonstrated experimentally, in a bulk NMR system, that using a two bit phase damping detecting code, the distortion of the accepted output states can be largely removed. These experimental results also provided quantitative measures of the major imperfections in the system. The principle source of errors, RF field inhomogeneity, was studied and a numerical simulation was developed to model our data. Despite the imperfections, a net amount of error correction was observed, when comparing cases with and without coding, and including gate errors in both cases.

Our analysis also addresses several theoretical questions in quantum error correction in bulk samples such as the fidelity measures of deviation density matrices. In the following, we conclude with some observations regarding quantum error correction in bulk systems, including syndrome measurements, the equivalence between error correction and logical labeling [53, 29] (see also Section 6.5.2), the applicability and advantages of detecting codes and some issues in signal strength.

Projective syndrome measurements traditionally employed in the standard theory of quantum error correction are impossible in ensemble quantum computation. Measurements via the acquisition of the FID do not reduce individual quantum states and provide only "average syndromes". Moreover, the quantum states are destroyed after acquisition. However, the important point is that syndrome measurement is *not necessary* in error correction [92, 11].

In each molecule, the syndrome bits carry the error syndrome for that particular molecule after decoding. These bits can either be used in a controlled-operation to correct the error [92], or in the case of a detecting code, to "logically label" the correct and erroneous states. Conversely, logical labeling to obtain effective pure states can be considered as error detection: unsuitable initial states are "detected" and are labeled as "bad" to start with. Both processes involve ejecting the entropy of the system to the ancilla bits. Error detection and logical labeling are therefore closely related concepts.

The distinction between error correcting and detecting codes is blurred when using bulk systems. The objective of error correction is to achieve reliable data transmission or information processing with high probability of success. When information is encoded in single systems, encoding with an error detection scheme will fail to provide reliable output for two different reasons: accepting an erroneous state or losing the state upon the detection of an error. Therefore, coding schemes capable of *distinguishing and correcting* errors are necessary to improve the probability of successful data processing. In contrast, in a bulk sample, a large initial redundancy exists upon preparation, and the combined signal of all the accepted cases forms the output. Therefore, rejection of the erroneous cases results in a reduction of the signal strength in the improved accepted cases without necessarily causing a failure. Error detecting codes thus provide a tradeoff between probability of error-free computation and signal strength.

As suggested by this analysis, and in concert with our experimental results, it thus makes sense to use detecting codes instead of correcting codes in bulk quantum computing systems under certain circumstances. Fundamentally, it is valuable to be able to interchange resources depending on their relative costs. This is illustrated by the following simple example. Suppose a total pool of $m$ qubits is available for transmission, and one just wants to correct for single phase flip errors of probability $p$. Using a three-qubit code, one would obtain an aggregate signal strength of $m/3$, with fidelity $1 - 3p^2$, whereas with a two-qubit code, the accepted signal strength would be $m(1 - 2p)/2$, with fidelity $1 - p^2$. Therefore, when $p \leq 1/6$, the two-qubit code performs better in this model due to its higher rate.

Another example relevant to bulk computation arises when the encoding and decoding circuits fail with probability proportional to the number of elementary gates used. Although errors in consecutive gates can be made to cancel sometimes, this basic scenario is substantiated by our experiment, in which imperfect pulses contribute significantly to the net error. Assume now that we have $n$ molecules, which are either two or three-qubit systems. Let us compare the performance of the two and three-qubit codes, based on the strength of the correct output signal. Because the correcting code requires at least three times as many operations as the detecting code [6], the figures of merit obtained for the two schemes are $n(1 - 3p_g)$ and $n(1 - 2p)(1 - p_g)$ respectively, where $p_g$ is the gate failure probability. In this model, the detecting code performs better for $p \leq p_g/(1 - p_g)$ due to the simplicity of the coding operations.

A third example is the case of current state NMR quantum computation at room temperature, in which the intrinsic signal strength decreases exponentially with the number of qubits [53, 29, 71]. In this model, the initial signal strength of an effective pure state of $m$ qubits is approximately of the order $2^{-m}$, and thus, for an ensemble of $n$ molecules, the signal strengths of the outputs from the correcting and detecting codes are about $n/8$ and $n(1-2p)/4$ respectively. According to this measure of performance, the detecting code outperforms the correcting code for $p \leq 1/4$ ($p \leq 0.27$ in our experiments).

If signal strength indeed decreases exponentially with $m$, then some interesting generalizations can be made. For arbitrary qubit errors, a $t$-error detecting code has distance $d \geq t + 1$, while a $t$-error correcting code has distance $d \geq 2t + 1$ [56]. If one encodes $k$ bits in $l$, the extra number of qubits used, $l - k$, satisfies the singleton bound [72, 26, 24], $l - k \geq 2d - 2$. Therefore, the output signal strengths for the detecting and correcting codes would be approximately proportional to $(1 - pf(p))/2^{2t}$ and $1/2^{4t}$, where $f(p)$ is a polynomial. The detecting code is thus always better asymptotically in this model [7].

This chapter illustrates how a careful study of dynamics in bulk quantum systems can provide a valuable opportunity to demonstrate and test theories of quantum information and computation. The development of temporal, spatial, logical, and related labeling techniques opens a window allowing information about the dynamics of single quantum systems to be extracted from bulk systems. Furthermore, by systematically developing an experimental toolbox of quantum circuits and quantum error correcting and detecting codes, experiments which test multiple particle quantum behavior become increasingly accessible. With improvements in the initial polarization in the systems and new labeling algorithms which do not incur exponential signal loss [100], and with better methods to control the major source of error, the RF field inhomogeneity, we believe that further study of bulk quantum systems will complement the study of single quantum systems, provide new insights into the dynamical behavior of open quantum systems, and further the potential for quantum information processing.

# Appendix A

# Master Equation

## A.1 Master Equation formalism

The dynamics of an open quantum system are traditionally described by a "master equation" which governs evolution of the density matrix as a function of time. A master equation is generally derived in the following manner. The system $s$, described by $\rho(t)$, couples to an environment $e$, described by $\rho_e$, through an interaction Hamiltonian $H_I$. Due to the evolution, quantum information originally in the system is delocalized over both the system and the environment. Tracing over the inaccessible environmental degrees of freedom gives the reduced density matrix for the system alone. Assuming weak interactions and a memoryless environment (so that the Born and Markov approximations hold) the Schrödinger equation for the system state has the form [82, 52]

$$\dot{\rho}(t) = -\frac{1}{\hbar^2} \int_0^{\tau_c} dt' \mathrm{Tr}_e \left[ H_I(t), [H_I(t-t'), \rho(t) \otimes \rho_e]] \right], \tag{A.1}$$

where the operators are given in the interaction picture, and $\tau_c$ is the correlation time of the environment. When $\tau_c$ is much smaller than the time scales in which $\rho(t)$ or $H_I(t)$ changes significantly, Eq. (A.1) can be approximated to give

$$\dot{\rho}(t) = -\eta \frac{\tau_c}{\hbar^2} \mathrm{Tr}_e \left[ H_I, [H_I, \rho(t) \otimes \rho_e]] \right], \tag{A.2}$$

where $\eta$ is a prefactor resulting from the integration. We will call Eq. (A.2) the *simplified* master equation.

## A.1.1   Common noise processes

We study two common noise processes, amplitude damping and phase damping, using this master equation approach.

### Amplitude damping

Amplitude damping [82, 52] describes the energy loss from the system to a zero temperature environment, and is a good approximation to many physical systems. It can be studied by modeling the system as a simple harmonic oscillator (for simplicity, we dispense with self Hamiltonians). The energy exchange between the system and the environment is given by the interaction Hamiltonian in the Schrodinger picture:

$$H_I = \chi(a^\dagger b + b^\dagger a) \tag{A.3}$$

where $a$, $b$ are the annihilation operators of the system and the environment respectively. Here, a single mode (harmonic oscillator) for the environment is sufficient to model the dynamics of interest. $\chi$ is a coupling constant.

We can obtain a master equation by substituting Eq. (A.3) into the simplified master equation Eq. (A.2) in Section A.1:

$$\dot{\rho} = -\frac{\lambda}{2}(a^\dagger a\rho + \rho a^\dagger a - 2a\rho a^\dagger) - \lambda\bar{n}_b(a^\dagger a\rho + \rho a a^\dagger - 2a\rho a^\dagger - 2a^\dagger\rho a)\,, \tag{A.4}$$

where $\bar{n}_b = \langle b^\dagger b\rangle$, and $\lambda = 2\eta\chi^2\tau_c$. For amplitude damping with an environment at temperature $k_B T$ much smaller than the system's energy scale, we can set $\bar{n}_b = 0$, resulting in the master equation:

$$\dot{\rho} = -\frac{\lambda}{2}(a^\dagger a\rho + \rho a^\dagger a - 2a\rho a^\dagger)\,. \tag{A.5}$$

Writing $\rho(t) = \sum_{mn}\rho_{mn}|m\rangle\langle n|$ where the time dependence of $\rho_{mn}$ has been suppressed for simplicity, we obtain the system of equations:

$$\dot{\rho} = -\frac{\lambda}{2} \sum_{mn} \left[ \rho_{mn}(n+m) - 2\sqrt{m+1}\sqrt{n+1}\rho_{(m+1)(n+1)} \right] |m\rangle\langle n| \tag{A.6}$$

Equation (A.6) can be solved to obtain $\rho(t)$ which is a complete description of the dynamics. The result is best expressed in the operator sum representation, where $\rho(t) = \sum_k A_k(t)\rho(0)A_k^\dagger(t)$ and

$$A_k(t) = \sum_n \sqrt{\binom{n}{k}} \sqrt{(1-\gamma(t))^{n-k}\gamma(t)^k} \, |n-k\rangle\langle n| \,, \tag{A.7}$$

where $\gamma(t) = 1 - e^{-\lambda t}$.

## Phase damping

The interaction Hamiltonian is:

$$H_I = \chi a^\dagger a b^\dagger b \,, \tag{A.8}$$

Using the simplied master equation,

$$\dot{\rho} = -\frac{\lambda}{2}\langle n^2 \rangle_b \left[ (a^\dagger a)^2 \rho + \rho(a^\dagger a)^2 - 2a^\dagger a \rho a^\dagger a \right] \,. \tag{A.9}$$

where $\langle n^2 \rangle_b = \text{Tr} \left[ \rho_b (b^\dagger b)^2 \right]$. Substituting $\rho(t) = \sum_{mn} \rho_{mn}|m\rangle\langle n|$,

$$\sum_{mn} \dot{\rho}_{mn}|m\rangle\langle n| = -\frac{\lambda}{2} \sum_{mn} \rho_{mn}(m-n)^2|m\rangle\langle n| \tag{A.10}$$

which can be solved to obtain $\rho_{mn}(t) = e^{-\frac{\lambda t}{2}(m-n)^2}\rho_{mn}(0)$.

# Appendix B

# Unital processes

In this appendix, we investigate the possibility to represent unital processes by random unitary processes. We use the *linear representation* of unital quantum operations, and consider the complete positivity of such representation. We will show that unital processes on a qubit can be represented as random unitaries. For a higher dimensional system, we will show that there are unital processes that are not random unitaries.

The results were obtained in collaboration with A. Childs and X. Zhou (unpublished). Related results were independently reported in [64, 99].

## B.1   Definitions and Facts

*Definition 9*: A quantum operation $\mathcal{E}$ of the form $\mathcal{E}(\rho) = \sum_k p_k U_k \rho U_k^\dagger$, where $\sum_k p_k = 1$, is called a random unitary process.

*Definition 10*: A quantum operation $\mathcal{E}$ is unital if $\mathcal{E}(I) = I$. A trace-preserving unital quantum operation is called doubly stochastic.

A random unitary process is doubly stochastic. To investigate the converse, we consider the affine representation of quantum operations.

## B.2    Affine and linear representations of quantum operations

Since quantum operations are linear maps on density matrices, they can be represented by matrices and vectors respectively. These matrices and vectors are real if the basis for density matrices is hermitian. We consider an orthonormal hermitian basis for density matrices, with an element proportional to the identity, and all other elements traceless. We call the traceless component the *generalized Bloch vector*. When the identity component is chosen to be the first coordinate, the matrix representation of a quantum operation $\mathcal{E}$ is given by

$$\begin{bmatrix} M_0 & V_1 \\ V_2 & M \end{bmatrix} \tag{B.1}$$

where $M_0$ and $M$ are respectively $1 \times 1$ and $(d^2 - 1) \times (d^2 - 1)$ matrices for a $d$-dimensional system, $V_1$ and $V_2$ are $d^2 - 1$ row and column vectors. If $\mathcal{E}$ is trace-preserving, $M_0 = 1$ and $V_1 = 0$. If $\mathcal{E}$ is unital, $M_0 = 1$ and $V_2 = 0$. It follows that $\mathcal{E}$ is doubly stochastic if and only if $M_0 = 1$ and the generalized Bloch vectors form an invariant subspace under $\mathcal{E}$, in which case $M$ governs the dynamics, and we call $M$ a linear representation of $\mathcal{E}$. [1] The linear representation is unique. Composition of quantum operations corresponds to matrix multiplication in the linear representation. A matrix represents a (doubly stochastic) quantum operation if and only if it defines a completely positive map.

### B.2.1    Complete positivity in the linear representation

To-date, there is no simple characterization of complete positivity for the linear representation. In this section, we present a characterization in the qubit case. Before the discussion on complete positivity, we elaborate on the distinction between positive maps and completely positive maps.

---

[1] A non-unital trace-preserving process can be represented by an affine map on the generalized Bloch vector.

**Positivity vs complete positivity**

Recall from Section 2.3.1 that a mapping $\mathcal{E}$ is completely positive, if for any ancillary Hilbert space $R$, $\mathcal{E} \otimes \mathcal{I}_R \geq 0$, where $\mathcal{I}_R$ is the identity operation on $R$. Physically, $R$ is a reference system. We say that $\mathcal{E}$ is *incompletely positive* if $\mathcal{E}$ is positive but not completely positive; in other words, when there exists a reference system $R$ and a *witness* $\rho \geq 0$ in the joint system such that $(\mathcal{E} \otimes \mathcal{I}_R)(\rho) \ngeq 0$. We use "CP" and "ICP" as shorthand notations for "completely positive" and "incompletely positive". We first consider some examples in the qubit case:

- Any unitary operation is CP (since an operator sum representation exists).

- The depolarizing channel (see Section 2.4), defined by $\mathcal{D}_p(\rho) = (1-p)I\rho I + \frac{p}{3}(X\rho X + Y\rho Y + Z\rho Z)$ for $0 \leq p \leq 1$, is CP. The set of depolarizing channels is closed under composition. Denote $\frac{1}{2}(I + \vec{r} \cdot \vec{\sigma})$ by $\rho(\vec{r})$. It follows that

$$\mathcal{D}_p(\rho(\vec{r})) = \rho\left(\left(1 - \frac{4p}{3}\right)\vec{r}\right), \tag{B.2}$$

$$\mathcal{D}_{p_2} \circ \mathcal{D}_{p_1}(\rho(\vec{r})) = \rho\left(\left(1 - \frac{4p_2}{3}\right)\left(1 - \frac{4p_1}{3}\right)\vec{r}\right), \tag{B.3}$$

  and $\mathcal{D}_{p_2} \circ \mathcal{D}_{p_1} = \mathcal{D}_p$, where $p = p_1 + p_2 - \frac{4}{3}\,p_1 p_2$ is always in $[0, 1]$.

- The transposition, $\mathcal{T}(H) = H^T$, is ICP. Any inseparable state [2] is a witness [91].

- The inversion operation, $\mathcal{I}_v(\rho(\vec{r})) = \rho(-\vec{r})$, is ICP. It can be shown in two different ways. Note that $\mathcal{I}_v = \mathcal{T} \circ \mathcal{Y} = \mathcal{Y} \circ \mathcal{T}$ where $\mathcal{Y}(\rho) = Y\rho Y$. If $\rho$ is a witness for $\mathcal{T}$ to be ICP, $Y\rho Y$ is a witness for $\mathcal{T} \circ \mathcal{Y}$ because it is the preimage of $\rho$ under $\mathcal{Y}$. Alternatively, $\rho$ is a witness for $\mathcal{Y} \circ \mathcal{T}$ because $\mathcal{Y}$ preserves eigenvalues.

The last example can be generalized. If $\Lambda$ is ICP with witness $\rho$, and $\mathcal{U}$ is a unitary operation, $\mathcal{U} \circ \Lambda$ and $\Lambda \circ \mathcal{U}$ are both ICP, with respective witnesses $\rho$ and $U^\dagger \rho U$. This invites the question, are $\mathcal{E} \circ \Lambda$ and $\Lambda \circ \mathcal{E}$ ICP for a general quantum operation $\mathcal{E}$?

We consider the example $\mathcal{I}_v \circ \mathcal{D}_p (= \mathcal{D}_p \circ \mathcal{I}_v)$. As the range of $\mathcal{D}_p$ is smaller than the set of all possible states, it is unclear if all the witnesses of $\mathcal{I}_v$ are excluded in the range of $\mathcal{D}_p$, making $\mathcal{I}_v \circ \mathcal{D}_p$ CP. We now show that $\mathcal{I}_v \circ \mathcal{D}_p$ is ICP when $p < \frac{1}{2}$ and is CP when $p \geq \frac{1}{2}$.

---

[2] An inseparable state cannot be written as a convex sum of product states over the composite system.

Suppose $p < \frac{1}{2}$. Consider the Bell state $|\Phi^+\rangle = \frac{1}{\sqrt{2}}(|00\rangle + |11\rangle)$, with density matrix $|\Phi^+\rangle\langle\Phi^+| = \frac{1}{4}(II + XX - YY + ZZ)$.

$$
\begin{aligned}
((\mathcal{I}_v \circ \mathcal{D}_p) \otimes \mathcal{I}_R)(|\Phi^+\rangle\langle\Phi^+|) &= \frac{1}{4}\left[ II + \left(1 - \frac{4p}{3}\right)(-XX + YY - ZZ)\right] \quad &\text{(B.4)} \\
&= \frac{1}{4}\left[\frac{4p}{3}II + \left(1 - \frac{4p}{3}\right)(II - XX + YY - ZZ)\right] &\text{(B.5)}
\end{aligned}
$$

The minimum eigenvalue of $(II - XX + YY - ZZ)$ is $-2$. Therefore, the minimum eigenvalue of $((\mathcal{I}_v \circ \mathcal{D}_p) \otimes \mathcal{I}_R)(|\Phi^+\rangle\langle\Phi^+|)$ is $\frac{1}{4}[\frac{4p}{3} - 2(1 - \frac{4p}{3})] = p - \frac{1}{2} < 0$. Hence, $\mathcal{I}_v \circ \mathcal{D}_p$ is ICP.

Suppose $p \geq \frac{1}{2}$,

$$
(\mathcal{I}_v \circ \mathcal{D}_p)(\rho(\vec{r})) = \rho\left(-\left(1 - \frac{4p}{3}\right)\vec{r}\right) = \mathcal{D}_{p'}(\rho(\vec{r})), \tag{B.6}
$$

where $p' = \frac{3}{2} - p \in [\frac{1}{2}, 1]$. Therefore, $\mathcal{I}_v \circ \mathcal{D}_p$ is CP. It is interesting that "over-depolarization" ($p > \frac{3}{4}$) resembles "depolarization-inversion".

This counter-example shows that no general statement can be made on the composition of a CP map with an ICP map. We now proceed to the characterization of complete positivity in the linear representation.

## Necessary and sufficient condition for complete positivity

Let $\mathcal{D}_{\vec{q}}$ denote the *generalized random Pauli channel* (see Section 2.4)

$$
\mathcal{D}_{\vec{q}}(\rho) = q_0\rho + q_x X\rho X + q_y Y\rho Y + q_z Z\rho Z. \tag{B.7}
$$

$\mathcal{D}_{\vec{q}}$ is trace-preserving iff $q_0 + q_x + q_y + q_z = 1$ and $\mathcal{D}_{\vec{q}}$ is completely positive iff $q_i \geq 0$ for $i = 0, x, y, z$. To see why complete positivity of $\mathcal{D}_{\vec{q}}$ implies $q_i \geq 0$, consider

$$
\begin{aligned}
(\mathcal{D}_{\vec{q}} \otimes \mathcal{I}_R)(|\Phi^+\rangle\langle\Phi^+|) &= q_0|\Phi^+\rangle\langle\Phi^+| + q_x XI|\Phi^+\rangle\langle\Phi^+|XI \\
&+ q_y YI|\Phi^+\rangle\langle\Phi^+|YI + q_z ZI|\Phi^+\rangle\langle\Phi^+|ZI \tag{B.8}
\end{aligned}
$$

Since $|\Phi^+\rangle$, $XI|\Phi^+\rangle$, $YI|\Phi^+\rangle$, and $ZI|\Phi^+\rangle$ are orthonormal, the eigenvalues of $(\mathcal{D}_{\vec{q}} \otimes \mathcal{I}_R)(|\Phi^+\rangle\langle\Phi^+|)$ are precisely $q_i$ for $i = 0, x, y, z$, which have to be non-negative if $\mathcal{D}_{\vec{q}}$ is completely positive. When $\mathcal{D}_{\vec{q}}$ is trace-preserving, $\mathcal{D}_{\vec{q}}$ is completely positive iff $\vec{q} \in \triangle_q$, the simplex with vertices $(0, 0, 0)$, $(1, 0, 0)$, $(0, 1, 0)$, and $(0, 0, 1)$.

We now consider when a real $3\times 3$ matrix is a linear representation for a doubly stochastic process choosing $\{I, X, iY, Z\}$ to be the basis for density matrices. We first consider diagonal matrices and then general matrices.

*Lemma 4*: $D = \mathrm{Diag}(d_1, d_2, d_3)$ is the linear representation of a doubly stochastic quantum operation iff $(d_1, d_2, d_3) \in \triangle_d$, the simplex in $\mathrm{R}^3$ with vertices $(1, 1, 1)$, $(1, -1, -1)$, $(-1, 1, -1)$, and $(-1, -1, 1)$.

*Proof:* It can easily be verified that $D$ is the linear representation of a mapping given by $\mathcal{D}_{\vec{q}}$ in Eq. (B.7) where the parameters are related by the non-singular linear transformation

$$d_1 = 1 - 2q_y - 2q_z \qquad (B.9)$$

$$d_2 = 1 - 2q_x - 2q_z \qquad (B.10)$$

$$d_3 = 1 - 2q_x - 2q_y \qquad (B.11)$$

$\mathcal{D}_{\vec{q}}$ is completely positive iff $\vec{q} \in \triangle_q$ iff $(d_1, d_2, d_3) \in \triangle_d$.

*Theorem 8*: Let $M$ be a real $3 \times 3$ matrix. $M$ is the linear representation of a doubly stochastic process iff $M = O_1 D O_2$ for some $O_i \in SO(3)$ and $D = \mathrm{Diag}(d_1, d_2, d_3)$ with $(d_1, d_2, d_3) \in \triangle_d$.

*Proof:* Since $M$ is real, $M$ has a singular value decomposition $M = \tilde{O}_1 \tilde{D} \tilde{O}_2$ where each $\tilde{O}_i$ is orthogonal. We can rewrite $M = O_1 D O_2$ where $O_i = \det(\tilde{O}_i)\,\tilde{O}_i$ is in $SO(3)$ and $D = \det(\tilde{O}_1 \tilde{O}_2)\tilde{D}$. Each $O_i \in SO(3)$ represents a unitary operation $\mathcal{U}_i$. If $D \in \triangle_d$, it represents a quantum operation $\mathcal{D}_{\vec{q}}$, and $M$ represents $\mathcal{U}_1 \circ \mathcal{D}_{\vec{q}} \circ \mathcal{U}_2$ which is a quantum operation. Conversely, if $M$ represents a quantum operation, $D = O_1^\dagger M O_2^\dagger$ also represents a quantum operation. By lemma 4, $(d_1, d_2, d_3) \in \triangle_d$.

Note that $D$ is unique up to negating two entries simultaneously. One can verify that $\triangle_d$ is indeed invariant under the simultaneous negation of two entries.

## B.3   Unital processes as random unitary processes

The following result is now immediate:

*Theorem 9*: Any unital process $\mathcal{E}$ on a qubit is a random unitary process.

*Proof:* Let $M$ be the linear representation of $\mathcal{E}$. From Theorem 8, $M = O_1 D O_2$ where $O_i \in SO(3)$ and $D \in \triangle_d$. $O_i$ and $D$ respectively represent some unitary operations $\mathcal{U}_i(\rho) = U_i \rho U_i^\dagger$ and $\mathcal{D}_{\vec{q}}$. Therefore,

$$
\begin{aligned}
\mathcal{E}(\rho) & = \mathcal{U}_1 \circ \mathcal{D}_{\vec{q}} \circ \mathcal{U}_2(\rho) & (B.12) \\
& = q_0(U_1 U_2)\rho(U_1 U_2)^\dagger + q_x(U_1 X U_2)\rho(U_1 X U_2)^\dagger & (B.13) \\
& + q_y(U_1 Y U_2)\rho(U_1 Y U_2)^\dagger + q_z(U_1 Z U_2)\rho(U_1 Z U_2)^\dagger & (B.14)
\end{aligned}
$$

which is a random unitary process.

The above proof provides a constructive method to find the unitary operation elements, as well as a circuit for realizing the unital operation:

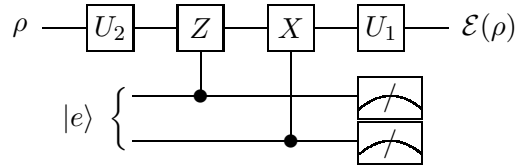

Figure B.1: A circuit model for an arbitrary unital quantum operation. The environmental state is given by $|e\rangle = \sqrt{q_0}|00\rangle + \sqrt{q_x}|01\rangle + \sqrt{q_z}|10\rangle + \sqrt{q_y}|11\rangle$.

## B.4   Unital processes for higher dimensional systems

Linear representations for quantum operations on a $d$-dimensional system are given by real $(d^2 - 1) \times (d^2 - 1)$ matrices. Most of the arguments for the qubit case can be generalized.

For example, one can generalize the random Pauli channel to $\mathcal{D}_{\vec{q}}(\rho) = \sum_k q_k E_k \rho E_k^\dagger$ where $\sum_k q_k = 1$, $E_k$ are unitary, and the states $(E_k \otimes I) \sum_i |i\rangle |i\rangle$ are orthogonal. For instance, $E_k$ can be chosen to be the nice error basis in [70]. The complete positivity of $\mathcal{D}_{\vec{q}}$ is still equivalent to the condition $q_k \geq 0 \ \forall k$. One can likewise deduce the criteria for a diagonal real $(d^2 - 1) \times (d^2 - 1)$ matrix to be a completely positive map. A general real $(d^2 - 1) \times (d^2 - 1)$ matrix still has a singular decomposition in terms of special orthogonal and diagonal components. However, a matrix in $SO(d^2 - 1)$ may not correspond to any unitary operation. In fact, there are $(d^2 - 1)(d^2 - 2)/2$ free parameters in $SO(d^2 - 1)$ but only $d^2 - 1$ free parameters in $SU(d)$, and a 1-1 correspondence between the two sets is impossible.

### B.4.1 Explicit counter-examples

We now construct unital processes that are *not* random unitary processes. The set of all doubly stochastic processes on a fixed Hilbert space is a convex set. [3] Likewise the set of all unital processes on a fixed Hilbert space is also convex. A point $x$ in a closed convex set $S$ is extreme if it cannot be expressed as a non-trivial convex combination of elements in $S$. [4] It follows that an extreme point which is not a unitary operation cannot be a random unitary process (which contradicts extremality). Moreover, a doubly stochastic process which is extreme in the set of all unital processes is extreme in the set of all doubly stochastic processes. The extreme points in the set of all unital processes are well characterized by theorem 5 in [28]. The special case for our purpose can be stated as:

*Theorem 10*: The set of completely positive unital linear maps on a $d$-dimensional system has extreme points of the form $\mathcal{E}(\rho) = \sum_k A_k \rho A_k^\dagger$ where $\{A_k A_l^\dagger\}$ is a linearly independent set.

*Corollary 1*: Consider the map on a qutrit:

$$\mathcal{E}(\rho) = (A_1 \rho A_1^\dagger + A_2 \rho A_2^\dagger + A_3 \rho A_3^\dagger) \tag{B.15}$$

---

[3] A set $S$ inside a vector space is convex if, $\forall x, y \in S$, $\alpha x + (1 - \alpha) y \in S \ \forall \ \alpha \in [0, 1]$.

[4] In other words, $x$ is extreme if $\forall \ 0 < \alpha < 1 \ y, z \in S$, $x = \alpha y + (1 - \alpha) z \Rightarrow x = y = z$.

where $A_i$ are related to the Gell-Mann matrices $\lambda_i$:

$$A_1 = \lambda_1 = \frac{1}{\sqrt{2}} \begin{bmatrix} 0 & 1 & 0 \\ 1 & 0 & 0 \\ 0 & 0 & 0 \end{bmatrix}, \ A_2 = \lambda_4 = \frac{1}{\sqrt{2}} \begin{bmatrix} 0 & 0 & 1 \\ 0 & 0 & 0 \\ 1 & 0 & 0 \end{bmatrix}, \ A_3 = \lambda_6 = \frac{1}{\sqrt{2}} \begin{bmatrix} 0 & 0 & 0 \\ 0 & 0 & 1 \\ 0 & 1 & 0 \end{bmatrix}$$

It is straightforward to verify that $\mathcal{E}$ is doubly stochastic and $\{A_k A_l^\dagger\}$ is linearly independent and therefore is not a random unitary process.

For a system of dimension $> 3$, a doubly stochastic process which affects a three dimensional subspace according to $\mathcal{E}$ and leaves the orthogonal complement invariant is likewise extreme, and cannot be a random unitary process.

# Appendix C

# Miscellaneous

## C.1 Upper bounds for $\overline{n}$

An argument for $c \approx 1$ for large $n$ is presented using Paley's construction (mentioned in Section 7.3.4), known results on primes in intervals and the prime number theorem for arithmetic progressions.

Let $\pi(x)$ be the number of primes $p$ which satisfy $2 \leq p \leq x$. For $x > 67$, $x/(\log x - 1/2) < \pi(x) < x/(\log x - 3/2)$ [98]. It follows that there exists a prime between $n$ and $n(1+\epsilon)$ for $\epsilon > 2/\log n$. Applying Paley's construction, $H(p+1)$ or $H(2(p+1))$ exists depending on whether $p \equiv 3 \mod 4$ or $p \equiv 1 \mod 4$. Therefore, $\overline{n} \leq n(1+\epsilon)+1$ or $\overline{n} \leq 2(n(1+\epsilon)+1)$ respectively.

The worse of the upper bounds $\overline{n} \leq 2(n(1+\epsilon)+1)$ resulting from $p \equiv 1 \mod 4$ can be improved. Note that there are at least $r$ primes between $n$ and $n(1+\epsilon)^r$. If the primes that equal 3 mod 4 and 1 mod 4 are randomly and uniformly distributed, the probability to find a prime which equals 3 mod 4 between $n$ and $n(1+\epsilon)^r$ is larger than $1-2^{-r}$. This assumption is true due to the prime number theorem for arithmetic progressions [43]. Let $\pi(x, a, q)$ denotes the number of primes $p$ in the arithmetic progression $\{a, a+q, a+2q, \ldots\}$ which satisfy $2 \leq p \leq x$. It is known that $\pi(x, 3, 4) \approx \pi(x, 1, 4)$. Therefore, with probability larger than $1-2^{-r}$, $\overline{n} \leq n(1+\epsilon)^r+1$, implying $c \leq (1+\epsilon)^r + 1/n \approx 1$ for large $n$.

## C.2   Mixed state description of the two-bit code

Recall that the initial state after the ancilla preparation is given by $\rho_0 = \sigma_z \otimes (I + \sigma_z)/2$ (see Eq. (8.29), with $\omega_a$ omitted). After $R_{ya}(\theta)$, the density matrix is given by

$$\rho_1 = (\cos\theta\sigma_z + \sin\theta\sigma_x) \otimes (I + \sigma_z)/2 \,. \tag{C.1}$$

Without coding, phase damping changes the density matrix to

$$\rho_5^{control} = \Big[ \cos\theta\sigma_z + (1 - 2p_a)\sin\theta\sigma_x \Big] \otimes (I + \sigma_z)/2 \,. \tag{C.2}$$

With coding, the encoding, phase damping, and decoding change the density matrix to $\rho_3$, $\rho_4$, and $\rho_5$:

$$
\begin{aligned}
\rho_3^{coded} &= \cos\theta(\sigma_z \otimes \sigma_z + \sigma_y \otimes \sigma_x)/2 \\
&+ \sin\theta(\sigma_x \otimes I + I \otimes \sigma_y)/2 \tag{C.3} \\
\rho_4^{coded} &= \cos\theta(\sigma_z \otimes \sigma_z + (1 - 2p_a)(1 - 2p_b)\sigma_y \otimes \sigma_x)/2 \\
&+ \sin\theta((1 - 2p_a)\sigma_x \otimes I + (1 - 2p_b)I \otimes \sigma_y)/2 \tag{C.4} \\
\rho_5^{coded} &= \cos\theta\sigma_z \otimes (I + (1 - 2p_a)(1 - 2p_b)\sigma_z)/2 \\
&+ \sin\theta\sigma_x \otimes ((1 - 2p_a)I + (1 - 2p_b)\sigma_z)/2 \tag{C.5} \\
&= \cos\theta\sigma_z \otimes \Big[ (1 - p_a - p_b + 2p_ap_b)(I + \sigma_z) + (p_a + p_b - 2p_ap_b)(I - \sigma_z) \Big]/2 \\
&+ \sin\theta\sigma_x \otimes \Big[ (1 - p_a - p_b)(I + \sigma_z) + (-p_a + p_b)(I - \sigma_z) \Big]/2 \tag{C.6} \\
&= \Big[ \cos\theta(1 - p_a - p_b + 2p_ap_b)\sigma_z + \sin\theta(1 - p_a - p_b)\sigma_x \Big] \otimes (I + \sigma_z)/2 \\
&+ \Big[ \cos\theta(p_a + p_b - 2p_ap_b)\sigma_z + \sin\theta(-p_a + p_b)\sigma_x \Big] \otimes (I - \sigma_z)/2 \,. \tag{C.7}
\end{aligned}
$$

## C.3 Two bit code experiment with very different $T_2$'s

While the case of equal $T_2$'s is interesting from a theoretical standpoint, different spins in a molecule typically have quite different $T_2$'s. To study the two-bit code in this regime, we performed experiments with carbon-13 labeled chloroform dissolved in acetone [8, 29]. All parameters were similar to the sodium formate sample, except for the relaxation time constants.

In the chloroform experiment, $T_1$'s were 16 s and 18.5 s and $T_2$'s were 7.5 s and 0.35 s for proton and carbon respectively. Separate experiments with the ancilla dephasing much slower or faster than the input were performed by interchanging the roles of proton and carbon. $T_2^*$'s and $t_d$'s were as listed in [9]. The ellipticities are shown in Fig. C.1.

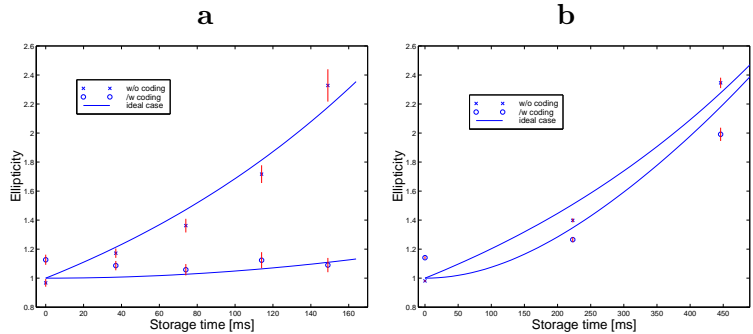

Figure C.1: Ellipticities obtained in the chloroform experiments, with (a) proton and (b) carbon as the ancilla. Carbon dephases much faster than proton. Error bars represent 95% confidence level.

From Fig. C.1 a, it is apparent that coding almost removes the distortion entirely when a much better ancilla is available. The question is, is coding advantageous over storing in the good ancilla alone? Theoretically, coding is always advantageous because the error probability is always reduced from $p_i$ ($i$ being the input spin) to $p_a p_b$. Fig. C.1 b shows that experimentally, such improvement is marginal, because the advantage of coding is offset by the noise introduced. Therefore, when the $T_2$'s are very different, the bottle neck is the dephasing of the bad qubit.